%% file: frex.tex
\PassOptionsToPackage{prologue,x11names}{xcolor}
\RequirePackage[bookmarksnumbered,unicode]{hyperref}
\RequirePackage{amssymb,hyperref}
\RequirePackage{placeins}
\documentclass[acmsmall]{acmart}
\setcopyright{cc}
\setcctype{by}
\acmJournal{PACMPL}
\acmYear{2025}
\acmVolume{9}
\acmNumber{ICFP}
\acmArticle{237}
\acmMonth{8}
\acmDOI{10.1145/3747506}
\received{2025-02-20}
\received[accepted]{2025-06-27}

\usepackage{booktabs}   
\usepackage{subcaption} 

\usepackage{graphicx}
\usepackage{multicol}
\usepackage[utf8]{inputenc}
\usepackage{newunicodechar}
\usepackage{enumitem}
\usepackage{agda}

\newunicodechar{⊎}{\ensuremath{\uplus}}
\newunicodechar{↔}{\ensuremath{\leftrightarrow}}
\newunicodechar{∎}{\ensuremath{\blacksquare}}
\newunicodechar{≡}{\ensuremath{\equiv}}
\newunicodechar{₁}{\ensuremath{_1}}
\newunicodechar{₂}{\ensuremath{_2}}
\newunicodechar{₃}{\ensuremath{_3}}
\newunicodechar{₄}{\ensuremath{_4}}
\newunicodechar{∀}{\ensuremath{\forall}}
\newunicodechar{ℕ}{\ensuremath{\mathbb{N}}}
\newunicodechar{λ}{\ensuremath{\lambda}}
\newunicodechar{•}{\ensuremath{\bullet}}
\newunicodechar{⟦}{\ensuremath{\llbracket}}
\newunicodechar{⟧}{\ensuremath{\rrbracket}}
\newunicodechar{ₑ}{\ensuremath{_e}}
\newunicodechar{ₐ}{\ensuremath{_a}}
\newunicodechar{ₛ}{\ensuremath{_s}}
\newunicodechar{Σ}{\ensuremath{\Sigma}}
\newunicodechar{Θ}{\ensuremath{\Theta}}
\newunicodechar{θ}{\ensuremath{\theta}}
\newunicodechar{ω}{\ensuremath{\omega}}
\newunicodechar{∥}{\ensuremath{\|}}
\newunicodechar{≈}{\ensuremath{\approx}}
\newunicodechar{⊔}{\ensuremath{\sqcup}}
\newunicodechar{ℓ}{\ensuremath{\ell}}
\newunicodechar{⟿}{\ensuremath{\rightsquigarrow}}
\newunicodechar{↝}{\ensuremath{\leadsto}}
\newunicodechar{∣}{\ensuremath{|}}
\newunicodechar{⃗}{\ensuremath{^{\rightarrow}}}
\newunicodechar{∘}{\ensuremath{\circ}}
\newunicodechar{⊙}{\ensuremath{\odot}}
\newunicodechar{⊥}{\ensuremath{\bot}}
\newunicodechar{≍}{\ensuremath{\asymp}}
\newunicodechar{≗}{\ensuremath{\circeq}}
\newunicodechar{⟨}{\ensuremath{\langle}}
\newunicodechar{⟩}{\ensuremath{\rangle}}
\newunicodechar{⊨}{\ensuremath{\vDash}}
\newunicodechar{≊}{\ensuremath{\approxeq}}
\newunicodechar{≋}{\ensuremath{\threesim}}
\newunicodechar{ˡ}{\ensuremath{^l}}
\newunicodechar{ʳ}{\ensuremath{^r}}
\newunicodechar{⃖}{\ensuremath{\leftarrow}}
\newunicodechar{≃}{\ensuremath{\simeq}}

\input{katla-preamble.tex}

\input{code-listings.tex}

\input{def}

\begin{document}

\title{Frex: Dependently Typed Algebraic Simplification}

\author{Guillaume Allais}
\orcid{0000-0002-4091-657X}
\affiliation{%
  \institution{University of Strathclyde}
  \city{Glasgow}
  \country{United Kingdom}
}
\email{guillaume.allais@ens-lyon.org}

\author{Edwin Brady}
\orcid{0000-0002-9734-367X}
\affiliation{%
  \institution{University of St. Andrews}
  \city{St. Andrews}
  \country{United Kingdom}
}
\email{ecb10@st-andrews.ac.uk}

\author{Nathan Corbyn}
\orcid{0000-0003-0513-8618}
\affiliation{%
  \institution{University of Oxford}
  \city{Oxford}
  \country{United Kingdom}
}
\email{nathan.corbyn@cs.ox.ac.uk}

\author{Ohad Kammar}
\orcid{0000-0002-2071-0929}
\affiliation{%
  \institution{University of Edinburgh}
  \city{Edinburgh}
  \country{United Kingdom}
}
\email{ohad.kammar@ed.ac.uk}

\author{Jeremy Yallop}
\orcid{0009-0002-1650-6340}
\affiliation{%
  \institution{University of Cambridge}
  \city{Cambridge}
  \country{United Kingdom}
}
\email{jeremy.yallop@cl.cam.ac.uk}

\begin{abstract}
  \input{abstract}
\end{abstract}

\input{ccsxml}


\maketitle

\input{new-intro}
\input{overview}
\input{background}
\input{frexlib}
\input{core}

\input{printing}
\input{reflection}
\input{evaluation}
\input{lessons}

\input{conclusion}

\newpage

\appendix
\input{automated-appendix}

\FloatBarrier
\begin{acks}                            
%
  Supported by the Engineering and Physical Sciences Research Council
  grant EP/T007265/1 and an Industrial CASE Studentship, a Royal
  Society University Research Fellowship, a Facebook Research Award,
  an Alan Turing Institute seed-funding grant, and UK Advanced
  Research and Innovation Agency (ARIA) as part of the project
  Qbs4Safety: Core Representation Underlying Safeguarded AI. An
  earlier, unpublished, outline of this work appeared as part of a
  short-abstract in TyDe'20~\cite{frex:tyde}.
  We are grateful to
  Jacques Carette,   %
  Donovan Crichton,  %
  Joey Eremondi,     %
  Sam Lindley,       %
  Conor McBride,     %
  James McKinna,     %
  Kasia Marek,       %
  Wojciech Nawrocki, %
  and                %
  Robert Wright      %
  for useful discussions and suggestions,
  and to the anonymous referees of the various iterations of this article who
  insisted we improve its presentation.
\end{acks}
\paragraph{Note.}
  For the purpose of Open Access the author(s) have applied a CC BY public
  copyright licence to any Author Accepted Manuscript version arising
  from this submission.

\bibliography{frex}

\end{document}

%% file: katla-preamble.tex
\usepackage{fancyvrb}
\usepackage[x11names]{xcolor}

\newcommand{\KatlaNewline}            {}
\newcommand{\KatlaSpace}              {\footnotesize\ }
\newcommand{\KatlaDash}               {\string-}
\newcommand{\KatlaUnderscore}         {\string_}
\newcommand{\KatlaTilde}              {\string~}
\newcommand{\IdrisHlightFont}         {\ttfamily\footnotesize}
\newcommand{\IdrisHlightStyleData}    {\bfseries}
\newcommand{\IdrisHlightStyleType}    {}
\newcommand{\IdrisHlightStyleBound}   {}
\newcommand{\IdrisHlightStyleFunction}{}
\newcommand{\IdrisHlightStyleKeyword} {}

\newcommand{\IdrisHlightStyleComment} {}

\newcommand{\IdrisHlightStyleModule}   {\itshape}

\newcommand{\IdrisHlightColourData}    {IndianRed1}
\newcommand{\IdrisHlightColourType}    {DeepSkyBlue3}
\newcommand{\IdrisHlightColourBound}   {DarkOrchid3}
\newcommand{\IdrisHlightColourFunction}{Chartreuse4}
\newcommand{\IdrisHlightColourKeyword} {black}

\newcommand{\IdrisHlightColourComment} {gray}

\newcommand{\IdrisHlightColourModule}   {Blue4}

\newcommand{\RawIdrisHighlight}[3]{{\textcolor{#1}{\IdrisHlightFont#2{#3}}}}

\newcommand{\IdrisData}[1]{\RawIdrisHighlight{\IdrisHlightColourData}{\IdrisHlightStyleData}{#1}}
\newcommand{\IdrisType}[1]{\RawIdrisHighlight{\IdrisHlightColourType}{\IdrisHlightStyleType}{#1}}
\newcommand{\IdrisBound}[1]{\RawIdrisHighlight{\IdrisHlightColourBound}{\IdrisHlightStyleBound}{#1}}
\newcommand{\IdrisFunction}[1]{\RawIdrisHighlight{\IdrisHlightColourFunction}{\IdrisHlightStyleFunction}{#1}}
\newcommand{\IdrisKeyword}[1]{\RawIdrisHighlight{\IdrisHlightColourKeyword}{\IdrisHlightStyleKeyword}{#1}}

\newcommand{\IdrisComment}[1]{\RawIdrisHighlight{\IdrisHlightColourComment}{\IdrisHlightStyleComment}{#1}}

\newcommand{\IdrisModule}[1]{\RawIdrisHighlight{\IdrisHlightColourModule}{\IdrisHlightStyleModule}{#1}}

\makeatletter
\let\FV@ProcessLine\relax
\makeatother

%% file: code-listings.tex
\input{introduction-listings}
\input{background-listings}
\input{frexlib-listings}
\input{reflection-listings}
\input{intro-printer}
\input{core-listings}
\input{certificates-listings.tex}
\input{evaluation-listings.tex}

%% file: introduction-listings.tex
\begin{SaveVerbatim}[commandchars=\\\{\}]{TypeAscription}
\IdrisFunction{the}\KatlaSpace{}\IdrisKeyword{:}\KatlaSpace{}\IdrisKeyword{(}\IdrisBound{a}\KatlaSpace{}\IdrisKeyword{:}\KatlaSpace{}\IdrisType{Type}\IdrisKeyword{)}\KatlaSpace{}\IdrisKeyword{\KatlaDash{}>}\KatlaSpace{}\IdrisBound{a}\KatlaSpace{}\IdrisKeyword{\KatlaDash{}>}\KatlaSpace{}\IdrisBound{a}\KatlaNewline{}
\IdrisFunction{the}\KatlaSpace{}\IdrisKeyword{\KatlaUnderscore{}}\KatlaSpace{}\IdrisBound{x}\KatlaSpace{}\IdrisKeyword{=}\KatlaSpace{}\IdrisBound{x}\KatlaNewline{}
\end{SaveVerbatim}
\newcommand\MonoidOperationIntro[1][]{\UseVerbatim[#1]{MonoidOperationIntro}}
\begin{SaveVerbatim}[commandchars=\\\{\}]{MonoidOperationIntro}
\IdrisKeyword{data}\KatlaSpace{}\IdrisType{Operation}\KatlaNewline{}
\KatlaSpace{}\KatlaSpace{}\KatlaSpace{}\KatlaSpace{}\IdrisKeyword{:}\KatlaSpace{}\IdrisType{Nat}\KatlaSpace{}\IdrisKeyword{\KatlaDash{}>}\KatlaSpace{}\IdrisType{Type}\KatlaSpace{}\IdrisKeyword{where}\KatlaNewline{}
\KatlaSpace{}\KatlaSpace{}\IdrisData{Neutral}\KatlaSpace{}\IdrisKeyword{:}\KatlaSpace{}\IdrisType{Operation}\KatlaSpace{}\IdrisData{0}\KatlaNewline{}
\KatlaSpace{}\KatlaSpace{}\IdrisData{Product}\KatlaSpace{}\IdrisKeyword{:}\KatlaSpace{}\IdrisType{Operation}\KatlaSpace{}\IdrisData{2}\KatlaNewline{}
\end{SaveVerbatim}

\begin{SaveVerbatim}[commandchars=\\\{\}]{ListAppend}
\IdrisFunction{(++)}\KatlaSpace{}\IdrisKeyword{:}\KatlaSpace{}\IdrisKeyword{(}\IdrisBound{xs}\IdrisKeyword{,}\KatlaSpace{}\IdrisBound{ys}\KatlaSpace{}\IdrisKeyword{:}\KatlaSpace{}\IdrisType{List}\KatlaSpace{}\IdrisBound{a}\IdrisKeyword{)}\KatlaSpace{}\IdrisKeyword{\KatlaDash{}>}\KatlaSpace{}\IdrisType{List}\KatlaSpace{}\IdrisBound{a}\KatlaNewline{}
\IdrisData{[]}\KatlaSpace{}\KatlaSpace{}\KatlaSpace{}\KatlaSpace{}\KatlaSpace{}\KatlaSpace{}\KatlaSpace{}\KatlaSpace{}\IdrisFunction{++}\KatlaSpace{}\IdrisBound{ys}\KatlaSpace{}\IdrisKeyword{=}\KatlaSpace{}\IdrisBound{ys}\KatlaNewline{}
\IdrisKeyword{(}\IdrisBound{x}\KatlaSpace{}\IdrisData{::}\KatlaSpace{}\IdrisBound{xs}\IdrisKeyword{)}\KatlaSpace{}\IdrisFunction{++}\KatlaSpace{}\IdrisBound{ys}\KatlaSpace{}\IdrisKeyword{=}\KatlaSpace{}\IdrisBound{x}\KatlaSpace{}\IdrisData{::}\KatlaSpace{}\IdrisKeyword{(}\IdrisBound{xs}\KatlaSpace{}\IdrisFunction{++}\KatlaSpace{}\IdrisBound{ys}\IdrisKeyword{)}\KatlaNewline{}
\KatlaNewline{}
\end{SaveVerbatim}

\begin{SaveVerbatim}[commandchars=\\\{\}]{ListMergeBy}
\IdrisFunction{mergeBy}\KatlaSpace{}\IdrisKeyword{:}\KatlaSpace{}\IdrisKeyword{(}\IdrisBound{isLT}\IdrisKeyword{:}\KatlaSpace{}\IdrisBound{a}\KatlaSpace{}\IdrisKeyword{\KatlaDash{}>}\KatlaSpace{}\IdrisBound{a}\KatlaSpace{}\IdrisKeyword{\KatlaDash{}>}\KatlaSpace{}\IdrisType{Bool}\IdrisKeyword{)}\KatlaSpace{}\IdrisKeyword{\KatlaDash{}>}\KatlaNewline{}
\KatlaSpace{}\KatlaSpace{}\IdrisKeyword{(}\IdrisBound{xs}\IdrisKeyword{,}\KatlaSpace{}\IdrisBound{ys}\KatlaSpace{}\IdrisKeyword{:}\KatlaSpace{}\IdrisType{List}\KatlaSpace{}\IdrisBound{a}\IdrisKeyword{)}\KatlaSpace{}\IdrisKeyword{\KatlaDash{}>}\KatlaSpace{}\IdrisType{List}\KatlaSpace{}\IdrisBound{a}\KatlaNewline{}
\IdrisFunction{mergeBy}\KatlaSpace{}\IdrisKeyword{\KatlaUnderscore{}}\KatlaSpace{}\IdrisData{[]}\KatlaSpace{}\IdrisBound{ys}\KatlaSpace{}\IdrisKeyword{=}\KatlaSpace{}\IdrisBound{ys}\KatlaNewline{}
\IdrisFunction{mergeBy}\KatlaSpace{}\IdrisKeyword{\KatlaUnderscore{}}\KatlaSpace{}\IdrisBound{xs}\KatlaSpace{}\IdrisData{[]}\KatlaSpace{}\IdrisKeyword{=}\KatlaNewline{}
\KatlaSpace{}\KatlaSpace{}\KatlaSpace{}\KatlaSpace{}\KatlaSpace{}\KatlaSpace{}\KatlaSpace{}\KatlaSpace{}\KatlaSpace{}\KatlaSpace{}\KatlaSpace{}\KatlaSpace{}\KatlaSpace{}\KatlaSpace{}\KatlaSpace{}\KatlaSpace{}\KatlaSpace{}\KatlaSpace{}\IdrisBound{xs}\KatlaNewline{}
\KatlaNewline{}
\IdrisFunction{mergeBy}\KatlaSpace{}\KatlaSpace{}\IdrisComment{\KatlaDash{}\KatlaDash{}\KatlaSpace{}@\KatlaSpace{}introduces\KatlaSpace{}an\KatlaSpace{}alias}\KatlaNewline{}
\KatlaSpace{}\KatlaSpace{}\IdrisBound{isLT}\KatlaSpace{}\IdrisBound{xs'}\IdrisKeyword{@(}\IdrisBound{x}\KatlaSpace{}\IdrisData{::}\KatlaSpace{}\IdrisBound{xs}\IdrisKeyword{)}\KatlaSpace{}\IdrisBound{ys'}\IdrisKeyword{@(}\IdrisBound{y}\KatlaSpace{}\IdrisData{::}\KatlaSpace{}\IdrisBound{ys}\IdrisKeyword{)}\KatlaSpace{}\IdrisKeyword{=}\KatlaNewline{}
\KatlaSpace{}\KatlaSpace{}\IdrisKeyword{if}\KatlaSpace{}\IdrisBound{isLT}\KatlaSpace{}\IdrisBound{x}\KatlaSpace{}\IdrisBound{y}\KatlaNewline{}
\KatlaSpace{}\KatlaSpace{}\IdrisKeyword{then}\KatlaSpace{}\IdrisBound{x}\KatlaSpace{}\IdrisData{::}\KatlaSpace{}\IdrisFunction{mergeBy}\KatlaSpace{}\IdrisBound{isLT}\KatlaSpace{}\IdrisBound{xs}\KatlaSpace{}\IdrisBound{ys'}\KatlaNewline{}
\KatlaSpace{}\KatlaSpace{}\IdrisKeyword{else}\KatlaNewline{}
\KatlaSpace{}\KatlaSpace{}\KatlaSpace{}\KatlaSpace{}\KatlaSpace{}\KatlaSpace{}\KatlaSpace{}\IdrisBound{y}\KatlaSpace{}\IdrisData{::}\KatlaSpace{}\IdrisFunction{mergeBy}\KatlaSpace{}\IdrisBound{isLT}\KatlaSpace{}\IdrisBound{xs'}\KatlaSpace{}\IdrisBound{ys}\KatlaNewline{}
\KatlaNewline{}
\KatlaNewline{}
\end{SaveVerbatim}

\begin{SaveVerbatim}[commandchars=\\\{\}]{VectAppend}
\IdrisFunction{(++)}\KatlaSpace{}\IdrisKeyword{:}\KatlaSpace{}\IdrisType{Vect}\KatlaSpace{}\IdrisBound{n}\KatlaSpace{}\IdrisBound{a}\KatlaSpace{}\IdrisKeyword{\KatlaDash{}>}\KatlaSpace{}\IdrisType{Vect}\KatlaSpace{}\IdrisBound{m}\KatlaSpace{}\IdrisBound{a}\KatlaSpace{}\IdrisKeyword{\KatlaDash{}>}\KatlaSpace{}\IdrisType{Vect}\KatlaSpace{}\IdrisKeyword{(}\IdrisBound{n}\IdrisFunction{+}\IdrisBound{m}\IdrisKeyword{)}\KatlaSpace{}\IdrisBound{a}\KatlaNewline{}
\IdrisData{[]}\KatlaSpace{}\KatlaSpace{}\KatlaSpace{}\KatlaSpace{}\KatlaSpace{}\KatlaSpace{}\KatlaSpace{}\KatlaSpace{}\IdrisFunction{++}\KatlaSpace{}\IdrisBound{ys}\KatlaSpace{}\IdrisKeyword{=}\KatlaSpace{}\IdrisBound{ys}\KatlaNewline{}
\IdrisKeyword{(}\IdrisBound{x}\KatlaSpace{}\IdrisData{::}\KatlaSpace{}\IdrisBound{xs}\IdrisKeyword{)}\KatlaSpace{}\IdrisFunction{++}\KatlaSpace{}\IdrisBound{ys}\KatlaSpace{}\IdrisKeyword{=}\KatlaSpace{}\IdrisBound{x}\KatlaSpace{}\IdrisData{::}\KatlaSpace{}\IdrisKeyword{(}\IdrisBound{xs}\KatlaSpace{}\IdrisFunction{++}\KatlaSpace{}\IdrisBound{ys}\IdrisKeyword{)}\KatlaNewline{}
\KatlaNewline{}
\end{SaveVerbatim}

\begin{SaveVerbatim}[commandchars=\\\{\}]{MergeBy}
\IdrisFunction{mergeBy}\KatlaSpace{}\IdrisKeyword{:}\KatlaSpace{}\IdrisKeyword{(}\IdrisBound{isLT}\IdrisKeyword{:}\KatlaSpace{}\IdrisBound{a}\KatlaSpace{}\IdrisKeyword{\KatlaDash{}>}\KatlaSpace{}\IdrisBound{a}\KatlaSpace{}\IdrisKeyword{\KatlaDash{}>}\KatlaSpace{}\IdrisType{Bool}\IdrisKeyword{)}\KatlaSpace{}\IdrisKeyword{\KatlaDash{}>}\KatlaNewline{}
\KatlaSpace{}\KatlaSpace{}\IdrisType{Vect}\KatlaSpace{}\IdrisBound{n}\KatlaSpace{}\IdrisBound{a}\KatlaSpace{}\IdrisKeyword{\KatlaDash{}>}\KatlaSpace{}\IdrisType{Vect}\KatlaSpace{}\IdrisBound{m}\KatlaSpace{}\IdrisBound{a}\KatlaSpace{}\IdrisKeyword{\KatlaDash{}>}\KatlaSpace{}\IdrisType{Vect}\KatlaSpace{}\IdrisKeyword{(}\IdrisBound{n}\KatlaSpace{}\IdrisFunction{+}\KatlaSpace{}\IdrisBound{m}\IdrisKeyword{)}\KatlaSpace{}\IdrisBound{a}\KatlaNewline{}
\IdrisFunction{mergeBy}\KatlaSpace{}\KatlaSpace{}\KatlaSpace{}\KatlaSpace{}\KatlaSpace{}\IdrisKeyword{\KatlaUnderscore{}}\KatlaSpace{}\IdrisData{[]}\KatlaSpace{}\IdrisBound{ys}\KatlaSpace{}\IdrisKeyword{=}\KatlaSpace{}\IdrisBound{ys}\KatlaNewline{}
\IdrisFunction{mergeBy}\KatlaSpace{}\IdrisKeyword{\{}\IdrisBound{n}\IdrisKeyword{\}}\KatlaSpace{}\IdrisKeyword{\KatlaUnderscore{}}\KatlaSpace{}\IdrisBound{xs}\KatlaSpace{}\IdrisData{[]}\KatlaSpace{}\IdrisKeyword{=}\KatlaSpace{}\IdrisKeyword{rewrite}\KatlaSpace{}\IdrisFunction{plusZeroRightNeutral}\KatlaSpace{}\IdrisBound{n}\KatlaSpace{}\IdrisKeyword{in}\KatlaNewline{}
\KatlaSpace{}\KatlaSpace{}\KatlaSpace{}\KatlaSpace{}\KatlaSpace{}\KatlaSpace{}\KatlaSpace{}\KatlaSpace{}\KatlaSpace{}\KatlaSpace{}\KatlaSpace{}\KatlaSpace{}\KatlaSpace{}\KatlaSpace{}\KatlaSpace{}\KatlaSpace{}\KatlaSpace{}\KatlaSpace{}\KatlaSpace{}\KatlaSpace{}\KatlaSpace{}\KatlaSpace{}\IdrisBound{xs}\KatlaNewline{}
\KatlaSpace{}\KatlaSpace{}\IdrisKeyword{where}\KatlaSpace{}\IdrisFunction{plusZeroRightNeutral}\KatlaSpace{}\IdrisKeyword{:}\KatlaSpace{}\IdrisKeyword{(}\IdrisBound{n}\KatlaSpace{}\IdrisKeyword{:}\KatlaSpace{}\IdrisType{Nat}\IdrisKeyword{)}\KatlaSpace{}\IdrisKeyword{\KatlaDash{}>}\KatlaSpace{}\IdrisBound{n}\KatlaSpace{}\IdrisFunction{+}\KatlaSpace{}\IdrisData{0}\KatlaSpace{}\IdrisKeyword{=}\KatlaSpace{}\IdrisBound{n}\KatlaNewline{}
\IdrisFunction{mergeBy}\KatlaSpace{}\IdrisKeyword{\{}\IdrisBound{n}\KatlaSpace{}\IdrisKeyword{=}\KatlaSpace{}\IdrisData{S}\KatlaSpace{}\IdrisBound{n}\IdrisKeyword{\}}\KatlaSpace{}\IdrisKeyword{\{}\IdrisBound{m}\KatlaSpace{}\IdrisKeyword{=}\KatlaSpace{}\IdrisData{S}\KatlaSpace{}\IdrisBound{m}\IdrisKeyword{\}}\KatlaNewline{}
\KatlaSpace{}\KatlaSpace{}\IdrisBound{isLT}\KatlaSpace{}\IdrisBound{xs'}\IdrisKeyword{@(}\IdrisBound{x}\KatlaSpace{}\IdrisData{::}\KatlaSpace{}\IdrisBound{xs}\IdrisKeyword{)}\KatlaSpace{}\IdrisBound{ys'}\IdrisKeyword{@(}\IdrisBound{y}\KatlaSpace{}\IdrisData{::}\KatlaSpace{}\IdrisBound{ys}\IdrisKeyword{)}\KatlaSpace{}\IdrisKeyword{=}\KatlaNewline{}
\KatlaSpace{}\KatlaSpace{}\IdrisKeyword{if}\KatlaSpace{}\IdrisBound{isLT}\KatlaSpace{}\IdrisBound{x}\KatlaSpace{}\IdrisBound{y}\KatlaNewline{}
\KatlaSpace{}\KatlaSpace{}\IdrisKeyword{then}\KatlaSpace{}\IdrisBound{x}\KatlaSpace{}\IdrisData{::}\KatlaSpace{}\IdrisFunction{mergeBy}\KatlaSpace{}\IdrisBound{isLT}\KatlaSpace{}\IdrisBound{xs}\KatlaSpace{}\IdrisBound{ys'}\KatlaNewline{}
\KatlaSpace{}\KatlaSpace{}\IdrisKeyword{else}\KatlaSpace{}\IdrisKeyword{rewrite}\KatlaSpace{}\IdrisFunction{sym}\KatlaSpace{}\IdrisKeyword{(}\IdrisFunction{plusSuccRightSucc}\KatlaSpace{}\IdrisBound{n}\KatlaSpace{}\IdrisBound{m}\IdrisKeyword{)}\KatlaSpace{}\IdrisKeyword{in}\KatlaNewline{}
\KatlaSpace{}\KatlaSpace{}\KatlaSpace{}\KatlaSpace{}\KatlaSpace{}\KatlaSpace{}\KatlaSpace{}\IdrisBound{y}\KatlaSpace{}\IdrisData{::}\KatlaSpace{}\IdrisFunction{mergeBy}\KatlaSpace{}\IdrisBound{isLT}\KatlaSpace{}\IdrisBound{xs'}\KatlaSpace{}\IdrisBound{ys}\KatlaNewline{}
\KatlaSpace{}\KatlaSpace{}\IdrisKeyword{where}\KatlaSpace{}\IdrisFunction{plusSuccRightSucc}\KatlaSpace{}\IdrisKeyword{:}\KatlaSpace{}\IdrisKeyword{(}\IdrisBound{n}\IdrisKeyword{,}\IdrisBound{m}\KatlaSpace{}\IdrisKeyword{:}\KatlaSpace{}\IdrisType{Nat}\IdrisKeyword{)}\KatlaSpace{}\IdrisKeyword{\KatlaDash{}>}\KatlaNewline{}
\KatlaSpace{}\KatlaSpace{}\KatlaSpace{}\KatlaSpace{}\KatlaSpace{}\KatlaSpace{}\KatlaSpace{}\KatlaSpace{}\KatlaSpace{}\KatlaSpace{}\IdrisData{S}\KatlaSpace{}\IdrisKeyword{(}\IdrisBound{n}\KatlaSpace{}\IdrisFunction{+}\KatlaSpace{}\IdrisBound{m}\IdrisKeyword{)}\KatlaSpace{}\IdrisKeyword{=}\KatlaSpace{}\IdrisBound{n}\KatlaSpace{}\IdrisFunction{+}\KatlaSpace{}\IdrisData{S}\KatlaSpace{}\IdrisBound{m}\KatlaNewline{}
\end{SaveVerbatim}

\begin{SaveVerbatim}[commandchars=\\\{\}]{xs}
\IdrisBound{xs}
\end{SaveVerbatim}

\begin{SaveVerbatim}[commandchars=\\\{\}]{ys}
\IdrisBound{ys}
\end{SaveVerbatim}

\begin{SaveVerbatim}[commandchars=\\\{\}]{DataVect}
\IdrisModule{Data.Vect}\KatlaNewline{}
\end{SaveVerbatim}

\begin{SaveVerbatim}[commandchars=\\\{\}]{DataNat}
\IdrisModule{Data.Nat}\KatlaNewline{}
\end{SaveVerbatim}

\begin{SaveVerbatim}[commandchars=\\\{\}]{VectRev}
\IdrisFunction{reverse}\KatlaSpace{}\IdrisKeyword{:}\KatlaSpace{}\IdrisType{Vect}\KatlaSpace{}\IdrisBound{k}\KatlaSpace{}\IdrisBound{a}\KatlaSpace{}\IdrisKeyword{\KatlaDash{}>}\KatlaSpace{}\IdrisType{Vect}\KatlaSpace{}\IdrisBound{k}\KatlaSpace{}\IdrisBound{a}\KatlaNewline{}
\IdrisFunction{reverse}\KatlaSpace{}\IdrisBound{xs}\KatlaSpace{}\IdrisKeyword{=}\KatlaSpace{}\IdrisFunction{go}\KatlaSpace{}\IdrisData{[]}\KatlaSpace{}\IdrisBound{xs}\KatlaSpace{}\IdrisKeyword{where}\KatlaNewline{}
\KatlaSpace{}\KatlaSpace{}\IdrisFunction{plusZeroRightNeutral}\KatlaSpace{}\IdrisKeyword{:}\KatlaSpace{}\IdrisKeyword{(}\IdrisBound{n}\KatlaSpace{}\IdrisKeyword{:}\KatlaSpace{}\IdrisType{Nat}\IdrisKeyword{)}\KatlaSpace{}\IdrisKeyword{\KatlaDash{}>}\KatlaSpace{}\IdrisBound{n}\KatlaSpace{}\IdrisFunction{+}\KatlaSpace{}\IdrisData{0}\KatlaSpace{}\KatlaSpace{}\KatlaSpace{}\IdrisKeyword{=}\KatlaSpace{}\IdrisBound{n}\KatlaNewline{}
\KatlaSpace{}\KatlaSpace{}\IdrisFunction{plusSuccRightSucc}\KatlaSpace{}\IdrisKeyword{:}\KatlaSpace{}\IdrisKeyword{(}\IdrisBound{n}\IdrisKeyword{,}\KatlaSpace{}\IdrisBound{m}\KatlaSpace{}\IdrisKeyword{:}\KatlaSpace{}\IdrisType{Nat}\IdrisKeyword{)}\KatlaSpace{}\IdrisKeyword{\KatlaDash{}>}\KatlaSpace{}\IdrisBound{n}\KatlaSpace{}\IdrisFunction{+}\KatlaSpace{}\IdrisData{S}\KatlaSpace{}\IdrisBound{m}\KatlaSpace{}\IdrisKeyword{=}\KatlaSpace{}\IdrisData{S}\KatlaSpace{}\IdrisKeyword{(}\IdrisBound{n}\KatlaSpace{}\IdrisFunction{+}\KatlaSpace{}\IdrisBound{m}\IdrisKeyword{)}\KatlaNewline{}
\KatlaNewline{}
\KatlaSpace{}\KatlaSpace{}\IdrisFunction{go}\KatlaSpace{}\IdrisKeyword{:}\KatlaSpace{}\IdrisType{Vect}\KatlaSpace{}\IdrisBound{n}\KatlaSpace{}\IdrisBound{a}\KatlaSpace{}\IdrisKeyword{\KatlaDash{}>}\KatlaSpace{}\IdrisType{Vect}\KatlaSpace{}\IdrisBound{m}\KatlaSpace{}\IdrisBound{a}\KatlaSpace{}\IdrisKeyword{\KatlaDash{}>}\KatlaSpace{}\IdrisType{Vect}\KatlaSpace{}\IdrisKeyword{(}\IdrisBound{n}\IdrisFunction{+}\IdrisBound{m}\IdrisKeyword{)}\KatlaSpace{}\IdrisBound{a}\KatlaNewline{}
\KatlaSpace{}\KatlaSpace{}\IdrisFunction{go}\KatlaSpace{}\IdrisKeyword{\{}\IdrisBound{n}\IdrisKeyword{\}}\KatlaSpace{}\KatlaSpace{}\KatlaSpace{}\KatlaSpace{}\KatlaSpace{}\KatlaSpace{}\KatlaSpace{}\KatlaSpace{}\KatlaSpace{}\IdrisBound{acc}\KatlaSpace{}\IdrisData{[]}\KatlaSpace{}\KatlaSpace{}\KatlaSpace{}\KatlaSpace{}\KatlaSpace{}\KatlaSpace{}\KatlaSpace{}\KatlaSpace{}\IdrisKeyword{=}\KatlaSpace{}\IdrisKeyword{rewrite}\KatlaSpace{}\IdrisFunction{plusZeroRightNeutral}\KatlaSpace{}\IdrisBound{n}\KatlaSpace{}\IdrisKeyword{in}\KatlaSpace{}\IdrisBound{acc}\KatlaNewline{}
\KatlaSpace{}\KatlaSpace{}\IdrisFunction{go}\KatlaSpace{}\IdrisKeyword{\{}\IdrisBound{n}\IdrisKeyword{\}}\KatlaSpace{}\IdrisKeyword{\{}\IdrisBound{m}\IdrisKeyword{=}\IdrisData{S}\KatlaSpace{}\IdrisBound{m}\IdrisKeyword{\}}\KatlaSpace{}\IdrisBound{acc}\KatlaSpace{}\IdrisKeyword{(}\IdrisBound{x}\KatlaSpace{}\IdrisData{::}\KatlaSpace{}\IdrisBound{xs}\IdrisKeyword{)}\KatlaSpace{}\IdrisKeyword{=}\KatlaSpace{}\IdrisKeyword{rewrite}\KatlaSpace{}\IdrisFunction{plusSuccRightSucc}\KatlaSpace{}\IdrisBound{n}\KatlaSpace{}\IdrisBound{m}\KatlaSpace{}\KatlaSpace{}\IdrisKeyword{in}\KatlaNewline{}
\KatlaSpace{}\KatlaSpace{}\KatlaSpace{}\KatlaSpace{}\KatlaSpace{}\KatlaSpace{}\KatlaSpace{}\KatlaSpace{}\KatlaSpace{}\KatlaSpace{}\KatlaSpace{}\KatlaSpace{}\KatlaSpace{}\KatlaSpace{}\KatlaSpace{}\KatlaSpace{}\KatlaSpace{}\KatlaSpace{}\KatlaSpace{}\KatlaSpace{}\KatlaSpace{}\KatlaSpace{}\KatlaSpace{}\KatlaSpace{}\KatlaSpace{}\KatlaSpace{}\KatlaSpace{}\KatlaSpace{}\KatlaSpace{}\KatlaSpace{}\KatlaSpace{}\KatlaSpace{}\KatlaSpace{}\IdrisFunction{go}\KatlaSpace{}\IdrisKeyword{(}\IdrisBound{x}\IdrisData{::}\IdrisBound{acc}\IdrisKeyword{)}\KatlaSpace{}\IdrisBound{xs}\KatlaNewline{}
\end{SaveVerbatim}

\begin{SaveVerbatim}[commandchars=\\\{\}]{VectAlternative}
\IdrisType{Vect}\KatlaSpace{}\IdrisKeyword{(}\IdrisBound{m}\IdrisFunction{+}\IdrisBound{n}\IdrisKeyword{)}\KatlaSpace{}\IdrisBound{a}\KatlaNewline{}
\end{SaveVerbatim}
\newcommand\nvar[1][]{\UseVerb[#1]{nvar}}
\begin{SaveVerbatim}[commandchars=\\\{\}]{nvar}
\IdrisBound{n}
\end{SaveVerbatim}
\newcommand\mvar[1][]{\UseVerb[#1]{mvar}}
\begin{SaveVerbatim}[commandchars=\\\{\}]{mvar}
\IdrisBound{m}
\end{SaveVerbatim}

\begin{SaveVerbatim}[commandchars=\\\{\}]{npm}
\IdrisBound{n}\KatlaSpace{}\IdrisFunction{+}\KatlaSpace{}\IdrisBound{m}
\end{SaveVerbatim}

\begin{SaveVerbatim}[commandchars=\\\{\}]{snm}
\IdrisKeyword{((}\IdrisData{1}\IdrisFunction{+}\IdrisBound{n}\IdrisKeyword{)}\IdrisFunction{+}\IdrisBound{m}\IdrisKeyword{)}
\end{SaveVerbatim}

\begin{SaveVerbatim}[commandchars=\\\{\}]{nsm}
\IdrisKeyword{(}\IdrisBound{n}\IdrisFunction{+}\IdrisKeyword{(}\IdrisData{1}\IdrisFunction{+}\IdrisBound{m}\IdrisKeyword{))}
\end{SaveVerbatim}

\begin{SaveVerbatim}[commandchars=\\\{\}]{nz}
\IdrisKeyword{(}\IdrisBound{n}\IdrisFunction{+}\IdrisData{0}\IdrisKeyword{)}
\end{SaveVerbatim}

\begin{SaveVerbatim}[commandchars=\\\{\}]{zm}
\IdrisData{0}\IdrisFunction{+}\IdrisBound{m}
\end{SaveVerbatim}

\begin{SaveVerbatim}[commandchars=\\\{\}]{snml}
\IdrisKeyword{(}\IdrisData{1}\IdrisFunction{+}\IdrisBound{n}\IdrisKeyword{)}\IdrisFunction{+}\IdrisBound{m}
\end{SaveVerbatim}

\begin{SaveVerbatim}[commandchars=\\\{\}]{snmr}
\IdrisData{1}\IdrisFunction{+}\IdrisKeyword{(}\IdrisBound{n}\IdrisFunction{+}\IdrisBound{m}\IdrisKeyword{)}
\end{SaveVerbatim}
\newcommand\Nat[1][]{\UseVerb[#1]{Nat}}
\begin{SaveVerbatim}[commandchars=\\\{\}]{Nat}
\IdrisType{Nat}
\end{SaveVerbatim}
\newcommand\Plus[1][]{\UseVerb[#1]{Plus}}
\begin{SaveVerbatim}[commandchars=\\\{\}]{Plus}
\IdrisFunction{+}
\end{SaveVerbatim}

\begin{SaveVerbatim}[commandchars=\\\{\}]{plusZeroRightNeutral}
\IdrisFunction{plusZeroRightNeutral}
\end{SaveVerbatim}

\begin{SaveVerbatim}[commandchars=\\\{\}]{plusSuccRightSucc}
\IdrisFunction{plusSuccRightSucc}
\end{SaveVerbatim}
\newcommand\MonoidTheory[1][]{\UseVerbatim[#1]{MonoidTheory}}
\begin{SaveVerbatim}[commandchars=\\\{\}]{MonoidTheory}
\IdrisFunction{MonoidTheory}\KatlaSpace{}\IdrisKeyword{:}\KatlaSpace{}\IdrisType{Presentation}\KatlaNewline{}
\IdrisFunction{MonoidTheory}\KatlaSpace{}\IdrisKeyword{=}\KatlaSpace{}\IdrisData{MkPresentation}\KatlaSpace{}\IdrisFunction{Theory.Signature}\KatlaSpace{}\IdrisType{Theory.Axiom}\KatlaSpace{}\$\KatlaSpace{}\IdrisKeyword{\textbackslash{}case}\KatlaNewline{}
\KatlaSpace{}\KatlaSpace{}\KatlaSpace{}\KatlaSpace{}\KatlaSpace{}\KatlaSpace{}\KatlaSpace{}\KatlaSpace{}\KatlaSpace{}\KatlaSpace{}\KatlaSpace{}\KatlaSpace{}\KatlaSpace{}\KatlaSpace{}\KatlaSpace{}\KatlaSpace{}\KatlaSpace{}\IdrisData{LftNeutrality}\KatlaSpace{}\IdrisKeyword{=>}\KatlaSpace{}\IdrisFunction{lftNeutrality}\KatlaSpace{}\IdrisData{Neutral}\KatlaSpace{}\IdrisData{Product}\KatlaNewline{}
\KatlaSpace{}\KatlaSpace{}\KatlaSpace{}\KatlaSpace{}\KatlaSpace{}\KatlaSpace{}\KatlaSpace{}\KatlaSpace{}\KatlaSpace{}\KatlaSpace{}\KatlaSpace{}\KatlaSpace{}\KatlaSpace{}\KatlaSpace{}\KatlaSpace{}\KatlaSpace{}\KatlaSpace{}\IdrisData{RgtNeutrality}\KatlaSpace{}\IdrisKeyword{=>}\KatlaSpace{}\IdrisFunction{rgtNeutrality}\KatlaSpace{}\IdrisData{Neutral}\KatlaSpace{}\IdrisData{Product}\KatlaNewline{}
\KatlaSpace{}\KatlaSpace{}\KatlaSpace{}\KatlaSpace{}\KatlaSpace{}\KatlaSpace{}\KatlaSpace{}\KatlaSpace{}\KatlaSpace{}\KatlaSpace{}\KatlaSpace{}\KatlaSpace{}\KatlaSpace{}\KatlaSpace{}\KatlaSpace{}\KatlaSpace{}\KatlaSpace{}\IdrisData{Associativity}\KatlaSpace{}\IdrisKeyword{=>}\KatlaSpace{}\IdrisFunction{associativity}\KatlaSpace{}\KatlaSpace{}\KatlaSpace{}\KatlaSpace{}\KatlaSpace{}\KatlaSpace{}\KatlaSpace{}\KatlaSpace{}\KatlaSpace{}\IdrisData{Product}\KatlaNewline{}
\end{SaveVerbatim}

\begin{SaveVerbatim}[commandchars=\\\{\}]{lftNeutrality}
\IdrisData{eutrality}\KatlaSpace{}\IdrisKeyword{=>}\KatlaSpace{}
\end{SaveVerbatim}

\begin{SaveVerbatim}[commandchars=\\\{\}]{monoidsTestUnits}
\IdrisFunction{units}\KatlaSpace{}\IdrisKeyword{:}\KatlaSpace{}\IdrisKeyword{\{}\IdrisBound{monoid}\KatlaSpace{}\IdrisKeyword{:}\KatlaSpace{}\IdrisFunction{Monoid}\IdrisKeyword{\}}\KatlaSpace{}\IdrisKeyword{\KatlaDash{}>}\KatlaSpace{}\IdrisKeyword{\{}\IdrisBound{a}\KatlaSpace{}\IdrisKeyword{:}\KatlaSpace{}\IdrisFunction{U}\KatlaSpace{}\IdrisBound{monoid}\IdrisKeyword{\}}\KatlaSpace{}\IdrisKeyword{\KatlaDash{}>}\KatlaNewline{}
\KatlaSpace{}\KatlaSpace{}\KatlaSpace{}\KatlaSpace{}\KatlaSpace{}\KatlaSpace{}\KatlaSpace{}\KatlaSpace{}\IdrisKeyword{(}\IdrisFunction{O1}\KatlaSpace{}\IdrisFunction{.+.}\KatlaSpace{}\IdrisKeyword{(}\IdrisBound{a}\KatlaSpace{}\IdrisFunction{.+.}\KatlaSpace{}\IdrisFunction{O1}\IdrisKeyword{))}\KatlaSpace{}\IdrisFunction{.+.}\KatlaSpace{}\IdrisFunction{O1}\KatlaSpace{}\IdrisFunction{\KatlaTilde{}\KatlaTilde{}}\KatlaSpace{}\IdrisBound{a}\KatlaNewline{}
\IdrisFunction{units}\KatlaSpace{}\IdrisKeyword{=}\KatlaSpace{}\IdrisFunction{solve}\KatlaSpace{}\IdrisData{1}\KatlaSpace{}\IdrisKeyword{(}\IdrisFunction{FreeMonoidOver}\KatlaSpace{}\IdrisKeyword{(}\IdrisFunction{cast}\KatlaSpace{}\$\KatlaSpace{}\IdrisType{Fin}\KatlaSpace{}\IdrisData{1}\IdrisKeyword{))}\KatlaNewline{}
\KatlaSpace{}\KatlaSpace{}\KatlaSpace{}\KatlaSpace{}\KatlaSpace{}\KatlaSpace{}\KatlaSpace{}\KatlaSpace{}\KatlaSpace{}\KatlaSpace{}\KatlaSpace{}\KatlaSpace{}\KatlaSpace{}\$\KatlaSpace{}\IdrisKeyword{(}\IdrisFunction{O1}\KatlaSpace{}\IdrisFunction{.+.}\KatlaSpace{}\IdrisKeyword{(}\IdrisFunction{X}\KatlaSpace{}\IdrisData{0}\KatlaSpace{}\IdrisFunction{.+.}\KatlaSpace{}\IdrisFunction{O1}\IdrisKeyword{))}\KatlaSpace{}\IdrisFunction{.+.}\KatlaSpace{}\IdrisFunction{O1}\KatlaSpace{}\IdrisFunction{=\KatlaDash{}=}\KatlaSpace{}\IdrisFunction{X}\KatlaSpace{}\IdrisData{0}\KatlaNewline{}
\KatlaNewline{}
\end{SaveVerbatim}

\begin{SaveVerbatim}[commandchars=\\\{\}]{exampleAlgebraicSyntax}
\IdrisKeyword{(}\IdrisFunction{O1}\KatlaSpace{}\IdrisFunction{.+.}\KatlaSpace{}\IdrisKeyword{(}\IdrisBound{a}\KatlaSpace{}\IdrisFunction{.+.}\KatlaSpace{}\IdrisFunction{O1}\IdrisKeyword{))}\KatlaSpace{}\IdrisFunction{.+.}\KatlaSpace{}\IdrisFunction{O1}\KatlaSpace{}\IdrisFunction{\KatlaTilde{}\KatlaTilde{}}\KatlaSpace{}\IdrisBound{a}\KatlaNewline{}
\end{SaveVerbatim}
\newcommand\solve[1][]{\UseVerb[#1]{solve}}
\begin{SaveVerbatim}[commandchars=\\\{\}]{solve}
\IdrisFunction{solve}
\end{SaveVerbatim}

\begin{SaveVerbatim}[commandchars=\\\{\}]{SolveArgCount}
\IdrisData{1}
\end{SaveVerbatim}

\begin{SaveVerbatim}[commandchars=\\\{\}]{FreeMonoidOver}
\IdrisFunction{FreeMonoidOver}
\end{SaveVerbatim}

\begin{SaveVerbatim}[commandchars=\\\{\}]{monoidsTestNotationHint}
\IdrisKeyword{\%hint}\KatlaNewline{}
\IdrisFunction{monoidNotation}\KatlaSpace{}\IdrisKeyword{:}\KatlaSpace{}\IdrisKeyword{(}\IdrisBound{a}\KatlaSpace{}\IdrisKeyword{:}\KatlaSpace{}\IdrisFunction{Monoid}\IdrisKeyword{)}\KatlaSpace{}\IdrisKeyword{\KatlaDash{}>}\KatlaSpace{}\IdrisFunction{NotationHint}\KatlaSpace{}\IdrisBound{a}\KatlaSpace{}\IdrisType{Additive1}\KatlaNewline{}
\IdrisFunction{monoidNotation}\KatlaSpace{}\IdrisBound{a}\KatlaSpace{}\IdrisKeyword{=}\KatlaSpace{}\IdrisBound{a}\IdrisFunction{.notationHint}\KatlaSpace{}\IdrisType{Additive1}\KatlaSpace{}\IdrisBound{a}\IdrisFunction{.Additive1}\KatlaNewline{}
\end{SaveVerbatim}

\begin{SaveVerbatim}[commandchars=\\\{\}]{monoidsTestNotationHintB}
\IdrisFunction{(\KatlaTilde{}\KatlaTilde{})}\KatlaSpace{}\IdrisKeyword{:}\KatlaSpace{}\IdrisKeyword{(}\IdrisBound{monoid}\KatlaSpace{}\IdrisKeyword{:}\KatlaSpace{}\IdrisFunction{Monoid}\IdrisKeyword{)}\KatlaSpace{}\IdrisKeyword{=>}\KatlaSpace{}\IdrisKeyword{(}\IdrisBound{lhs}\IdrisKeyword{,}\KatlaSpace{}\IdrisBound{rhs}\KatlaSpace{}\IdrisKeyword{:}\KatlaSpace{}\IdrisFunction{U}\KatlaSpace{}\IdrisBound{monoid}\IdrisKeyword{)}\KatlaSpace{}\IdrisKeyword{\KatlaDash{}>}\KatlaSpace{}\IdrisType{Type}\KatlaNewline{}
\IdrisFunction{(\KatlaTilde{}\KatlaTilde{})}\KatlaSpace{}\IdrisKeyword{=}\KatlaSpace{}\IdrisBound{monoid}\IdrisFunction{.equivalence.relation}\KatlaNewline{}
\end{SaveVerbatim}

\begin{SaveVerbatim}[commandchars=\\\{\}]{zeroQuantity}
\IdrisFunction{(}
\end{SaveVerbatim}

\begin{SaveVerbatim}[commandchars=\\\{\}]{setoidEqualityRelation}
\IdrisFunction{\KatlaTilde{})}\KatlaSpace{}\IdrisKeyword{=}
\end{SaveVerbatim}
\newcommand\setoidEqualityRelationInfix[1][]{\UseVerb[#1]{setoidEqualityRelationInfix}}
\begin{SaveVerbatim}[commandchars=\\\{\}]{setoidEqualityRelationInfix}
\IdrisFunction{)}\KatlaSpace{}
\end{SaveVerbatim}

\begin{SaveVerbatim}[commandchars=\\\{\}]{monoidsTestIIUnits}
\IdrisFunction{Units}\KatlaSpace{}\IdrisKeyword{:}\KatlaSpace{}\IdrisType{Lemma}\KatlaSpace{}\IdrisFunction{MonoidTheory}\KatlaNewline{}
\IdrisFunction{Units}\KatlaSpace{}\IdrisKeyword{=}\KatlaSpace{}\IdrisFunction{mkLemma}\KatlaSpace{}\IdrisKeyword{(}\IdrisFunction{FreeMonoidOver}\KatlaSpace{}\$\KatlaSpace{}\IdrisFunction{cast}\KatlaSpace{}\$\KatlaSpace{}\IdrisType{Fin}\KatlaSpace{}\IdrisData{1}\IdrisKeyword{)}\KatlaNewline{}
\KatlaSpace{}\KatlaSpace{}\KatlaSpace{}\KatlaSpace{}\KatlaSpace{}\KatlaSpace{}\$\KatlaSpace{}\IdrisKeyword{(}\IdrisFunction{O1}\KatlaSpace{}\IdrisFunction{.+.}\KatlaSpace{}\IdrisKeyword{(}\IdrisFunction{X}\KatlaSpace{}\IdrisData{0}\KatlaSpace{}\IdrisFunction{.+.}\KatlaSpace{}\IdrisFunction{O1}\IdrisKeyword{))}\KatlaSpace{}\IdrisFunction{.+.}\KatlaSpace{}\IdrisFunction{O1}\KatlaSpace{}\IdrisFunction{=\KatlaDash{}=}\KatlaSpace{}\IdrisFunction{X}\KatlaSpace{}\IdrisData{0}\KatlaNewline{}
\end{SaveVerbatim}

\begin{SaveVerbatim}[commandchars=\\\{\}]{commutativeMonoidsTestNSM}
\IdrisFunction{simplify}\KatlaSpace{}\IdrisKeyword{:}\KatlaSpace{}\IdrisKeyword{(}\IdrisBound{n}\IdrisKeyword{,}\KatlaSpace{}\IdrisBound{m}\IdrisKeyword{,}\KatlaSpace{}\IdrisBound{k}\KatlaSpace{}\IdrisKeyword{:}\KatlaSpace{}\IdrisType{Nat}\IdrisKeyword{)}\KatlaSpace{}\IdrisKeyword{\KatlaDash{}>}\KatlaNewline{}
\KatlaSpace{}\KatlaSpace{}\IdrisKeyword{(}\IdrisBound{n}\KatlaSpace{}\IdrisFunction{+}\KatlaSpace{}\IdrisData{6}\IdrisKeyword{)}\KatlaSpace{}\IdrisFunction{+}\KatlaSpace{}\IdrisKeyword{(}\IdrisBound{k}\KatlaSpace{}\IdrisFunction{+}\KatlaSpace{}\IdrisBound{n}\IdrisKeyword{)}\KatlaSpace{}\IdrisFunction{+}\KatlaSpace{}\IdrisKeyword{(}\IdrisBound{m}\KatlaSpace{}\IdrisFunction{+}\KatlaSpace{}\IdrisData{2}\IdrisKeyword{)}\KatlaSpace{}\IdrisKeyword{=}\KatlaSpace{}\IdrisBound{k}\KatlaSpace{}\IdrisFunction{+}\KatlaSpace{}\IdrisData{2}\IdrisFunction{*}\IdrisBound{n}\KatlaSpace{}\IdrisFunction{+}\KatlaSpace{}\IdrisBound{m}\KatlaSpace{}\IdrisFunction{+}\KatlaSpace{}\IdrisData{8}\KatlaNewline{}
\IdrisFunction{simplify}\KatlaSpace{}\IdrisBound{n}\KatlaSpace{}\IdrisBound{m}\KatlaSpace{}\IdrisBound{k}\KatlaSpace{}\IdrisKeyword{=}\KatlaSpace{}\IdrisFunction{solve}\KatlaSpace{}\IdrisData{3}\KatlaSpace{}\IdrisKeyword{(}\IdrisFunction{Monoid.Commutative.Frex}\KatlaSpace{}\IdrisFunction{Nat.Additive}\IdrisKeyword{)}\KatlaNewline{}
\KatlaSpace{}\KatlaSpace{}\$\KatlaSpace{}\IdrisKeyword{(}\IdrisFunction{Dyn}\KatlaSpace{}\IdrisData{0}\KatlaSpace{}\IdrisFunction{.+.}\KatlaSpace{}\IdrisFunction{Sta}\KatlaSpace{}\IdrisData{6}\IdrisKeyword{)}\KatlaSpace{}\IdrisFunction{.+.}\KatlaSpace{}\IdrisKeyword{(}\IdrisFunction{Dyn}\KatlaSpace{}\IdrisData{1}\KatlaSpace{}\IdrisFunction{.+.}\KatlaSpace{}\IdrisFunction{Dyn}\KatlaSpace{}\IdrisData{0}\IdrisKeyword{)}\KatlaSpace{}\IdrisFunction{.+.}\KatlaSpace{}\IdrisKeyword{(}\IdrisFunction{Dyn}\KatlaSpace{}\IdrisData{2}\KatlaSpace{}\IdrisFunction{.+.}\KatlaSpace{}\IdrisFunction{Sta}\KatlaSpace{}\IdrisData{2}\IdrisKeyword{)}\KatlaSpace{}\IdrisFunction{=\KatlaDash{}=}\KatlaNewline{}
\KatlaSpace{}\KatlaSpace{}\IdrisFunction{Dyn}\KatlaSpace{}\IdrisData{1}\KatlaSpace{}\IdrisFunction{.+.}\KatlaSpace{}\IdrisKeyword{((}\IdrisFunction{the}\KatlaSpace{}\IdrisType{Nat}\KatlaSpace{}\IdrisData{2}\IdrisKeyword{)}\KatlaSpace{}\IdrisFunction{*.}\KatlaSpace{}\IdrisFunction{Dyn}\KatlaSpace{}\IdrisData{0}\IdrisKeyword{)}\KatlaSpace{}\IdrisFunction{.+.}\KatlaSpace{}\IdrisFunction{Dyn}\KatlaSpace{}\IdrisData{2}\KatlaSpace{}\IdrisFunction{.+.}\KatlaSpace{}\IdrisFunction{Sta}\KatlaSpace{}\IdrisData{8}\KatlaNewline{}
\end{SaveVerbatim}

\begin{SaveVerbatim}[commandchars=\\\{\}]{Dyn}
\IdrisFunction{Dyn}
\end{SaveVerbatim}

\begin{SaveVerbatim}[commandchars=\\\{\}]{Sta}
\IdrisFunction{Sta}
\end{SaveVerbatim}

\begin{SaveVerbatim}[commandchars=\\\{\}]{FralByMonoidFrex}
\IdrisFunction{FreeMonoidOver}\KatlaSpace{}\IdrisKeyword{:}\KatlaSpace{}\IdrisKeyword{(}\IdrisBound{s}\KatlaSpace{}\IdrisKeyword{:}\KatlaSpace{}\IdrisType{Setoid}\IdrisKeyword{)}\KatlaSpace{}\IdrisKeyword{\KatlaDash{}>}\KatlaSpace{}\IdrisType{Free}\KatlaSpace{}\IdrisFunction{MonoidTheory}\KatlaSpace{}\IdrisBound{s}\KatlaNewline{}
\IdrisFunction{FreeMonoidOver}\KatlaSpace{}\IdrisBound{s}\KatlaSpace{}\IdrisKeyword{=}\KatlaSpace{}\IdrisFunction{ByFrex}\KatlaSpace{}\IdrisFunction{FreeMonoidVoid}\KatlaNewline{}
\KatlaSpace{}\KatlaSpace{}\KatlaSpace{}\KatlaSpace{}\KatlaSpace{}\KatlaSpace{}\KatlaSpace{}\KatlaSpace{}\KatlaSpace{}\KatlaSpace{}\KatlaSpace{}\KatlaSpace{}\KatlaSpace{}\KatlaSpace{}\KatlaSpace{}\KatlaSpace{}\KatlaSpace{}\KatlaSpace{}\KatlaSpace{}\KatlaSpace{}\KatlaSpace{}\KatlaSpace{}\KatlaSpace{}\KatlaSpace{}\KatlaSpace{}\KatlaSpace{}\IdrisKeyword{(}\IdrisFunction{MonoidFrex}\KatlaSpace{}\IdrisFunction{TrivialMonoid}\KatlaSpace{}\IdrisBound{s}\IdrisKeyword{)}\KatlaNewline{}
\end{SaveVerbatim}

\begin{SaveVerbatim}[commandchars=\\\{\}]{ByFrexType}
\IdrisFunction{ByFrex}\KatlaSpace{}\IdrisKeyword{:}\KatlaSpace{}\IdrisKeyword{(}\IdrisBound{initial}\KatlaSpace{}\IdrisKeyword{:}\KatlaSpace{}\IdrisType{Free}\KatlaSpace{}\IdrisBound{pres}\KatlaSpace{}\IdrisKeyword{(}\IdrisFunction{cast}\KatlaSpace{}\IdrisType{Void}\IdrisKeyword{))}\KatlaSpace{}\IdrisKeyword{\KatlaDash{}>}\KatlaNewline{}
\KatlaSpace{}\KatlaSpace{}\IdrisType{Frex}\KatlaSpace{}\IdrisBound{initial}\IdrisFunction{.Data.Model}\KatlaSpace{}\IdrisBound{s}\KatlaSpace{}\IdrisKeyword{\KatlaDash{}>}\KatlaSpace{}\IdrisType{Free}\KatlaSpace{}\IdrisBound{pres}\KatlaSpace{}\IdrisBound{s}\KatlaNewline{}
\end{SaveVerbatim}

\begin{SaveVerbatim}[commandchars=\\\{\}]{ByFrex}
\IdrisFunction{ByFrex}
\end{SaveVerbatim}

\begin{SaveVerbatim}[commandchars=\\\{\}]{InvFrexByMonoidFrex}
\IdrisFunction{MonoidFrex}\KatlaSpace{}\IdrisKeyword{(}\IdrisFunction{cast}\KatlaSpace{}\IdrisBound{a}\IdrisKeyword{)}\KatlaSpace{}\IdrisKeyword{(}\IdrisFunction{cast}\KatlaSpace{}\IdrisType{Bool}\KatlaSpace{}\IdrisFunction{`Pair`}\KatlaSpace{}\IdrisBound{s}\IdrisKeyword{)}\KatlaNewline{}
\end{SaveVerbatim}
\newcommand\Certificate[1][]{\UseVerbatim[#1]{Certificate}}
\begin{SaveVerbatim}[commandchars=\\\{\}]{Certificate}
\KatlaSpace{}\KatlaSpace{}\IdrisFunction{units}\KatlaSpace{}\IdrisKeyword{:}\KatlaSpace{}\IdrisKeyword{(}\IdrisBound{x}\KatlaSpace{}\IdrisKeyword{:}\KatlaSpace{}\IdrisFunction{U}\KatlaSpace{}\IdrisBound{m}\IdrisKeyword{)}\KatlaSpace{}\IdrisKeyword{\KatlaDash{}>}\KatlaSpace{}\IdrisFunction{O1}\KatlaSpace{}\IdrisFunction{.+.}\KatlaSpace{}\IdrisKeyword{(}\IdrisBound{x}\KatlaSpace{}\IdrisFunction{.+.}\KatlaSpace{}\IdrisFunction{O1}\IdrisKeyword{)}\KatlaSpace{}\IdrisFunction{.+.}\KatlaSpace{}\IdrisFunction{O1}\KatlaSpace{}\IdrisFunction{=\KatlaTilde{}=}\KatlaSpace{}\IdrisBound{x}\KatlaNewline{}
\KatlaSpace{}\KatlaSpace{}\IdrisFunction{units}\KatlaSpace{}\IdrisBound{x}\KatlaSpace{}\IdrisKeyword{=}\KatlaSpace{}\IdrisFunction{CalcWith}\KatlaSpace{}\IdrisKeyword{(}\IdrisFunction{cast}\KatlaSpace{}\IdrisBound{m}\IdrisKeyword{)}\KatlaSpace{}\$\KatlaNewline{}
\KatlaSpace{}\KatlaSpace{}\KatlaSpace{}\KatlaSpace{}\IdrisData{|\KatlaTilde{}}\KatlaSpace{}\IdrisFunction{O1}\KatlaSpace{}\IdrisFunction{.+.}\KatlaSpace{}\IdrisKeyword{(}\IdrisBound{x}\KatlaSpace{}\IdrisFunction{.+.}\KatlaSpace{}\IdrisFunction{O1}\IdrisKeyword{)}\KatlaSpace{}\IdrisFunction{.+.}\KatlaSpace{}\IdrisFunction{O1}\KatlaNewline{}
\KatlaSpace{}\KatlaSpace{}\KatlaSpace{}\KatlaSpace{}\IdrisData{\KatlaTilde{}\KatlaTilde{}}\KatlaSpace{}\IdrisFunction{O1}\KatlaSpace{}\IdrisFunction{.+.}\KatlaSpace{}\IdrisKeyword{(}\IdrisFunction{O1}\KatlaSpace{}\IdrisFunction{.+.}\KatlaSpace{}\IdrisBound{x}\KatlaSpace{}\IdrisFunction{.+.}\KatlaSpace{}\IdrisFunction{O1}\IdrisKeyword{)}\KatlaSpace{}\IdrisFunction{.+.}\KatlaSpace{}\IdrisFunction{O1}\KatlaNewline{}
\KatlaSpace{}\KatlaSpace{}\KatlaSpace{}\KatlaSpace{}\KatlaSpace{}\KatlaSpace{}\IdrisFunction{..<}\IdrisKeyword{(}\KatlaSpace{}\IdrisFunction{Cong}\KatlaSpace{}\IdrisKeyword{(\textbackslash{}}\KatlaSpace{}\IdrisBound{focus}\KatlaSpace{}\IdrisKeyword{=>}\KatlaSpace{}\IdrisFunction{O2}\KatlaSpace{}\IdrisFunction{:+:}\KatlaSpace{}\IdrisKeyword{(}\IdrisBound{focus}\KatlaSpace{}\IdrisFunction{:+:}\KatlaSpace{}\IdrisFunction{O2}\IdrisKeyword{)}\KatlaSpace{}\IdrisFunction{:+:}\KatlaSpace{}\IdrisFunction{O2}\IdrisKeyword{)}\KatlaSpace{}\$\KatlaSpace{}\IdrisFunction{lftNeutrality}\KatlaSpace{}\IdrisBound{x}\KatlaSpace{}\IdrisKeyword{)}\KatlaNewline{}
\KatlaSpace{}\KatlaSpace{}\KatlaSpace{}\KatlaSpace{}\IdrisData{\KatlaTilde{}\KatlaTilde{}}\KatlaSpace{}\IdrisFunction{O1}\KatlaSpace{}\IdrisFunction{.+.}\KatlaSpace{}\IdrisKeyword{(}\IdrisFunction{O1}\KatlaSpace{}\IdrisFunction{.+.}\KatlaSpace{}\IdrisKeyword{(}\IdrisBound{x}\KatlaSpace{}\IdrisFunction{.+.}\KatlaSpace{}\IdrisFunction{O1}\IdrisKeyword{))}\KatlaSpace{}\IdrisFunction{.+.}\KatlaSpace{}\IdrisFunction{O1}\KatlaNewline{}
\KatlaSpace{}\KatlaSpace{}\KatlaSpace{}\KatlaSpace{}\KatlaSpace{}\KatlaSpace{}\IdrisFunction{..<}\IdrisKeyword{(}\KatlaSpace{}\IdrisFunction{Cong}\KatlaSpace{}\IdrisKeyword{(\textbackslash{}}\KatlaSpace{}\IdrisBound{focus}\KatlaSpace{}\IdrisKeyword{=>}\KatlaSpace{}\IdrisFunction{O2}\KatlaSpace{}\IdrisFunction{:+:}\KatlaSpace{}\IdrisBound{focus}\KatlaSpace{}\IdrisFunction{:+:}\KatlaSpace{}\IdrisFunction{O2}\IdrisKeyword{)}\KatlaSpace{}\$\KatlaSpace{}\IdrisFunction{associativity}\KatlaSpace{}\IdrisFunction{O1}\KatlaSpace{}\IdrisBound{x}\KatlaSpace{}\IdrisFunction{O1}\KatlaSpace{}\IdrisKeyword{)}\KatlaNewline{}
\KatlaSpace{}\KatlaSpace{}\KatlaSpace{}\KatlaSpace{}\IdrisData{\KatlaTilde{}\KatlaTilde{}}\KatlaSpace{}\IdrisFunction{O1}\KatlaSpace{}\IdrisFunction{.+.}\KatlaSpace{}\IdrisFunction{O1}\KatlaSpace{}\IdrisFunction{.+.}\KatlaSpace{}\IdrisKeyword{(}\IdrisBound{x}\KatlaSpace{}\IdrisFunction{.+.}\KatlaSpace{}\IdrisFunction{O1}\IdrisKeyword{)}\KatlaSpace{}\IdrisFunction{.+.}\KatlaSpace{}\IdrisFunction{O1}\KatlaNewline{}
\KatlaSpace{}\KatlaSpace{}\KatlaSpace{}\KatlaSpace{}\KatlaSpace{}\KatlaSpace{}\IdrisData{...}\IdrisKeyword{(}\KatlaSpace{}\IdrisFunction{Cong}\KatlaSpace{}\IdrisKeyword{(\textbackslash{}}\KatlaSpace{}\IdrisBound{focus}\KatlaSpace{}\IdrisKeyword{=>}\KatlaSpace{}\IdrisBound{focus}\KatlaSpace{}\IdrisFunction{:+:}\KatlaSpace{}\IdrisFunction{O2}\IdrisKeyword{)}\KatlaSpace{}\$\KatlaSpace{}\IdrisFunction{associativity}\KatlaSpace{}\IdrisFunction{O1}\KatlaSpace{}\IdrisFunction{O1}\KatlaSpace{}\IdrisKeyword{(}\IdrisBound{x}\KatlaSpace{}\IdrisFunction{.+.}\KatlaSpace{}\IdrisFunction{O1}\IdrisKeyword{)}\KatlaSpace{}\IdrisKeyword{)}\KatlaNewline{}
\KatlaSpace{}\KatlaSpace{}\KatlaSpace{}\KatlaSpace{}\IdrisData{\KatlaTilde{}\KatlaTilde{}}\KatlaSpace{}\IdrisFunction{O1}\KatlaSpace{}\IdrisFunction{.+.}\KatlaSpace{}\IdrisFunction{O1}\KatlaSpace{}\IdrisFunction{.+.}\KatlaSpace{}\IdrisBound{x}\KatlaSpace{}\IdrisFunction{.+.}\KatlaSpace{}\IdrisFunction{O1}\KatlaSpace{}\IdrisFunction{.+.}\KatlaSpace{}\IdrisFunction{O1}\KatlaNewline{}
\KatlaSpace{}\KatlaSpace{}\KatlaSpace{}\KatlaSpace{}\KatlaSpace{}\KatlaSpace{}\IdrisData{...}\IdrisKeyword{(}\KatlaSpace{}\IdrisFunction{Cong}\KatlaSpace{}\IdrisKeyword{(\textbackslash{}}\KatlaSpace{}\IdrisBound{focus}\KatlaSpace{}\IdrisKeyword{=>}\KatlaSpace{}\IdrisBound{focus}\KatlaSpace{}\IdrisFunction{:+:}\KatlaSpace{}\IdrisFunction{O2}\IdrisKeyword{)}\KatlaSpace{}\$\KatlaSpace{}\IdrisFunction{associativity}\KatlaSpace{}\IdrisKeyword{(}\IdrisFunction{O1}\KatlaSpace{}\IdrisFunction{.+.}\KatlaSpace{}\IdrisFunction{O1}\IdrisKeyword{)}\KatlaSpace{}\IdrisBound{x}\KatlaSpace{}\IdrisFunction{O1}\KatlaSpace{}\IdrisKeyword{)}\KatlaNewline{}
\KatlaSpace{}\KatlaSpace{}\KatlaSpace{}\KatlaSpace{}\IdrisData{\KatlaTilde{}\KatlaTilde{}}\KatlaSpace{}\IdrisFunction{O1}\KatlaSpace{}\IdrisFunction{.+.}\KatlaSpace{}\IdrisBound{x}\KatlaSpace{}\IdrisFunction{.+.}\KatlaSpace{}\IdrisFunction{O1}\KatlaSpace{}\IdrisFunction{.+.}\KatlaSpace{}\IdrisFunction{O1}\KatlaNewline{}
\KatlaSpace{}\KatlaSpace{}\KatlaSpace{}\KatlaSpace{}\KatlaSpace{}\KatlaSpace{}\IdrisData{...}\IdrisKeyword{(}\KatlaSpace{}\IdrisFunction{Cong}\KatlaSpace{}\IdrisKeyword{(\textbackslash{}}\KatlaSpace{}\IdrisBound{focus}\KatlaSpace{}\IdrisKeyword{=>}\KatlaSpace{}\IdrisBound{focus}\KatlaSpace{}\IdrisFunction{:+:}\KatlaSpace{}\IdrisFunction{Val}\KatlaSpace{}\IdrisBound{x}\KatlaSpace{}\IdrisFunction{:+:}\KatlaSpace{}\IdrisFunction{O2}\KatlaSpace{}\IdrisFunction{:+:}\KatlaSpace{}\IdrisFunction{O2}\IdrisKeyword{)}\KatlaSpace{}\$\KatlaSpace{}\IdrisFunction{lftNeutrality}\KatlaSpace{}\IdrisFunction{O1}\KatlaSpace{}\IdrisKeyword{)}\KatlaNewline{}
\KatlaSpace{}\KatlaSpace{}\KatlaSpace{}\KatlaSpace{}\IdrisData{\KatlaTilde{}\KatlaTilde{}}\KatlaSpace{}\IdrisFunction{O1}\KatlaSpace{}\IdrisFunction{.+.}\KatlaSpace{}\IdrisBound{x}\KatlaSpace{}\IdrisFunction{.+.}\KatlaSpace{}\IdrisKeyword{(}\IdrisFunction{O1}\KatlaSpace{}\IdrisFunction{.+.}\KatlaSpace{}\IdrisFunction{O1}\IdrisKeyword{)}\KatlaNewline{}
\KatlaSpace{}\KatlaSpace{}\KatlaSpace{}\KatlaSpace{}\KatlaSpace{}\KatlaSpace{}\IdrisFunction{..<}\IdrisKeyword{(}\KatlaSpace{}\IdrisFunction{associativity}\KatlaSpace{}\IdrisKeyword{(}\IdrisFunction{O1}\KatlaSpace{}\IdrisFunction{.+.}\KatlaSpace{}\IdrisBound{x}\IdrisKeyword{)}\KatlaSpace{}\IdrisFunction{O1}\KatlaSpace{}\IdrisFunction{O1}\KatlaSpace{}\IdrisKeyword{)}\KatlaNewline{}
\KatlaSpace{}\KatlaSpace{}\KatlaSpace{}\KatlaSpace{}\IdrisData{\KatlaTilde{}\KatlaTilde{}}\KatlaSpace{}\IdrisFunction{O1}\KatlaSpace{}\IdrisFunction{.+.}\KatlaSpace{}\IdrisBound{x}\KatlaSpace{}\IdrisFunction{.+.}\KatlaSpace{}\IdrisFunction{O1}\KatlaNewline{}
\KatlaSpace{}\KatlaSpace{}\KatlaSpace{}\KatlaSpace{}\KatlaSpace{}\KatlaSpace{}\IdrisData{...}\IdrisKeyword{(}\KatlaSpace{}\IdrisFunction{Cong}\KatlaSpace{}\IdrisKeyword{(\textbackslash{}}\KatlaSpace{}\IdrisBound{focus}\KatlaSpace{}\IdrisKeyword{=>}\KatlaSpace{}\IdrisFunction{O2}\KatlaSpace{}\IdrisFunction{:+:}\KatlaSpace{}\IdrisFunction{Val}\KatlaSpace{}\IdrisBound{x}\KatlaSpace{}\IdrisFunction{:+:}\KatlaSpace{}\IdrisBound{focus}\IdrisKeyword{)}\KatlaSpace{}\$\KatlaSpace{}\IdrisFunction{lftNeutrality}\KatlaSpace{}\IdrisFunction{O1}\KatlaSpace{}\IdrisKeyword{)}\KatlaNewline{}
\KatlaSpace{}\KatlaSpace{}\KatlaSpace{}\KatlaSpace{}\IdrisData{\KatlaTilde{}\KatlaTilde{}}\KatlaSpace{}\IdrisFunction{O1}\KatlaSpace{}\IdrisFunction{.+.}\KatlaSpace{}\IdrisBound{x}\KatlaNewline{}
\KatlaSpace{}\KatlaSpace{}\KatlaSpace{}\KatlaSpace{}\KatlaSpace{}\KatlaSpace{}\IdrisData{...}\IdrisKeyword{(}\KatlaSpace{}\IdrisFunction{rgtNeutrality}\KatlaSpace{}\IdrisKeyword{(}\IdrisFunction{O1}\KatlaSpace{}\IdrisFunction{.+.}\KatlaSpace{}\IdrisBound{x}\IdrisKeyword{)}\KatlaSpace{}\IdrisKeyword{)}\KatlaNewline{}
\KatlaSpace{}\KatlaSpace{}\KatlaSpace{}\KatlaSpace{}\IdrisData{\KatlaTilde{}\KatlaTilde{}}\KatlaSpace{}\IdrisBound{x}\KatlaNewline{}
\KatlaSpace{}\KatlaSpace{}\KatlaSpace{}\KatlaSpace{}\KatlaSpace{}\KatlaSpace{}\IdrisData{...}\IdrisKeyword{(}\KatlaSpace{}\IdrisFunction{lftNeutrality}\KatlaSpace{}\IdrisBound{x}\KatlaSpace{}\IdrisKeyword{)}\KatlaNewline{}
\end{SaveVerbatim}
\newcommand\Um[1][]{\UseVerb[#1]{Um}}
\begin{SaveVerbatim}[commandchars=\\\{\}]{Um}
\IdrisFunction{U}\KatlaSpace{}\IdrisBound{m}
\end{SaveVerbatim}
\newcommand\OI[1][]{\UseVerb[#1]{OI}}
\begin{SaveVerbatim}[commandchars=\\\{\}]{OI}
\IdrisFunction{O1}
\end{SaveVerbatim}
\newcommand\DotPlus[1][]{\UseVerb[#1]{DotPlus}}
\begin{SaveVerbatim}[commandchars=\\\{\}]{DotPlus}
\IdrisFunction{.+.}
\end{SaveVerbatim}

\begin{SaveVerbatim}[commandchars=\\\{\}]{IdrisCertificatePrinter}
\IdrisFunction{idris}\KatlaSpace{}\IdrisKeyword{:}\KatlaSpace{}\IdrisType{List}\KatlaSpace{}\IdrisKeyword{(}\IdrisType{String,}\KatlaSpace{}\IdrisType{Lemma}\KatlaSpace{}\IdrisFunction{MonoidTheory}\IdrisKeyword{)}\KatlaSpace{}\IdrisKeyword{\KatlaDash{}>}\KatlaSpace{}\IdrisType{String}\KatlaNewline{}
\end{SaveVerbatim}

\begin{SaveVerbatim}[commandchars=\\\{\}]{IntroTerm}
\KatlaDash{}\IdrisData{6}\KatlaSpace{}\IdrisFunction{+}\KatlaSpace{}\IdrisKeyword{(}\IdrisBound{x}\KatlaSpace{}\IdrisFunction{+}\KatlaSpace{}\IdrisData{3}\IdrisKeyword{)}\KatlaSpace{}\IdrisFunction{+}\KatlaSpace{}\IdrisKeyword{(}\IdrisBound{y}\KatlaSpace{}\IdrisFunction{+}\KatlaSpace{}\IdrisBound{x}\IdrisKeyword{)}\KatlaNewline{}
\end{SaveVerbatim}

\begin{SaveVerbatim}[commandchars=\\\{\}]{IntroTermNF}
\KatlaDash{}\IdrisData{3}\KatlaSpace{}\IdrisFunction{+}\KatlaSpace{}\IdrisData{2}\IdrisFunction{*}\IdrisBound{x}\KatlaSpace{}\IdrisFunction{+}\KatlaSpace{}\IdrisData{1}\IdrisFunction{*}\IdrisBound{y}\KatlaNewline{}
\end{SaveVerbatim}

\begin{SaveVerbatim}[commandchars=\\\{\}]{IntroFrexType}
\IdrisKeyword{(}\IdrisType{Integer,}\KatlaSpace{}\IdrisType{Vect}\KatlaSpace{}\IdrisData{2}\KatlaSpace{}\IdrisType{Nat}\IdrisKeyword{)}
\end{SaveVerbatim}

\begin{SaveVerbatim}[commandchars=\\\{\}]{IntroFrexNF}
\IdrisKeyword{(}\KatlaDash{}\IdrisData{3,}\KatlaSpace{}\IdrisData{[2,}\KatlaSpace{}\IdrisData{1]}\IdrisKeyword{)}\KatlaNewline{}
\end{SaveVerbatim}

\begin{SaveVerbatim}[commandchars=\\\{\}]{ListReversal}
\IdrisKeyword{(}\IdrisFunction{reverse}\KatlaSpace{}\IdrisKeyword{(}\IdrisFunction{reverse}\KatlaSpace{}\IdrisBound{ys}\KatlaSpace{}\IdrisFunction{++}\KatlaSpace{}\IdrisKeyword{(}\IdrisData{[}\IdrisBound{x3}\IdrisData{,}\KatlaSpace{}\IdrisBound{x2}\IdrisData{,}\KatlaSpace{}\IdrisBound{x1}\IdrisData{]}\KatlaSpace{}\IdrisFunction{++}\KatlaSpace{}\IdrisFunction{reverse}\KatlaSpace{}\IdrisBound{xs}\IdrisKeyword{)))}\KatlaSpace{}\IdrisKeyword{=}\KatlaSpace{}\IdrisKeyword{(}\IdrisBound{xs}\KatlaSpace{}\IdrisFunction{++}\KatlaSpace{}\IdrisData{[}\IdrisBound{x1}\IdrisData{,}\KatlaSpace{}\IdrisBound{x2}\IdrisData{,}\KatlaSpace{}\IdrisBound{x3}\IdrisData{]}\KatlaSpace{}\IdrisFunction{++}\KatlaSpace{}\IdrisBound{ys}\IdrisKeyword{)}\KatlaNewline{}
\end{SaveVerbatim}

%% file: background-listings.tex
\newcommand\SetoidDef[1][]{\UseVerbatim[#1]{SetoidDef}}
\begin{SaveVerbatim}[commandchars=\\\{\}]{SetoidDef}
\IdrisKeyword{record}\KatlaSpace{}\IdrisType{Setoid}\KatlaSpace{}\IdrisKeyword{where}
\KatlaSpace{}\KatlaSpace{}constructor\KatlaSpace{}\IdrisData{MkSetoid}
\KatlaSpace{}\KatlaSpace{}\IdrisKeyword{0}\KatlaSpace{}\IdrisFunction{U}\KatlaSpace{}\IdrisKeyword{:}\KatlaSpace{}\IdrisType{Type}
\KatlaSpace{}\KatlaSpace{}\IdrisFunction{equivalence}\KatlaSpace{}\IdrisKeyword{:}\KatlaSpace{}\IdrisType{Equivalence}\KatlaSpace{}\IdrisBound{U}
\end{SaveVerbatim}
\newcommand\SetoidHomo[1][]{\UseVerbatim[#1]{SetoidHomo}}
\begin{SaveVerbatim}[commandchars=\\\{\}]{SetoidHomo}
\IdrisFunction{SetoidHomomorphism}\KatlaSpace{}\IdrisKeyword{:}\KatlaSpace{}\IdrisKeyword{(}\IdrisBound{a}\IdrisKeyword{,}\IdrisBound{b}\KatlaSpace{}\IdrisKeyword{:}\KatlaSpace{}\IdrisType{Setoid}\IdrisKeyword{)}
\KatlaSpace{}\KatlaSpace{}\IdrisKeyword{->}\KatlaSpace{}\IdrisKeyword{(}\IdrisBound{f}\KatlaSpace{}\IdrisKeyword{:}\KatlaSpace{}\IdrisFunction{U}\KatlaSpace{}\IdrisBound{a}\KatlaSpace{}\IdrisKeyword{->}\KatlaSpace{}\IdrisFunction{U}\KatlaSpace{}\IdrisBound{b}\IdrisKeyword{)}\KatlaSpace{}\IdrisKeyword{->}\KatlaSpace{}\IdrisType{Type}
\IdrisFunction{SetoidHomomorphism}\KatlaSpace{}\IdrisBound{a}\KatlaSpace{}\IdrisBound{b}\KatlaSpace{}\IdrisBound{f}
\KatlaSpace{}\KatlaSpace{}\IdrisKeyword{=}\KatlaSpace{}\IdrisKeyword{(}\IdrisBound{x}\IdrisKeyword{,}\IdrisBound{y}\KatlaSpace{}\IdrisKeyword{:}\KatlaSpace{}\IdrisFunction{U}\KatlaSpace{}\IdrisBound{a}\IdrisKeyword{)}\KatlaSpace{}\IdrisKeyword{->}\KatlaSpace{}\IdrisBound{a}\IdrisFunction{.equivalence.relation}\KatlaSpace{}\IdrisBound{x}\KatlaSpace{}\IdrisBound{y}
\KatlaSpace{}\KatlaSpace{}\IdrisKeyword{->}\KatlaSpace{}\IdrisBound{b}\IdrisFunction{.equivalence.relation}\KatlaSpace{}\IdrisKeyword{(}\IdrisBound{f}\KatlaSpace{}\IdrisBound{x}\IdrisKeyword{)}\KatlaSpace{}\IdrisKeyword{(}\IdrisBound{f}\KatlaSpace{}\IdrisBound{y}\IdrisKeyword{)}
\end{SaveVerbatim}
\newcommand\SetoidArrow[1][]{\UseVerbatim[#1]{SetoidArrow}}
\begin{SaveVerbatim}[commandchars=\\\{\}]{SetoidArrow}
\IdrisKeyword{record}\KatlaSpace{}\IdrisType{(~>)}\KatlaSpace{}\IdrisKeyword{(}\IdrisBound{A}\IdrisKeyword{,}\IdrisBound{B}\KatlaSpace{}\IdrisKeyword{:}\KatlaSpace{}\IdrisType{Setoid}\IdrisKeyword{)}\KatlaSpace{}\IdrisKeyword{where}
\KatlaSpace{}\KatlaSpace{}constructor\KatlaSpace{}\IdrisData{MkSetoidHomomorphism}
\KatlaSpace{}\KatlaSpace{}\IdrisFunction{H}\KatlaSpace{}\IdrisKeyword{:}\KatlaSpace{}\IdrisFunction{U}\KatlaSpace{}\IdrisBound{A}\KatlaSpace{}\IdrisKeyword{->}\KatlaSpace{}\IdrisFunction{U}\KatlaSpace{}\IdrisBound{B}
\KatlaSpace{}\KatlaSpace{}\IdrisFunction{homomorphic}\KatlaSpace{}\IdrisKeyword{:}\KatlaSpace{}\IdrisFunction{SetoidHomomorphism}\KatlaSpace{}\IdrisBound{A}\KatlaSpace{}\IdrisBound{B}\KatlaSpace{}\IdrisBound{H}
\end{SaveVerbatim}
\newcommand\SetoidArrowInline[1][]{\UseVerb[#1]{SetoidArrowInline}}
\begin{SaveVerbatim}[commandchars=\\\{\}]{SetoidArrowInline}
\IdrisType{~>}
\end{SaveVerbatim}

\begin{SaveVerbatim}[commandchars=\\\{\}]{EquivalenceName}
\IdrisType{Equivalence}
\end{SaveVerbatim}

\begin{SaveVerbatim}[commandchars=\\\{\}]{EquivalenceType}
\KatlaSpace{}\IdrisKeyword{(}\IdrisBound{A}\KatlaSpace{}\IdrisKeyword{:}\KatlaSpace{}\IdrisType{Type}\IdrisKeyword{)}\KatlaSpace{}\IdrisKeyword{where}
\end{SaveVerbatim}

\begin{SaveVerbatim}[commandchars=\\\{\}]{relationField}
\IdrisKeyword{0}\KatlaSpace{}
\end{SaveVerbatim}
\newcommand\EquivalenceDef[1][]{\UseVerbatim[#1]{EquivalenceDef}}
\begin{SaveVerbatim}[commandchars=\\\{\}]{EquivalenceDef}
\IdrisKeyword{record}\KatlaSpace{}\IdrisType{Equivalence}\KatlaSpace{}\IdrisKeyword{(}\IdrisBound{A}\KatlaSpace{}\IdrisKeyword{:}\KatlaSpace{}\IdrisType{Type}\IdrisKeyword{)}\KatlaSpace{}\IdrisKeyword{where}
\KatlaSpace{}\KatlaSpace{}constructor\KatlaSpace{}\IdrisData{MkEquivalence}
\KatlaSpace{}\KatlaSpace{}\IdrisKeyword{0}\KatlaSpace{}\IdrisFunction{relation}\IdrisKeyword{:}\KatlaSpace{}\IdrisFunction{Rel}\KatlaSpace{}\IdrisBound{A}
\KatlaSpace{}\KatlaSpace{}\IdrisFunction{reflexive}\KatlaSpace{}\IdrisKeyword{:}\KatlaSpace{}\IdrisKeyword{(}\IdrisBound{x}\KatlaSpace{}\KatlaSpace{}\KatlaSpace{}\KatlaSpace{}\KatlaSpace{}\KatlaSpace{}\KatlaSpace{}\IdrisKeyword{:}\KatlaSpace{}\IdrisBound{A}\IdrisKeyword{)}\KatlaSpace{}\IdrisKeyword{->}\KatlaSpace{}\IdrisBound{relation}\KatlaSpace{}\IdrisBound{x}\KatlaSpace{}\IdrisBound{x}
\KatlaSpace{}\KatlaSpace{}\IdrisFunction{symmetric}\KatlaSpace{}\IdrisKeyword{:}\KatlaSpace{}\IdrisKeyword{(}\IdrisBound{x}\IdrisKeyword{,}\KatlaSpace{}\IdrisBound{y}\KatlaSpace{}\KatlaSpace{}\KatlaSpace{}\KatlaSpace{}\IdrisKeyword{:}\KatlaSpace{}\IdrisBound{A}\IdrisKeyword{)}\KatlaSpace{}\IdrisKeyword{->}\KatlaSpace{}\IdrisBound{relation}\KatlaSpace{}\IdrisBound{x}\KatlaSpace{}\IdrisBound{y}\KatlaSpace{}\IdrisKeyword{->}\KatlaSpace{}\IdrisBound{relation}\KatlaSpace{}\IdrisBound{y}\KatlaSpace{}\IdrisBound{x}
\KatlaSpace{}\KatlaSpace{}\IdrisFunction{transitive}\IdrisKeyword{:}\KatlaSpace{}\IdrisKeyword{(}\IdrisBound{x}\IdrisKeyword{,}\KatlaSpace{}\IdrisBound{y}\IdrisKeyword{,}\KatlaSpace{}\IdrisBound{z}\KatlaSpace{}\IdrisKeyword{:}\KatlaSpace{}\IdrisBound{A}\IdrisKeyword{)}\KatlaSpace{}\IdrisKeyword{->}\KatlaSpace{}\IdrisBound{relation}\KatlaSpace{}\IdrisBound{x}\KatlaSpace{}\IdrisBound{y}\KatlaSpace{}\IdrisKeyword{->}\KatlaSpace{}\IdrisBound{relation}\KatlaSpace{}\IdrisBound{y}\KatlaSpace{}\IdrisBound{z}
\KatlaSpace{}\KatlaSpace{}\KatlaSpace{}\KatlaSpace{}\KatlaSpace{}\KatlaSpace{}\KatlaSpace{}\KatlaSpace{}\KatlaSpace{}\KatlaSpace{}\KatlaSpace{}\KatlaSpace{}\KatlaSpace{}\KatlaSpace{}\KatlaSpace{}\KatlaSpace{}\KatlaSpace{}\KatlaSpace{}\KatlaSpace{}\KatlaSpace{}\KatlaSpace{}\KatlaSpace{}\KatlaSpace{}\KatlaSpace{}\KatlaSpace{}\KatlaSpace{}\KatlaSpace{}\KatlaSpace{}\IdrisKeyword{->}\KatlaSpace{}\IdrisBound{relation}\KatlaSpace{}\IdrisBound{x}\KatlaSpace{}\IdrisBound{z}
\end{SaveVerbatim}
\newcommand\SetoidSugar[1][]{\UseVerbatim[#1]{SetoidSugar}}
\begin{SaveVerbatim}[commandchars=\\\{\}]{SetoidSugar}
\IdrisKeyword{data}\KatlaSpace{}\IdrisType{Setoid}\KatlaSpace{}\IdrisKeyword{:}\KatlaSpace{}\IdrisType{Type}\KatlaSpace{}\IdrisKeyword{where}
\KatlaSpace{}\KatlaSpace{}\IdrisData{MkSetoid}\KatlaSpace{}\IdrisKeyword{:}\KatlaSpace{}\IdrisKeyword{(0}\KatlaSpace{}\IdrisBound{U}\KatlaSpace{}\IdrisKeyword{:}\KatlaSpace{}\IdrisType{Type}\IdrisKeyword{)}\KatlaSpace{}\IdrisKeyword{->}
\KatlaSpace{}\KatlaSpace{}\KatlaSpace{}\KatlaSpace{}\IdrisKeyword{(}\IdrisBound{equivalence}
\KatlaSpace{}\KatlaSpace{}\KatlaSpace{}\KatlaSpace{}\KatlaSpace{}\KatlaSpace{}\IdrisKeyword{:}\KatlaSpace{}\IdrisType{Equivalence}\KatlaSpace{}\IdrisBound{U}\IdrisKeyword{)}\KatlaSpace{}\IdrisKeyword{->}\KatlaSpace{}\IdrisType{Setoid}

\end{SaveVerbatim}
\newcommand\SetoidCarrierSugar[1][]{\UseVerbatim[#1]{SetoidCarrierSugar}}
\begin{SaveVerbatim}[commandchars=\\\{\}]{SetoidCarrierSugar}
\IdrisKeyword{0}
\IdrisFunction{U}\KatlaSpace{}\IdrisKeyword{:}\KatlaSpace{}\IdrisType{Setoid}\KatlaSpace{}\IdrisKeyword{->}\KatlaSpace{}\IdrisType{Type}
\IdrisFunction{U}\KatlaSpace{}\IdrisKeyword{(}\IdrisData{MkSetoid}\KatlaSpace{}\IdrisBound{x}\KatlaSpace{}\IdrisKeyword{_)}\KatlaSpace{}\IdrisKeyword{=}\KatlaSpace{}\IdrisBound{x}
\end{SaveVerbatim}
\newcommand\SetoidRelationSugar[1][]{\UseVerbatim[#1]{SetoidRelationSugar}}
\begin{SaveVerbatim}[commandchars=\\\{\}]{SetoidRelationSugar}
\IdrisFunction{equivalence}\KatlaSpace{}\IdrisKeyword{:}\KatlaSpace{}\IdrisKeyword{(}\IdrisBound{s}\KatlaSpace{}\IdrisKeyword{:}\KatlaSpace{}\IdrisType{Setoid}\IdrisKeyword{)}\KatlaSpace{}\IdrisKeyword{->}
\KatlaSpace{}\KatlaSpace{}\KatlaSpace{}\KatlaSpace{}\KatlaSpace{}\KatlaSpace{}\KatlaSpace{}\KatlaSpace{}\KatlaSpace{}\KatlaSpace{}\KatlaSpace{}\KatlaSpace{}\KatlaSpace{}\KatlaSpace{}\IdrisType{Equivalence}\KatlaSpace{}\IdrisKeyword{(}\IdrisFunction{U}\KatlaSpace{}\IdrisBound{s}\IdrisKeyword{)}
\IdrisFunction{equivalence}\KatlaSpace{}\IdrisKeyword{(}\IdrisData{MkSetoid}\KatlaSpace{}\IdrisKeyword{_}\KatlaSpace{}\IdrisBound{y}\IdrisKeyword{)}\KatlaSpace{}\IdrisKeyword{=}\KatlaSpace{}\IdrisBound{y}
\end{SaveVerbatim}

\begin{SaveVerbatim}[commandchars=\\\{\}]{SetoidCarrierSugarPostfix}
\IdrisKeyword{0}
\IdrisFunction{(.U)}\KatlaSpace{}\IdrisKeyword{:}\KatlaSpace{}\IdrisType{Setoid}\KatlaSpace{}\IdrisKeyword{->}\KatlaSpace{}\IdrisType{Type}
\IdrisKeyword{(}\IdrisData{MkSetoid}\KatlaSpace{}\IdrisBound{x}\KatlaSpace{}\IdrisKeyword{_)}\IdrisFunction{.U}\KatlaSpace{}\IdrisKeyword{=}\KatlaSpace{}\IdrisBound{x}
\end{SaveVerbatim}

\begin{SaveVerbatim}[commandchars=\\\{\}]{SetoidRelationSugarPostfix}
\IdrisFunction{(.equivalence)}\KatlaSpace{}\IdrisKeyword{:}\KatlaSpace{}\IdrisKeyword{(}\IdrisBound{s}\KatlaSpace{}\IdrisKeyword{:}\KatlaSpace{}\IdrisType{Setoid}\IdrisKeyword{)}\KatlaSpace{}\IdrisKeyword{->}
\KatlaSpace{}\KatlaSpace{}\IdrisType{Equivalence}\KatlaSpace{}\IdrisBound{s}\IdrisFunction{.U}
\IdrisKeyword{(}\IdrisData{MkSetoid}\KatlaSpace{}\IdrisKeyword{_}\KatlaSpace{}\IdrisBound{y}\IdrisKeyword{)}\IdrisFunction{.equivalence}\KatlaSpace{}\IdrisKeyword{=}\KatlaSpace{}\IdrisBound{y}
\end{SaveVerbatim}
\newcommand\QuotientDef[1][]{\UseVerbatim[#1]{QuotientDef}}
\begin{SaveVerbatim}[commandchars=\\\{\}]{QuotientDef}
\IdrisFunction{Quotient}\KatlaSpace{}\IdrisKeyword{:}\KatlaSpace{}\IdrisKeyword{(}\IdrisBound{b}\KatlaSpace{}\IdrisKeyword{:}\KatlaSpace{}\IdrisType{Setoid}\IdrisKeyword{)}\KatlaSpace{}\IdrisKeyword{->}\KatlaSpace{}\IdrisKeyword{(}\IdrisBound{a}\KatlaSpace{}\IdrisKeyword{->}\KatlaSpace{}\IdrisFunction{U}\KatlaSpace{}\IdrisBound{b}\IdrisKeyword{)}
\KatlaSpace{}\KatlaSpace{}\IdrisKeyword{->}\KatlaSpace{}\IdrisType{Setoid}
\IdrisFunction{Quotient}\KatlaSpace{}\IdrisBound{b}\KatlaSpace{}\IdrisBound{q}\KatlaSpace{}\IdrisKeyword{=}\KatlaSpace{}\IdrisData{MkSetoid}\KatlaSpace{}\IdrisBound{a}\KatlaSpace{}$
\KatlaSpace{}\KatlaSpace{}\IdrisKeyword{let}\KatlaSpace{}\IdrisKeyword{0}\KatlaSpace{}\IdrisFunction{relation}\KatlaSpace{}\IdrisKeyword{:}\KatlaSpace{}\IdrisBound{a}\KatlaSpace{}\IdrisKeyword{->}\KatlaSpace{}\IdrisBound{a}\KatlaSpace{}\IdrisKeyword{->}\KatlaSpace{}\IdrisType{Type}
\KatlaSpace{}\KatlaSpace{}\KatlaSpace{}\KatlaSpace{}\KatlaSpace{}\KatlaSpace{}\IdrisFunction{relation}\KatlaSpace{}\IdrisBound{x}\KatlaSpace{}\IdrisBound{y}\KatlaSpace{}\IdrisKeyword{=}
\KatlaSpace{}\KatlaSpace{}\KatlaSpace{}\KatlaSpace{}\KatlaSpace{}\KatlaSpace{}\KatlaSpace{}\KatlaSpace{}\IdrisBound{b}\IdrisFunction{.equivalence.relation}\KatlaSpace{}\IdrisKeyword{(}\IdrisBound{q}\KatlaSpace{}\IdrisBound{x}\IdrisKeyword{)}\KatlaSpace{}\IdrisKeyword{(}\IdrisBound{q}\KatlaSpace{}\IdrisBound{y}\IdrisKeyword{)}
\KatlaSpace{}\KatlaSpace{}\IdrisKeyword{in}\KatlaSpace{}\IdrisData{MkEquivalence}
\KatlaSpace{}\KatlaSpace{}\KatlaSpace{}\KatlaSpace{}\IdrisKeyword{\{}\KatlaSpace{}\IdrisBound{relation}\KatlaSpace{}\IdrisKeyword{=}\KatlaSpace{}\IdrisFunction{relation}
\KatlaSpace{}\KatlaSpace{}\KatlaSpace{}\KatlaSpace{}\IdrisKeyword{,}\KatlaSpace{}\IdrisBound{reflexive}\KatlaSpace{}\KatlaSpace{}\IdrisKeyword{=}\KatlaSpace{}\IdrisKeyword{\textbackslash{}}\IdrisBound{x}\KatlaSpace{}\KatlaSpace{}\KatlaSpace{}\KatlaSpace{}\KatlaSpace{}\KatlaSpace{}\IdrisKeyword{=>}
\KatlaSpace{}\KatlaSpace{}\KatlaSpace{}\KatlaSpace{}\KatlaSpace{}\KatlaSpace{}\KatlaSpace{}\KatlaSpace{}\IdrisBound{b}\IdrisFunction{.equivalence.reflexive}\KatlaSpace{}\KatlaSpace{}\IdrisKeyword{(}\IdrisBound{q}\KatlaSpace{}\IdrisBound{x}\IdrisKeyword{)}
\KatlaSpace{}\KatlaSpace{}\KatlaSpace{}\KatlaSpace{}\IdrisKeyword{,}\KatlaSpace{}\IdrisBound{symmetric}\KatlaSpace{}\KatlaSpace{}\IdrisKeyword{=}\KatlaSpace{}\IdrisKeyword{\textbackslash{}}\IdrisBound{x}\IdrisKeyword{,}\IdrisBound{y}\KatlaSpace{}\KatlaSpace{}\KatlaSpace{}\KatlaSpace{}\IdrisKeyword{=>}
\KatlaSpace{}\KatlaSpace{}\KatlaSpace{}\KatlaSpace{}\KatlaSpace{}\KatlaSpace{}\KatlaSpace{}\KatlaSpace{}\IdrisBound{b}\IdrisFunction{.equivalence.symmetric}\KatlaSpace{}\KatlaSpace{}\IdrisKeyword{(}\IdrisBound{q}\KatlaSpace{}\IdrisBound{x}\IdrisKeyword{)}\KatlaSpace{}\IdrisKeyword{(}\IdrisBound{q}\KatlaSpace{}\IdrisBound{y}\IdrisKeyword{)}
\KatlaSpace{}\KatlaSpace{}\KatlaSpace{}\KatlaSpace{}\IdrisKeyword{,}\KatlaSpace{}\IdrisBound{transitive}\KatlaSpace{}\IdrisKeyword{=}\KatlaSpace{}\IdrisKeyword{\textbackslash{}}\IdrisBound{x}\IdrisKeyword{,}\IdrisBound{y}\IdrisKeyword{,}\IdrisBound{z}\KatlaSpace{}\KatlaSpace{}\IdrisKeyword{=>}
\KatlaSpace{}\KatlaSpace{}\KatlaSpace{}\KatlaSpace{}\KatlaSpace{}\KatlaSpace{}\KatlaSpace{}\KatlaSpace{}\IdrisBound{b}\IdrisFunction{.equivalence.transitive}
\KatlaSpace{}\KatlaSpace{}\KatlaSpace{}\KatlaSpace{}\KatlaSpace{}\KatlaSpace{}\KatlaSpace{}\KatlaSpace{}\KatlaSpace{}\KatlaSpace{}\IdrisKeyword{(}\IdrisBound{q}\KatlaSpace{}\IdrisBound{x}\IdrisKeyword{)}\KatlaSpace{}\IdrisKeyword{(}\IdrisBound{q}\KatlaSpace{}\IdrisBound{y}\IdrisKeyword{)}\KatlaSpace{}\IdrisKeyword{(}\IdrisBound{q}\KatlaSpace{}\IdrisBound{z}\IdrisKeyword{)}
\KatlaSpace{}\KatlaSpace{}\KatlaSpace{}\KatlaSpace{}\IdrisKeyword{\}}
\end{SaveVerbatim}
\newcommand\QuotientingMap[1][]{\UseVerb[#1]{QuotientingMap}}
\begin{SaveVerbatim}[commandchars=\\\{\}]{QuotientingMap}
\IdrisBound{q}
\end{SaveVerbatim}
\newcommand\QuotientName[1][]{\UseVerb[#1]{QuotientName}}
\begin{SaveVerbatim}[commandchars=\\\{\}]{QuotientName}
\IdrisFunction{Quotient}
\end{SaveVerbatim}
\newcommand\ImplicitArg[1][]{\UseVerb[#1]{ImplicitArg}}
\begin{SaveVerbatim}[commandchars=\\\{\}]{ImplicitArg}
\IdrisBound{a}
\end{SaveVerbatim}
\newcommand\FunSpaceDef[1][]{\UseVerbatim[#1]{FunSpaceDef}}
\begin{SaveVerbatim}[commandchars=\\\{\}]{FunSpaceDef}
\IdrisFunction{(~~>)}\KatlaSpace{}\IdrisKeyword{:}\KatlaSpace{}\IdrisKeyword{(}\IdrisBound{a}\IdrisKeyword{,}\IdrisBound{b}\KatlaSpace{}\IdrisKeyword{:}\KatlaSpace{}\IdrisType{Setoid}\IdrisKeyword{)}\KatlaSpace{}\IdrisKeyword{->}\KatlaSpace{}\IdrisType{Setoid}
\IdrisFunction{(~~>)}\KatlaSpace{}\IdrisBound{a}\KatlaSpace{}\IdrisBound{b}\KatlaSpace{}\IdrisKeyword{=}\KatlaSpace{}\IdrisData{MkSetoid}\KatlaSpace{}\IdrisKeyword{(}\IdrisBound{a}\KatlaSpace{}\IdrisType{~>}\KatlaSpace{}\IdrisBound{b}\IdrisKeyword{)}\KatlaSpace{}$
\KatlaSpace{}\KatlaSpace{}\IdrisKeyword{let}\KatlaSpace{}\IdrisKeyword{0}\KatlaSpace{}\IdrisFunction{relation}\KatlaSpace{}\IdrisKeyword{:}\KatlaSpace{}\IdrisKeyword{(}\IdrisBound{f}\IdrisKeyword{,}\KatlaSpace{}\IdrisBound{g}\KatlaSpace{}\IdrisKeyword{:}\KatlaSpace{}\IdrisBound{a}\KatlaSpace{}\IdrisType{~>}\KatlaSpace{}\IdrisBound{b}\IdrisKeyword{)}\KatlaSpace{}\IdrisKeyword{->}\KatlaSpace{}\IdrisType{Type}
\KatlaSpace{}\KatlaSpace{}\KatlaSpace{}\KatlaSpace{}\KatlaSpace{}\KatlaSpace{}\IdrisFunction{relation}\KatlaSpace{}\IdrisBound{f}\KatlaSpace{}\IdrisBound{g}\KatlaSpace{}\IdrisKeyword{=}\KatlaSpace{}\IdrisKeyword{(}\IdrisBound{x}\KatlaSpace{}\IdrisKeyword{:}\KatlaSpace{}\IdrisFunction{U}\KatlaSpace{}\IdrisBound{a}\IdrisKeyword{)}\KatlaSpace{}\IdrisKeyword{->}
\KatlaSpace{}\KatlaSpace{}\KatlaSpace{}\KatlaSpace{}\KatlaSpace{}\KatlaSpace{}\KatlaSpace{}\KatlaSpace{}\IdrisBound{b}\IdrisFunction{.equivalence.relation}\KatlaSpace{}\IdrisKeyword{(}\IdrisBound{f}\IdrisFunction{.H}\KatlaSpace{}\IdrisBound{x}\IdrisKeyword{)}\KatlaSpace{}\IdrisKeyword{(}\IdrisBound{g}\IdrisFunction{.H}\KatlaSpace{}\IdrisBound{x}\IdrisKeyword{)}
\KatlaSpace{}\KatlaSpace{}\IdrisKeyword{in}\KatlaSpace{}\IdrisData{MkEquivalence}
\KatlaSpace{}\KatlaSpace{}\IdrisKeyword{\{}\KatlaSpace{}\IdrisFunction{relation}
\KatlaSpace{}\KatlaSpace{}\IdrisKeyword{,}\KatlaSpace{}\IdrisBound{reflexive}\KatlaSpace{}\IdrisKeyword{=}\KatlaSpace{}\IdrisKeyword{\textbackslash{}}\IdrisBound{f}\IdrisKeyword{,}\IdrisBound{v}\KatlaSpace{}\KatlaSpace{}\KatlaSpace{}\KatlaSpace{}\KatlaSpace{}\KatlaSpace{}\KatlaSpace{}\IdrisKeyword{=>}
\KatlaSpace{}\KatlaSpace{}\KatlaSpace{}\KatlaSpace{}\KatlaSpace{}\KatlaSpace{}\IdrisBound{b}\IdrisFunction{.equivalence.reflexive}\KatlaSpace{}\IdrisKeyword{(}\IdrisBound{f}\IdrisFunction{.H}\KatlaSpace{}\IdrisBound{v}\IdrisKeyword{)}
\KatlaSpace{}\KatlaSpace{}\IdrisKeyword{,}\KatlaSpace{}\IdrisBound{symmetric}\KatlaSpace{}\IdrisKeyword{=}\KatlaSpace{}\IdrisKeyword{\textbackslash{}}\IdrisBound{f}\IdrisKeyword{,}\IdrisBound{g}\IdrisKeyword{,}\IdrisBound{prf}\IdrisKeyword{,}\IdrisBound{w}\KatlaSpace{}\IdrisKeyword{=>}
\KatlaSpace{}\KatlaSpace{}\KatlaSpace{}\KatlaSpace{}\KatlaSpace{}\KatlaSpace{}\IdrisBound{b}\IdrisFunction{.equivalence.symmetric}\KatlaSpace{}\IdrisKeyword{_}\KatlaSpace{}\IdrisKeyword{_}\KatlaSpace{}\IdrisKeyword{(}\IdrisBound{prf}\KatlaSpace{}\IdrisBound{w}\IdrisKeyword{)}
\KatlaSpace{}\KatlaSpace{}\IdrisKeyword{,}\KatlaSpace{}\IdrisBound{transitive}\KatlaSpace{}\IdrisKeyword{=}\KatlaSpace{}\IdrisKeyword{\textbackslash{}}\IdrisBound{f}\IdrisKeyword{,}\IdrisBound{g}\IdrisKeyword{,}\IdrisBound{h}\IdrisKeyword{,}\IdrisBound{f_eq_g}\IdrisKeyword{,}\KatlaSpace{}\IdrisBound{g_eq_h}\IdrisKeyword{,}\KatlaSpace{}\IdrisBound{q}\KatlaSpace{}\IdrisKeyword{=>}
\KatlaSpace{}\KatlaSpace{}\KatlaSpace{}\KatlaSpace{}\KatlaSpace{}\KatlaSpace{}\IdrisBound{b}\IdrisFunction{.equivalence.transitive}\KatlaSpace{}\IdrisKeyword{_}\KatlaSpace{}\IdrisKeyword{_}\KatlaSpace{}\IdrisKeyword{_}
\KatlaSpace{}\KatlaSpace{}\KatlaSpace{}\KatlaSpace{}\KatlaSpace{}\KatlaSpace{}\KatlaSpace{}\KatlaSpace{}\IdrisKeyword{(}\IdrisBound{f_eq_g}\KatlaSpace{}\IdrisBound{q}\IdrisKeyword{)}\KatlaSpace{}\IdrisKeyword{(}\IdrisBound{g_eq_h}\KatlaSpace{}\IdrisBound{q}\IdrisKeyword{)}
\KatlaSpace{}\KatlaSpace{}\IdrisKeyword{\}}
\end{SaveVerbatim}

\begin{SaveVerbatim}[commandchars=\\\{\}]{implicitArg}
\IdrisKeyword{(}\IdrisBound{a}\KatlaSpace{}\IdrisKeyword{->}\KatlaSpace{}\IdrisFunction{U}\KatlaSpace{}\IdrisBound{b}\IdrisKeyword{)}\KatlaSpace{}\IdrisKeyword{-}
\end{SaveVerbatim}
\newcommand\VectEquality[1][]{\UseVerbatim[#1]{VectEquality}}
\begin{SaveVerbatim}[commandchars=\\\{\}]{VectEquality}
\IdrisKeyword{0}\KatlaSpace{}\IdrisFunction{(.VectEquality)}\IdrisKeyword{:}\IdrisKeyword{(}\IdrisBound{a}\KatlaSpace{}\IdrisKeyword{:}\KatlaSpace{}\IdrisType{Setoid}\IdrisKeyword{)}\KatlaSpace{}\IdrisKeyword{->}\KatlaSpace{}\IdrisFunction{Rel}\KatlaSpace{}\IdrisKeyword{(}\IdrisType{Vect}\KatlaSpace{}\IdrisBound{n}\KatlaSpace{}\IdrisKeyword{(}\IdrisFunction{U}\KatlaSpace{}\IdrisBound{a}\IdrisKeyword{))}
\IdrisBound{a}\IdrisFunction{.VectEquality}\KatlaSpace{}\IdrisBound{xs}\KatlaSpace{}\IdrisBound{ys}\KatlaSpace{}\IdrisKeyword{=}\KatlaSpace{}\IdrisKeyword{(}\IdrisBound{i}\KatlaSpace{}\IdrisKeyword{:}\KatlaSpace{}\IdrisType{Fin}\KatlaSpace{}\IdrisBound{n}\IdrisKeyword{)}\KatlaSpace{}\IdrisKeyword{->}
\KatlaSpace{}\KatlaSpace{}\IdrisBound{a}\IdrisFunction{.equivalence.relation}\KatlaSpace{}\IdrisKeyword{(}\IdrisFunction{index}\KatlaSpace{}\IdrisBound{i}\KatlaSpace{}\IdrisBound{xs}\IdrisKeyword{)}\KatlaSpace{}\IdrisKeyword{(}\IdrisFunction{index}\KatlaSpace{}\IdrisBound{i}\KatlaSpace{}\IdrisBound{ys}\IdrisKeyword{)}
\end{SaveVerbatim}
\newcommand\VectSetoid[1][]{\UseVerbatim[#1]{VectSetoid}}
\begin{SaveVerbatim}[commandchars=\\\{\}]{VectSetoid}
\IdrisFunction{VectSetoid}\KatlaSpace{}\IdrisKeyword{:}\KatlaSpace{}\IdrisKeyword{(}\IdrisBound{n}\KatlaSpace{}\IdrisKeyword{:}\KatlaSpace{}\IdrisType{Nat}\IdrisKeyword{)}\KatlaSpace{}\IdrisKeyword{->}\KatlaSpace{}\IdrisKeyword{(}\IdrisBound{a}\KatlaSpace{}\IdrisKeyword{:}\KatlaSpace{}\IdrisType{Setoid}\IdrisKeyword{)}\KatlaSpace{}\IdrisKeyword{->}\KatlaSpace{}\IdrisType{Setoid}
\IdrisFunction{VectSetoid}\KatlaSpace{}\IdrisBound{n}\KatlaSpace{}\IdrisBound{a}\KatlaSpace{}\IdrisKeyword{=}\KatlaSpace{}\IdrisData{MkSetoid}\KatlaSpace{}\IdrisKeyword{(}\IdrisType{Vect}\KatlaSpace{}\IdrisBound{n}\KatlaSpace{}\IdrisKeyword{(}\IdrisFunction{U}\KatlaSpace{}\IdrisBound{a}\IdrisKeyword{))}
\KatlaSpace{}\KatlaSpace{}$\KatlaSpace{}\IdrisData{MkEquivalence}
\KatlaSpace{}\KatlaSpace{}\IdrisKeyword{\{}\KatlaSpace{}\IdrisBound{relation}\KatlaSpace{}\KatlaSpace{}\KatlaSpace{}\IdrisKeyword{=}\KatlaSpace{}\IdrisFunction{(.VectEquality)}\KatlaSpace{}\IdrisBound{a}
\KatlaSpace{}\KatlaSpace{}\IdrisKeyword{,}\KatlaSpace{}\IdrisBound{reflexive}\KatlaSpace{}\KatlaSpace{}\IdrisKeyword{=}\KatlaSpace{}\IdrisKeyword{\textbackslash{}}\IdrisBound{xs}\KatlaSpace{}\KatlaSpace{}\KatlaSpace{}\KatlaSpace{}\KatlaSpace{}\KatlaSpace{}\KatlaSpace{}\KatlaSpace{}\KatlaSpace{}\KatlaSpace{}\KatlaSpace{}\KatlaSpace{}\KatlaSpace{}\KatlaSpace{}\KatlaSpace{}\KatlaSpace{}\KatlaSpace{}\KatlaSpace{}\KatlaSpace{}\KatlaSpace{}\IdrisKeyword{,}\KatlaSpace{}\IdrisBound{i}\KatlaSpace{}\IdrisKeyword{=>}
\KatlaSpace{}\KatlaSpace{}\KatlaSpace{}\KatlaSpace{}\KatlaSpace{}\KatlaSpace{}\IdrisBound{a}\IdrisFunction{.equivalence.reflexive}\KatlaSpace{}\KatlaSpace{}\IdrisKeyword{_}
\KatlaSpace{}\KatlaSpace{}\IdrisKeyword{,}\KatlaSpace{}\IdrisBound{symmetric}\KatlaSpace{}\KatlaSpace{}\IdrisKeyword{=}\KatlaSpace{}\IdrisKeyword{\textbackslash{}}\IdrisBound{xs}\IdrisKeyword{,}\IdrisBound{ys}\IdrisKeyword{,}\KatlaSpace{}\IdrisBound{prf}\KatlaSpace{}\KatlaSpace{}\KatlaSpace{}\KatlaSpace{}\KatlaSpace{}\KatlaSpace{}\KatlaSpace{}\KatlaSpace{}\KatlaSpace{}\KatlaSpace{}\KatlaSpace{}\KatlaSpace{}\IdrisKeyword{,}\KatlaSpace{}\IdrisBound{i}\KatlaSpace{}\IdrisKeyword{=>}
\KatlaSpace{}\KatlaSpace{}\KatlaSpace{}\KatlaSpace{}\KatlaSpace{}\KatlaSpace{}\IdrisBound{a}\IdrisFunction{.equivalence.symmetric}\KatlaSpace{}\KatlaSpace{}\IdrisKeyword{_}\KatlaSpace{}\IdrisKeyword{_}\KatlaSpace{}\KatlaSpace{}\KatlaSpace{}\IdrisKeyword{(}\IdrisBound{prf}\KatlaSpace{}\KatlaSpace{}\IdrisBound{i}\IdrisKeyword{)}
\KatlaSpace{}\KatlaSpace{}\IdrisKeyword{,}\KatlaSpace{}\IdrisBound{transitive}\KatlaSpace{}\IdrisKeyword{=}\KatlaSpace{}\IdrisKeyword{\textbackslash{}}\IdrisBound{xs}\IdrisKeyword{,}\KatlaSpace{}\IdrisBound{ys}\IdrisKeyword{,}\KatlaSpace{}\IdrisBound{zs}\IdrisKeyword{,}\KatlaSpace{}\IdrisBound{prf1}\IdrisKeyword{,}\KatlaSpace{}\IdrisBound{prf2}\IdrisKeyword{,}\KatlaSpace{}\IdrisBound{i}\KatlaSpace{}\IdrisKeyword{=>}
\KatlaSpace{}\KatlaSpace{}\KatlaSpace{}\KatlaSpace{}\KatlaSpace{}\KatlaSpace{}\IdrisBound{a}\IdrisFunction{.equivalence.transitive}\KatlaSpace{}\IdrisKeyword{_}\KatlaSpace{}\IdrisKeyword{_}\KatlaSpace{}\IdrisKeyword{_}\KatlaSpace{}\IdrisKeyword{(}\IdrisBound{prf1}\KatlaSpace{}\IdrisBound{i}\IdrisKeyword{)}\IdrisKeyword{(}\IdrisBound{prf2}\KatlaSpace{}\IdrisBound{i}\IdrisKeyword{)}
\KatlaSpace{}\KatlaSpace{}\IdrisKeyword{\}}
\end{SaveVerbatim}
\newcommand\VectMap[1][]{\UseVerbatim[#1]{VectMap}}
\begin{SaveVerbatim}[commandchars=\\\{\}]{VectMap}
\IdrisFunction{VectMap}\KatlaSpace{}\IdrisKeyword{:}\KatlaSpace{}\IdrisKeyword{\{}\IdrisBound{a}\IdrisKeyword{,}\KatlaSpace{}\IdrisBound{b}\KatlaSpace{}\IdrisKeyword{:}\KatlaSpace{}\IdrisType{Setoid}\IdrisKeyword{\}}\KatlaSpace{}\IdrisKeyword{->}\KatlaSpace{}\IdrisKeyword{(}\IdrisBound{a}\KatlaSpace{}\IdrisFunction{~~>}\KatlaSpace{}\IdrisBound{b}\IdrisKeyword{)}\KatlaSpace{}\IdrisType{~>}
\KatlaSpace{}\KatlaSpace{}\IdrisKeyword{(}\IdrisFunction{VectSetoid}\KatlaSpace{}\IdrisBound{n}\KatlaSpace{}\IdrisBound{a}\KatlaSpace{}\IdrisFunction{~~>}\KatlaSpace{}\IdrisFunction{VectSetoid}\KatlaSpace{}\IdrisBound{n}\KatlaSpace{}\IdrisBound{b}\IdrisKeyword{)}
\IdrisFunction{VectMap}\KatlaSpace{}\IdrisKeyword{=}\KatlaSpace{}\IdrisData{MkSetoidHomomorphism}
\KatlaSpace{}\KatlaSpace{}\IdrisKeyword{(\textbackslash{}}\IdrisBound{f}\KatlaSpace{}\IdrisKeyword{=>}\KatlaSpace{}\IdrisData{MkSetoidHomomorphism}
\KatlaSpace{}\KatlaSpace{}\KatlaSpace{}\KatlaSpace{}\IdrisKeyword{(\textbackslash{}}\IdrisBound{xs}\KatlaSpace{}\IdrisKeyword{=>}\KatlaSpace{}\IdrisFunction{map}\KatlaSpace{}\IdrisBound{f}\IdrisFunction{.H}\KatlaSpace{}\IdrisBound{xs}\IdrisKeyword{)}
\KatlaSpace{}\KatlaSpace{}\KatlaSpace{}\KatlaSpace{}$\KatlaSpace{}\IdrisKeyword{\textbackslash{}}\IdrisBound{xs}\IdrisKeyword{,}\KatlaSpace{}\IdrisBound{ys}\IdrisKeyword{,}\KatlaSpace{}\IdrisBound{prf}\IdrisKeyword{,}\KatlaSpace{}\IdrisBound{i}\KatlaSpace{}\KatlaSpace{}\IdrisKeyword{=>}\KatlaSpace{}\IdrisFunction{CalcWith}\KatlaSpace{}\IdrisBound{b}\KatlaSpace{}$
\KatlaSpace{}\KatlaSpace{}\KatlaSpace{}\KatlaSpace{}\IdrisData{|~}\KatlaSpace{}\IdrisFunction{index}\KatlaSpace{}\IdrisBound{i}\KatlaSpace{}\IdrisKeyword{(}\IdrisFunction{map}\KatlaSpace{}\IdrisBound{f}\IdrisFunction{.H}\KatlaSpace{}\IdrisBound{xs}\IdrisKeyword{)}
\KatlaSpace{}\KatlaSpace{}\KatlaSpace{}\KatlaSpace{}\IdrisData{~~}\KatlaSpace{}\IdrisBound{f}\IdrisFunction{.H}\IdrisKeyword{(}\IdrisFunction{index}\KatlaSpace{}\IdrisBound{i}\KatlaSpace{}\IdrisBound{xs}\IdrisKeyword{)}
\KatlaSpace{}\KatlaSpace{}\KatlaSpace{}\KatlaSpace{}\KatlaSpace{}\KatlaSpace{}\KatlaSpace{}\KatlaSpace{}\KatlaSpace{}\KatlaSpace{}\KatlaSpace{}\KatlaSpace{}\KatlaSpace{}\KatlaSpace{}\KatlaSpace{}\KatlaSpace{}\KatlaSpace{}\KatlaSpace{}\KatlaSpace{}\KatlaSpace{}\KatlaSpace{}\KatlaSpace{}\KatlaSpace{}\IdrisFunction{.=.}\IdrisKeyword{(}\IdrisFunction{indexNaturality}\KatlaSpace{}\IdrisKeyword{_}\KatlaSpace{}\IdrisKeyword{_}\KatlaSpace{}\IdrisKeyword{_)}
\KatlaSpace{}\KatlaSpace{}\KatlaSpace{}\KatlaSpace{}\IdrisData{~~}\KatlaSpace{}\IdrisBound{f}\IdrisFunction{.H}\IdrisKeyword{(}\IdrisFunction{index}\KatlaSpace{}\IdrisBound{i}\KatlaSpace{}\IdrisBound{ys}\IdrisKeyword{)}\IdrisData{...}\IdrisKeyword{(}\IdrisBound{f}\IdrisFunction{.homomorphic}\KatlaSpace{}\IdrisKeyword{_}\KatlaSpace{}\IdrisKeyword{_}\KatlaSpace{}$\KatlaSpace{}\IdrisBound{prf}\KatlaSpace{}\IdrisBound{i}\IdrisKeyword{)}
\KatlaSpace{}\KatlaSpace{}\KatlaSpace{}\KatlaSpace{}\IdrisData{~~}\KatlaSpace{}\IdrisFunction{index}\KatlaSpace{}\IdrisBound{i}\KatlaSpace{}\IdrisKeyword{(}\IdrisFunction{map}\KatlaSpace{}\IdrisBound{f}\IdrisFunction{.H}\KatlaSpace{}\IdrisBound{ys}\IdrisKeyword{)}
\KatlaSpace{}\KatlaSpace{}\KatlaSpace{}\KatlaSpace{}\KatlaSpace{}\KatlaSpace{}\KatlaSpace{}\KatlaSpace{}\KatlaSpace{}\KatlaSpace{}\KatlaSpace{}\KatlaSpace{}\KatlaSpace{}\KatlaSpace{}\KatlaSpace{}\KatlaSpace{}\KatlaSpace{}\KatlaSpace{}\KatlaSpace{}\KatlaSpace{}\KatlaSpace{}\KatlaSpace{}\KatlaSpace{}\KatlaSpace{}\KatlaSpace{}\IdrisFunction{.=<}\IdrisKeyword{(}\IdrisFunction{indexNaturality}\KatlaSpace{}\IdrisKeyword{_}\KatlaSpace{}\IdrisKeyword{_}\KatlaSpace{}\IdrisKeyword{_))}
\KatlaSpace{}\KatlaSpace{}$\KatlaSpace{}\IdrisKeyword{\textbackslash{}}\IdrisBound{f}\IdrisKeyword{,}\IdrisBound{g}\IdrisKeyword{,}\IdrisBound{prf}\IdrisKeyword{,}\IdrisBound{xs}\IdrisKeyword{,}\IdrisBound{i}\KatlaSpace{}\IdrisKeyword{=>}\KatlaSpace{}\IdrisFunction{CalcWith}\KatlaSpace{}\IdrisBound{b}\KatlaSpace{}$
\KatlaSpace{}\KatlaSpace{}\IdrisData{|~}\KatlaSpace{}\IdrisFunction{index}\KatlaSpace{}\IdrisBound{i}\KatlaSpace{}\IdrisKeyword{(}\IdrisFunction{map}\KatlaSpace{}\IdrisBound{f}\IdrisFunction{.H}\KatlaSpace{}\IdrisBound{xs}\IdrisKeyword{)}
\KatlaSpace{}\KatlaSpace{}\IdrisData{~~}\KatlaSpace{}\IdrisBound{f}\IdrisFunction{.H}\KatlaSpace{}\IdrisKeyword{(}\IdrisFunction{index}\KatlaSpace{}\IdrisBound{i}\KatlaSpace{}\IdrisBound{xs}\IdrisKeyword{)}\KatlaSpace{}\IdrisFunction{.=.}\IdrisKeyword{(}\IdrisFunction{indexNaturality}\KatlaSpace{}\IdrisKeyword{_}\KatlaSpace{}\IdrisKeyword{_}\KatlaSpace{}\IdrisKeyword{_)}
\KatlaSpace{}\KatlaSpace{}\IdrisData{~~}\KatlaSpace{}\IdrisBound{g}\IdrisFunction{.H}\KatlaSpace{}\IdrisKeyword{(}\IdrisFunction{index}\KatlaSpace{}\IdrisBound{i}\KatlaSpace{}\IdrisBound{xs}\IdrisKeyword{)}\KatlaSpace{}\IdrisData{...}\IdrisKeyword{(}\IdrisBound{prf}\KatlaSpace{}\IdrisKeyword{_)}
\KatlaSpace{}\KatlaSpace{}\IdrisData{~~}\KatlaSpace{}\IdrisFunction{index}\KatlaSpace{}\IdrisBound{i}\KatlaSpace{}\IdrisKeyword{(}\IdrisFunction{map}\KatlaSpace{}\IdrisBound{g}\IdrisFunction{.H}\KatlaSpace{}\IdrisBound{xs}\IdrisKeyword{)}\KatlaSpace{}\IdrisFunction{.=<}\IdrisKeyword{(}\IdrisFunction{indexNaturality}\KatlaSpace{}\IdrisKeyword{_}\KatlaSpace{}\IdrisKeyword{_}\KatlaSpace{}\IdrisKeyword{_)}
\end{SaveVerbatim}
\newcommand\VectMapName[1][]{\UseVerb[#1]{VectMapName}}
\begin{SaveVerbatim}[commandchars=\\\{\}]{VectMapName}
\IdrisFunction{VectMap}
\end{SaveVerbatim}
\newcommand\VectMapF[1][]{\UseVerb[#1]{VectMapF}}
\begin{SaveVerbatim}[commandchars=\\\{\}]{VectMapF}
\IdrisFunction{map}\KatlaSpace{}\IdrisBound{f}\IdrisFunction{.H}
\end{SaveVerbatim}
\newcommand\BasicThoughtBubble[1][]{\UseVerb[#1]{BasicThoughtBubble}}
\begin{SaveVerbatim}[commandchars=\\\{\}]{BasicThoughtBubble}
\IdrisData{(...)}
\end{SaveVerbatim}
\newcommand\EqualThoughtBubble[1][]{\UseVerb[#1]{EqualThoughtBubble}}
\begin{SaveVerbatim}[commandchars=\\\{\}]{EqualThoughtBubble}
\IdrisFunction{(.=.)}
\end{SaveVerbatim}
\newcommand\EqualSymThoughtBubble[1][]{\UseVerb[#1]{EqualSymThoughtBubble}}
\begin{SaveVerbatim}[commandchars=\\\{\}]{EqualSymThoughtBubble}
\IdrisFunction{(.=<)}
\end{SaveVerbatim}

%% file: frexlib-listings.tex
\newcommand\SignatureName[1][]{\UseVerb[#1]{SignatureName}}
\begin{SaveVerbatim}[commandchars=\\\{\}]{SignatureName}
\IdrisType{Signature}
\end{SaveVerbatim}
\newcommand\ArityName[1][]{\UseVerb[#1]{ArityName}}
\begin{SaveVerbatim}[commandchars=\\\{\}]{ArityName}
\IdrisFunction{arity}
\end{SaveVerbatim}
\newcommand\SignaturesPartA[1][]{\UseVerbatim[#1]{SignaturesPartA}}
\begin{SaveVerbatim}[commandchars=\\\{\}]{SignaturesPartA}
\IdrisKeyword{record}\KatlaSpace{}\IdrisType{Signature}\KatlaSpace{}\IdrisKeyword{where}\KatlaNewline{}
\KatlaSpace{}\KatlaSpace{}\IdrisKeyword{constructor}\KatlaSpace{}\IdrisData{MkSignature}\KatlaNewline{}
\end{SaveVerbatim}
\newcommand\SignaturesPartB[1][]{\UseVerbatim[#1]{SignaturesPartB}}
\begin{SaveVerbatim}[commandchars=\\\{\}]{SignaturesPartB}
\KatlaSpace{}\KatlaSpace{}\IdrisFunction{OpWithArity}\KatlaSpace{}\IdrisKeyword{:}\KatlaSpace{}\IdrisType{Nat}\KatlaSpace{}\IdrisKeyword{\KatlaDash{}>}\KatlaSpace{}\IdrisType{Type}\KatlaNewline{}
\end{SaveVerbatim}
\newcommand\OpWithArityName[1][]{\UseVerb[#1]{OpWithArityName}}
\begin{SaveVerbatim}[commandchars=\\\{\}]{OpWithArityName}
\IdrisFunction{OpWithArity}
\end{SaveVerbatim}
\newcommand\OpDecl[1][]{\UseVerbatim[#1]{OpDecl}}
\begin{SaveVerbatim}[commandchars=\\\{\}]{OpDecl}
\IdrisKeyword{record}\KatlaSpace{}\IdrisType{Op}\KatlaSpace{}\IdrisKeyword{(}\IdrisBound{sig}\KatlaSpace{}\IdrisKeyword{:}\KatlaSpace{}\IdrisType{Signature}\IdrisKeyword{)}\KatlaSpace{}\IdrisKeyword{where}\KatlaNewline{}
\KatlaSpace{}\KatlaSpace{}\IdrisKeyword{constructor}\KatlaSpace{}\IdrisData{MkOp}\KatlaNewline{}
\KatlaSpace{}\KatlaSpace{}\IdrisKeyword{\{}\IdrisFunction{arity}\KatlaSpace{}\IdrisKeyword{:}\KatlaSpace{}\IdrisType{Nat}\IdrisKeyword{\}}\KatlaNewline{}
\KatlaSpace{}\KatlaSpace{}\IdrisFunction{snd}\KatlaSpace{}\IdrisKeyword{:}\KatlaSpace{}\IdrisBound{sig}\IdrisFunction{.OpWithArity}\KatlaSpace{}\IdrisBound{arity}\KatlaNewline{}
\end{SaveVerbatim}

\begin{SaveVerbatim}[commandchars=\\\{\}]{MonoidOperation}
\IdrisKeyword{data}\KatlaSpace{}\IdrisType{Operation}\KatlaSpace{}\IdrisKeyword{:}\KatlaSpace{}\IdrisType{Nat}\KatlaSpace{}\IdrisKeyword{\KatlaDash{}>}\KatlaSpace{}\IdrisType{Type}\KatlaSpace{}\IdrisKeyword{where}\KatlaNewline{}
\KatlaSpace{}\KatlaSpace{}\IdrisData{Neutral}\KatlaSpace{}\IdrisKeyword{:}\KatlaSpace{}\IdrisType{Operation}\KatlaSpace{}\IdrisData{0}\KatlaNewline{}
\KatlaSpace{}\KatlaSpace{}\IdrisData{Product}\KatlaSpace{}\IdrisKeyword{:}\KatlaSpace{}\IdrisType{Operation}\KatlaSpace{}\IdrisData{2}\KatlaNewline{}
\end{SaveVerbatim}
\newcommand\MonoidSig[1][]{\UseVerb[#1]{MonoidSig}}
\begin{SaveVerbatim}[commandchars=\\\{\}]{MonoidSig}
\IdrisData{MkSignature}\KatlaSpace{}\IdrisType{Operation}\KatlaNewline{}
\end{SaveVerbatim}
\newcommand\Tuple[1][]{\UseVerbatim[#1]{Tuple}}
\begin{SaveVerbatim}[commandchars=\\\{\}]{Tuple}
\IdrisFunction{(^)}\KatlaSpace{}\IdrisKeyword{:}\KatlaSpace{}\IdrisType{Type}\KatlaSpace{}\IdrisKeyword{\KatlaDash{}>}\KatlaSpace{}\IdrisType{Nat}\KatlaSpace{}\IdrisKeyword{\KatlaDash{}>}\KatlaSpace{}\IdrisType{Type}\KatlaNewline{}
\IdrisFunction{(^)}\KatlaSpace{}\IdrisBound{a}\KatlaSpace{}\IdrisBound{n}\KatlaSpace{}\IdrisKeyword{=}\KatlaSpace{}\IdrisType{Vect}\KatlaSpace{}\IdrisBound{n}\KatlaSpace{}\IdrisBound{a}\KatlaNewline{}
\end{SaveVerbatim}
\newcommand\algebraOverDecl[1][]{\UseVerbatim[#1]{algebraOverDecl}}
\begin{SaveVerbatim}[commandchars=\\\{\}]{algebraOverDecl}
\IdrisFunction{algebraOver}\KatlaSpace{}\IdrisKeyword{:}\KatlaSpace{}\IdrisKeyword{(}\IdrisBound{sig}\KatlaSpace{}\IdrisKeyword{:}\KatlaSpace{}\IdrisType{Signature}\IdrisKeyword{)}\KatlaNewline{}
\KatlaSpace{}\KatlaSpace{}\IdrisKeyword{\KatlaDash{}>}\KatlaSpace{}\IdrisKeyword{(}\IdrisBound{a}\KatlaSpace{}\IdrisKeyword{:}\KatlaSpace{}\IdrisType{Type}\IdrisKeyword{)}\KatlaSpace{}\IdrisKeyword{\KatlaDash{}>}\KatlaSpace{}\IdrisType{Type}\KatlaNewline{}
\IdrisBound{sig}\KatlaSpace{}\IdrisFunction{`algebraOver`}\KatlaSpace{}\IdrisBound{a}\KatlaSpace{}\IdrisKeyword{=}\KatlaNewline{}
\KatlaSpace{}\KatlaSpace{}\IdrisKeyword{(}\IdrisBound{f}\KatlaSpace{}\IdrisKeyword{:}\KatlaSpace{}\IdrisType{Op}\KatlaSpace{}\IdrisBound{sig}\IdrisKeyword{)}\KatlaSpace{}\IdrisKeyword{\KatlaDash{}>}\KatlaSpace{}\IdrisBound{a}\KatlaSpace{}\IdrisFunction{^}\KatlaSpace{}\IdrisKeyword{(}\IdrisFunction{arity}\KatlaSpace{}\IdrisBound{f}\IdrisKeyword{)}\KatlaSpace{}\IdrisKeyword{\KatlaDash{}>}\KatlaSpace{}\IdrisBound{a}\KatlaNewline{}
\end{SaveVerbatim}
\newcommand\AlgebraDecl[1][]{\UseVerbatim[#1]{AlgebraDecl}}
\begin{SaveVerbatim}[commandchars=\\\{\}]{AlgebraDecl}
\IdrisKeyword{record}\KatlaSpace{}\IdrisType{Algebra}\KatlaSpace{}\IdrisKeyword{(}\IdrisBound{Sig}\KatlaSpace{}\IdrisKeyword{:}\KatlaSpace{}\IdrisType{Signature}\IdrisKeyword{)}\KatlaSpace{}\IdrisKeyword{where}\KatlaNewline{}
\KatlaSpace{}\KatlaSpace{}\IdrisKeyword{constructor}\KatlaSpace{}\IdrisData{MakeAlgebra}\KatlaNewline{}
\KatlaSpace{}\KatlaSpace{}\IdrisKeyword{0}\KatlaSpace{}\IdrisFunction{U}\KatlaSpace{}\KatlaSpace{}\KatlaSpace{}\IdrisKeyword{:}\KatlaSpace{}\IdrisType{Type}\KatlaNewline{}
\KatlaSpace{}\KatlaSpace{}\IdrisFunction{Semantics}\KatlaSpace{}\IdrisKeyword{:}\KatlaSpace{}\IdrisBound{Sig}\KatlaSpace{}\IdrisFunction{`algebraOver`}\KatlaSpace{}\IdrisBound{U}\KatlaNewline{}
\end{SaveVerbatim}
\newcommand\CongruenceWRT[1][]{\UseVerbatim[#1]{CongruenceWRT}}
\begin{SaveVerbatim}[commandchars=\\\{\}]{CongruenceWRT}
\IdrisFunction{CongruenceWRT}\KatlaSpace{}\IdrisKeyword{:}\KatlaSpace{}\IdrisKeyword{\{}\IdrisBound{n}\KatlaSpace{}\IdrisKeyword{:}\KatlaSpace{}\IdrisType{Nat}\IdrisKeyword{\}}\KatlaSpace{}\IdrisKeyword{\KatlaDash{}>}\KatlaSpace{}\IdrisKeyword{(}\IdrisBound{a}\KatlaSpace{}\IdrisKeyword{:}\KatlaSpace{}\IdrisType{Setoid}\IdrisKeyword{)}\KatlaSpace{}\IdrisKeyword{\KatlaDash{}>}\KatlaNewline{}
\KatlaSpace{}\KatlaSpace{}\IdrisKeyword{(}\IdrisBound{f}\KatlaSpace{}\IdrisKeyword{:}\KatlaSpace{}\IdrisKeyword{(}\IdrisFunction{U}\KatlaSpace{}\IdrisBound{a}\IdrisKeyword{)}\KatlaSpace{}\IdrisFunction{^}\KatlaSpace{}\IdrisBound{n}\KatlaSpace{}\IdrisKeyword{\KatlaDash{}>}\KatlaSpace{}\IdrisFunction{U}\KatlaSpace{}\IdrisBound{a}\IdrisKeyword{)}\KatlaSpace{}\IdrisKeyword{\KatlaDash{}>}\KatlaSpace{}\IdrisType{Type}\KatlaNewline{}
\IdrisFunction{CongruenceWRT}\KatlaSpace{}\IdrisBound{a}\KatlaSpace{}\IdrisBound{f}\KatlaSpace{}\IdrisKeyword{=}\KatlaSpace{}\IdrisFunction{SetoidHomomorphism}\KatlaSpace{}\IdrisKeyword{(}\IdrisFunction{VectSetoid}\KatlaSpace{}\IdrisBound{n}\KatlaSpace{}\IdrisBound{a}\IdrisKeyword{)}\KatlaSpace{}\IdrisBound{a}\KatlaSpace{}\IdrisBound{f}\KatlaNewline{}
\end{SaveVerbatim}
\newcommand\SetoidAlgebra[1][]{\UseVerbatim[#1]{SetoidAlgebra}}
\begin{SaveVerbatim}[commandchars=\\\{\}]{SetoidAlgebra}
\IdrisKeyword{record}\KatlaSpace{}\IdrisType{SetoidAlgebra}\KatlaSpace{}\IdrisKeyword{(}\IdrisBound{Sig}\KatlaSpace{}\IdrisKeyword{:}\KatlaSpace{}\IdrisType{Signature}\IdrisKeyword{)}\KatlaSpace{}\IdrisKeyword{where}\KatlaNewline{}
\KatlaSpace{}\KatlaSpace{}\IdrisKeyword{constructor}\KatlaSpace{}\IdrisData{MkSetoidAlgebra}\KatlaNewline{}
\KatlaSpace{}\KatlaSpace{}\IdrisFunction{algebra}\KatlaSpace{}\IdrisKeyword{:}\KatlaSpace{}\IdrisType{Algebra}\KatlaSpace{}\IdrisBound{Sig}\KatlaNewline{}
\KatlaSpace{}\KatlaSpace{}\IdrisFunction{equivalence}\KatlaSpace{}\IdrisKeyword{:}\KatlaSpace{}\IdrisType{Equivalence}\KatlaSpace{}\IdrisKeyword{(}\IdrisFunction{U}\KatlaSpace{}\IdrisBound{algebra}\IdrisKeyword{)}\KatlaNewline{}
\KatlaSpace{}\KatlaSpace{}\IdrisFunction{congruence}\KatlaSpace{}\IdrisKeyword{:}\KatlaSpace{}\IdrisKeyword{(}\IdrisBound{f\KatlaSpace{}}\IdrisKeyword{:}\IdrisBound{\KatlaSpace{}}\IdrisType{Op}\IdrisBound{\KatlaSpace{}Sig}\IdrisKeyword{)}\IdrisBound{\KatlaSpace{}}\IdrisKeyword{\KatlaDash{}>}\KatlaNewline{}
\IdrisBound{\KatlaSpace{}\KatlaSpace{}\KatlaSpace{}\KatlaSpace{}}\IdrisKeyword{(}\IdrisData{MkSetoid}\IdrisBound{\KatlaSpace{}}\IdrisKeyword{(}\IdrisFunction{U}\IdrisBound{\KatlaSpace{}algebra}\IdrisKeyword{)}\IdrisBound{\KatlaSpace{}equivalence}\IdrisKeyword{)}\KatlaNewline{}
\IdrisBound{\KatlaSpace{}\KatlaSpace{}\KatlaSpace{}\KatlaSpace{}\KatlaSpace{}\KatlaSpace{}\KatlaSpace{}}\IdrisFunction{`CongruenceWRT`}\IdrisBound{\KatlaSpace{}}\IdrisKeyword{(}\IdrisBound{algebra}\IdrisFunction{.Sem}\IdrisBound{\KatlaSpace{}f)}\KatlaNewline{}
\end{SaveVerbatim}
\newcommand\NatAdditive[1][]{\UseVerbatim[#1]{NatAdditive}}
\begin{SaveVerbatim}[commandchars=\\\{\}]{NatAdditive}
\IdrisFunction{Additive}\KatlaSpace{}\IdrisKeyword{:}\KatlaSpace{}\IdrisType{Algebra}\KatlaSpace{}\IdrisFunction{Monoid.Theory.Signature}\KatlaNewline{}
\IdrisFunction{Additive}\KatlaSpace{}\IdrisKeyword{=}\KatlaSpace{}\KatlaSpace{}\IdrisFunction{MkAlgebra}\KatlaSpace{}\IdrisKeyword{\{}\IdrisBound{U}\KatlaSpace{}\IdrisKeyword{=}\KatlaSpace{}\IdrisType{Nat}\IdrisKeyword{,}\KatlaSpace{}\IdrisBound{Sem}\KatlaSpace{}\IdrisKeyword{=}\KatlaSpace{}\IdrisKeyword{\textbackslash{}case}\KatlaNewline{}
\KatlaSpace{}\KatlaSpace{}\IdrisData{Neutral}\KatlaSpace{}\IdrisKeyword{=>}\KatlaSpace{}\IdrisData{0}\KatlaNewline{}
\KatlaSpace{}\KatlaSpace{}\IdrisData{Product}\KatlaSpace{}\IdrisKeyword{=>}\KatlaSpace{}\IdrisFunction{plus}\IdrisKeyword{\}}\KatlaNewline{}
\KatlaNewline{}
\end{SaveVerbatim}
\newcommand\NatMultitive[1][]{\UseVerbatim[#1]{NatMultitive}}
\begin{SaveVerbatim}[commandchars=\\\{\}]{NatMultitive}
\IdrisFunction{Multiplicative}\KatlaSpace{}\IdrisKeyword{:}\KatlaSpace{}\IdrisFunction{Monoid}\KatlaNewline{}
\IdrisFunction{Multiplicative}\KatlaSpace{}\IdrisKeyword{=}\KatlaSpace{}\IdrisData{MkModel}\KatlaNewline{}
\KatlaSpace{}\KatlaSpace{}\IdrisKeyword{\{}\KatlaSpace{}\IdrisBound{Algebra}\KatlaSpace{}\IdrisKeyword{=}\KatlaSpace{}\IdrisFunction{cast}\KatlaSpace{}\IdrisKeyword{\{}\IdrisBound{from}\KatlaSpace{}\IdrisKeyword{=}\KatlaSpace{}\IdrisType{Algebra}\KatlaSpace{}\IdrisFunction{Signature}\IdrisKeyword{\}}\KatlaSpace{}\$\KatlaNewline{}
\KatlaSpace{}\KatlaSpace{}\KatlaSpace{}\KatlaSpace{}\KatlaSpace{}\KatlaSpace{}\IdrisFunction{MkAlgebra}\KatlaSpace{}\IdrisKeyword{\{}\IdrisBound{U}\KatlaSpace{}\IdrisKeyword{=}\KatlaSpace{}\IdrisType{Nat}\IdrisKeyword{,}\KatlaSpace{}\IdrisBound{Sem}\KatlaSpace{}\IdrisKeyword{=}\KatlaSpace{}\IdrisKeyword{\textbackslash{}case}\KatlaSpace{}\IdrisData{Neutral}\KatlaSpace{}\IdrisKeyword{=>}\KatlaSpace{}\IdrisData{1}\KatlaNewline{}
\KatlaSpace{}\KatlaSpace{}\KatlaSpace{}\KatlaSpace{}\KatlaSpace{}\KatlaSpace{}\KatlaSpace{}\KatlaSpace{}\KatlaSpace{}\KatlaSpace{}\KatlaSpace{}\KatlaSpace{}\KatlaSpace{}\KatlaSpace{}\KatlaSpace{}\KatlaSpace{}\KatlaSpace{}\KatlaSpace{}\KatlaSpace{}\KatlaSpace{}\KatlaSpace{}\KatlaSpace{}\KatlaSpace{}\KatlaSpace{}\KatlaSpace{}\KatlaSpace{}\KatlaSpace{}\KatlaSpace{}\KatlaSpace{}\KatlaSpace{}\KatlaSpace{}\KatlaSpace{}\KatlaSpace{}\KatlaSpace{}\KatlaSpace{}\KatlaSpace{}\KatlaSpace{}\KatlaSpace{}\IdrisData{Product}\KatlaSpace{}\IdrisKeyword{=>}\KatlaSpace{}\IdrisFunction{mult}\IdrisKeyword{\}}\KatlaNewline{}
\KatlaSpace{}\KatlaSpace{}\IdrisKeyword{,}\KatlaSpace{}\IdrisBound{Validate}\KatlaSpace{}\IdrisKeyword{=}\KatlaSpace{}\IdrisKeyword{\textbackslash{}case}\KatlaNewline{}
\KatlaSpace{}\KatlaSpace{}\KatlaSpace{}\KatlaSpace{}\KatlaSpace{}\KatlaSpace{}\IdrisData{LftNeutrality}\KatlaSpace{}\IdrisKeyword{=>}\KatlaSpace{}\IdrisKeyword{\textbackslash{}}\IdrisBound{env}\KatlaSpace{}\IdrisKeyword{=>}\KatlaSpace{}\IdrisFunction{plusZeroRightNeutral}\KatlaSpace{}\IdrisKeyword{\KatlaUnderscore{}}\KatlaNewline{}
\KatlaSpace{}\KatlaSpace{}\KatlaSpace{}\KatlaSpace{}\KatlaSpace{}\KatlaSpace{}\IdrisData{RgtNeutrality}\KatlaSpace{}\IdrisKeyword{=>}\KatlaSpace{}\IdrisKeyword{\textbackslash{}}\IdrisBound{env}\KatlaSpace{}\IdrisKeyword{=>}\KatlaSpace{}\IdrisFunction{multOneRightNeutral}\KatlaSpace{}\IdrisKeyword{\KatlaUnderscore{}}\KatlaNewline{}
\KatlaSpace{}\KatlaSpace{}\KatlaSpace{}\KatlaSpace{}\KatlaSpace{}\KatlaSpace{}\IdrisData{Associativity}\KatlaSpace{}\IdrisKeyword{=>}\KatlaSpace{}\IdrisKeyword{\textbackslash{}}\IdrisBound{env}\KatlaSpace{}\IdrisKeyword{=>}\KatlaSpace{}\IdrisFunction{multAssociative}\KatlaSpace{}\IdrisKeyword{\KatlaUnderscore{}}\KatlaSpace{}\IdrisKeyword{\KatlaUnderscore{}}\KatlaSpace{}\IdrisKeyword{\KatlaUnderscore{}}\KatlaNewline{}
\KatlaSpace{}\KatlaSpace{}\IdrisKeyword{\}}\KatlaNewline{}
\end{SaveVerbatim}
\newcommand\TermDecl[1][]{\UseVerbatim[#1]{TermDecl}}
\begin{SaveVerbatim}[commandchars=\\\{\}]{TermDecl}
\IdrisKeyword{data}\KatlaSpace{}\IdrisType{Term}\KatlaSpace{}\IdrisKeyword{:}\KatlaSpace{}\IdrisKeyword{(0}\KatlaSpace{}\IdrisBound{sig}\KatlaSpace{}\IdrisKeyword{:}\KatlaSpace{}\IdrisType{Signature}\IdrisKeyword{)}\KatlaSpace{}\IdrisKeyword{\KatlaDash{}>}\KatlaSpace{}\IdrisType{Type}\KatlaSpace{}\IdrisKeyword{\KatlaDash{}>}\KatlaSpace{}\IdrisType{Type}\KatlaSpace{}\IdrisKeyword{where}\KatlaNewline{}
\KatlaSpace{}\KatlaSpace{}\IdrisComment{|||\KatlaSpace{}A\KatlaSpace{}variable\KatlaSpace{}with\KatlaSpace{}the\KatlaSpace{}given\KatlaSpace{}index}\KatlaNewline{}
\KatlaSpace{}\KatlaSpace{}\IdrisData{Done}\KatlaSpace{}\IdrisKeyword{:}\KatlaSpace{}\IdrisKeyword{\{0}\KatlaSpace{}\IdrisBound{sig}\KatlaSpace{}\IdrisKeyword{:}\KatlaSpace{}\IdrisType{Signature}\IdrisKeyword{\}}\KatlaSpace{}\IdrisKeyword{\KatlaDash{}>}\KatlaSpace{}\IdrisBound{a}\KatlaSpace{}\IdrisKeyword{\KatlaDash{}>}\KatlaSpace{}\IdrisType{Term}\KatlaSpace{}\IdrisBound{sig}\KatlaSpace{}\IdrisBound{a}\KatlaNewline{}
\KatlaSpace{}\KatlaSpace{}\IdrisComment{|||\KatlaSpace{}An\KatlaSpace{}operator,\KatlaSpace{}applied\KatlaSpace{}to\KatlaSpace{}a\KatlaSpace{}vector\KatlaSpace{}of\KatlaSpace{}sub\KatlaDash{}terms}\KatlaNewline{}
\KatlaSpace{}\KatlaSpace{}\IdrisData{Call}\KatlaSpace{}\IdrisKeyword{:}\KatlaSpace{}\IdrisKeyword{\{0}\KatlaSpace{}\IdrisBound{sig}\KatlaSpace{}\IdrisKeyword{:}\KatlaSpace{}\IdrisType{Signature}\IdrisKeyword{\}}\KatlaSpace{}\IdrisKeyword{\KatlaDash{}>}\KatlaSpace{}\IdrisKeyword{(}\IdrisBound{f}\KatlaSpace{}\IdrisKeyword{:}\KatlaSpace{}\IdrisType{Op}\KatlaSpace{}\IdrisBound{sig}\IdrisKeyword{)}\KatlaSpace{}\IdrisKeyword{\KatlaDash{}>}\KatlaNewline{}
\KatlaSpace{}\KatlaSpace{}\KatlaSpace{}\KatlaSpace{}\KatlaSpace{}\KatlaSpace{}\KatlaSpace{}\KatlaSpace{}\KatlaSpace{}\IdrisType{Vect}\KatlaSpace{}\IdrisKeyword{(}\IdrisFunction{arity}\KatlaSpace{}\IdrisBound{f}\IdrisKeyword{)}\KatlaSpace{}\IdrisKeyword{(}\IdrisType{Term}\KatlaSpace{}\IdrisBound{sig}\KatlaSpace{}\IdrisBound{a}\IdrisKeyword{)}\KatlaSpace{}\IdrisKeyword{\KatlaDash{}>}\KatlaSpace{}\IdrisType{Term}\KatlaSpace{}\IdrisBound{sig}\KatlaSpace{}\IdrisBound{a}\KatlaNewline{}
\end{SaveVerbatim}
\newcommand\Term[1][]{\UseVerb[#1]{Term}}
\begin{SaveVerbatim}[commandchars=\\\{\}]{Term}
\IdrisType{Term}\KatlaSpace{}
\end{SaveVerbatim}
\newcommand\FreeDecl[1][]{\UseVerbatim[#1]{FreeDecl}}
\begin{SaveVerbatim}[commandchars=\\\{\}]{FreeDecl}
\IdrisFunction{Free}\KatlaSpace{}\IdrisKeyword{:}\KatlaSpace{}\IdrisKeyword{(0}\KatlaSpace{}\IdrisBound{sig}\KatlaSpace{}\IdrisKeyword{:}\KatlaSpace{}\IdrisType{Signature}\IdrisKeyword{)}\KatlaSpace{}\IdrisKeyword{\KatlaDash{}>}\KatlaSpace{}\IdrisKeyword{(0}\KatlaSpace{}\IdrisBound{x}\KatlaSpace{}\IdrisKeyword{:}\KatlaSpace{}\IdrisType{Type}\IdrisKeyword{)}\KatlaSpace{}\IdrisKeyword{\KatlaDash{}>}\KatlaSpace{}\IdrisType{Algebra}\KatlaSpace{}\IdrisBound{sig}\KatlaNewline{}
\IdrisFunction{Free}\KatlaSpace{}\IdrisBound{sig}\KatlaSpace{}\IdrisBound{x}\KatlaSpace{}\IdrisKeyword{=}\KatlaSpace{}\IdrisData{MakeAlgebra}\KatlaSpace{}\IdrisKeyword{(}\IdrisType{Term}\KatlaSpace{}\IdrisBound{sig}\KatlaSpace{}\IdrisBound{x}\IdrisKeyword{)}\KatlaSpace{}\IdrisData{Call}\KatlaNewline{}
\end{SaveVerbatim}
\newcommand\OpPreservation[1][]{\UseVerbatim[#1]{OpPreservation}}
\begin{SaveVerbatim}[commandchars=\\\{\}]{OpPreservation}
\IdrisFunction{Preserves}\KatlaSpace{}\IdrisKeyword{:}\KatlaSpace{}\IdrisKeyword{\{}\IdrisBound{sig}\KatlaSpace{}\IdrisKeyword{:}\KatlaSpace{}\IdrisType{Signature}\IdrisKeyword{\}}\KatlaNewline{}
\KatlaSpace{}\KatlaSpace{}\IdrisKeyword{\KatlaDash{}>}\KatlaSpace{}\IdrisKeyword{(}\IdrisBound{a}\IdrisKeyword{,}\KatlaSpace{}\IdrisBound{b}\KatlaSpace{}\IdrisKeyword{:}\KatlaSpace{}\IdrisType{SetoidAlgebra}\KatlaSpace{}\IdrisBound{sig}\IdrisKeyword{)}\KatlaNewline{}
\KatlaSpace{}\KatlaSpace{}\IdrisKeyword{\KatlaDash{}>}\KatlaSpace{}\IdrisKeyword{(}\IdrisBound{h}\KatlaSpace{}\IdrisKeyword{:}\KatlaSpace{}\IdrisFunction{U}\KatlaSpace{}\IdrisBound{a}\KatlaSpace{}\IdrisKeyword{\KatlaDash{}>}\KatlaSpace{}\IdrisFunction{U}\KatlaSpace{}\IdrisBound{b}\IdrisKeyword{)}\KatlaNewline{}
\KatlaSpace{}\KatlaSpace{}\IdrisKeyword{\KatlaDash{}>}\KatlaSpace{}\IdrisKeyword{(}\IdrisBound{f}\KatlaSpace{}\IdrisKeyword{:}\KatlaSpace{}\IdrisType{Op}\KatlaSpace{}\IdrisBound{sig}\IdrisKeyword{)}\KatlaSpace{}\IdrisKeyword{\KatlaDash{}>}\KatlaSpace{}\IdrisType{Type}\KatlaNewline{}
\IdrisFunction{Preserves}\KatlaSpace{}\IdrisKeyword{\{}\IdrisBound{sig}\IdrisKeyword{\}}\KatlaSpace{}\IdrisBound{a}\KatlaSpace{}\IdrisBound{b}\KatlaSpace{}\IdrisBound{h}\KatlaSpace{}\IdrisBound{f}\KatlaNewline{}
\KatlaSpace{}\KatlaSpace{}\IdrisKeyword{=}\KatlaSpace{}\IdrisKeyword{(}\IdrisBound{xs}\KatlaSpace{}\IdrisKeyword{:}\KatlaSpace{}\IdrisType{Vect}\KatlaSpace{}\IdrisKeyword{(}\IdrisFunction{arity}\KatlaSpace{}\IdrisBound{f}\IdrisKeyword{)}\KatlaSpace{}\IdrisKeyword{(}\IdrisFunction{U}\KatlaSpace{}\IdrisBound{a}\IdrisKeyword{))}\KatlaNewline{}
\KatlaSpace{}\KatlaSpace{}\KatlaSpace{}\KatlaSpace{}\IdrisKeyword{\KatlaDash{}>}\KatlaSpace{}\IdrisBound{b}\IdrisFunction{.equivalence.relation}\KatlaNewline{}
\KatlaSpace{}\KatlaSpace{}\KatlaSpace{}\KatlaSpace{}\KatlaSpace{}\KatlaSpace{}\KatlaSpace{}\KatlaSpace{}\KatlaSpace{}\IdrisKeyword{(}\IdrisBound{h}\KatlaSpace{}\$\KatlaSpace{}\IdrisBound{a}\IdrisFunction{.Sem}\KatlaSpace{}\IdrisBound{f}\KatlaSpace{}\IdrisBound{xs}\IdrisKeyword{)}\KatlaNewline{}
\KatlaSpace{}\KatlaSpace{}\KatlaSpace{}\KatlaSpace{}\KatlaSpace{}\KatlaSpace{}\KatlaSpace{}\KatlaSpace{}\KatlaSpace{}\IdrisKeyword{(}\IdrisBound{b}\IdrisFunction{.Sem}\KatlaSpace{}\IdrisBound{f}\KatlaSpace{}\IdrisKeyword{(}\IdrisFunction{map}\KatlaSpace{}\IdrisBound{h}\KatlaSpace{}\IdrisBound{xs}\IdrisKeyword{))}\KatlaNewline{}
\end{SaveVerbatim}
\newcommand\IsAlgHomomorphism[1][]{\UseVerbatim[#1]{IsAlgHomomorphism}}
\begin{SaveVerbatim}[commandchars=\\\{\}]{IsAlgHomomorphism}
\IdrisFunction{Homomorphism}\KatlaSpace{}\IdrisKeyword{:}\KatlaSpace{}\IdrisKeyword{\{}\IdrisBound{sig}\KatlaSpace{}\IdrisKeyword{:}\KatlaSpace{}\IdrisType{Signature}\IdrisKeyword{\}}\KatlaNewline{}
\KatlaSpace{}\KatlaSpace{}\IdrisKeyword{\KatlaDash{}>}\KatlaSpace{}\IdrisKeyword{(}\IdrisBound{a}\IdrisKeyword{,}\KatlaSpace{}\IdrisBound{b}\KatlaSpace{}\IdrisKeyword{:}\KatlaSpace{}\IdrisType{SetoidAlgebra}\KatlaSpace{}\IdrisBound{sig}\IdrisKeyword{)}\KatlaSpace{}\IdrisKeyword{\KatlaDash{}>}\KatlaSpace{}\IdrisKeyword{(}\IdrisBound{h}\KatlaSpace{}\IdrisKeyword{:}\KatlaSpace{}\IdrisFunction{U}\KatlaSpace{}\IdrisBound{a}\KatlaSpace{}\IdrisKeyword{\KatlaDash{}>}\KatlaSpace{}\IdrisFunction{U}\KatlaSpace{}\IdrisBound{b}\IdrisKeyword{)}\KatlaSpace{}\IdrisKeyword{\KatlaDash{}>}\KatlaSpace{}\IdrisType{Type}\KatlaNewline{}
\IdrisFunction{Homomorphism}\KatlaSpace{}\IdrisBound{a}\KatlaSpace{}\IdrisBound{b}\KatlaSpace{}\IdrisBound{h}\KatlaSpace{}\IdrisKeyword{=}\KatlaSpace{}\IdrisKeyword{(}\IdrisBound{f}\KatlaSpace{}\IdrisKeyword{:}\KatlaSpace{}\IdrisType{Op}\KatlaSpace{}\IdrisBound{sig}\IdrisKeyword{)}\KatlaSpace{}\IdrisKeyword{\KatlaDash{}>}\KatlaSpace{}\IdrisFunction{Preserves}\KatlaSpace{}\IdrisBound{a}\KatlaSpace{}\IdrisBound{b}\KatlaSpace{}\IdrisBound{h}\KatlaSpace{}\IdrisBound{f}\KatlaNewline{}
\KatlaNewline{}
\end{SaveVerbatim}
\newcommand\AlgHomomorphism[1][]{\UseVerbatim[#1]{AlgHomomorphism}}
\begin{SaveVerbatim}[commandchars=\\\{\}]{AlgHomomorphism}
\IdrisKeyword{record}\KatlaSpace{}\IdrisType{(\KatlaTilde{}>)}\KatlaSpace{}\IdrisKeyword{\{}\IdrisBound{Sig}\KatlaSpace{}\IdrisKeyword{:}\KatlaSpace{}\IdrisType{Signature}\IdrisKeyword{\}}\KatlaSpace{}\IdrisKeyword{(}\IdrisBound{a}\IdrisKeyword{,}\KatlaSpace{}\IdrisBound{b}\KatlaSpace{}\IdrisKeyword{:}\KatlaSpace{}\IdrisType{SetoidAlgebra}\KatlaSpace{}\IdrisBound{Sig}\IdrisKeyword{)}\KatlaSpace{}\IdrisKeyword{where}\KatlaNewline{}
\KatlaSpace{}\KatlaSpace{}\IdrisKeyword{constructor}\KatlaSpace{}\IdrisData{MkSetoidHomomorphism}\KatlaNewline{}
\KatlaSpace{}\KatlaSpace{}\IdrisFunction{H}\KatlaSpace{}\IdrisKeyword{:}\KatlaSpace{}\IdrisFunction{cast\KatlaSpace{}}\IdrisKeyword{\{}\IdrisBound{to}\IdrisFunction{\KatlaSpace{}}\IdrisKeyword{=}\IdrisFunction{\KatlaSpace{}}\IdrisType{Setoid}\IdrisKeyword{\}}\IdrisFunction{\KatlaSpace{}a}\KatlaSpace{}\IdrisType{\KatlaTilde{}>}\KatlaSpace{}\IdrisFunction{cast\KatlaSpace{}b}\KatlaNewline{}
\KatlaSpace{}\KatlaSpace{}\IdrisFunction{preserves}\KatlaSpace{}\IdrisKeyword{:}\KatlaSpace{}\IdrisFunction{Homomorphism}\KatlaSpace{}\IdrisBound{a}\KatlaSpace{}\IdrisBound{b}\KatlaSpace{}\IdrisKeyword{(}\IdrisFunction{.H}\KatlaSpace{}\IdrisBound{H}\IdrisKeyword{)}\KatlaNewline{}
\KatlaNewline{}
\end{SaveVerbatim}

\begin{SaveVerbatim}[commandchars=\\\{\}]{lftNeutralityEq}
\KatlaNewline{}
\end{SaveVerbatim}
\newcommand\EquationDecl[1][]{\UseVerbatim[#1]{EquationDecl}}
\begin{SaveVerbatim}[commandchars=\\\{\}]{EquationDecl}
\IdrisKeyword{record}\KatlaSpace{}\IdrisType{Equation}\KatlaNewline{}
\KatlaSpace{}\KatlaSpace{}\IdrisKeyword{(}\IdrisBound{Sig}\KatlaSpace{}\IdrisKeyword{:}\KatlaSpace{}\IdrisType{Signature}\IdrisKeyword{)}\KatlaSpace{}\IdrisKeyword{where}\KatlaNewline{}
\KatlaSpace{}\KatlaSpace{}\IdrisKeyword{constructor}\KatlaSpace{}\IdrisData{MkEq}\KatlaNewline{}
\KatlaSpace{}\KatlaSpace{}\IdrisFunction{support}\KatlaSpace{}\KatlaSpace{}\IdrisKeyword{:}\KatlaSpace{}\IdrisType{Nat}\KatlaNewline{}
\KatlaSpace{}\KatlaSpace{}\IdrisFunction{lhs}\IdrisKeyword{,}\KatlaSpace{}\IdrisFunction{rhs}\KatlaSpace{}\IdrisKeyword{:}\KatlaNewline{}
\KatlaSpace{}\KatlaSpace{}\KatlaSpace{}\KatlaSpace{}\IdrisType{Term}\KatlaSpace{}\IdrisBound{Sig}\KatlaSpace{}\IdrisKeyword{(}\IdrisType{Fin}\KatlaSpace{}\IdrisBound{support}\IdrisKeyword{)}\KatlaNewline{}
\end{SaveVerbatim}
\newcommand\PresentationDecl[1][]{\UseVerbatim[#1]{PresentationDecl}}
\begin{SaveVerbatim}[commandchars=\\\{\}]{PresentationDecl}
\IdrisKeyword{record}\KatlaSpace{}\IdrisType{Presentation}\KatlaSpace{}\IdrisKeyword{where}\KatlaNewline{}
\KatlaSpace{}\KatlaSpace{}\IdrisKeyword{constructor}\KatlaSpace{}\IdrisData{MkPresentation}\KatlaNewline{}
\KatlaSpace{}\KatlaSpace{}\IdrisFunction{signature}\KatlaSpace{}\IdrisKeyword{:}\KatlaSpace{}\IdrisType{Signature}\KatlaNewline{}
\KatlaSpace{}\KatlaSpace{}\IdrisKeyword{0}\KatlaSpace{}\IdrisFunction{Axiom}\KatlaSpace{}\KatlaSpace{}\KatlaSpace{}\IdrisKeyword{:}\KatlaSpace{}\IdrisType{Type}\KatlaNewline{}
\KatlaSpace{}\KatlaSpace{}\IdrisFunction{axiom}\KatlaSpace{}\KatlaSpace{}\KatlaSpace{}\KatlaSpace{}\KatlaSpace{}\IdrisKeyword{:}\KatlaSpace{}\IdrisKeyword{(}\IdrisBound{ax}\KatlaSpace{}\IdrisKeyword{:}\KatlaSpace{}\IdrisBound{Axiom}\IdrisKeyword{)}\KatlaSpace{}\IdrisKeyword{\KatlaDash{}>}\KatlaNewline{}
\KatlaSpace{}\KatlaSpace{}\KatlaSpace{}\KatlaSpace{}\IdrisType{Equation}\KatlaSpace{}\IdrisBound{signature}\KatlaNewline{}
\end{SaveVerbatim}
\newcommand\AssociativityScheme[1][]{\UseVerbatim[#1]{AssociativityScheme}}
\begin{SaveVerbatim}[commandchars=\\\{\}]{AssociativityScheme}
\IdrisFunction{associativity}\KatlaSpace{}\IdrisKeyword{:}\KatlaSpace{}\IdrisKeyword{\{}\IdrisBound{sig}\KatlaSpace{}\IdrisKeyword{:}\KatlaSpace{}\IdrisType{Signature}\IdrisKeyword{\}}\KatlaNewline{}
\KatlaSpace{}\KatlaSpace{}\IdrisKeyword{\KatlaDash{}>}\KatlaSpace{}\IdrisFunction{EqSpec}\KatlaSpace{}\IdrisBound{sig}\KatlaSpace{}\IdrisData{[2]}\KatlaNewline{}
\IdrisFunction{associativity}\KatlaSpace{}\IdrisBound{product}\KatlaSpace{}\IdrisKeyword{=}\KatlaNewline{}
\KatlaSpace{}\KatlaSpace{}\IdrisKeyword{let}\KatlaSpace{}\IdrisBound{(+)}\KatlaSpace{}\IdrisKeyword{=}\KatlaSpace{}\IdrisFunction{call}\KatlaSpace{}\IdrisBound{product}\KatlaSpace{}\IdrisKeyword{in}\KatlaNewline{}
\KatlaSpace{}\KatlaSpace{}\IdrisFunction{MkEquation}\KatlaSpace{}\IdrisData{3}\KatlaSpace{}\$\KatlaSpace{}\IdrisFunction{X}\KatlaSpace{}\IdrisData{0}\KatlaSpace{}\IdrisBound{+}\KatlaSpace{}\IdrisKeyword{(}\IdrisFunction{X}\KatlaSpace{}\IdrisData{1}\KatlaSpace{}\IdrisBound{+}\KatlaSpace{}\KatlaSpace{}\IdrisFunction{X}\KatlaSpace{}\IdrisData{2}\IdrisKeyword{)}\KatlaNewline{}
\KatlaSpace{}\KatlaSpace{}\KatlaSpace{}\KatlaSpace{}\KatlaSpace{}\KatlaSpace{}\KatlaSpace{}\KatlaSpace{}\KatlaSpace{}\KatlaSpace{}\KatlaSpace{}\KatlaSpace{}\IdrisFunction{=\KatlaDash{}=}\KatlaSpace{}\IdrisKeyword{(}\IdrisFunction{X}\KatlaSpace{}\IdrisData{0}\KatlaSpace{}\IdrisBound{+}\KatlaSpace{}\KatlaSpace{}\IdrisFunction{X}\KatlaSpace{}\IdrisData{1}\IdrisKeyword{)}\KatlaSpace{}\IdrisBound{+}\KatlaSpace{}\IdrisFunction{X}\KatlaSpace{}\IdrisData{2}\KatlaNewline{}
\end{SaveVerbatim}
\newcommand\SchemeOps[1][]{\UseVerb[#1]{SchemeOps}}
\begin{SaveVerbatim}[commandchars=\\\{\}]{SchemeOps}
\IdrisFunction{EqSpec}\KatlaSpace{}\IdrisBound{sig}\KatlaSpace{}\IdrisData{[2]}\KatlaNewline{}
\end{SaveVerbatim}
\newcommand\MkEquation[1][]{\UseVerb[#1]{MkEquation}}
\begin{SaveVerbatim}[commandchars=\\\{\}]{MkEquation}
\IdrisFunction{MkEquation}\KatlaSpace{}\IdrisData{3}
\end{SaveVerbatim}
\newcommand\MonoidAxiom[1][]{\UseVerbatim[#1]{MonoidAxiom}}
\begin{SaveVerbatim}[commandchars=\\\{\}]{MonoidAxiom}
\IdrisKeyword{data}\KatlaSpace{}\IdrisType{Axiom}\KatlaSpace{}\IdrisKeyword{=}\KatlaSpace{}\IdrisData{LftNeutrality}\KatlaNewline{}
\KatlaSpace{}\KatlaSpace{}\KatlaSpace{}\KatlaSpace{}\KatlaSpace{}\KatlaSpace{}\KatlaSpace{}\KatlaSpace{}\KatlaSpace{}\KatlaSpace{}\KatlaSpace{}\IdrisKeyword{|}\KatlaSpace{}\IdrisData{RgtNeutrality}\KatlaNewline{}
\KatlaSpace{}\KatlaSpace{}\KatlaSpace{}\KatlaSpace{}\KatlaSpace{}\KatlaSpace{}\KatlaSpace{}\KatlaSpace{}\KatlaSpace{}\KatlaSpace{}\KatlaSpace{}\IdrisKeyword{|}\KatlaSpace{}\IdrisData{Associativity}\KatlaNewline{}
\end{SaveVerbatim}
\newcommand\MonoidStructure[1][]{\UseVerbatim[#1]{MonoidStructure}}
\begin{SaveVerbatim}[commandchars=\\\{\}]{MonoidStructure}
\IdrisFunction{MonoidStructure}\KatlaSpace{}\IdrisKeyword{:}\KatlaSpace{}\IdrisType{Type}\KatlaNewline{}
\IdrisFunction{MonoidStructure}\KatlaSpace{}\IdrisKeyword{=}\KatlaSpace{}\IdrisType{SetoidAlgebra}\KatlaSpace{}\IdrisFunction{Signature}\KatlaNewline{}
\end{SaveVerbatim}
\newcommand\MonoidModel[1][]{\UseVerbatim[#1]{MonoidModel}}
\begin{SaveVerbatim}[commandchars=\\\{\}]{MonoidModel}
\IdrisFunction{Monoid}\KatlaSpace{}\IdrisKeyword{:}\KatlaSpace{}\IdrisType{Type}\KatlaNewline{}
\IdrisFunction{Monoid}\KatlaSpace{}\IdrisKeyword{=}\KatlaSpace{}\IdrisType{Model}\KatlaSpace{}\IdrisFunction{MonoidTheory}\KatlaNewline{}
\end{SaveVerbatim}

\begin{SaveVerbatim}[commandchars=\\\{\}]{AdditiveNat}
\IdrisFunction{Additive}\KatlaSpace{}\IdrisKeyword{:}\KatlaSpace{}\IdrisFunction{Monoid}\KatlaNewline{}
\IdrisFunction{Additive}\KatlaSpace{}\IdrisKeyword{=}\KatlaSpace{}\IdrisData{MkModel}\KatlaNewline{}
\KatlaSpace{}\KatlaSpace{}\IdrisKeyword{\{}\KatlaSpace{}\IdrisBound{Algebra}\KatlaSpace{}\IdrisKeyword{=}\KatlaSpace{}\IdrisFunction{cast}\KatlaSpace{}\IdrisKeyword{\{}\IdrisBound{from}\KatlaSpace{}\IdrisKeyword{=}\KatlaSpace{}\IdrisType{Algebra}\KatlaSpace{}\IdrisFunction{Signature}\IdrisKeyword{\}}\KatlaSpace{}\$\KatlaNewline{}
\KatlaSpace{}\KatlaSpace{}\KatlaSpace{}\KatlaSpace{}\KatlaSpace{}\KatlaSpace{}\KatlaSpace{}\KatlaSpace{}\KatlaSpace{}\KatlaSpace{}\KatlaSpace{}\KatlaSpace{}\KatlaSpace{}\KatlaSpace{}\IdrisFunction{MkAlgebra}\KatlaSpace{}\IdrisKeyword{\{}\KatlaSpace{}\IdrisBound{U}\KatlaSpace{}\IdrisKeyword{=}\KatlaSpace{}\IdrisType{Nat}\KatlaNewline{}
\KatlaSpace{}\KatlaSpace{}\KatlaSpace{}\KatlaSpace{}\KatlaSpace{}\KatlaSpace{}\KatlaSpace{}\KatlaSpace{}\KatlaSpace{}\KatlaSpace{}\KatlaSpace{}\KatlaSpace{}\KatlaSpace{}\KatlaSpace{}\KatlaSpace{}\KatlaSpace{}\KatlaSpace{}\KatlaSpace{}\KatlaSpace{}\KatlaSpace{}\KatlaSpace{}\KatlaSpace{}\KatlaSpace{}\KatlaSpace{}\IdrisKeyword{,}\KatlaSpace{}\IdrisBound{Sem}\KatlaSpace{}\IdrisKeyword{=}\KatlaSpace{}\IdrisKeyword{\textbackslash{}case}\KatlaNewline{}
\KatlaSpace{}\KatlaSpace{}\KatlaSpace{}\KatlaSpace{}\KatlaSpace{}\KatlaSpace{}\KatlaSpace{}\KatlaSpace{}\KatlaSpace{}\KatlaSpace{}\KatlaSpace{}\KatlaSpace{}\KatlaSpace{}\KatlaSpace{}\KatlaSpace{}\KatlaSpace{}\KatlaSpace{}\KatlaSpace{}\KatlaSpace{}\KatlaSpace{}\KatlaSpace{}\KatlaSpace{}\KatlaSpace{}\KatlaSpace{}\KatlaSpace{}\KatlaSpace{}\KatlaSpace{}\IdrisData{Neutral}\KatlaSpace{}\IdrisKeyword{=>}\KatlaSpace{}\IdrisData{0}\KatlaNewline{}
\KatlaSpace{}\KatlaSpace{}\KatlaSpace{}\KatlaSpace{}\KatlaSpace{}\KatlaSpace{}\KatlaSpace{}\KatlaSpace{}\KatlaSpace{}\KatlaSpace{}\KatlaSpace{}\KatlaSpace{}\KatlaSpace{}\KatlaSpace{}\KatlaSpace{}\KatlaSpace{}\KatlaSpace{}\KatlaSpace{}\KatlaSpace{}\KatlaSpace{}\KatlaSpace{}\KatlaSpace{}\KatlaSpace{}\KatlaSpace{}\KatlaSpace{}\KatlaSpace{}\KatlaSpace{}\IdrisData{Product}\KatlaSpace{}\IdrisKeyword{=>}\KatlaSpace{}\IdrisFunction{plus}\KatlaNewline{}
\KatlaSpace{}\KatlaSpace{}\KatlaSpace{}\KatlaSpace{}\KatlaSpace{}\KatlaSpace{}\KatlaSpace{}\KatlaSpace{}\KatlaSpace{}\KatlaSpace{}\KatlaSpace{}\KatlaSpace{}\KatlaSpace{}\KatlaSpace{}\KatlaSpace{}\KatlaSpace{}\KatlaSpace{}\KatlaSpace{}\KatlaSpace{}\KatlaSpace{}\KatlaSpace{}\KatlaSpace{}\KatlaSpace{}\KatlaSpace{}\IdrisKeyword{\}}\KatlaNewline{}
\KatlaSpace{}\KatlaSpace{}\IdrisKeyword{,}\KatlaSpace{}\IdrisBound{Validate}\KatlaSpace{}\IdrisKeyword{=}\KatlaSpace{}\IdrisKeyword{\textbackslash{}case}\KatlaNewline{}
\KatlaSpace{}\KatlaSpace{}\KatlaSpace{}\KatlaSpace{}\KatlaSpace{}\KatlaSpace{}\IdrisData{LftNeutrality}\KatlaSpace{}\IdrisKeyword{=>}\KatlaSpace{}\IdrisKeyword{\textbackslash{}}\IdrisBound{env}\KatlaSpace{}\IdrisKeyword{=>}\KatlaSpace{}\IdrisData{Refl}\KatlaNewline{}
\KatlaSpace{}\KatlaSpace{}\KatlaSpace{}\KatlaSpace{}\KatlaSpace{}\KatlaSpace{}\IdrisData{RgtNeutrality}\KatlaSpace{}\IdrisKeyword{=>}\KatlaSpace{}\IdrisKeyword{\textbackslash{}}\IdrisBound{env}\KatlaSpace{}\IdrisKeyword{=>}\KatlaSpace{}\IdrisFunction{plusZeroRightNeutral}\KatlaSpace{}\IdrisKeyword{\KatlaUnderscore{}}\KatlaNewline{}
\KatlaSpace{}\KatlaSpace{}\KatlaSpace{}\KatlaSpace{}\KatlaSpace{}\KatlaSpace{}\IdrisData{Associativity}\KatlaSpace{}\IdrisKeyword{=>}\KatlaSpace{}\IdrisKeyword{\textbackslash{}}\IdrisBound{env}\KatlaSpace{}\IdrisKeyword{=>}\KatlaSpace{}\IdrisFunction{plusAssociative}\KatlaSpace{}\IdrisKeyword{\KatlaUnderscore{}}\KatlaSpace{}\IdrisKeyword{\KatlaUnderscore{}}\KatlaSpace{}\IdrisKeyword{\KatlaUnderscore{}}\KatlaNewline{}
\KatlaSpace{}\KatlaSpace{}\IdrisKeyword{\}}\KatlaNewline{}
\end{SaveVerbatim}
\newcommand\SigAlgebra[1][]{\UseVerb[#1]{SigAlgebra}}
\begin{SaveVerbatim}[commandchars=\\\{\}]{SigAlgebra}
\IdrisBound{a}\KatlaSpace{}\IdrisKeyword{:}\KatlaSpace{}\IdrisType{Algebra}\KatlaSpace{}\IdrisBound{sig}
\end{SaveVerbatim}
\newcommand\AlgBindType[1][]{\UseVerb[#1]{AlgBindType}}
\begin{SaveVerbatim}[commandchars=\\\{\}]{AlgBindType}
\KatlaSpace{}\IdrisKeyword{:}\KatlaSpace{}\IdrisType{Term}\KatlaSpace{}\IdrisBound{sig}\KatlaSpace{}\IdrisBound{x}\KatlaSpace{}\IdrisKeyword{\KatlaDash{}>}\KatlaSpace{}\IdrisKeyword{(}\IdrisBound{x}\KatlaSpace{}\IdrisKeyword{\KatlaDash{}>}\KatlaSpace{}\IdrisFunction{U}\KatlaSpace{}\IdrisBound{a}\IdrisKeyword{)}\KatlaSpace{}\IdrisKeyword{\KatlaDash{}>}\KatlaSpace{}\IdrisFunction{U}\KatlaSpace{}\IdrisBound{a}\KatlaNewline{}
\end{SaveVerbatim}
\newcommand\AlgBind[1][]{\UseVerb[#1]{AlgBind}}
\begin{SaveVerbatim}[commandchars=\\\{\}]{AlgBind}
\IdrisBound{a}\IdrisFunction{.Sem}\KatlaNewline{}
\end{SaveVerbatim}
\newcommand\BindEx[1][]{\UseVerb[#1]{BindEx}}
\begin{SaveVerbatim}[commandchars=\\\{\}]{BindEx}
\IdrisKeyword{(}\IdrisFunction{Nat.Additive}\IdrisKeyword{)}\IdrisFunction{.Sem}\KatlaSpace{}\IdrisKeyword{(}\IdrisFunction{X}\KatlaSpace{}\IdrisData{0}\IdrisFunction{.+.O1.+.X}\KatlaSpace{}\IdrisData{1}\IdrisKeyword{)}\KatlaSpace{}\IdrisKeyword{(\textbackslash{}case}\KatlaSpace{}\{\IdrisData{0}\IdrisKeyword{=>}\IdrisData{5};\KatlaSpace{}\IdrisData{1}\IdrisKeyword{=>}\IdrisData{7}\}\IdrisKeyword{)}\KatlaNewline{}
\end{SaveVerbatim}
\newcommand\BindExResult[1][]{\UseVerb[#1]{BindExResult}}
\begin{SaveVerbatim}[commandchars=\\\{\}]{BindExResult}
\IdrisData{5}\IdrisFunction{+}\IdrisData{0}\IdrisFunction{+}\IdrisData{7}\KatlaNewline{}
\end{SaveVerbatim}
\newcommand\Models[1][]{\UseVerbatim[#1]{Models}}
\begin{SaveVerbatim}[commandchars=\\\{\}]{Models}
\IdrisFunction{models}\KatlaSpace{}\IdrisKeyword{:}\KatlaSpace{}\IdrisKeyword{\{}\IdrisBound{sig}\KatlaSpace{}\IdrisKeyword{:}\KatlaSpace{}\IdrisType{Signature}\IdrisKeyword{\}}\KatlaSpace{}\IdrisKeyword{\KatlaDash{}>}\KatlaSpace{}\IdrisKeyword{(}\IdrisBound{a}\KatlaSpace{}\IdrisKeyword{:}\KatlaSpace{}\IdrisType{SetoidAlgebra}\KatlaSpace{}\IdrisBound{sig}\IdrisKeyword{)}\KatlaSpace{}\IdrisKeyword{\KatlaDash{}>}\KatlaSpace{}\IdrisKeyword{(}\IdrisBound{eq}\KatlaSpace{}\IdrisKeyword{:}\KatlaSpace{}\IdrisType{Equation}\KatlaSpace{}\IdrisBound{sig}\IdrisKeyword{)}\KatlaSpace{}\IdrisKeyword{\KatlaDash{}>}\KatlaNewline{}
\KatlaSpace{}\KatlaSpace{}\KatlaSpace{}\KatlaSpace{}\KatlaSpace{}\KatlaSpace{}\KatlaSpace{}\KatlaSpace{}\KatlaSpace{}\IdrisKeyword{(}\IdrisBound{env}\KatlaSpace{}\IdrisKeyword{:}\KatlaSpace{}\IdrisType{Fin}\KatlaSpace{}\IdrisBound{eq}\IdrisFunction{.support}\KatlaSpace{}\IdrisKeyword{\KatlaDash{}>}\KatlaSpace{}\IdrisFunction{U}\KatlaSpace{}\IdrisBound{a}\IdrisFunction{.algebra}\IdrisKeyword{)}\KatlaSpace{}\IdrisKeyword{\KatlaDash{}>}\KatlaSpace{}\IdrisType{Type}\KatlaNewline{}
\IdrisFunction{models}\KatlaSpace{}\IdrisBound{a}\KatlaSpace{}\IdrisBound{eq}\KatlaSpace{}\IdrisBound{env}\KatlaSpace{}\IdrisKeyword{=}\KatlaSpace{}\IdrisBound{a}\IdrisFunction{.equivalence.relation}\KatlaSpace{}\IdrisKeyword{(}\IdrisBound{a}\IdrisFunction{.Sem}\KatlaSpace{}\IdrisBound{eq}\IdrisFunction{.lhs}\KatlaSpace{}\IdrisBound{env}\IdrisKeyword{)}\KatlaSpace{}\IdrisKeyword{(}\IdrisBound{a}\IdrisFunction{.Sem}\KatlaSpace{}\IdrisBound{eq}\IdrisFunction{.rhs}\KatlaSpace{}\IdrisBound{env}\IdrisKeyword{)}\KatlaNewline{}
\end{SaveVerbatim}
\newcommand\Entails[1][]{\UseVerbatim[#1]{Entails}}
\begin{SaveVerbatim}[commandchars=\\\{\}]{Entails}
\IdrisFunction{(=|)}\KatlaSpace{}\IdrisKeyword{:}\KatlaSpace{}\IdrisKeyword{\{}\IdrisBound{sig}\KatlaSpace{}\IdrisKeyword{:}\KatlaSpace{}\IdrisType{Signature}\IdrisKeyword{\}}\KatlaSpace{}\IdrisKeyword{\KatlaDash{}>}\KatlaSpace{}\IdrisKeyword{(}\IdrisBound{eq}\KatlaSpace{}\IdrisKeyword{:}\KatlaSpace{}\IdrisType{Equation}\KatlaSpace{}\IdrisBound{sig}\IdrisKeyword{)}\KatlaSpace{}\IdrisKeyword{\KatlaDash{}>}\KatlaNewline{}
\KatlaSpace{}\KatlaSpace{}\KatlaSpace{}\KatlaSpace{}\KatlaSpace{}\KatlaSpace{}\KatlaSpace{}\IdrisType{(}\IdrisBound{a}\KatlaSpace{}\IdrisKeyword{:}\KatlaSpace{}\IdrisType{SetoidAlgebra}\KatlaSpace{}\IdrisBound{sig}\KatlaSpace{}\IdrisType{**}\KatlaSpace{}\IdrisType{Fin}\KatlaSpace{}\IdrisBound{eq}\IdrisFunction{.support}\KatlaSpace{}\IdrisKeyword{\KatlaDash{}>}\KatlaSpace{}\IdrisFunction{U}\KatlaSpace{}\IdrisBound{a}\IdrisFunction{.algebra}\IdrisType{)}\KatlaSpace{}\IdrisKeyword{\KatlaDash{}>}\KatlaSpace{}\IdrisType{Type}\KatlaNewline{}
\IdrisBound{eq}\KatlaSpace{}\IdrisFunction{=|}\KatlaSpace{}(\IdrisBound{a}\KatlaSpace{}\IdrisData{**}\KatlaSpace{}\IdrisBound{env})\KatlaSpace{}\IdrisKeyword{=}\KatlaSpace{}\IdrisFunction{models}\KatlaSpace{}\IdrisBound{a}\KatlaSpace{}\IdrisBound{eq}\KatlaSpace{}\IdrisBound{env}\KatlaNewline{}
\end{SaveVerbatim}
\newcommand\ValidatesEquation[1][]{\UseVerbatim[#1]{ValidatesEquation}}
\begin{SaveVerbatim}[commandchars=\\\{\}]{ValidatesEquation}
\IdrisFunction{ValidatesEquation}\KatlaSpace{}\IdrisKeyword{:}\KatlaSpace{}\IdrisKeyword{(}\IdrisBound{eq}\KatlaSpace{}\IdrisKeyword{:}\KatlaSpace{}\IdrisType{Equation}\KatlaSpace{}\IdrisBound{sig}\IdrisKeyword{)}\KatlaSpace{}\IdrisKeyword{\KatlaDash{}>}\KatlaSpace{}\IdrisKeyword{(}\IdrisBound{a}\KatlaSpace{}\IdrisKeyword{:}\KatlaSpace{}\IdrisType{SetoidAlgebra}\KatlaSpace{}\IdrisBound{sig}\IdrisKeyword{)}\KatlaSpace{}\IdrisKeyword{\KatlaDash{}>}\KatlaSpace{}\IdrisType{Type}\KatlaNewline{}
\IdrisFunction{ValidatesEquation}\KatlaSpace{}\IdrisBound{eq}\KatlaSpace{}\IdrisBound{a}\KatlaSpace{}\IdrisKeyword{=}\KatlaSpace{}\IdrisKeyword{(}\IdrisBound{env}\KatlaSpace{}\IdrisKeyword{:}\KatlaSpace{}\IdrisType{Fin}\KatlaSpace{}\IdrisBound{eq}\IdrisFunction{.support}\KatlaSpace{}\IdrisKeyword{\KatlaDash{}>}\KatlaSpace{}\IdrisFunction{U}\KatlaSpace{}\IdrisBound{a}\IdrisFunction{.algebra}\IdrisKeyword{)}\KatlaSpace{}\IdrisKeyword{\KatlaDash{}>}\KatlaSpace{}\IdrisBound{eq}\KatlaSpace{}\IdrisFunction{=|}\KatlaSpace{}(\IdrisBound{a}\KatlaSpace{}\IdrisData{**}\KatlaSpace{}\IdrisBound{env})\KatlaNewline{}
\end{SaveVerbatim}
\newcommand\Validates[1][]{\UseVerbatim[#1]{Validates}}
\begin{SaveVerbatim}[commandchars=\\\{\}]{Validates}
\IdrisFunction{Validates}\KatlaSpace{}\IdrisKeyword{:}\KatlaSpace{}\IdrisKeyword{(}\IdrisBound{pres}\KatlaSpace{}\IdrisKeyword{:}\KatlaSpace{}\IdrisType{Presentation}\IdrisKeyword{)}\KatlaSpace{}\IdrisKeyword{\KatlaDash{}>}\KatlaSpace{}\IdrisKeyword{(}\IdrisBound{a}\KatlaSpace{}\IdrisKeyword{:}\KatlaSpace{}\IdrisType{SetoidAlgebra}\KatlaSpace{}\IdrisBound{pres}\IdrisFunction{.signature}\IdrisKeyword{)}\KatlaSpace{}\IdrisKeyword{\KatlaDash{}>}\KatlaSpace{}\IdrisType{Type}\KatlaNewline{}
\IdrisFunction{Validates}\KatlaSpace{}\IdrisBound{pres}\KatlaSpace{}\IdrisBound{a}\KatlaSpace{}\IdrisKeyword{=}\KatlaSpace{}\IdrisKeyword{(}\IdrisBound{ax}\KatlaSpace{}\IdrisKeyword{:}\KatlaSpace{}\IdrisBound{pres}\IdrisFunction{.Axiom}\IdrisKeyword{)}\KatlaSpace{}\IdrisKeyword{\KatlaDash{}>}\KatlaSpace{}\IdrisFunction{ValidatesEquation}\KatlaSpace{}\IdrisKeyword{(}\IdrisBound{pres}\IdrisFunction{.axiom}\KatlaSpace{}\IdrisBound{ax}\IdrisKeyword{)}\KatlaSpace{}\IdrisBound{a}\KatlaNewline{}
\end{SaveVerbatim}
\newcommand\ModelDecl[1][]{\UseVerbatim[#1]{ModelDecl}}
\begin{SaveVerbatim}[commandchars=\\\{\}]{ModelDecl}
\IdrisKeyword{record}\KatlaSpace{}\IdrisType{Model}\KatlaSpace{}\IdrisKeyword{(}\IdrisBound{Pres}\KatlaSpace{}\IdrisKeyword{:}\KatlaSpace{}\IdrisType{Presentation}\IdrisKeyword{)}\KatlaSpace{}\IdrisKeyword{where}\KatlaNewline{}
\KatlaSpace{}\KatlaSpace{}\IdrisKeyword{constructor}\KatlaSpace{}\IdrisData{MkModel}\KatlaNewline{}
\KatlaSpace{}\KatlaSpace{}\IdrisFunction{Algebra}\KatlaSpace{}\KatlaSpace{}\IdrisKeyword{:}\KatlaSpace{}\IdrisType{SetoidAlgebra}\KatlaSpace{}\IdrisKeyword{(}\IdrisBound{Pres}\IdrisKeyword{)}\IdrisFunction{.signature}\KatlaNewline{}
\KatlaSpace{}\KatlaSpace{}\IdrisFunction{Validate}\KatlaSpace{}\IdrisKeyword{:}\KatlaSpace{}\IdrisFunction{Validates}\KatlaSpace{}\IdrisBound{Pres}\KatlaSpace{}\IdrisBound{Algebra}\KatlaNewline{}
\end{SaveVerbatim}
\newcommand\AdditiveNatValid[1][]{\UseVerbatim[#1]{AdditiveNatValid}}
\begin{SaveVerbatim}[commandchars=\\\{\}]{AdditiveNatValid}
\IdrisFunction{IsMonoid}\KatlaSpace{}\IdrisKeyword{:}\KatlaSpace{}\IdrisFunction{Validates}\KatlaSpace{}\IdrisFunction{MonoidTheory}\KatlaSpace{}\IdrisFunction{NatAdditive}\KatlaNewline{}
\IdrisFunction{IsMonoid}\KatlaSpace{}\IdrisData{LftNeutrality}\KatlaSpace{}\IdrisBound{env}\KatlaSpace{}\IdrisKeyword{=}\KatlaSpace{}\IdrisData{Refl}\KatlaNewline{}
\IdrisFunction{IsMonoid}\KatlaSpace{}\IdrisData{RgtNeutrality}\KatlaSpace{}\IdrisBound{env}\KatlaSpace{}\IdrisKeyword{=}\KatlaSpace{}\IdrisFunction{plusZeroRightNeutral}\KatlaSpace{}\IdrisKeyword{\KatlaUnderscore{}}\KatlaNewline{}
\IdrisFunction{IsMonoid}\KatlaSpace{}\IdrisData{Associativity}\KatlaSpace{}\IdrisBound{env}\KatlaSpace{}\IdrisKeyword{=}\KatlaSpace{}\IdrisFunction{plusAssociative}\KatlaSpace{}\IdrisKeyword{\KatlaUnderscore{}}\KatlaSpace{}\IdrisKeyword{\KatlaUnderscore{}}\KatlaSpace{}\IdrisKeyword{\KatlaUnderscore{}}\KatlaNewline{}
\end{SaveVerbatim}
\newcommand\ModelEnv[1][]{\UseVerb[#1]{ModelEnv}}
\begin{SaveVerbatim}[commandchars=\\\{\}]{ModelEnv}
\KatlaNewline{}
\end{SaveVerbatim}
\newcommand\ListInvMonoid[1][]{\UseVerbatim[#1]{ListInvMonoid}}
\begin{SaveVerbatim}[commandchars=\\\{\}]{ListInvMonoid}
\IdrisFunction{ListInvMonoid}\KatlaSpace{}\IdrisKeyword{:}\KatlaSpace{}\IdrisKeyword{\{0}\KatlaSpace{}\IdrisBound{a}\KatlaSpace{}\IdrisKeyword{:}\KatlaSpace{}\IdrisType{Type}\IdrisKeyword{\}}\KatlaSpace{}\IdrisKeyword{\KatlaDash{}>}\KatlaSpace{}\IdrisFunction{InvolutiveMonoid}\KatlaNewline{}
\IdrisFunction{ListInvMonoid}\KatlaSpace{}\IdrisKeyword{=}\KatlaSpace{}\IdrisData{MkModel}\KatlaNewline{}
\KatlaSpace{}\KatlaSpace{}\IdrisKeyword{\{}\KatlaSpace{}\IdrisBound{Algebra}\KatlaSpace{}\IdrisKeyword{=}\KatlaSpace{}\IdrisFunction{cast}\KatlaSpace{}\$\KatlaSpace{}\IdrisFunction{MkAlgebra}\KatlaNewline{}
\KatlaSpace{}\KatlaSpace{}\KatlaSpace{}\KatlaSpace{}\KatlaSpace{}\KatlaSpace{}\IdrisKeyword{\{}\IdrisBound{sig}\KatlaSpace{}\IdrisKeyword{=}\KatlaSpace{}\IdrisFunction{Monoid.Involutive.Theory.Signature}\IdrisKeyword{\}}\KatlaNewline{}
\KatlaSpace{}\KatlaSpace{}\KatlaSpace{}\KatlaSpace{}\KatlaSpace{}\KatlaSpace{}\IdrisKeyword{\{}\KatlaSpace{}\IdrisBound{U}\KatlaSpace{}\IdrisKeyword{=}\KatlaSpace{}\IdrisType{List}\KatlaSpace{}\IdrisBound{a}\KatlaSpace{}\KatlaSpace{}\KatlaSpace{}\KatlaSpace{}\KatlaSpace{}\KatlaSpace{}\KatlaSpace{}\KatlaSpace{}\KatlaSpace{}\KatlaSpace{}\KatlaSpace{}\KatlaSpace{}\KatlaSpace{}\KatlaSpace{}\KatlaSpace{}\KatlaSpace{}\KatlaSpace{}\KatlaSpace{}\KatlaSpace{}\KatlaSpace{}\KatlaSpace{}\KatlaSpace{}\KatlaSpace{}\KatlaSpace{}\KatlaSpace{}\KatlaSpace{}\KatlaSpace{}\KatlaSpace{}\KatlaSpace{}\KatlaSpace{}\KatlaSpace{}\KatlaSpace{}\KatlaSpace{}\KatlaSpace{}\KatlaSpace{}\KatlaSpace{}\KatlaSpace{}\KatlaSpace{}\KatlaSpace{}\KatlaSpace{}\KatlaSpace{}\KatlaSpace{}\KatlaSpace{}\KatlaSpace{}\IdrisComment{\KatlaDash{}\KatlaDash{}\KatlaSpace{}Carrier}\KatlaNewline{}
\KatlaSpace{}\KatlaSpace{}\KatlaSpace{}\KatlaSpace{}\KatlaSpace{}\KatlaSpace{}\IdrisKeyword{,}\KatlaSpace{}\IdrisBound{Sem}\KatlaSpace{}\IdrisKeyword{=}\KatlaSpace{}\IdrisKeyword{\textbackslash{}case}\KatlaSpace{}\KatlaSpace{}\KatlaSpace{}\KatlaSpace{}\KatlaSpace{}\KatlaSpace{}\KatlaSpace{}\KatlaSpace{}\KatlaSpace{}\KatlaSpace{}\KatlaSpace{}\KatlaSpace{}\KatlaSpace{}\KatlaSpace{}\KatlaSpace{}\KatlaSpace{}\KatlaSpace{}\KatlaSpace{}\KatlaSpace{}\KatlaSpace{}\KatlaSpace{}\KatlaSpace{}\KatlaSpace{}\KatlaSpace{}\KatlaSpace{}\KatlaSpace{}\KatlaSpace{}\KatlaSpace{}\KatlaSpace{}\KatlaSpace{}\KatlaSpace{}\KatlaSpace{}\KatlaSpace{}\KatlaSpace{}\KatlaSpace{}\KatlaSpace{}\KatlaSpace{}\KatlaSpace{}\KatlaSpace{}\KatlaSpace{}\KatlaSpace{}\KatlaSpace{}\KatlaSpace{}\IdrisComment{\KatlaDash{}\KatlaDash{}\KatlaSpace{}Operations}\KatlaNewline{}
\KatlaSpace{}\KatlaSpace{}\KatlaSpace{}\KatlaSpace{}\KatlaSpace{}\KatlaSpace{}\KatlaSpace{}\KatlaSpace{}\KatlaSpace{}\KatlaSpace{}\IdrisData{Mono}\KatlaSpace{}\IdrisBound{monoidOp}\KatlaSpace{}\IdrisKeyword{=>}\KatlaSpace{}\IdrisKeyword{case}\KatlaSpace{}\IdrisBound{monoidOp}\KatlaSpace{}\IdrisKeyword{of}\KatlaSpace{}\KatlaSpace{}\KatlaSpace{}\KatlaSpace{}\KatlaSpace{}\KatlaSpace{}\KatlaSpace{}\KatlaSpace{}\KatlaSpace{}\KatlaSpace{}\KatlaSpace{}\KatlaSpace{}\KatlaSpace{}\KatlaSpace{}\KatlaSpace{}\KatlaSpace{}\KatlaSpace{}\KatlaSpace{}\KatlaSpace{}\IdrisComment{\KatlaDash{}\KatlaDash{}\KatlaSpace{}\KatlaSpace{}\KatlaSpace{}Inherited\KatlaSpace{}from\KatlaSpace{}monoids}\KatlaNewline{}
\KatlaSpace{}\KatlaSpace{}\KatlaSpace{}\KatlaSpace{}\KatlaSpace{}\KatlaSpace{}\KatlaSpace{}\KatlaSpace{}\KatlaSpace{}\KatlaSpace{}\KatlaSpace{}\KatlaSpace{}\KatlaSpace{}\KatlaSpace{}\KatlaSpace{}\IdrisData{Neutral}\KatlaSpace{}\IdrisKeyword{=>}\KatlaSpace{}\IdrisData{[]}\KatlaNewline{}
\KatlaSpace{}\KatlaSpace{}\KatlaSpace{}\KatlaSpace{}\KatlaSpace{}\KatlaSpace{}\KatlaSpace{}\KatlaSpace{}\KatlaSpace{}\KatlaSpace{}\KatlaSpace{}\KatlaSpace{}\KatlaSpace{}\KatlaSpace{}\KatlaSpace{}\IdrisData{Product}\KatlaSpace{}\IdrisKeyword{=>}\KatlaSpace{}\IdrisFunction{(++)}\KatlaNewline{}
\KatlaSpace{}\KatlaSpace{}\KatlaSpace{}\KatlaSpace{}\KatlaSpace{}\KatlaSpace{}\KatlaSpace{}\KatlaSpace{}\KatlaSpace{}\KatlaSpace{}\IdrisData{Involution}\KatlaSpace{}\IdrisKeyword{=>}\KatlaSpace{}\IdrisFunction{reverse}\KatlaNewline{}
\KatlaSpace{}\KatlaSpace{}\KatlaSpace{}\KatlaSpace{}\KatlaSpace{}\KatlaSpace{}\IdrisKeyword{\}}\KatlaNewline{}
\KatlaSpace{}\KatlaSpace{}\IdrisKeyword{,}\KatlaSpace{}\IdrisBound{Validate}\KatlaSpace{}\IdrisKeyword{=}\KatlaSpace{}\IdrisKeyword{\textbackslash{}case}\KatlaSpace{}\KatlaSpace{}\KatlaSpace{}\KatlaSpace{}\KatlaSpace{}\KatlaSpace{}\KatlaSpace{}\KatlaSpace{}\KatlaSpace{}\KatlaSpace{}\KatlaSpace{}\KatlaSpace{}\KatlaSpace{}\KatlaSpace{}\KatlaSpace{}\KatlaSpace{}\KatlaSpace{}\KatlaSpace{}\KatlaSpace{}\KatlaSpace{}\KatlaSpace{}\KatlaSpace{}\KatlaSpace{}\KatlaSpace{}\KatlaSpace{}\KatlaSpace{}\KatlaSpace{}\KatlaSpace{}\KatlaSpace{}\KatlaSpace{}\KatlaSpace{}\KatlaSpace{}\KatlaSpace{}\KatlaSpace{}\KatlaSpace{}\KatlaSpace{}\KatlaSpace{}\KatlaSpace{}\KatlaSpace{}\KatlaSpace{}\KatlaSpace{}\KatlaSpace{}\IdrisComment{\KatlaDash{}\KatlaDash{}\KatlaSpace{}Validate\KatlaSpace{}equations}\KatlaNewline{}
\KatlaSpace{}\KatlaSpace{}\KatlaSpace{}\KatlaSpace{}\KatlaSpace{}\KatlaSpace{}\IdrisData{Mon}\KatlaSpace{}\IdrisData{LftNeutrality}\KatlaSpace{}\KatlaSpace{}\IdrisKeyword{=>}\KatlaSpace{}\IdrisKeyword{\textbackslash{}}\IdrisBound{env}\KatlaSpace{}\IdrisKeyword{=>}\KatlaSpace{}\IdrisData{Refl}\KatlaSpace{}\KatlaSpace{}\KatlaSpace{}\KatlaSpace{}\KatlaSpace{}\KatlaSpace{}\KatlaSpace{}\KatlaSpace{}\KatlaSpace{}\KatlaSpace{}\KatlaSpace{}\KatlaSpace{}\KatlaSpace{}\KatlaSpace{}\KatlaSpace{}\KatlaSpace{}\KatlaSpace{}\KatlaSpace{}\KatlaSpace{}\KatlaSpace{}\KatlaSpace{}\KatlaSpace{}\IdrisComment{\KatlaDash{}\KatlaDash{}\KatlaSpace{}Directly,\KatlaSpace{}or}\KatlaNewline{}
\KatlaSpace{}\KatlaSpace{}\KatlaSpace{}\KatlaSpace{}\KatlaSpace{}\KatlaSpace{}\IdrisData{Mon}\KatlaSpace{}\IdrisData{RgtNeutrality}\KatlaSpace{}\KatlaSpace{}\IdrisKeyword{=>}\KatlaSpace{}\IdrisKeyword{\textbackslash{}}\IdrisBound{env}\KatlaSpace{}\IdrisKeyword{=>}\KatlaSpace{}\IdrisFunction{appendNilRightNeutral}\KatlaSpace{}\IdrisKeyword{\KatlaUnderscore{}}\KatlaSpace{}\KatlaSpace{}\KatlaSpace{}\IdrisComment{\KatlaDash{}\KatlaDash{}\KatlaSpace{}use\KatlaSpace{}existing\KatlaSpace{}standard}\KatlaNewline{}
\KatlaSpace{}\KatlaSpace{}\KatlaSpace{}\KatlaSpace{}\KatlaSpace{}\KatlaSpace{}\IdrisData{Mon}\KatlaSpace{}\IdrisData{Associativity}\KatlaSpace{}\KatlaSpace{}\IdrisKeyword{=>}\KatlaSpace{}\IdrisKeyword{\textbackslash{}}\IdrisBound{env}\KatlaSpace{}\IdrisKeyword{=>}\KatlaSpace{}\IdrisFunction{appendAssociative}\KatlaSpace{}\IdrisKeyword{\KatlaUnderscore{}}\KatlaSpace{}\IdrisKeyword{\KatlaUnderscore{}}\KatlaSpace{}\IdrisKeyword{\KatlaUnderscore{}}\KatlaSpace{}\KatlaSpace{}\KatlaSpace{}\IdrisComment{\KatlaDash{}\KatlaDash{}\KatlaSpace{}library\KatlaSpace{}functions}\KatlaNewline{}
\KatlaSpace{}\KatlaSpace{}\KatlaSpace{}\KatlaSpace{}\KatlaSpace{}\KatlaSpace{}\IdrisData{Involutivity}\KatlaSpace{}\KatlaSpace{}\KatlaSpace{}\KatlaSpace{}\KatlaSpace{}\KatlaSpace{}\KatlaSpace{}\IdrisKeyword{=>}\KatlaSpace{}\IdrisKeyword{\textbackslash{}}\IdrisBound{env}\KatlaSpace{}\IdrisKeyword{=>}\KatlaSpace{}\IdrisFunction{reverseInvolutive}\KatlaSpace{}\IdrisKeyword{\KatlaUnderscore{}}\KatlaNewline{}
\KatlaSpace{}\KatlaSpace{}\KatlaSpace{}\KatlaSpace{}\KatlaSpace{}\KatlaSpace{}\IdrisData{Antidistributivity}\KatlaSpace{}\IdrisKeyword{=>}\KatlaSpace{}\IdrisKeyword{\textbackslash{}}\IdrisBound{env}\KatlaSpace{}\IdrisKeyword{=>}\KatlaSpace{}\IdrisFunction{sym}\KatlaSpace{}\IdrisKeyword{(}\IdrisFunction{revAppend}\KatlaSpace{}\IdrisKeyword{\KatlaUnderscore{}}\KatlaSpace{}\IdrisKeyword{\KatlaUnderscore{})}\KatlaNewline{}
\KatlaSpace{}\KatlaSpace{}\IdrisKeyword{\}}\KatlaNewline{}
\end{SaveVerbatim}
\newcommand\InvMonoidFrexExample[1][]{\UseVerbatim[#1]{InvMonoidFrexExample}}
\begin{SaveVerbatim}[commandchars=\\\{\}]{InvMonoidFrexExample}
\IdrisFunction{lemma}\KatlaSpace{}\IdrisKeyword{:}\KatlaSpace{}\IdrisKeyword{\{}\IdrisBound{x}\KatlaSpace{}\IdrisKeyword{:}\KatlaSpace{}\IdrisType{List}\KatlaSpace{}\IdrisBound{a}\IdrisKeyword{\}}\KatlaSpace{}\IdrisKeyword{\KatlaDash{}>}\KatlaSpace{}\IdrisKeyword{(}\IdrisBound{i}\IdrisKeyword{,}\IdrisBound{j}\KatlaSpace{}\IdrisKeyword{:}\KatlaSpace{}\IdrisBound{a}\IdrisKeyword{)}\KatlaSpace{}\IdrisKeyword{\KatlaDash{}>}\KatlaNewline{}
\KatlaSpace{}\KatlaSpace{}\KatlaSpace{}\KatlaSpace{}\KatlaSpace{}\IdrisFunction{reverse}\KatlaSpace{}\IdrisKeyword{(}\IdrisData{[}\IdrisBound{i}\IdrisData{]}\KatlaSpace{}\IdrisFunction{++}\KatlaSpace{}\IdrisFunction{reverse}\KatlaSpace{}\IdrisBound{x}\KatlaSpace{}\IdrisFunction{++}\KatlaSpace{}\IdrisData{[]}\IdrisKeyword{)}\KatlaSpace{}\IdrisFunction{++}\KatlaSpace{}\IdrisData{[}\IdrisBound{j}\IdrisData{]}\KatlaNewline{}
\KatlaSpace{}\KatlaSpace{}\KatlaSpace{}\KatlaSpace{}\KatlaSpace{}\IdrisKeyword{=}\KatlaNewline{}
\KatlaSpace{}\KatlaSpace{}\KatlaSpace{}\KatlaSpace{}\KatlaSpace{}\IdrisBound{x}\KatlaSpace{}\IdrisFunction{++}\KatlaSpace{}\IdrisData{[}\IdrisBound{i}\IdrisData{,}\KatlaSpace{}\IdrisBound{j}\IdrisData{]}\KatlaNewline{}
\IdrisFunction{lemma}\KatlaSpace{}\IdrisBound{i}\KatlaSpace{}\IdrisBound{j}\KatlaSpace{}\IdrisKeyword{=}\KatlaSpace{}\IdrisFunction{solve}\KatlaSpace{}\IdrisData{1}\KatlaSpace{}\IdrisKeyword{(}\IdrisFunction{Involutive.Frex.Frex}\KatlaSpace{}\IdrisFunction{ListInvMonoid}\IdrisKeyword{)}\KatlaSpace{}\$\KatlaNewline{}
\KatlaSpace{}\KatlaSpace{}\KatlaSpace{}\KatlaSpace{}\KatlaSpace{}\IdrisKeyword{((}\IdrisFunction{Sta}\KatlaSpace{}\IdrisData{[}\IdrisBound{i}\IdrisData{]}\KatlaSpace{}\IdrisFunction{.*.}\KatlaSpace{}\IdrisKeyword{(}\IdrisFunction{Dyn}\KatlaSpace{}\IdrisData{0}\IdrisKeyword{)}\KatlaSpace{}\IdrisFunction{.inv}\KatlaSpace{}\IdrisFunction{.*.}\KatlaSpace{}\IdrisFunction{I1}\IdrisKeyword{)}\KatlaSpace{}\IdrisFunction{.inv}\IdrisKeyword{)}\KatlaSpace{}\IdrisFunction{.*.}\KatlaSpace{}\IdrisFunction{Sta}\KatlaSpace{}\IdrisData{[}\IdrisBound{j}\IdrisData{]}\KatlaNewline{}
\KatlaSpace{}\KatlaSpace{}\KatlaSpace{}\KatlaSpace{}\KatlaSpace{}\IdrisFunction{=\KatlaDash{}=}\KatlaNewline{}
\KatlaSpace{}\KatlaSpace{}\KatlaSpace{}\KatlaSpace{}\KatlaSpace{}\IdrisFunction{Dyn}\KatlaSpace{}\IdrisData{0}\KatlaSpace{}\IdrisFunction{.*.}\KatlaSpace{}\IdrisFunction{Sta}\KatlaSpace{}\IdrisData{[}\IdrisBound{i}\IdrisData{,}\KatlaSpace{}\IdrisBound{j}\IdrisData{]}\KatlaNewline{}
\end{SaveVerbatim}
\newcommand\InvMonoidFralExample[1][]{\UseVerbatim[#1]{InvMonoidFralExample}}
\begin{SaveVerbatim}[commandchars=\\\{\}]{InvMonoidFralExample}
\IdrisFunction{ExampleFral}\KatlaSpace{}\IdrisKeyword{:}\KatlaSpace{}\IdrisKeyword{\{}\IdrisBound{a}\KatlaSpace{}\IdrisKeyword{:}\KatlaSpace{}\IdrisFunction{InvolutiveMonoid}\IdrisKeyword{\}}\KatlaSpace{}\IdrisKeyword{\KatlaDash{}>}\KatlaSpace{}\IdrisKeyword{(}\IdrisBound{x}\IdrisKeyword{,}\IdrisBound{y}\IdrisKeyword{,}\IdrisBound{z}\KatlaSpace{}\IdrisKeyword{:}\KatlaSpace{}\IdrisFunction{U}\KatlaSpace{}\IdrisBound{a}\IdrisKeyword{)}\KatlaNewline{}
\KatlaSpace{}\KatlaSpace{}\IdrisKeyword{\KatlaDash{}>}\KatlaSpace{}\IdrisKeyword{let}\KatlaSpace{}\IdrisKeyword{\%hint}\KatlaSpace{}\IdrisFunction{notation}\KatlaSpace{}\IdrisKeyword{:}\KatlaSpace{}?\KatlaSpace{}\KatlaSpace{}\KatlaSpace{}\KatlaSpace{}\KatlaSpace{}\KatlaSpace{}\KatlaSpace{}\KatlaSpace{}\KatlaSpace{}\KatlaSpace{}\KatlaSpace{}\KatlaSpace{}\KatlaSpace{}\KatlaSpace{}\KatlaSpace{}\KatlaSpace{}\KatlaSpace{}\KatlaSpace{}\KatlaSpace{}\KatlaSpace{}\KatlaSpace{}\KatlaSpace{}\IdrisComment{\KatlaDash{}\KatlaDash{}\KatlaSpace{}Open\KatlaSpace{}notation\KatlaSpace{}hints\KatlaSpace{}for\KatlaSpace{}the\KatlaSpace{}monoid}\KatlaNewline{}
\KatlaSpace{}\KatlaSpace{}\KatlaSpace{}\KatlaSpace{}\KatlaSpace{}\KatlaSpace{}\KatlaSpace{}\KatlaSpace{}\KatlaSpace{}\IdrisFunction{notation}\KatlaSpace{}\IdrisKeyword{=}\KatlaSpace{}\IdrisBound{a}\IdrisFunction{.Notation1}\KatlaSpace{}\KatlaSpace{}\KatlaSpace{}\KatlaSpace{}\KatlaSpace{}\KatlaSpace{}\KatlaSpace{}\KatlaSpace{}\KatlaSpace{}\KatlaSpace{}\KatlaSpace{}\KatlaSpace{}\KatlaSpace{}\KatlaSpace{}\KatlaSpace{}\KatlaSpace{}\KatlaSpace{}\KatlaSpace{}\IdrisComment{\KatlaDash{}\KatlaDash{}\KatlaSpace{}for\KatlaSpace{}infix\KatlaSpace{}operator\KatlaSpace{}(.*.)\KatlaSpace{}and}\KatlaNewline{}
\KatlaSpace{}\KatlaSpace{}\KatlaSpace{}\KatlaSpace{}\KatlaSpace{}\IdrisKeyword{in}\KatlaSpace{}\IdrisBound{a}\IdrisFunction{.rel}\KatlaSpace{}\KatlaSpace{}\KatlaSpace{}\KatlaSpace{}\KatlaSpace{}\KatlaSpace{}\KatlaSpace{}\KatlaSpace{}\KatlaSpace{}\KatlaSpace{}\KatlaSpace{}\KatlaSpace{}\KatlaSpace{}\KatlaSpace{}\KatlaSpace{}\KatlaSpace{}\KatlaSpace{}\KatlaSpace{}\KatlaSpace{}\KatlaSpace{}\KatlaSpace{}\KatlaSpace{}\KatlaSpace{}\KatlaSpace{}\KatlaSpace{}\KatlaSpace{}\KatlaSpace{}\KatlaSpace{}\KatlaSpace{}\KatlaSpace{}\KatlaSpace{}\KatlaSpace{}\KatlaSpace{}\KatlaSpace{}\KatlaSpace{}\KatlaSpace{}\IdrisComment{\KatlaDash{}\KatlaDash{}\KatlaSpace{}postfix\KatlaSpace{}operator\KatlaSpace{}(.inv)}\KatlaNewline{}
\KatlaSpace{}\KatlaSpace{}\KatlaSpace{}\KatlaSpace{}\KatlaSpace{}\KatlaSpace{}\KatlaSpace{}\IdrisKeyword{(}\IdrisBound{x}\KatlaSpace{}\IdrisFunction{.*.}\KatlaSpace{}\IdrisBound{y}\IdrisFunction{.inv}\KatlaSpace{}\IdrisFunction{.*.}\KatlaSpace{}\IdrisBound{z}\IdrisKeyword{)}\IdrisFunction{.inv}\KatlaNewline{}
\KatlaSpace{}\KatlaSpace{}\KatlaSpace{}\KatlaSpace{}\KatlaSpace{}\KatlaSpace{}\KatlaSpace{}\IdrisKeyword{(}\IdrisBound{z}\IdrisFunction{.inv}\KatlaSpace{}\IdrisFunction{.*.}\KatlaSpace{}\IdrisBound{y}\KatlaSpace{}\IdrisFunction{.*.}\KatlaSpace{}\IdrisBound{x}\IdrisFunction{.inv}\IdrisKeyword{)}\KatlaNewline{}
\IdrisFunction{ExampleFral}\KatlaSpace{}\IdrisBound{x}\KatlaSpace{}\IdrisBound{y}\KatlaSpace{}\IdrisBound{z}\KatlaSpace{}\IdrisKeyword{=}\KatlaNewline{}
\KatlaSpace{}\KatlaSpace{}\IdrisKeyword{let}\KatlaSpace{}\IdrisKeyword{\%hint}\KatlaSpace{}\IdrisFunction{notation}\KatlaSpace{}\IdrisKeyword{:}\KatlaSpace{}?\KatlaSpace{}\KatlaSpace{}\KatlaSpace{}\KatlaSpace{}\KatlaSpace{}\KatlaSpace{}\KatlaSpace{}\KatlaSpace{}\KatlaSpace{}\KatlaSpace{}\KatlaSpace{}\KatlaSpace{}\KatlaSpace{}\KatlaSpace{}\KatlaSpace{}\KatlaSpace{}\KatlaSpace{}\KatlaSpace{}\KatlaSpace{}\KatlaSpace{}\KatlaSpace{}\KatlaSpace{}\KatlaSpace{}\KatlaSpace{}\KatlaSpace{}\IdrisComment{\KatlaDash{}\KatlaDash{}\KatlaSpace{}ditto,\KatlaSpace{}but\KatlaSpace{}for\KatlaSpace{}terms}\KatlaNewline{}
\KatlaSpace{}\KatlaSpace{}\KatlaSpace{}\KatlaSpace{}\KatlaSpace{}\KatlaSpace{}\IdrisFunction{notation}\KatlaSpace{}\IdrisKeyword{=}\KatlaSpace{}\IdrisFunction{Involutive.Notation.multiplicative1}\KatlaNewline{}
\KatlaSpace{}\KatlaSpace{}\IdrisKeyword{in}\KatlaSpace{}\IdrisFunction{solve}\KatlaSpace{}\IdrisData{3}\KatlaSpace{}\IdrisKeyword{(}\IdrisFunction{Involutive.Free.FreeInvolutiveMonoidOver}\KatlaSpace{}\IdrisData{3}\IdrisKeyword{)}\KatlaSpace{}\$\KatlaNewline{}
\KatlaSpace{}\KatlaSpace{}\IdrisKeyword{(}\IdrisFunction{X}\KatlaSpace{}\IdrisData{0}\KatlaSpace{}\IdrisFunction{.*.}\KatlaSpace{}\IdrisKeyword{(}\IdrisFunction{X}\KatlaSpace{}\IdrisData{1}\IdrisKeyword{)}\IdrisFunction{.inv}\KatlaSpace{}\IdrisFunction{.*.}\KatlaSpace{}\IdrisFunction{X}\KatlaSpace{}\IdrisData{2}\IdrisKeyword{)}\IdrisFunction{.inv}\KatlaSpace{}\IdrisFunction{=\KatlaDash{}=}\KatlaSpace{}\IdrisKeyword{(}\IdrisFunction{X}\KatlaSpace{}\IdrisData{2}\IdrisKeyword{)}\IdrisFunction{.inv}\KatlaSpace{}\IdrisFunction{.*.}\KatlaSpace{}\IdrisFunction{X}\KatlaSpace{}\IdrisData{1}\KatlaSpace{}\IdrisFunction{.*.}\KatlaSpace{}\IdrisKeyword{(}\IdrisFunction{X}\KatlaSpace{}\IdrisData{0}\IdrisKeyword{)}\IdrisFunction{.inv}\KatlaNewline{}
\end{SaveVerbatim}

%% file: reflection-listings.tex
\newcommand\ExOne[1][]{\UseVerb[#1]{ExOne}}
\begin{SaveVerbatim}[commandchars=\\\{\}]{ExOne}
\IdrisKeyword{(}\IdrisBound{x}\KatlaSpace{}\IdrisFunction{+}\KatlaSpace{}\IdrisData{1}\IdrisKeyword{)}\KatlaSpace{}\IdrisFunction{+}\KatlaSpace{}\IdrisBound{y}\KatlaSpace{}\IdrisKeyword{=}\KatlaSpace{}\IdrisBound{x}\KatlaSpace{}\IdrisFunction{+}\KatlaSpace{}\IdrisKeyword{(}\IdrisData{1}\KatlaSpace{}\IdrisFunction{+}\KatlaSpace{}\IdrisBound{y}\IdrisKeyword{)}
\end{SaveVerbatim}
\newcommand\ExTwo[1][]{\UseVerb[#1]{ExTwo}}
\begin{SaveVerbatim}[commandchars=\\\{\}]{ExTwo}
\IdrisKeyword{(}\IdrisBound{x}\KatlaSpace{}\IdrisFunction{+}\KatlaSpace{}\IdrisData{1}\IdrisKeyword{)}\KatlaSpace{}\IdrisFunction{+}\KatlaSpace{}\IdrisBound{y}\KatlaSpace{}\IdrisKeyword{=}\KatlaSpace{}\IdrisBound{x}\KatlaSpace{}\IdrisFunction{+}\KatlaSpace{}\IdrisData{S}\KatlaSpace{}\IdrisBound{y}
\end{SaveVerbatim}
\newcommand\ReflectionY[1][]{\UseVerb[#1]{ReflectionY}}
\begin{SaveVerbatim}[commandchars=\\\{\}]{ReflectionY}
\IdrisBound{y}
\end{SaveVerbatim}
\newcommand\ReflectionSY[1][]{\UseVerb[#1]{ReflectionSY}}
\begin{SaveVerbatim}[commandchars=\\\{\}]{ReflectionSY}
\IdrisData{S}\KatlaSpace{}\IdrisBound{y}
\end{SaveVerbatim}
\newcommand\ExThree[1][]{\UseVerb[#1]{ExThree}}
\begin{SaveVerbatim}[commandchars=\\\{\}]{ExThree}
\IdrisFunction{Dyn}\KatlaSpace{}\IdrisData{0}\KatlaSpace{}\IdrisFunction{.+.}\KatlaSpace{}\IdrisFunction{Sta}\KatlaSpace{}\IdrisData{1}\KatlaSpace{}\IdrisFunction{.+.}\KatlaSpace{}\IdrisFunction{Dyn}\KatlaSpace{}\IdrisData{1}\KatlaSpace{}\IdrisFunction{=-=}\KatlaSpace{}\IdrisFunction{Dyn}\KatlaSpace{}\IdrisData{0}\KatlaSpace{}\IdrisFunction{.+.}\KatlaSpace{}\IdrisFunction{Dyn}\KatlaSpace{}\IdrisData{2}
\end{SaveVerbatim}
\newcommand\MagicExOne[1][]{\UseVerbatim[#1]{MagicExOne}}
\begin{SaveVerbatim}[commandchars=\\\{\}]{MagicExOne}
\IdrisFunction{units}\KatlaSpace{}\IdrisKeyword{:}\KatlaSpace{}\IdrisKeyword{\{}\IdrisBound{a}\IdrisKeyword{,}\KatlaSpace{}\IdrisBound{b}\KatlaSpace{}\IdrisKeyword{:}\KatlaSpace{}\IdrisType{Nat}\IdrisKeyword{\}}\KatlaSpace{}\IdrisKeyword{->}\KatlaSpace{}\IdrisKeyword{(}\IdrisData{0}\IdrisFunction{\KatlaSpace{}+\KatlaSpace{}}\IdrisKeyword{(}\IdrisBound{a}\IdrisFunction{\KatlaSpace{}+\KatlaSpace{}0)}\IdrisKeyword{)}\IdrisFunction{\KatlaSpace{}+\KatlaSpace{}}\IdrisBound{b}\IdrisFunction{\KatlaSpace{}+\KatlaSpace{}}\IdrisData{0}\IdrisFunction{\KatlaSpace{}}\IdrisKeyword{=}\IdrisFunction{\KatlaSpace{}}\IdrisBound{a}\IdrisFunction{\KatlaSpace{}+\KatlaSpace{}b}
\IdrisFunction{units}\IdrisBound{\KatlaSpace{}}\IdrisKeyword{=}\IdrisBound{\KatlaSpace{}}\IdrisKeyword{\%runElab}\IdrisBound{\KatlaSpace{}}\IdrisFunction{frexMagic}\IdrisBound{\KatlaSpace{}}\IdrisFunction{MonoidFrexlet}\IdrisBound{\KatlaSpace{}Additive}

\end{SaveVerbatim}

%% file: intro-printer.tex
\usepackage{amsmath}
\usepackage{mathtools}
\usepackage{amssymb}
\usepackage{newunicodechar}
\usepackage{newunicodechar}
\newunicodechar{ε}{\ensuremath{\varepsilon}}
\newunicodechar{•}{\ensuremath{\bullet}}
\newunicodechar{⟨}{\ensuremath{\langle}}
\newunicodechar{⟩}{\ensuremath{\rangle}}
\usepackage{ifthen}
\usepackage{mathpartir}
\newboolean{explanation}
\setboolean{explanation}{false}
\newcommand\explainabove[2]{\overset{\overset{%
\clap{\text{\scriptsize #2}}}%
{\downarrow}}{#1}}
\newcommand\explainbelow[2]{\underset{\underset{%
\clap{\text{\scriptsize #2}}}%
{\uparrow}}{#1}}
\newcommand\negate[1]{%
  \ifthenelse{\boolean{#1}}{%
    \setboolean{#1}{false}%
    }{%
    \setboolean{#1}{true}%
    }%
}
\newcommand\explain[2]{%
  \ifthenelse{\boolean{explanation}}{%
    \explainabove{#1}{#2}%
  }{%
    \explainbelow{#1}{#2}%
  }
  \negate{explanation}
}
\usepackage{suffix}
\WithSuffix\newcommand\explain*[2]{\explain{#1}{\llap{#2}}}
\newcommand\IntroPrintedListing{
\begin{mathpar}\mprset{flushleft}
(x • 3) • [2]\explain={⟨ Right neutrality ]}
(x • 3) • 2 • [ε]\explain={⟨ Left neutrality ]}
(x • [3]) • 2 • ε • ε\explain={⟨ Right neutrality ]}
(x • 3 • [ε]) • 2 • ε • ε\explain={⟨ Left neutrality ]}
([x] • 3 • ε • ε) • 2 • ε • ε\explain*={⟨ Right neutrality ]}
([x • ε] • 3 • ε • ε) • 2 • ε • ε\explain={⟨ Right neutrality ]}
([(x • ε) • ε] • 3 • ε • ε) • 2 • ε • ε\explain={⟨ Left neutrality ]}
(([ε] • (x • ε) • ε) • 3 • ε • ε) • 2 • ε • ε\explain*={[ Evaluate ⟩}
[(0 • (x • ε) • ε) • 3 • ε • ε] • 2 • ε • ε\explain={⟨ Associativity ]}
(0 • [((x • ε) • ε) • 3 • ε • ε]) • 2 • ε • ε\explain={[ Associativity ⟩}
(0 • [((x • ε) • ε) • 3] • ε • ε) • 2 • ε • ε\explain*={[ Commutativity ⟩}
(0 • [(3 • (x • ε) • ε) • ε • ε]) • 2 • ε • ε\explain={⟨ Associativity ]}
[0 • 3 • ((x • ε) • ε) • ε • ε] • 2 • ε • ε\explain={[ Associativity ⟩}
((0 • 3) • ((x • ε) • ε) • [ε • ε]) • 2 • ε • ε\explain*={[ Left neutrality ⟩}
((0 • 3) • [(x • ε) • ε] • ε) • 2 • ε • ε\explain={⟨ Associativity ]}
((0 • 3) • (x • [ε • ε]) • ε) • 2 • ε • ε\explain={[ Left neutrality ⟩}
([0 • 3] • (x • ε) • ε) • 2 • ε • ε\explain*={[ Evaluate ⟩}
(3 • (x • ε) • ε) • 2 • ε • ε\explain={⟨ Associativity ]}
3 • [((x • ε) • ε) • 2 • ε • ε]\explain={[ Associativity ⟩}
3 • [((x • ε) • ε) • 2] • ε • ε\explain*={[ Commutativity ⟩}
3 • [(2 • (x • ε) • ε) • ε • ε]\explain={⟨ Associativity ]}
3 • 2 • ((x • ε) • ε) • ε • ε\explain={[ Associativity ⟩}
(3 • 2) • ((x • ε) • ε) • [ε • ε]\explain*={[ Left neutrality ⟩}
(3 • 2) • [(x • ε) • ε] • ε\explain={⟨ Associativity ]}
(3 • 2) • (x • [ε • ε]) • ε\explain={[ Left neutrality ⟩}
[3 • 2] • (x • ε) • ε\explain={[ Evaluate ⟩}
5 • [(x • ε) • ε]\explain={[ Right neutrality ⟩}
5 • [x • ε]\explain*={[ Right neutrality ⟩}
5 • x
\end{mathpar}
}

\begin{SaveVerbatim}[commandchars=\\\{\}]{ExtractedProof}
\IdrisFunction{extractedProof}\KatlaSpace{}\IdrisKeyword{:}\KatlaSpace{}\IdrisKeyword{((}\IdrisFunction{Dyn'}\KatlaSpace{}\IdrisData{0}\KatlaSpace{}\IdrisFunction{:+:}\KatlaSpace{}\IdrisFunction{Sta'}\KatlaSpace{}\IdrisData{3}\IdrisKeyword{)}\KatlaSpace{}\IdrisFunction{:+:}\KatlaSpace{}\IdrisFunction{Sta'}\KatlaSpace{}\IdrisData{2}\KatlaSpace{}\IdrisFunction{~~}\KatlaSpace{}\IdrisFunction{Sta'}\KatlaSpace{}\IdrisData{5}\KatlaSpace{}\IdrisFunction{:+:}\KatlaSpace{}\IdrisFunction{Dyn'}\KatlaSpace{}\IdrisData{0}\IdrisKeyword{)}\KatlaSpace{}\IdrisKeyword{\{}\IdrisBound{vars}\KatlaSpace{}\IdrisKeyword{=}\KatlaSpace{}\IdrisType{Fin}\KatlaSpace{}\IdrisData{1}\IdrisKeyword{\}}
\IdrisFunction{extractedProof}
\KatlaSpace{}\KatlaSpace{}\IdrisKeyword{=}\KatlaSpace{}\IdrisFunction{Frex.prove}\KatlaSpace{}\IdrisKeyword{_}\KatlaSpace{}\IdrisKeyword{(}\IdrisFunction{Monoid.Commutative.Frex}\KatlaSpace{}\IdrisFunction{Nat.Additive}\IdrisKeyword{)}
\KatlaSpace{}\KatlaSpace{}$\KatlaSpace{}\IdrisKeyword{((}\IdrisFunction{Dyn}\KatlaSpace{}\IdrisData{0}\KatlaSpace{}\IdrisFunction{:+:}\KatlaSpace{}\IdrisFunction{Sta}\KatlaSpace{}\IdrisData{3}\IdrisKeyword{)}\KatlaSpace{}\IdrisFunction{:+:}\KatlaSpace{}\IdrisFunction{Sta}\KatlaSpace{}\IdrisData{2}\KatlaSpace{}\IdrisFunction{=-=}\KatlaSpace{}\IdrisFunction{Sta}\KatlaSpace{}\IdrisData{5}\KatlaSpace{}\IdrisFunction{:+:}\KatlaSpace{}\IdrisFunction{Dyn}\KatlaSpace{}\IdrisData{0}\IdrisKeyword{)}

\end{SaveVerbatim}

%% file: core-listings.tex
\newcommand\Visibility[1][]{\UseVerbatim[#1]{Visibility}}
\begin{SaveVerbatim}[commandchars=\\\{\}]{Visibility}
\IdrisKeyword{data}\KatlaSpace{}\IdrisType{Visibility}\KatlaSpace{}\IdrisKeyword{=}\KatlaSpace{}\IdrisData{Visible}\KatlaSpace{}\IdrisKeyword{|}\KatlaSpace{}\IdrisData{Hidden}\KatlaSpace{}\IdrisKeyword{|}\KatlaSpace{}\IdrisData{Auto}
\end{SaveVerbatim}
\newcommand\metaPi[1][]{\UseVerbatim[#1]{metaPi}}
\begin{SaveVerbatim}[commandchars=\\\{\}]{metaPi}
\IdrisFunction{Pi}\KatlaSpace{}\IdrisKeyword{:}\KatlaSpace{}\IdrisType{Visibility}\KatlaSpace{}\IdrisKeyword{->}\KatlaSpace{}\IdrisKeyword{(}\IdrisBound{a}\KatlaSpace{}\IdrisKeyword{:}\KatlaSpace{}\IdrisType{Type}\IdrisKeyword{)}\KatlaSpace{}\IdrisKeyword{->}\KatlaSpace{}\IdrisKeyword{(}\IdrisBound{a}\KatlaSpace{}\IdrisKeyword{->}\KatlaSpace{}\IdrisType{Type}\IdrisKeyword{)}\KatlaSpace{}\IdrisKeyword{->}\KatlaSpace{}\IdrisType{Type}
\IdrisFunction{Pi}\KatlaSpace{}\IdrisData{Visible}\KatlaSpace{}\IdrisBound{a}\KatlaSpace{}\IdrisBound{b}\KatlaSpace{}\IdrisKeyword{=}\KatlaSpace{}\IdrisKeyword{(}\IdrisBound{x}\KatlaSpace{}\IdrisKeyword{:}\KatlaSpace{}\IdrisBound{a}\IdrisKeyword{)}\KatlaSpace{}\IdrisKeyword{->}\KatlaSpace{}\IdrisBound{b}\KatlaSpace{}\IdrisBound{x}
\IdrisFunction{Pi}\KatlaSpace{}\IdrisData{Hidden}\KatlaSpace{}\KatlaSpace{}\IdrisBound{a}\KatlaSpace{}\IdrisBound{b}\KatlaSpace{}\IdrisKeyword{=}\KatlaSpace{}\IdrisKeyword{\{}\IdrisBound{x}\KatlaSpace{}\IdrisKeyword{:}\KatlaSpace{}\IdrisBound{a}\IdrisKeyword{\}}\KatlaSpace{}\IdrisKeyword{->}\KatlaSpace{}\IdrisBound{b}\KatlaSpace{}\IdrisBound{x}
\IdrisFunction{Pi}\KatlaSpace{}\IdrisData{Auto}\KatlaSpace{}\KatlaSpace{}\KatlaSpace{}\KatlaSpace{}\IdrisBound{a}\KatlaSpace{}\IdrisBound{b}\KatlaSpace{}\IdrisKeyword{=}\KatlaSpace{}\IdrisKeyword{\{auto}\KatlaSpace{}\IdrisBound{x}\KatlaSpace{}\IdrisKeyword{:}\KatlaSpace{}\IdrisBound{a}\IdrisKeyword{\}}\KatlaSpace{}\IdrisKeyword{->}\KatlaSpace{}\IdrisBound{b}\KatlaSpace{}\IdrisBound{x}
\end{SaveVerbatim}
\newcommand\metaPI[1][]{\UseVerbatim[#1]{metaPI}}
\begin{SaveVerbatim}[commandchars=\\\{\}]{metaPI}
\IdrisFunction{PI}\KatlaSpace{}\IdrisKeyword{:}\KatlaSpace{}\IdrisKeyword{(}\IdrisBound{n}\KatlaSpace{}\IdrisKeyword{:}\KatlaSpace{}\IdrisType{Nat}\IdrisKeyword{)}\KatlaSpace{}\IdrisKeyword{->}\KatlaSpace{}\IdrisType{Visibility}\KatlaSpace{}\IdrisKeyword{->}\KatlaSpace{}\IdrisKeyword{(}\IdrisBound{a}\KatlaSpace{}\IdrisKeyword{:}\KatlaSpace{}\IdrisType{Type}\IdrisKeyword{)}\KatlaSpace{}\IdrisKeyword{->}\KatlaSpace{}\IdrisKeyword{(}\IdrisType{Vect}\KatlaSpace{}\IdrisBound{n}\KatlaSpace{}\IdrisBound{a}\KatlaSpace{}\IdrisKeyword{->}\KatlaSpace{}\IdrisType{Type}\IdrisKeyword{)}\KatlaSpace{}\IdrisKeyword{->}\KatlaSpace{}\IdrisType{Type}
\IdrisFunction{PI}\KatlaSpace{}\IdrisData{Z}\KatlaSpace{}\KatlaSpace{}\KatlaSpace{}\KatlaSpace{}\KatlaSpace{}\IdrisBound{vis}\KatlaSpace{}\IdrisBound{a}\KatlaSpace{}\IdrisBound{b}\KatlaSpace{}\IdrisKeyword{=}\KatlaSpace{}\IdrisBound{b}\KatlaSpace{}\IdrisData{[]}
\IdrisFunction{PI}\KatlaSpace{}\IdrisKeyword{(}\IdrisData{S}\KatlaSpace{}\IdrisBound{n}\IdrisKeyword{)}\KatlaSpace{}\IdrisBound{vis}\KatlaSpace{}\IdrisBound{a}\KatlaSpace{}\IdrisBound{b}\KatlaSpace{}\IdrisKeyword{=}\KatlaSpace{}\IdrisFunction{Pi}\KatlaSpace{}\IdrisBound{vis}\KatlaSpace{}\IdrisBound{a}\KatlaSpace{}\IdrisKeyword{(\textbackslash{}}\KatlaSpace{}\IdrisBound{x}\KatlaSpace{}\IdrisKeyword{=>}\KatlaSpace{}\IdrisFunction{PI}\KatlaSpace{}\IdrisBound{n}\KatlaSpace{}\IdrisBound{vis}\KatlaSpace{}\IdrisBound{a}\KatlaSpace{}\IdrisKeyword{(}\IdrisBound{b}\KatlaSpace{}\IdrisFunction{.}\KatlaSpace{}\IdrisKeyword{(}\IdrisBound{x}\KatlaSpace{}\IdrisData{::}\IdrisKeyword{)))}
\end{SaveVerbatim}
\newcommand\setoidEither[1][]{\UseVerbatim[#1]{setoidEither}}
\begin{SaveVerbatim}[commandchars=\\\{\}]{setoidEither}
\IdrisFunction{either}\KatlaSpace{}\IdrisKeyword{:}\KatlaSpace{}\IdrisKeyword{\{0}\KatlaSpace{}\IdrisBound{a}\IdrisKeyword{,}\KatlaSpace{}\IdrisBound{b}\IdrisKeyword{,}\KatlaSpace{}\IdrisBound{c}\KatlaSpace{}\IdrisKeyword{:}\KatlaSpace{}\IdrisType{Setoid}\IdrisKeyword{\}}\KatlaSpace{}\IdrisKeyword{->}\KatlaSpace{}\IdrisKeyword{(}\IdrisBound{a}\KatlaSpace{}\IdrisType{~>}\KatlaSpace{}\IdrisBound{c}\IdrisKeyword{)}\KatlaSpace{}\IdrisKeyword{->}\KatlaSpace{}\IdrisKeyword{(}\IdrisBound{b}\KatlaSpace{}\IdrisType{~>}\KatlaSpace{}\IdrisBound{c}\IdrisKeyword{)}\KatlaSpace{}\IdrisKeyword{->}\KatlaSpace{}\IdrisKeyword{(}\IdrisBound{a}\KatlaSpace{}\IdrisFunction{`Either`}\KatlaSpace{}\IdrisBound{b}\IdrisKeyword{)}\KatlaSpace{}\IdrisType{~>}\KatlaSpace{}\IdrisBound{c}
\end{SaveVerbatim}
\newcommand\extEnv[1][]{\UseVerbatim[#1]{extEnv}}
\begin{SaveVerbatim}[commandchars=\\\{\}]{extEnv}
\IdrisFunction{extEnv}\KatlaSpace{}\IdrisKeyword{:}\KatlaSpace{}\IdrisKeyword{\{}\IdrisBound{a}\KatlaSpace{}\IdrisKeyword{:}\KatlaSpace{}\IdrisType{Model}\KatlaSpace{}\IdrisBound{pres}\IdrisKeyword{\}}\KatlaSpace{}\IdrisKeyword{->}\KatlaSpace{}\IdrisKeyword{\{}\IdrisBound{x}\KatlaSpace{}\IdrisKeyword{:}\KatlaSpace{}\IdrisType{Setoid}\IdrisKeyword{\}}\KatlaSpace{}\IdrisKeyword{->}\KatlaSpace{}\IdrisKeyword{(}\IdrisBound{ext}\KatlaSpace{}\IdrisKeyword{:}\KatlaSpace{}\IdrisType{Extension}\KatlaSpace{}\IdrisBound{a}\KatlaSpace{}\IdrisBound{x}\IdrisKeyword{)}\KatlaSpace{}\IdrisKeyword{->}
\KatlaSpace{}\KatlaSpace{}\IdrisFunction{Either}\KatlaSpace{}\IdrisKeyword{(}\IdrisFunction{cast}\KatlaSpace{}\IdrisBound{a}\IdrisKeyword{)}\KatlaSpace{}\IdrisBound{x}\KatlaSpace{}\IdrisType{~>}\KatlaSpace{}\IdrisFunction{cast}\KatlaSpace{}\IdrisBound{ext}\IdrisFunction{.Model}
\IdrisFunction{extEnv}\KatlaSpace{}\IdrisBound{ext}\KatlaSpace{}\IdrisKeyword{=}\KatlaSpace{}\IdrisFunction{either}\KatlaSpace{}\IdrisBound{ext}\IdrisFunction{.Embed.H}\KatlaSpace{}\IdrisBound{ext}\IdrisFunction{.Var}
\end{SaveVerbatim}
\newcommand\irrelevantCast[1][]{\UseVerbatim[#1]{irrelevantCast}}
\begin{SaveVerbatim}[commandchars=\\\{\}]{irrelevantCast}
\IdrisFunction{irrelevantCast}\KatlaSpace{}\IdrisKeyword{:}\KatlaSpace{}\IdrisKeyword{(0}\KatlaSpace{}\IdrisBound{a}\KatlaSpace{}\IdrisKeyword{:}\KatlaSpace{}\IdrisType{Type}\IdrisKeyword{)}\KatlaSpace{}\IdrisKeyword{->}\KatlaSpace{}\IdrisType{Setoid}
\end{SaveVerbatim}
\newcommand\relevantCast[1][]{\UseVerbatim[#1]{relevantCast}}
\begin{SaveVerbatim}[commandchars=\\\{\}]{relevantCast}
\IdrisFunction{cast}\KatlaSpace{}\IdrisKeyword{:}\KatlaSpace{}\IdrisKeyword{(}\IdrisBound{a}\KatlaSpace{}\IdrisKeyword{:}\KatlaSpace{}\IdrisType{Type}\IdrisKeyword{)}\KatlaSpace{}\IdrisKeyword{->}\KatlaSpace{}\IdrisType{Setoid}
\end{SaveVerbatim}
\newcommand\solveVectType[1][]{\UseVerbatim[#1]{solveVectType}}
\begin{SaveVerbatim}[commandchars=\\\{\}]{solveVectType}
\IdrisFunction{solveVect}\KatlaSpace{}\IdrisKeyword{:}\KatlaSpace{}\IdrisKeyword{\{0}\KatlaSpace{}\IdrisBound{n}\KatlaSpace{}\IdrisKeyword{:}\KatlaSpace{}\IdrisType{Nat}\IdrisKeyword{\}}\KatlaSpace{}\IdrisKeyword{->}\KatlaSpace{}\IdrisKeyword{\{}\IdrisBound{pres}\KatlaSpace{}\IdrisKeyword{:}\KatlaSpace{}\IdrisType{Presentation}\IdrisKeyword{\}}\KatlaSpace{}\IdrisKeyword{->}\KatlaSpace{}\IdrisKeyword{\{}\IdrisBound{a}\KatlaSpace{}\IdrisKeyword{:}\KatlaSpace{}\IdrisType{Model}\KatlaSpace{}\IdrisBound{pres}\IdrisKeyword{\}}\KatlaSpace{}\IdrisKeyword{->}
\KatlaSpace{}\KatlaSpace{}\IdrisKeyword{(}\IdrisBound{frex}\KatlaSpace{}\IdrisKeyword{:}\KatlaSpace{}\IdrisType{Frex}\KatlaSpace{}\IdrisBound{a}\KatlaSpace{}\IdrisKeyword{(}\IdrisFunction{irrelevantCast}\KatlaSpace{}$\KatlaSpace{}\IdrisType{Fin}\KatlaSpace{}\IdrisBound{n}\IdrisKeyword{))}\KatlaSpace{}\IdrisKeyword{->}\KatlaSpace{}\IdrisKeyword{(}\IdrisBound{env}\KatlaSpace{}\IdrisKeyword{:}\KatlaSpace{}\IdrisType{Vect}\KatlaSpace{}\IdrisBound{n}\KatlaSpace{}\IdrisKeyword{(}\IdrisFunction{U}\KatlaSpace{}\IdrisBound{a}\IdrisKeyword{))}\KatlaSpace{}\IdrisKeyword{->}
\KatlaSpace{}\KatlaSpace{}\IdrisKeyword{(}\IdrisBound{eq}\KatlaSpace{}\IdrisKeyword{:}\KatlaSpace{}\IdrisKeyword{(}\KatlaSpace{}\IdrisType{Term}\KatlaSpace{}\IdrisBound{pres}\IdrisFunction{.signature}\KatlaSpace{}\IdrisKeyword{(}\IdrisFunction{U}\KatlaSpace{}\IdrisBound{a}\KatlaSpace{}\IdrisType{`Either`}\KatlaSpace{}\IdrisType{Fin}\KatlaSpace{}\IdrisBound{n}\IdrisKeyword{)}
\KatlaSpace{}\KatlaSpace{}\KatlaSpace{}\KatlaSpace{}\KatlaSpace{}\KatlaSpace{}\KatlaSpace{}\KatlaSpace{}\IdrisType{,}\KatlaSpace{}\IdrisType{Term}\KatlaSpace{}\IdrisBound{pres}\IdrisFunction{.signature}\KatlaSpace{}\IdrisKeyword{(}\IdrisFunction{U}\KatlaSpace{}\IdrisBound{a}\KatlaSpace{}\IdrisType{`Either`}\KatlaSpace{}\IdrisType{Fin}\KatlaSpace{}\IdrisBound{n}\IdrisKeyword{)))}\KatlaSpace{}\IdrisKeyword{->}
\KatlaSpace{}\KatlaSpace{}\IdrisKeyword{\{auto}\KatlaSpace{}\IdrisBound{prf}\KatlaSpace{}\IdrisKeyword{:}\KatlaSpace{}\IdrisBound{frex}\IdrisFunction{.Data.Model.rel}
\KatlaSpace{}\KatlaSpace{}\KatlaSpace{}\KatlaSpace{}\KatlaSpace{}\IdrisKeyword{(}\IdrisBound{frex}\IdrisFunction{.Sem}\KatlaSpace{}\IdrisKeyword{(}\IdrisFunction{fst}\KatlaSpace{}\IdrisBound{eq}\IdrisKeyword{)}\KatlaSpace{}\IdrisKeyword{(}\IdrisFunction{frexEnv}\KatlaSpace{}\IdrisKeyword{\{}\IdrisBound{x}\KatlaSpace{}\IdrisKeyword{=}\KatlaSpace{}\IdrisFunction{cast}\KatlaSpace{}$\KatlaSpace{}\IdrisType{Fin}\KatlaSpace{}\IdrisBound{n}\IdrisKeyword{\}}\KatlaSpace{}\IdrisBound{frex}\IdrisKeyword{)}\IdrisFunction{.H}\IdrisKeyword{)}
\KatlaSpace{}\KatlaSpace{}\KatlaSpace{}\KatlaSpace{}\KatlaSpace{}\IdrisKeyword{(}\IdrisBound{frex}\IdrisFunction{.Sem}\KatlaSpace{}\IdrisKeyword{(}\IdrisFunction{snd}\KatlaSpace{}\IdrisBound{eq}\IdrisKeyword{)}\KatlaSpace{}\IdrisKeyword{(}\IdrisFunction{frexEnv}\KatlaSpace{}\IdrisKeyword{\{}\IdrisBound{x}\KatlaSpace{}\IdrisKeyword{=}\KatlaSpace{}\IdrisFunction{cast}\KatlaSpace{}$\KatlaSpace{}\IdrisType{Fin}\KatlaSpace{}\IdrisBound{n}\IdrisKeyword{\}}\KatlaSpace{}\IdrisBound{frex}\IdrisKeyword{)}\IdrisFunction{.H}\IdrisKeyword{)\}}
\KatlaSpace{}\KatlaSpace{}\KatlaSpace{}\KatlaSpace{}\KatlaSpace{}\IdrisKeyword{->}
\KatlaSpace{}\KatlaSpace{}\IdrisBound{a}\IdrisFunction{.rel}\KatlaSpace{}\IdrisKeyword{(}\IdrisBound{a}\IdrisFunction{.Sem}\KatlaSpace{}\IdrisKeyword{(}\IdrisFunction{fst}\KatlaSpace{}\IdrisBound{eq}\IdrisKeyword{)}\KatlaSpace{}\IdrisKeyword{(}\IdrisFunction{either}\KatlaSpace{}\IdrisFunction{Prelude.id}\KatlaSpace{}\IdrisKeyword{(}\IdrisFunction{flip}\KatlaSpace{}\IdrisFunction{Vect.index}\KatlaSpace{}\IdrisBound{env}\IdrisKeyword{)))}
\KatlaSpace{}\KatlaSpace{}\KatlaSpace{}\KatlaSpace{}\KatlaSpace{}\KatlaSpace{}\KatlaSpace{}\KatlaSpace{}\IdrisKeyword{(}\IdrisBound{a}\IdrisFunction{.Sem}\KatlaSpace{}\IdrisKeyword{(}\IdrisFunction{snd}\KatlaSpace{}\IdrisBound{eq}\IdrisKeyword{)}\KatlaSpace{}\IdrisKeyword{(}\IdrisFunction{either}\KatlaSpace{}\IdrisFunction{Prelude.id}\KatlaSpace{}\IdrisKeyword{(}\IdrisFunction{flip}\KatlaSpace{}\IdrisFunction{Vect.index}\KatlaSpace{}\IdrisBound{env}\IdrisKeyword{)))}
\end{SaveVerbatim}
\newcommand\solveType[1][]{\UseVerbatim[#1]{solveType}}
\begin{SaveVerbatim}[commandchars=\\\{\}]{solveType}
\IdrisFunction{solve}\KatlaSpace{}\IdrisKeyword{:}\KatlaSpace{}\IdrisKeyword{(}\IdrisBound{n}\KatlaSpace{}\IdrisKeyword{:}\KatlaSpace{}\IdrisType{Nat}\IdrisKeyword{)}\KatlaSpace{}\IdrisKeyword{->}\KatlaSpace{}\IdrisKeyword{\{}\IdrisBound{pres}\KatlaSpace{}\IdrisKeyword{:}\KatlaSpace{}\IdrisType{Presentation}\IdrisKeyword{\}}\KatlaSpace{}\IdrisKeyword{->}\KatlaSpace{}\IdrisKeyword{\{}\IdrisBound{a}\KatlaSpace{}\IdrisKeyword{:}\KatlaSpace{}\IdrisType{Model}\KatlaSpace{}\IdrisBound{pres}\IdrisKeyword{\}}\KatlaSpace{}\IdrisKeyword{->}
\KatlaSpace{}\KatlaSpace{}\IdrisKeyword{(}\IdrisBound{frex}\KatlaSpace{}\IdrisKeyword{:}\KatlaSpace{}\IdrisType{Frex}\KatlaSpace{}\IdrisBound{a}\KatlaSpace{}\IdrisKeyword{(}\IdrisFunction{cast}\KatlaSpace{}$\KatlaSpace{}\IdrisType{Fin}\KatlaSpace{}\IdrisBound{n}\IdrisKeyword{))}\KatlaSpace{}\IdrisKeyword{->}
\KatlaSpace{}\KatlaSpace{}\IdrisFunction{PI}\KatlaSpace{}\IdrisBound{n}\KatlaSpace{}\IdrisData{Hidden}\KatlaSpace{}\IdrisKeyword{(}\IdrisFunction{U}\KatlaSpace{}\IdrisBound{a}\IdrisKeyword{)}\KatlaSpace{}$\KatlaSpace{}\IdrisKeyword{(\textbackslash{}}\KatlaSpace{}\IdrisBound{env}\KatlaSpace{}\IdrisKeyword{=>}
\KatlaSpace{}\KatlaSpace{}\IdrisKeyword{(}\IdrisBound{eq}\KatlaSpace{}\IdrisKeyword{:}\KatlaSpace{}\IdrisKeyword{(}\KatlaSpace{}\IdrisType{Term}\KatlaSpace{}\IdrisBound{pres}\IdrisFunction{.signature}\KatlaSpace{}\IdrisKeyword{(}\IdrisFunction{U}\KatlaSpace{}\IdrisBound{a}\KatlaSpace{}\IdrisType{`Either`}\KatlaSpace{}\IdrisType{Fin}\KatlaSpace{}\IdrisBound{n}\IdrisKeyword{)}
\KatlaSpace{}\KatlaSpace{}\KatlaSpace{}\KatlaSpace{}\KatlaSpace{}\KatlaSpace{}\KatlaSpace{}\KatlaSpace{}\IdrisType{,}\KatlaSpace{}\IdrisType{Term}\KatlaSpace{}\IdrisBound{pres}\IdrisFunction{.signature}\KatlaSpace{}\IdrisKeyword{(}\IdrisFunction{U}\KatlaSpace{}\IdrisBound{a}\KatlaSpace{}\IdrisType{`Either`}\KatlaSpace{}\IdrisType{Fin}\KatlaSpace{}\IdrisBound{n}\IdrisKeyword{)))}\KatlaSpace{}\IdrisKeyword{->}
\KatlaSpace{}\KatlaSpace{}\IdrisKeyword{\{auto}\KatlaSpace{}\IdrisBound{prf}\KatlaSpace{}\IdrisKeyword{:}\KatlaSpace{}\IdrisBound{frex}\IdrisFunction{.Data.Model.rel}
\KatlaSpace{}\KatlaSpace{}\KatlaSpace{}\KatlaSpace{}\KatlaSpace{}\IdrisKeyword{(}\IdrisBound{frex}\IdrisFunction{.Sem}\KatlaSpace{}\IdrisKeyword{(}\IdrisFunction{fst}\KatlaSpace{}\IdrisBound{eq}\IdrisKeyword{)}\KatlaSpace{}\IdrisKeyword{(}\IdrisFunction{frexEnv}\KatlaSpace{}\IdrisBound{frex}\IdrisKeyword{)}\IdrisFunction{.H}\IdrisKeyword{)}
\KatlaSpace{}\KatlaSpace{}\KatlaSpace{}\KatlaSpace{}\KatlaSpace{}\IdrisKeyword{(}\IdrisBound{frex}\IdrisFunction{.Sem}\KatlaSpace{}\IdrisKeyword{(}\IdrisFunction{snd}\KatlaSpace{}\IdrisBound{eq}\IdrisKeyword{)}\KatlaSpace{}\IdrisKeyword{(}\IdrisFunction{frexEnv}\KatlaSpace{}\IdrisBound{frex}\IdrisKeyword{)}\IdrisFunction{.H}\IdrisKeyword{)\}}
\KatlaSpace{}\KatlaSpace{}\KatlaSpace{}\KatlaSpace{}\KatlaSpace{}\IdrisKeyword{->}
\KatlaSpace{}\KatlaSpace{}\IdrisBound{a}\IdrisFunction{.rel}\KatlaSpace{}\IdrisKeyword{(}\IdrisBound{a}\IdrisFunction{.Sem}\KatlaSpace{}\IdrisKeyword{(}\IdrisFunction{fst}\KatlaSpace{}\IdrisBound{eq}\IdrisKeyword{)}\KatlaSpace{}\IdrisKeyword{(}\IdrisFunction{either}\KatlaSpace{}\IdrisFunction{Prelude.id}\KatlaSpace{}\IdrisKeyword{(}\IdrisFunction{flip}\KatlaSpace{}\IdrisFunction{Vect.index}\KatlaSpace{}\IdrisBound{env}\IdrisKeyword{)))}
\KatlaSpace{}\KatlaSpace{}\KatlaSpace{}\KatlaSpace{}\KatlaSpace{}\KatlaSpace{}\KatlaSpace{}\KatlaSpace{}\IdrisKeyword{(}\IdrisBound{a}\IdrisFunction{.Sem}\KatlaSpace{}\IdrisKeyword{(}\IdrisFunction{snd}\KatlaSpace{}\IdrisBound{eq}\IdrisKeyword{)}\KatlaSpace{}\IdrisKeyword{(}\IdrisFunction{either}\KatlaSpace{}\IdrisFunction{Prelude.id}\KatlaSpace{}\IdrisKeyword{(}\IdrisFunction{flip}\KatlaSpace{}\IdrisFunction{Vect.index}\KatlaSpace{}\IdrisBound{env}\IdrisKeyword{))))}
\end{SaveVerbatim}
\newcommand\helloStr[1][]{\UseVerb[#1]{helloStr}}
\begin{SaveVerbatim}[commandchars=\\\{\}]{helloStr}
\IdrisData{"hello"}
\end{SaveVerbatim}
\newcommand\ollehStr[1][]{\UseVerb[#1]{ollehStr}}
\begin{SaveVerbatim}[commandchars=\\\{\}]{ollehStr}
\IdrisData{"olleh"}
\end{SaveVerbatim}
\newcommand\EmptyStr[1][]{\UseVerb[#1]{EmptyStr}}
\begin{SaveVerbatim}[commandchars=\\\{\}]{EmptyStr}
\IdrisData{""}
\end{SaveVerbatim}
\newcommand\String[1][]{\UseVerb[#1]{String}}
\begin{SaveVerbatim}[commandchars=\\\{\}]{String}
\IdrisType{String}
\end{SaveVerbatim}
\newcommand\ModelOver[1][]{\UseVerbatim[#1]{ModelOver}}
\begin{SaveVerbatim}[commandchars=\\\{\}]{ModelOver}
\IdrisKeyword{record}\KatlaSpace{}\IdrisType{ModelOver}
\KatlaSpace{}\KatlaSpace{}\KatlaSpace{}\KatlaSpace{}\IdrisKeyword{(}\IdrisBound{Pres}\KatlaSpace{}\IdrisKeyword{:}\KatlaSpace{}\IdrisType{Presentation}\IdrisKeyword{)}
\KatlaSpace{}\KatlaSpace{}\KatlaSpace{}\KatlaSpace{}\IdrisKeyword{(}\IdrisBound{X}\KatlaSpace{}\IdrisKeyword{:}\KatlaSpace{}\IdrisType{Setoid}\IdrisKeyword{)}\KatlaSpace{}\IdrisKeyword{where}
\KatlaSpace{}\KatlaSpace{}constructor\KatlaSpace{}\IdrisData{MkModelOver}
\KatlaSpace{}\KatlaSpace{}\IdrisFunction{Model}\KatlaSpace{}\IdrisKeyword{:}\KatlaSpace{}\IdrisType{Model}\KatlaSpace{}\IdrisBound{Pres}
\KatlaSpace{}\KatlaSpace{}\IdrisFunction{Env}\KatlaSpace{}\IdrisKeyword{:}\KatlaSpace{}\IdrisBound{X}\KatlaSpace{}\IdrisType{~>}\KatlaSpace{}\IdrisFunction{cast\KatlaSpace{}Model}
\end{SaveVerbatim}
\newcommand\PreservesEnvDecl[1][]{\UseVerbatim[#1]{PreservesEnvDecl}}
\begin{SaveVerbatim}[commandchars=\\\{\}]{PreservesEnvDecl}
\IdrisFunction{PreservesEnv}\KatlaSpace{}\IdrisKeyword{:}\KatlaSpace{}\IdrisKeyword{\{}\IdrisBound{Pres}\KatlaSpace{}\IdrisKeyword{:}\KatlaSpace{}\IdrisType{Presentation}\IdrisKeyword{\}}
\KatlaSpace{}\KatlaSpace{}\KatlaSpace{}\KatlaSpace{}\IdrisKeyword{->}\KatlaSpace{}\IdrisKeyword{\{}\IdrisBound{X}\KatlaSpace{}\IdrisKeyword{:}\KatlaSpace{}\IdrisType{Setoid}\IdrisKeyword{\}}
\KatlaSpace{}\KatlaSpace{}\KatlaSpace{}\KatlaSpace{}\IdrisKeyword{->}\KatlaSpace{}\IdrisKeyword{(}\IdrisBound{a}\IdrisKeyword{,}\KatlaSpace{}\IdrisBound{b}\KatlaSpace{}\IdrisKeyword{:}\KatlaSpace{}\IdrisBound{Pres}\KatlaSpace{}\IdrisType{`ModelOver`}\KatlaSpace{}\IdrisBound{X}\IdrisKeyword{)}\KatlaSpace{}\IdrisKeyword{->}
\KatlaSpace{}\KatlaSpace{}\KatlaSpace{}\KatlaSpace{}\IdrisKeyword{(}\IdrisFunction{cast}\KatlaSpace{}\IdrisKeyword{\{}\IdrisBound{to}\KatlaSpace{}\IdrisKeyword{=}\KatlaSpace{}\IdrisType{Setoid}\IdrisKeyword{\}}\KatlaSpace{}\IdrisBound{a}\IdrisFunction{.Model}
\KatlaSpace{}\KatlaSpace{}\KatlaSpace{}\KatlaSpace{}\KatlaSpace{}\KatlaSpace{}\IdrisType{~>}\KatlaSpace{}\IdrisFunction{cast}\KatlaSpace{}\IdrisBound{b}\IdrisFunction{.Model}\IdrisKeyword{)}\KatlaSpace{}\IdrisKeyword{->}\KatlaSpace{}\IdrisType{Type}
\IdrisFunction{PreservesEnv}\KatlaSpace{}\IdrisBound{a}\KatlaSpace{}\IdrisBound{b}\KatlaSpace{}\IdrisBound{h}\KatlaSpace{}\IdrisKeyword{=}
\KatlaSpace{}\IdrisKeyword{(}\IdrisBound{X}\KatlaSpace{}\IdrisFunction{~~>}\KatlaSpace{}\IdrisFunction{cast}\KatlaSpace{}\IdrisBound{b}\IdrisFunction{.Model}\IdrisKeyword{)}\IdrisFunction{.equivalence.relation}
\KatlaSpace{}\KatlaSpace{}\KatlaSpace{}\KatlaSpace{}\IdrisKeyword{(}\IdrisBound{h}\KatlaSpace{}\IdrisFunction{.}\KatlaSpace{}\IdrisBound{a}\IdrisFunction{.Env}\IdrisKeyword{)}\KatlaSpace{}\IdrisBound{b}\IdrisFunction{.Env}
\end{SaveVerbatim}
\newcommand\ModelOverHomo[1][]{\UseVerbatim[#1]{ModelOverHomo}}
\begin{SaveVerbatim}[commandchars=\\\{\}]{ModelOverHomo}
\IdrisKeyword{record}\KatlaSpace{}\IdrisType{(~>)}
\KatlaSpace{}\KatlaSpace{}\KatlaSpace{}\KatlaSpace{}\IdrisKeyword{\{}\IdrisBound{Pres}\KatlaSpace{}\IdrisKeyword{:}\KatlaSpace{}\IdrisType{Presentation}\IdrisKeyword{\}}\KatlaSpace{}\IdrisKeyword{\{}\IdrisBound{X}\KatlaSpace{}\IdrisKeyword{:}\KatlaSpace{}\IdrisType{Setoid}\IdrisKeyword{\}}
\KatlaSpace{}\KatlaSpace{}\KatlaSpace{}\KatlaSpace{}\IdrisKeyword{(}\IdrisBound{A}\IdrisKeyword{,}\KatlaSpace{}\IdrisBound{B}\KatlaSpace{}\IdrisKeyword{:}\KatlaSpace{}\IdrisBound{Pres}\KatlaSpace{}\IdrisType{`ModelOver`}\KatlaSpace{}\IdrisBound{X}\IdrisKeyword{)}\KatlaSpace{}\IdrisKeyword{where}
\KatlaSpace{}\KatlaSpace{}constructor\KatlaSpace{}\IdrisData{MkHomomorphism}
\KatlaSpace{}\KatlaSpace{}\IdrisFunction{H}\KatlaSpace{}\IdrisKeyword{:}\KatlaSpace{}\IdrisKeyword{(}\IdrisBound{A}\KatlaSpace{}\IdrisFunction{.Model}\IdrisKeyword{)}\KatlaSpace{}\IdrisFunction{~>}\KatlaSpace{}\IdrisKeyword{(}\IdrisBound{B}\KatlaSpace{}\IdrisFunction{.Model}\IdrisKeyword{)}
\KatlaSpace{}\KatlaSpace{}\IdrisFunction{preserves}\KatlaSpace{}\IdrisKeyword{:}\KatlaSpace{}\IdrisFunction{PreservesEnv}\KatlaSpace{}\IdrisBound{A}\KatlaSpace{}\IdrisBound{B}\KatlaSpace{}\IdrisKeyword{(}\IdrisBound{H}\KatlaSpace{}\IdrisFunction{.H}\IdrisKeyword{)}
\end{SaveVerbatim}
\newcommand\Extension[1][]{\UseVerbatim[#1]{Extension}}
\begin{SaveVerbatim}[commandchars=\\\{\}]{Extension}
\IdrisKeyword{record}\KatlaSpace{}\IdrisType{Extension}\KatlaSpace{}\IdrisKeyword{\{}\IdrisBound{Pres}\KatlaSpace{}\IdrisKeyword{:}\KatlaSpace{}\IdrisType{Presentation}\IdrisKeyword{\}}
\KatlaSpace{}\KatlaSpace{}\KatlaSpace{}\KatlaSpace{}\IdrisKeyword{(}\IdrisBound{A}\KatlaSpace{}\IdrisKeyword{:}\KatlaSpace{}\IdrisType{Model}\KatlaSpace{}\IdrisBound{Pres}\IdrisKeyword{)(}\IdrisBound{X}\KatlaSpace{}\IdrisKeyword{:}\KatlaSpace{}\IdrisType{Setoid}\IdrisKeyword{)}\KatlaSpace{}\IdrisKeyword{where}
\KatlaSpace{}\KatlaSpace{}constructor\KatlaSpace{}\IdrisData{MkExtension}
\KatlaSpace{}\KatlaSpace{}\IdrisFunction{Model}\KatlaSpace{}\IdrisKeyword{:}\KatlaSpace{}\IdrisType{Model}\KatlaSpace{}\IdrisBound{Pres}
\KatlaSpace{}\KatlaSpace{}\IdrisFunction{Embed}\KatlaSpace{}\IdrisKeyword{:}\KatlaSpace{}\IdrisBound{A}\KatlaSpace{}\IdrisFunction{~>}\KatlaSpace{}\IdrisBound{Model}
\KatlaSpace{}\KatlaSpace{}\IdrisFunction{Var}\KatlaSpace{}\KatlaSpace{}\KatlaSpace{}\IdrisKeyword{:}\KatlaSpace{}\IdrisBound{X}\KatlaSpace{}\IdrisType{~>}\KatlaSpace{}\IdrisFunction{cast\KatlaSpace{}Model}
\end{SaveVerbatim}
\newcommand\ExtensionHomo[1][]{\UseVerbatim[#1]{ExtensionHomo}}
\begin{SaveVerbatim}[commandchars=\\\{\}]{ExtensionHomo}
\IdrisKeyword{record}\KatlaSpace{}\IdrisType{(~>)}\KatlaSpace{}\IdrisKeyword{\{}\IdrisBound{Pres}\KatlaSpace{}\IdrisKeyword{:}\KatlaSpace{}\IdrisType{Presentation}\IdrisKeyword{\}}
\KatlaSpace{}\KatlaSpace{}\KatlaSpace{}\KatlaSpace{}\KatlaSpace{}\KatlaSpace{}\KatlaSpace{}\IdrisKeyword{\{}\IdrisBound{A}\KatlaSpace{}\IdrisKeyword{:}\KatlaSpace{}\IdrisType{Model}\KatlaSpace{}\IdrisBound{Pres}\IdrisKeyword{\}}\KatlaSpace{}\IdrisKeyword{\{}\IdrisBound{X}\KatlaSpace{}\IdrisKeyword{:}\KatlaSpace{}\IdrisType{Setoid}\IdrisKeyword{\}}
\KatlaSpace{}\KatlaSpace{}\KatlaSpace{}\KatlaSpace{}\IdrisKeyword{(}\IdrisBound{Extension1}\IdrisKeyword{,}\KatlaSpace{}\IdrisBound{Extension2}\KatlaSpace{}\IdrisKeyword{:}\KatlaSpace{}\IdrisType{Extension}\KatlaSpace{}\IdrisBound{A}\KatlaSpace{}\IdrisBound{X}\IdrisKeyword{)}\KatlaSpace{}\IdrisKeyword{where}
\KatlaSpace{}\KatlaSpace{}constructor\KatlaSpace{}\IdrisData{MkExtensionMorphism}
\KatlaSpace{}\KatlaSpace{}\IdrisFunction{H}\KatlaSpace{}\IdrisKeyword{:}\KatlaSpace{}\IdrisKeyword{(}\IdrisBound{Extension1}\IdrisKeyword{)}\IdrisFunction{.Model}\KatlaSpace{}\IdrisFunction{~>}\KatlaSpace{}\IdrisKeyword{(}\IdrisBound{Extension2}\IdrisKeyword{)}\IdrisFunction{.Model}
\KatlaSpace{}\KatlaSpace{}\IdrisFunction{PreserveEmbed}\KatlaSpace{}\IdrisKeyword{:}
\KatlaSpace{}\KatlaSpace{}\KatlaSpace{}\KatlaSpace{}\IdrisKeyword{(}\IdrisFunction{cast\KatlaSpace{}A}\KatlaSpace{}\IdrisFunction{~~>}\KatlaSpace{}\IdrisKeyword{(}\IdrisBound{Extension2}\IdrisKeyword{)}\IdrisFunction{.Model}\IdrisKeyword{)}
\KatlaSpace{}\KatlaSpace{}\KatlaSpace{}\KatlaSpace{}\KatlaSpace{}\KatlaSpace{}\IdrisFunction{.equivalence.relation}
\KatlaSpace{}\KatlaSpace{}\KatlaSpace{}\KatlaSpace{}\KatlaSpace{}\KatlaSpace{}\KatlaSpace{}\KatlaSpace{}\IdrisKeyword{(}\IdrisBound{H}\KatlaSpace{}\IdrisFunction{.}\KatlaSpace{}\IdrisKeyword{(}\IdrisBound{Extension1}\IdrisKeyword{)}\IdrisFunction{.Embed}\IdrisKeyword{)}
\KatlaSpace{}\KatlaSpace{}\KatlaSpace{}\KatlaSpace{}\KatlaSpace{}\KatlaSpace{}\KatlaSpace{}\KatlaSpace{}\KatlaSpace{}\KatlaSpace{}\KatlaSpace{}\KatlaSpace{}\KatlaSpace{}\IdrisKeyword{(}\IdrisBound{Extension2}\IdrisKeyword{)}\IdrisFunction{.Embed}
\KatlaSpace{}\KatlaSpace{}\IdrisFunction{PreserveVar}\KatlaSpace{}\KatlaSpace{}\KatlaSpace{}\IdrisKeyword{:}
\KatlaSpace{}\KatlaSpace{}\KatlaSpace{}\KatlaSpace{}\IdrisKeyword{(}\IdrisBound{X}\KatlaSpace{}\IdrisFunction{~~>}\KatlaSpace{}\IdrisFunction{cast\KatlaSpace{}}\IdrisKeyword{(}\IdrisBound{Extension2}\IdrisKeyword{)}\IdrisFunction{.Model}\IdrisKeyword{)}
\KatlaSpace{}\KatlaSpace{}\KatlaSpace{}\KatlaSpace{}\KatlaSpace{}\KatlaSpace{}\IdrisFunction{.equivalence.relation}
\KatlaSpace{}\KatlaSpace{}\KatlaSpace{}\KatlaSpace{}\KatlaSpace{}\KatlaSpace{}\KatlaSpace{}\KatlaSpace{}\IdrisKeyword{((}\IdrisBound{H}\IdrisKeyword{)}\IdrisFunction{.H}\KatlaSpace{}\IdrisFunction{.}\KatlaSpace{}\IdrisKeyword{(}\IdrisBound{Extension1}\IdrisKeyword{)}\IdrisFunction{.Var}\IdrisKeyword{)}
\KatlaSpace{}\KatlaSpace{}\KatlaSpace{}\KatlaSpace{}\KatlaSpace{}\KatlaSpace{}\KatlaSpace{}\KatlaSpace{}\KatlaSpace{}\KatlaSpace{}\KatlaSpace{}\KatlaSpace{}\KatlaSpace{}\KatlaSpace{}\KatlaSpace{}\KatlaSpace{}\KatlaSpace{}\IdrisKeyword{(}\IdrisBound{Extension2}\IdrisKeyword{)}\IdrisFunction{.Var}
\end{SaveVerbatim}
\newcommand\avar[1][]{\UseVerb[#1]{avar}}
\begin{SaveVerbatim}[commandchars=\\\{\}]{avar}
\IdrisBound{a}
\end{SaveVerbatim}
\newcommand\bvar[1][]{\UseVerb[#1]{bvar}}
\begin{SaveVerbatim}[commandchars=\\\{\}]{bvar}
\IdrisBound{b}
\end{SaveVerbatim}
\newcommand\aModel[1][]{\UseVerb[#1]{aModel}}
\begin{SaveVerbatim}[commandchars=\\\{\}]{aModel}
\IdrisBound{a}\IdrisFunction{.Model}
\end{SaveVerbatim}
\newcommand\bModel[1][]{\UseVerb[#1]{bModel}}
\begin{SaveVerbatim}[commandchars=\\\{\}]{bModel}
\IdrisBound{b}\IdrisFunction{.Model}
\end{SaveVerbatim}
\newcommand\aEnv[1][]{\UseVerb[#1]{aEnv}}
\begin{SaveVerbatim}[commandchars=\\\{\}]{aEnv}
\IdrisFunction{Env}\KatlaSpace{}\IdrisBound{a}
\end{SaveVerbatim}

\begin{SaveVerbatim}[commandchars=\\\{\}]{bEnv}
\KatlaSpace{}\IdrisBound{h}\KatlaSpace{}\IdrisKeyword{=}
\end{SaveVerbatim}
\newcommand\EnvName[1][]{\UseVerb[#1]{EnvName}}
\begin{SaveVerbatim}[commandchars=\\\{\}]{EnvName}
\IdrisFunction{Env}
\end{SaveVerbatim}
\newcommand\hName[1][]{\UseVerb[#1]{hName}}
\begin{SaveVerbatim}[commandchars=\\\{\}]{hName}
\IdrisBound{h}
\end{SaveVerbatim}

\begin{SaveVerbatim}[commandchars=\\\{\}]{EmbedName}
\IdrisFunction{Embed}
\end{SaveVerbatim}

\begin{SaveVerbatim}[commandchars=\\\{\}]{VarName}
\IdrisFunction{Var}\KatlaSpace{}
\end{SaveVerbatim}

\begin{SaveVerbatim}[commandchars=\\\{\}]{GenEmbed}
\IdrisFunction{Embed.H}
\end{SaveVerbatim}
\newcommand\aEmbed[1][]{\UseVerb[#1]{aEmbed}}
\begin{SaveVerbatim}[commandchars=\\\{\}]{aEmbed}
\IdrisBound{a}\IdrisFunction{.Embed.H}
\end{SaveVerbatim}
\newcommand\aVar[1][]{\UseVerb[#1]{aVar}}
\begin{SaveVerbatim}[commandchars=\\\{\}]{aVar}
\IdrisBound{a}\IdrisFunction{.Var}
\end{SaveVerbatim}

\begin{SaveVerbatim}[commandchars=\\\{\}]{bEmbed}
\IdrisBound{b}\IdrisFunction{.Embed.H}
\end{SaveVerbatim}

\begin{SaveVerbatim}[commandchars=\\\{\}]{bVar}
\IdrisBound{b}\IdrisFunction{.Var}
\end{SaveVerbatim}
\newcommand\Finn[1][]{\UseVerb[#1]{Finn}}
\begin{SaveVerbatim}[commandchars=\\\{\}]{Finn}
\IdrisType{Fin}\KatlaSpace{}\IdrisBound{n}
\end{SaveVerbatim}
\newcommand\ModelCarrier[1][]{\UseVerbatim[#1]{ModelCarrier}}
\begin{SaveVerbatim}[commandchars=\\\{\}]{ModelCarrier}
\IdrisFunction{Carrier}\KatlaSpace{}\IdrisKeyword{:}\KatlaSpace{}\IdrisKeyword{(}\IdrisBound{n}\KatlaSpace{}\IdrisKeyword{:}\KatlaSpace{}\IdrisType{Nat}\IdrisKeyword{)}\KatlaSpace{}\IdrisKeyword{->}\KatlaSpace{}\IdrisType{Setoid}
\IdrisFunction{Carrier}\KatlaSpace{}\IdrisBound{n}\KatlaSpace{}\IdrisKeyword{=}\KatlaSpace{}\IdrisFunction{VectSetoid}\KatlaSpace{}\IdrisBound{n}
\KatlaSpace{}\KatlaSpace{}\KatlaSpace{}\KatlaSpace{}\KatlaSpace{}\KatlaSpace{}\KatlaSpace{}\KatlaSpace{}\KatlaSpace{}\KatlaSpace{}\KatlaSpace{}\KatlaSpace{}\KatlaSpace{}\KatlaSpace{}\KatlaSpace{}\KatlaSpace{}\KatlaSpace{}\KatlaSpace{}\KatlaSpace{}\IdrisKeyword{(}\IdrisFunction{cast}\KatlaSpace{}\IdrisType{Nat}\IdrisKeyword{)}
\end{SaveVerbatim}
\newcommand\ModelCarrierName[1][]{\UseVerb[#1]{ModelCarrierName}}
\begin{SaveVerbatim}[commandchars=\\\{\}]{ModelCarrierName}
\IdrisFunction{Carrier}\KatlaSpace{}\IdrisBound{n}
\end{SaveVerbatim}
\newcommand\ReplicateZero[1][]{\UseVerb[#1]{ReplicateZero}}
\begin{SaveVerbatim}[commandchars=\\\{\}]{ReplicateZero}
\IdrisFunction{replicate}\KatlaSpace{}\IdrisBound{n}\KatlaSpace{}\IdrisData{0}
\end{SaveVerbatim}
\newcommand\ReplicateZeroPartA[1][]{\UseVerb[#1]{ReplicateZeroPartA}}
\begin{SaveVerbatim}[commandchars=\\\{\}]{ReplicateZeroPartA}
\IdrisData{[0,}
\end{SaveVerbatim}
\newcommand\ReplicateZeroPartB[1][]{\UseVerb[#1]{ReplicateZeroPartB}}
\begin{SaveVerbatim}[commandchars=\\\{\}]{ReplicateZeroPartB}
\IdrisData{,0]}
\end{SaveVerbatim}
\newcommand\OpenBracket[1][]{\UseVerb[#1]{OpenBracket}}
\begin{SaveVerbatim}[commandchars=\\\{\}]{OpenBracket}
\IdrisData{[}
\end{SaveVerbatim}
\newcommand\CloseBracket[1][]{\UseVerb[#1]{CloseBracket}}
\begin{SaveVerbatim}[commandchars=\\\{\}]{CloseBracket}
\IdrisData{]}
\end{SaveVerbatim}
\newcommand\Comma[1][]{\UseVerb[#1]{Comma}}
\begin{SaveVerbatim}[commandchars=\\\{\}]{Comma}
\IdrisData{,}
\end{SaveVerbatim}
\newcommand\ZipIt[1][]{\UseVerb[#1]{ZipIt}}
\begin{SaveVerbatim}[commandchars=\\\{\}]{ZipIt}
\IdrisFunction{map}\KatlaSpace{}\IdrisKeyword{(}\IdrisFunction{uncurry}\KatlaSpace{}\IdrisFunction{(+)}\IdrisKeyword{)}\KatlaSpace{}\IdrisKeyword{(}\IdrisFunction{zip}\KatlaSpace{}\IdrisBound{as}\KatlaSpace{}\IdrisBound{bs}\IdrisKeyword{)}
\end{SaveVerbatim}

\begin{SaveVerbatim}[commandchars=\\\{\}]{aModelA}
\IdrisBound{a}\IdrisFunction{.Model}\KatlaSpace{}\IdrisFunction{~~>}\KatlaSpace{}\IdrisBound{A}
\end{SaveVerbatim}

\begin{SaveVerbatim}[commandchars=\\\{\}]{bModelA}
\IdrisBound{b}\IdrisFunction{.Model}\KatlaSpace{}\IdrisFunction{~~>}\KatlaSpace{}\IdrisBound{A}
\end{SaveVerbatim}
\newcommand\aEval[1][]{\UseVerb[#1]{aEval}}
\begin{SaveVerbatim}[commandchars=\\\{\}]{aEval}
\IdrisBound{a}\IdrisFunction{.Eval}
\end{SaveVerbatim}

\begin{SaveVerbatim}[commandchars=\\\{\}]{bEval}
\IdrisBound{b}\IdrisFunction{.Eval}
\end{SaveVerbatim}
\newcommand\PowerConstruction[1][]{\UseVerb[#1]{PowerConstruction}}
\begin{SaveVerbatim}[commandchars=\\\{\}]{PowerConstruction}
\IdrisBound{X}\KatlaSpace{}\IdrisFunction{~~>}\KatlaSpace{}\IdrisBound{a}\IdrisFunction{.Model}
\end{SaveVerbatim}
\newcommand\KroneckerMiddle[1][]{\UseVerb[#1]{KroneckerMiddle}}
\begin{SaveVerbatim}[commandchars=\\\{\}]{KroneckerMiddle}
\IdrisData{,0,1,0,}
\end{SaveVerbatim}
\newcommand\tabulateDirac[1][]{\UseVerb[#1]{tabulateDirac}}
\begin{SaveVerbatim}[commandchars=\\\{\}]{tabulateDirac}
\IdrisFunction{tabulate}\KatlaSpace{}$\KatlaSpace{}\IdrisFunction{dirac}\KatlaSpace{}\IdrisBound{i}
\end{SaveVerbatim}
\newcommand\dirac[1][]{\UseVerb[#1]{dirac}}
\begin{SaveVerbatim}[commandchars=\\\{\}]{dirac}
\IdrisFunction{dirac}
\end{SaveVerbatim}
\newcommand\diracij[1][]{\UseVerb[#1]{diracij}}
\begin{SaveVerbatim}[commandchars=\\\{\}]{diracij}
\IdrisFunction{dirac}\KatlaSpace{}\IdrisBound{i}\KatlaSpace{}\IdrisBound{j}
\end{SaveVerbatim}
\newcommand\diracsi[1][]{\UseVerb[#1]{diracsi}}
\begin{SaveVerbatim}[commandchars=\\\{\}]{diracsi}
\IdrisBound{i}
\end{SaveVerbatim}
\newcommand\diracsj[1][]{\UseVerb[#1]{diracsj}}
\begin{SaveVerbatim}[commandchars=\\\{\}]{diracsj}
\IdrisBound{j}
\end{SaveVerbatim}
\newcommand\AppliedFreeCMonoidModel[1][]{\UseVerb[#1]{AppliedFreeCMonoidModel}}
\begin{SaveVerbatim}[commandchars=\\\{\}]{AppliedFreeCMonoidModel}
\IdrisFunction{Model}\KatlaSpace{}\IdrisBound{n}
\end{SaveVerbatim}
\newcommand\AppliedFreeCMonoidModelType[1][]{\UseVerb[#1]{AppliedFreeCMonoidModelType}}
\begin{SaveVerbatim}[commandchars=\\\{\}]{AppliedFreeCMonoidModelType}
\IdrisFunction{CommutativeMonoid}
\end{SaveVerbatim}

\begin{SaveVerbatim}[commandchars=\\\{\}]{AppliedCMonoidNormalForm}
\IdrisFunction{normalForm}\KatlaSpace{}\IdrisBound{n}\KatlaSpace{}\IdrisBound{xs}
\end{SaveVerbatim}
\newcommand\AppliedCMonoidNormalFormTypePartA[1][]{\UseVerb[#1]{AppliedCMonoidNormalFormTypePartA}}
\begin{SaveVerbatim}[commandchars=\\\{\}]{AppliedCMonoidNormalFormTypePartA}
\IdrisKeyword{(}\IdrisFunction{Model}\KatlaSpace{}\IdrisBound{n}\IdrisKeyword{)}\IdrisFunction{.sum}\KatlaSpace{}\IdrisKeyword{(}\IdrisFunction{tabulate}\KatlaSpace{}$\KatlaSpace{}\IdrisKeyword{\textbackslash{}}\IdrisBound{i}\KatlaSpace{}\IdrisKeyword{=>}\KatlaSpace{}\IdrisKeyword{(}\IdrisFunction{index}\KatlaSpace{}\IdrisBound{i}\KatlaSpace{}\IdrisBound{xs}\IdrisKeyword{)}\KatlaSpace{}\IdrisFunction{*.}\KatlaSpace{}\IdrisKeyword{(}\IdrisFunction{unit}\KatlaSpace{}\IdrisBound{n}\KatlaSpace{}\IdrisBound{i}\IdrisKeyword{))}
\end{SaveVerbatim}
\newcommand\AppliedCMonoidNormalFormTypePartB[1][]{\UseVerb[#1]{AppliedCMonoidNormalFormTypePartB}}
\begin{SaveVerbatim}[commandchars=\\\{\}]{AppliedCMonoidNormalFormTypePartB}
\IdrisFunction{normalForm}\KatlaSpace{}\IdrisKeyword{:}\KatlaSpace{}\IdrisKeyword{(}\IdrisBound{n}\KatlaSpace{}\IdrisKeyword{:}\KatlaSpace{}\IdrisType{Nat}\IdrisKeyword{)}\KatlaSpace{}\IdrisKeyword{->}\KatlaSpace{}\IdrisKeyword{(}\IdrisBound{xs}\KatlaSpace{}\IdrisKeyword{:}\KatlaSpace{}\IdrisFunction{U}\KatlaSpace{}\IdrisKeyword{(}\IdrisFunction{Model}\KatlaSpace{}\IdrisBound{n}\IdrisKeyword{))}\KatlaSpace{}\IdrisKeyword{->}
\end{SaveVerbatim}
\newcommand\AppliedCMonoidNormalFormTypePartC[1][]{\UseVerb[#1]{AppliedCMonoidNormalFormTypePartC}}
\begin{SaveVerbatim}[commandchars=\\\{\}]{AppliedCMonoidNormalFormTypePartC}
\KatlaSpace{}\IdrisBound{xs}\KatlaSpace{}\IdrisKeyword{=}
\end{SaveVerbatim}
\newcommand\FreeCMonoidUnit[1][]{\UseVerb[#1]{FreeCMonoidUnit}}
\begin{SaveVerbatim}[commandchars=\\\{\}]{FreeCMonoidUnit}
\IdrisFunction{unit}
\end{SaveVerbatim}
\newcommand\CMonoidUniqueMap[1][]{\UseVerb[#1]{CMonoidUniqueMap}}
\begin{SaveVerbatim}[commandchars=\\\{\}]{CMonoidUniqueMap}
\IdrisBound{a}\IdrisFunction{.Model.sum}\KatlaSpace{}\IdrisKeyword{(}\IdrisFunction{mapWithPos}\KatlaSpace{}\IdrisKeyword{(\textbackslash{}}\IdrisBound{i}\IdrisKeyword{,}\IdrisBound{k}\KatlaSpace{}\IdrisKeyword{=>}\KatlaSpace{}\IdrisBound{k}\KatlaSpace{}\IdrisFunction{*.}\KatlaSpace{}\IdrisBound{a}\IdrisFunction{.Env.H}\KatlaSpace{}\IdrisBound{i}\IdrisKeyword{)}\KatlaSpace{}\IdrisBound{xs}\IdrisKeyword{)}
\end{SaveVerbatim}
\newcommand\FreeName[1][]{\UseVerb[#1]{FreeName}}
\begin{SaveVerbatim}[commandchars=\\\{\}]{FreeName}
\IdrisType{Free}
\end{SaveVerbatim}
\newcommand\aSink[1][]{\UseVerb[#1]{aSink}}
\begin{SaveVerbatim}[commandchars=\\\{\}]{aSink}
\IdrisBound{a}\IdrisFunction{.Sink}
\end{SaveVerbatim}
\newcommand\aLft[1][]{\UseVerb[#1]{aLft}}
\begin{SaveVerbatim}[commandchars=\\\{\}]{aLft}
\IdrisBound{a}\IdrisFunction{.Lft}
\end{SaveVerbatim}
\newcommand\Lft[1][]{\UseVerb[#1]{Lft}}
\begin{SaveVerbatim}[commandchars=\\\{\}]{Lft}
\IdrisFunction{Lft}
\end{SaveVerbatim}
\newcommand\aRgt[1][]{\UseVerb[#1]{aRgt}}
\begin{SaveVerbatim}[commandchars=\\\{\}]{aRgt}
\IdrisBound{a}\IdrisFunction{.Rgt}
\end{SaveVerbatim}
\newcommand\Rgt[1][]{\UseVerb[#1]{Rgt}}
\begin{SaveVerbatim}[commandchars=\\\{\}]{Rgt}
\IdrisFunction{Rgt}
\end{SaveVerbatim}
\newcommand\bSink[1][]{\UseVerb[#1]{bSink}}
\begin{SaveVerbatim}[commandchars=\\\{\}]{bSink}
\IdrisBound{b}\IdrisFunction{.Sink}
\end{SaveVerbatim}

\begin{SaveVerbatim}[commandchars=\\\{\}]{bLft}
\IdrisBound{b}\IdrisFunction{.Lft}
\end{SaveVerbatim}

\begin{SaveVerbatim}[commandchars=\\\{\}]{bRgt}
\IdrisBound{b}\IdrisFunction{.Rgt}
\end{SaveVerbatim}
\newcommand\FrexByCoprodPartA[1][]{\UseVerb[#1]{FrexByCoprodPartA}}
\begin{SaveVerbatim}[commandchars=\\\{\}]{FrexByCoprodPartA}
\IdrisFunction{CoproductAlgebraWithFree}\KatlaSpace{}\IdrisBound{pres}\KatlaSpace{}\IdrisBound{a}\KatlaSpace{}\IdrisBound{x}
\end{SaveVerbatim}
\newcommand\FrexByCoprodPartB[1][]{\UseVerb[#1]{FrexByCoprodPartB}}
\begin{SaveVerbatim}[commandchars=\\\{\}]{FrexByCoprodPartB}
\KatlaSpace{}\IdrisKeyword{(}\IdrisBound{free}\KatlaSpace{}\IdrisKeyword{:}\KatlaSpace{}\IdrisType{Free}\KatlaSpace{}\IdrisBound{pres}\KatlaSpace{}\IdrisBound{x}\IdrisKeyword{)}\KatlaSpace{}\IdrisKeyword{->}
\end{SaveVerbatim}
\newcommand\FrexByCoprodPartC[1][]{\UseVerb[#1]{FrexByCoprodPartC}}
\begin{SaveVerbatim}[commandchars=\\\{\}]{FrexByCoprodPartC}
\KatlaSpace{}\KatlaSpace{}\IdrisKeyword{(}\IdrisBound{coprod}\KatlaSpace{}\IdrisKeyword{:}\KatlaSpace{}\IdrisType{Coproduct}\KatlaSpace{}\IdrisBound{a}\KatlaSpace{}\IdrisBound{free}\IdrisFunction{.Data.Model}\IdrisKeyword{)}\KatlaSpace{}\IdrisKeyword{->}\KatlaSpace{}\IdrisType{Frex}\KatlaSpace{}\IdrisBound{a}\KatlaSpace{}\IdrisBound{x}
\end{SaveVerbatim}
\newcommand\CMonoidFrexCarrier[1][]{\UseVerb[#1]{CMonoidFrexCarrier}}
\begin{SaveVerbatim}[commandchars=\\\{\}]{CMonoidFrexCarrier}
\IdrisKeyword{(}\IdrisFunction{U}\KatlaSpace{}\IdrisBound{A}\IdrisType{,}\KatlaSpace{}\IdrisType{Vect}\KatlaSpace{}\IdrisBound{n}\KatlaSpace{}\IdrisType{Nat}\IdrisKeyword{)}
\end{SaveVerbatim}
\newcommand\ByFrexFull[1][]{\UseVerbatim[#1]{ByFrexFull}}
\begin{SaveVerbatim}[commandchars=\\\{\}]{ByFrexFull}
\IdrisFunction{ByFrex}\KatlaSpace{}\IdrisKeyword{:}\KatlaSpace{}\IdrisKeyword{(}\IdrisBound{initial}\KatlaSpace{}\IdrisKeyword{:}\KatlaSpace{}\IdrisType{Free}\KatlaSpace{}\IdrisBound{pres}\KatlaSpace{}\IdrisKeyword{(}\IdrisFunction{cast}\KatlaSpace{}\IdrisType{Void}\IdrisKeyword{))}\KatlaSpace{}\IdrisKeyword{->}
\KatlaSpace{}\KatlaSpace{}\IdrisType{Frex}\KatlaSpace{}\IdrisBound{initial}\IdrisFunction{.Data.Model}\KatlaSpace{}\IdrisBound{s}\KatlaSpace{}\IdrisKeyword{->}\KatlaSpace{}\IdrisType{Free}\KatlaSpace{}\IdrisBound{pres}\KatlaSpace{}\IdrisBound{s}
\end{SaveVerbatim}
\newcommand\BoolX[1][]{\UseVerb[#1]{BoolX}}
\begin{SaveVerbatim}[commandchars=\\\{\}]{BoolX}
\IdrisKeyword{(}\IdrisType{Bool}\IdrisData{,}\IdrisFunction{X}\IdrisKeyword{)}
\end{SaveVerbatim}

%% file: certificates-listings.tex
\newcommand\RTList[1][]{\UseVerbatim[#1]{RTList}}
\begin{SaveVerbatim}[commandchars=\\\{\}]{RTList}
\IdrisKeyword{data}\KatlaSpace{}\IdrisType{RTList}\KatlaSpace{}\IdrisKeyword{:}\KatlaSpace{}\IdrisFunction{Rel}\KatlaSpace{}\IdrisBound{a}\KatlaSpace{}\IdrisKeyword{->}\KatlaSpace{}\IdrisFunction{Rel}\KatlaSpace{}\IdrisBound{a}\KatlaSpace{}\IdrisKeyword{where}
\KatlaSpace{}\KatlaSpace{}\IdrisData{Nil}\KatlaSpace{}\KatlaSpace{}\IdrisKeyword{:}\KatlaSpace{}\IdrisType{RTList}\KatlaSpace{}\IdrisBound{r}\KatlaSpace{}\IdrisBound{x}\KatlaSpace{}\IdrisBound{x}
\KatlaSpace{}\KatlaSpace{}\IdrisData{(::)}\KatlaSpace{}\IdrisKeyword{:}\KatlaSpace{}\IdrisKeyword{\{0}\KatlaSpace{}\IdrisBound{r}\KatlaSpace{}\IdrisKeyword{:}\KatlaSpace{}\IdrisFunction{Rel}\KatlaSpace{}\IdrisBound{a}\IdrisKeyword{\}}\KatlaSpace{}\IdrisKeyword{->}\KatlaSpace{}\IdrisKeyword{\{}\IdrisBound{y}\KatlaSpace{}\IdrisKeyword{:}\KatlaSpace{}\IdrisBound{a}\IdrisKeyword{\}}
\KatlaSpace{}\KatlaSpace{}\KatlaSpace{}\KatlaSpace{}\KatlaSpace{}\KatlaSpace{}\KatlaSpace{}\KatlaSpace{}\KatlaSpace{}\IdrisKeyword{->}\KatlaSpace{}\IdrisBound{r}\KatlaSpace{}\IdrisBound{x}\KatlaSpace{}\IdrisBound{y}\KatlaSpace{}\IdrisKeyword{->}\KatlaSpace{}\IdrisType{RTList}\KatlaSpace{}\IdrisBound{r}\KatlaSpace{}\IdrisBound{y}\KatlaSpace{}\IdrisBound{z}
\KatlaSpace{}\KatlaSpace{}\KatlaSpace{}\KatlaSpace{}\KatlaSpace{}\KatlaSpace{}\KatlaSpace{}\KatlaSpace{}\KatlaSpace{}\IdrisKeyword{->}\KatlaSpace{}\IdrisType{RTList}\KatlaSpace{}\IdrisBound{r}\KatlaSpace{}\IdrisBound{x}\KatlaSpace{}\IdrisBound{z}
\end{SaveVerbatim}
\newcommand\Symmetrise[1][]{\UseVerbatim[#1]{Symmetrise}}
\begin{SaveVerbatim}[commandchars=\\\{\}]{Symmetrise}
\IdrisKeyword{data}\KatlaSpace{}\IdrisType{Symmetrise}\KatlaSpace{}\IdrisKeyword{:}\KatlaSpace{}\IdrisFunction{Rel}\KatlaSpace{}\IdrisBound{a}\KatlaSpace{}\IdrisKeyword{->}\KatlaSpace{}\IdrisFunction{Rel}\KatlaSpace{}\IdrisBound{a}\KatlaSpace{}\IdrisKeyword{where}
\KatlaSpace{}\KatlaSpace{}\IdrisData{Fwd}\KatlaSpace{}\IdrisKeyword{:}\KatlaSpace{}\IdrisKeyword{\{0}\KatlaSpace{}\IdrisBound{r}\KatlaSpace{}\IdrisKeyword{:}\KatlaSpace{}\IdrisFunction{Rel}\KatlaSpace{}\IdrisBound{a}\IdrisKeyword{\}}\KatlaSpace{}\IdrisKeyword{->}\KatlaSpace{}\IdrisBound{r}\KatlaSpace{}\IdrisBound{x}\KatlaSpace{}\IdrisBound{y}
\KatlaSpace{}\KatlaSpace{}\KatlaSpace{}\KatlaSpace{}\KatlaSpace{}\KatlaSpace{}\KatlaSpace{}\KatlaSpace{}\KatlaSpace{}\KatlaSpace{}\KatlaSpace{}\IdrisKeyword{->}\KatlaSpace{}\IdrisType{Symmetrise}\KatlaSpace{}\IdrisBound{r}\KatlaSpace{}\IdrisBound{x}\KatlaSpace{}\IdrisBound{y}
\KatlaSpace{}\KatlaSpace{}\IdrisData{Bwd}\KatlaSpace{}\IdrisKeyword{:}\KatlaSpace{}\IdrisKeyword{\{0}\KatlaSpace{}\IdrisBound{r}\KatlaSpace{}\IdrisKeyword{:}\KatlaSpace{}\IdrisFunction{Rel}\KatlaSpace{}\IdrisBound{a}\IdrisKeyword{\}}\KatlaSpace{}\IdrisKeyword{->}\KatlaSpace{}\IdrisBound{r}\KatlaSpace{}\IdrisBound{x}\KatlaSpace{}\IdrisBound{y}
\KatlaSpace{}\KatlaSpace{}\KatlaSpace{}\KatlaSpace{}\KatlaSpace{}\KatlaSpace{}\KatlaSpace{}\KatlaSpace{}\KatlaSpace{}\KatlaSpace{}\KatlaSpace{}\IdrisKeyword{->}\KatlaSpace{}\IdrisType{Symmetrise}\KatlaSpace{}\IdrisBound{r}\KatlaSpace{}\IdrisBound{y}\KatlaSpace{}\IdrisBound{x}
\end{SaveVerbatim}
\newcommand\Step[1][]{\UseVerbatim[#1]{Step}}
\begin{SaveVerbatim}[commandchars=\\\{\}]{Step}
\IdrisKeyword{data}\KatlaSpace{}\IdrisType{Step}\KatlaSpace{}\IdrisKeyword{:}\KatlaSpace{}\IdrisKeyword{(}\IdrisBound{pres}\KatlaSpace{}\IdrisKeyword{:}\KatlaSpace{}\IdrisType{Presentation}\IdrisKeyword{)}
\KatlaSpace{}\KatlaSpace{}\KatlaSpace{}\KatlaSpace{}\KatlaSpace{}\KatlaSpace{}\KatlaSpace{}\KatlaSpace{}\KatlaSpace{}\KatlaSpace{}\KatlaSpace{}\KatlaSpace{}\IdrisKeyword{->}\KatlaSpace{}\IdrisKeyword{(}\IdrisBound{a}\KatlaSpace{}\IdrisKeyword{:}\KatlaSpace{}\IdrisType{PresetoidAlgebra}\KatlaSpace{}\IdrisBound{pres}\IdrisFunction{.signature}\IdrisKeyword{)}
\KatlaSpace{}\KatlaSpace{}\KatlaSpace{}\KatlaSpace{}\KatlaSpace{}\KatlaSpace{}\KatlaSpace{}\KatlaSpace{}\KatlaSpace{}\KatlaSpace{}\KatlaSpace{}\KatlaSpace{}\IdrisKeyword{->}\KatlaSpace{}\IdrisFunction{Rel}\KatlaSpace{}\IdrisKeyword{(}\IdrisFunction{U}\KatlaSpace{}\IdrisBound{a}\IdrisKeyword{)}\KatlaSpace{}\IdrisKeyword{where}
\KatlaSpace{}\KatlaSpace{}\IdrisData{Include}\KatlaSpace{}\IdrisKeyword{:}\KatlaSpace{}\IdrisKeyword{\{}\IdrisBound{x}\IdrisKeyword{,}\KatlaSpace{}\IdrisBound{y}\KatlaSpace{}\IdrisKeyword{:}\KatlaSpace{}\IdrisFunction{U}\KatlaSpace{}\IdrisBound{a}\IdrisKeyword{\}}\KatlaSpace{}\IdrisKeyword{->}\KatlaSpace{}\IdrisBound{a}\IdrisFunction{.relation}\KatlaSpace{}\IdrisBound{x}\KatlaSpace{}\IdrisBound{y}
\KatlaSpace{}\KatlaSpace{}\KatlaSpace{}\KatlaSpace{}\KatlaSpace{}\KatlaSpace{}\KatlaSpace{}\KatlaSpace{}\KatlaSpace{}\KatlaSpace{}\KatlaSpace{}\KatlaSpace{}\IdrisKeyword{->}\KatlaSpace{}\IdrisType{Step}\KatlaSpace{}\IdrisBound{pres}\KatlaSpace{}\IdrisBound{a}\KatlaSpace{}\IdrisBound{x}\KatlaSpace{}\IdrisBound{y}
\KatlaSpace{}\KatlaSpace{}\IdrisData{ByAxiom}\KatlaSpace{}\IdrisKeyword{:}\KatlaSpace{}\IdrisKeyword{\{0}\KatlaSpace{}\IdrisBound{a}\KatlaSpace{}\IdrisKeyword{:}\KatlaSpace{}\IdrisType{PresetoidAlgebra}\KatlaSpace{}\IdrisBound{pres}\IdrisFunction{.signature}\IdrisKeyword{\}}
\KatlaSpace{}\KatlaSpace{}\KatlaSpace{}\KatlaSpace{}\KatlaSpace{}\KatlaSpace{}\KatlaSpace{}\KatlaSpace{}\KatlaSpace{}\KatlaSpace{}\KatlaSpace{}\KatlaSpace{}\IdrisKeyword{->}\KatlaSpace{}\IdrisKeyword{(}\IdrisBound{eq}\KatlaSpace{}\IdrisKeyword{:}\KatlaSpace{}\IdrisFunction{Axiom}\KatlaSpace{}\IdrisBound{pres}\IdrisKeyword{)}
\KatlaSpace{}\KatlaSpace{}\KatlaSpace{}\KatlaSpace{}\KatlaSpace{}\KatlaSpace{}\KatlaSpace{}\KatlaSpace{}\KatlaSpace{}\KatlaSpace{}\KatlaSpace{}\KatlaSpace{}\IdrisKeyword{->}\KatlaSpace{}\IdrisKeyword{(}\IdrisBound{env}\KatlaSpace{}\IdrisKeyword{:}\KatlaSpace{}\IdrisType{Fin}\KatlaSpace{}\IdrisKeyword{(}\IdrisBound{pres}\IdrisFunction{.axiom}\KatlaSpace{}\IdrisBound{eq}\IdrisKeyword{)}\IdrisFunction{.support}\KatlaSpace{}\IdrisKeyword{->}\KatlaSpace{}\IdrisFunction{U}\KatlaSpace{}\IdrisBound{a}\IdrisKeyword{)}
\KatlaSpace{}\KatlaSpace{}\KatlaSpace{}\KatlaSpace{}\KatlaSpace{}\KatlaSpace{}\KatlaSpace{}\KatlaSpace{}\KatlaSpace{}\KatlaSpace{}\KatlaSpace{}\KatlaSpace{}\IdrisKeyword{->}\KatlaSpace{}\IdrisType{Step}\KatlaSpace{}\IdrisBound{pres}\KatlaSpace{}\IdrisBound{a}
\KatlaSpace{}\KatlaSpace{}\KatlaSpace{}\KatlaSpace{}\KatlaSpace{}\KatlaSpace{}\KatlaSpace{}\KatlaSpace{}\KatlaSpace{}\KatlaSpace{}\KatlaSpace{}\KatlaSpace{}\KatlaSpace{}\KatlaSpace{}\KatlaSpace{}\KatlaSpace{}\KatlaSpace{}\IdrisKeyword{(}\IdrisBound{a}\KatlaSpace{}\IdrisFunction{.bindTerm}\KatlaSpace{}\IdrisKeyword{(}\IdrisBound{pres}\IdrisFunction{.axiom}\KatlaSpace{}\IdrisBound{eq}\IdrisKeyword{)}\IdrisFunction{.lhs}\KatlaSpace{}\IdrisBound{env}\IdrisKeyword{)}
\KatlaSpace{}\KatlaSpace{}\KatlaSpace{}\KatlaSpace{}\KatlaSpace{}\KatlaSpace{}\KatlaSpace{}\KatlaSpace{}\KatlaSpace{}\KatlaSpace{}\KatlaSpace{}\KatlaSpace{}\KatlaSpace{}\KatlaSpace{}\KatlaSpace{}\KatlaSpace{}\KatlaSpace{}\IdrisKeyword{(}\IdrisBound{a}\KatlaSpace{}\IdrisFunction{.bindTerm}\KatlaSpace{}\IdrisKeyword{(}\IdrisBound{pres}\IdrisFunction{.axiom}\KatlaSpace{}\IdrisBound{eq}\IdrisKeyword{)}\IdrisFunction{.rhs}\KatlaSpace{}\IdrisBound{env}\IdrisKeyword{)}
\end{SaveVerbatim}
\newcommand\Locate[1][]{\UseVerbatim[#1]{Locate}}
\begin{SaveVerbatim}[commandchars=\\\{\}]{Locate}
\IdrisKeyword{data}\KatlaSpace{}\IdrisType{Locate}\KatlaSpace{}\IdrisKeyword{:}\KatlaSpace{}\IdrisKeyword{(}\IdrisBound{sig}\KatlaSpace{}\IdrisKeyword{:}\KatlaSpace{}\IdrisType{Signature}\IdrisKeyword{)}\KatlaSpace{}\IdrisKeyword{->}\KatlaSpace{}\IdrisKeyword{(}\IdrisBound{a}\KatlaSpace{}\IdrisKeyword{:}\KatlaSpace{}\IdrisType{Algebra}\KatlaSpace{}\IdrisBound{sig}\IdrisKeyword{)}\KatlaSpace{}\IdrisKeyword{->}
\KatlaSpace{}\KatlaSpace{}\KatlaSpace{}\KatlaSpace{}\KatlaSpace{}\KatlaSpace{}\KatlaSpace{}\KatlaSpace{}\KatlaSpace{}\KatlaSpace{}\KatlaSpace{}\KatlaSpace{}\KatlaSpace{}\KatlaSpace{}\IdrisFunction{Rel}\KatlaSpace{}\IdrisKeyword{(}\IdrisFunction{U}\KatlaSpace{}\IdrisBound{a}\IdrisKeyword{)}\KatlaSpace{}\IdrisKeyword{->}\KatlaSpace{}\IdrisFunction{Rel}\KatlaSpace{}\IdrisKeyword{(}\IdrisFunction{U}\KatlaSpace{}\IdrisBound{a}\IdrisKeyword{)}\KatlaSpace{}\IdrisKeyword{where}
\KatlaSpace{}\KatlaSpace{}\IdrisComment{|||\KatlaSpace{}We\KatlaSpace{}prove\KatlaSpace{}the\KatlaSpace{}equality\KatlaSpace{}by\KatlaSpace{}invoking\KatlaSpace{}a\KatlaSpace{}rule\KatlaSpace{}at\KatlaSpace{}the}
\IdrisComment{\KatlaSpace{}\KatlaSpace{}|||\KatlaSpace{}toplevel}
\KatlaSpace{}\KatlaSpace{}\IdrisData{Here}\KatlaSpace{}\IdrisKeyword{:}\KatlaSpace{}\IdrisKeyword{\{0}\KatlaSpace{}\IdrisBound{r}\KatlaSpace{}\IdrisKeyword{:}\KatlaSpace{}\IdrisFunction{Rel}\KatlaSpace{}\IdrisKeyword{(}\IdrisFunction{U}\KatlaSpace{}\IdrisBound{a}\IdrisKeyword{)\}}\KatlaSpace{}\IdrisKeyword{->}\KatlaSpace{}\IdrisBound{r}\KatlaSpace{}\IdrisBound{x}\KatlaSpace{}\IdrisBound{y}
\KatlaSpace{}\KatlaSpace{}\KatlaSpace{}\KatlaSpace{}\KatlaSpace{}\KatlaSpace{}\KatlaSpace{}\KatlaSpace{}\KatlaSpace{}\KatlaSpace{}\KatlaSpace{}\KatlaSpace{}\KatlaSpace{}\KatlaSpace{}\IdrisKeyword{->}\KatlaSpace{}\IdrisType{Locate}\KatlaSpace{}\IdrisBound{sig}\KatlaSpace{}\IdrisBound{a}\KatlaSpace{}\IdrisBound{r}\KatlaSpace{}\IdrisBound{x}\KatlaSpace{}\IdrisBound{y}
\KatlaSpace{}\KatlaSpace{}\IdrisComment{|||\KatlaSpace{}We\KatlaSpace{}focus\KatlaSpace{}on\KatlaSpace{}a\KatlaSpace{}subterm\KatlaSpace{}`lhs`\KatlaSpace{}that\KatlaSpace{}may\KatlaSpace{}appear\KatlaSpace{}in}
\IdrisComment{\KatlaSpace{}\KatlaSpace{}|||\KatlaSpace{}multiple\KatlaSpace{}locations\KatlaSpace{}and\KatlaSpace{}rewrite\KatlaSpace{}it\KatlaSpace{}to\KatlaSpace{}`rhs`\KatlaSpace{}using\KatlaSpace{}a}
\IdrisComment{\KatlaSpace{}\KatlaSpace{}|||\KatlaSpace{}specific\KatlaSpace{}rule.}
\KatlaSpace{}\KatlaSpace{}\IdrisData{Cong}\KatlaSpace{}\IdrisKeyword{:}\KatlaSpace{}\IdrisKeyword{\{0}\KatlaSpace{}\IdrisBound{r}\KatlaSpace{}\IdrisKeyword{:}\KatlaSpace{}\IdrisFunction{Rel}\KatlaSpace{}\IdrisKeyword{(}\IdrisFunction{U}\KatlaSpace{}\IdrisBound{a}\IdrisKeyword{)\}}\KatlaSpace{}\IdrisKeyword{->}
\KatlaSpace{}\KatlaSpace{}\KatlaSpace{}\KatlaSpace{}\KatlaSpace{}\KatlaSpace{}\KatlaSpace{}\KatlaSpace{}\KatlaSpace{}\IdrisKeyword{(}\IdrisBound{t}\KatlaSpace{}\IdrisKeyword{:}\KatlaSpace{}\IdrisType{Term}\KatlaSpace{}\IdrisBound{sig}\KatlaSpace{}\IdrisKeyword{(}\IdrisType{Maybe}\KatlaSpace{}\IdrisKeyword{(}\IdrisFunction{U}\KatlaSpace{}\IdrisBound{a}\IdrisKeyword{)))}\KatlaSpace{}\IdrisKeyword{->}
\KatlaSpace{}\KatlaSpace{}\KatlaSpace{}\KatlaSpace{}\KatlaSpace{}\KatlaSpace{}\KatlaSpace{}\KatlaSpace{}\KatlaSpace{}\IdrisKeyword{\{}\IdrisBound{lhs}\IdrisKeyword{,}\KatlaSpace{}\IdrisBound{rhs}\KatlaSpace{}\IdrisKeyword{:}\KatlaSpace{}\IdrisFunction{U}\KatlaSpace{}\IdrisBound{a}\IdrisKeyword{\}}\KatlaSpace{}\IdrisKeyword{->}\KatlaSpace{}\IdrisBound{r}\KatlaSpace{}\IdrisBound{lhs}\KatlaSpace{}\IdrisBound{rhs}\KatlaSpace{}\IdrisKeyword{->}
\KatlaSpace{}\KatlaSpace{}\KatlaSpace{}\KatlaSpace{}\KatlaSpace{}\KatlaSpace{}\KatlaSpace{}\KatlaSpace{}\KatlaSpace{}\IdrisType{Locate}\KatlaSpace{}\IdrisBound{sig}\KatlaSpace{}\IdrisBound{a}\KatlaSpace{}\IdrisBound{r}\KatlaSpace{}\IdrisKeyword{(}\IdrisFunction{plug}\KatlaSpace{}\IdrisBound{a}\KatlaSpace{}\IdrisBound{t}\KatlaSpace{}\IdrisBound{lhs}\IdrisKeyword{)}\KatlaSpace{}\IdrisKeyword{(}\IdrisFunction{plug}\KatlaSpace{}\IdrisBound{a}\KatlaSpace{}\IdrisBound{t}\KatlaSpace{}\IdrisBound{rhs}\IdrisKeyword{)}
\end{SaveVerbatim}
\newcommand\Derivation[1][]{\UseVerbatim[#1]{Derivation}}
\begin{SaveVerbatim}[commandchars=\\\{\}]{Derivation}
\IdrisFunction{Derivation}\KatlaSpace{}\IdrisKeyword{:}\KatlaSpace{}\IdrisKeyword{(}\IdrisBound{p}\KatlaSpace{}\IdrisKeyword{:}\KatlaSpace{}\IdrisType{Presentation}\IdrisKeyword{)}
\KatlaSpace{}\KatlaSpace{}\IdrisKeyword{->}\KatlaSpace{}\IdrisKeyword{(}\IdrisBound{a}\KatlaSpace{}\IdrisKeyword{:}\KatlaSpace{}\IdrisType{PresetoidAlgebra}
\KatlaSpace{}\KatlaSpace{}\KatlaSpace{}\KatlaSpace{}\KatlaSpace{}\KatlaSpace{}\KatlaSpace{}\KatlaSpace{}\KatlaSpace{}\KatlaSpace{}\KatlaSpace{}\KatlaSpace{}\IdrisBound{p}\IdrisFunction{.signature}\IdrisKeyword{)}
\KatlaSpace{}\KatlaSpace{}\IdrisKeyword{->}\KatlaSpace{}\IdrisFunction{Rel}\KatlaSpace{}\IdrisKeyword{(}\IdrisFunction{U}\KatlaSpace{}\IdrisBound{a}\IdrisKeyword{)}
\IdrisFunction{Derivation}\KatlaSpace{}\IdrisBound{p}\KatlaSpace{}\IdrisBound{a}
\KatlaSpace{}\KatlaSpace{}\IdrisKeyword{=}\KatlaSpace{}\IdrisType{RTList}\KatlaSpace{}\KatlaSpace{}\KatlaSpace{}\KatlaSpace{}\KatlaSpace{}\IdrisComment{--\KatlaSpace{}Reflexive,\KatlaSpace{}Transitive}
\KatlaSpace{}\KatlaSpace{}$\KatlaSpace{}\IdrisType{Symmetrise}\KatlaSpace{}\IdrisComment{--\KatlaSpace{}Symmetric}
\KatlaSpace{}\KatlaSpace{}\KatlaSpace{}\KatlaSpace{}\KatlaSpace{}\KatlaSpace{}\KatlaSpace{}\KatlaSpace{}\KatlaSpace{}\KatlaSpace{}\KatlaSpace{}\KatlaSpace{}\KatlaSpace{}\KatlaSpace{}\KatlaSpace{}\IdrisComment{--\KatlaSpace{}Congruence}
\KatlaSpace{}\KatlaSpace{}$\KatlaSpace{}\IdrisType{Locate}\KatlaSpace{}\IdrisBound{p}\IdrisFunction{.signature}\KatlaSpace{}\IdrisBound{a}\IdrisFunction{.algebra}
\KatlaSpace{}\KatlaSpace{}$\KatlaSpace{}\IdrisType{Step}\KatlaSpace{}\IdrisBound{p}\KatlaSpace{}\IdrisBound{a}\KatlaSpace{}\KatlaSpace{}\KatlaSpace{}\IdrisComment{--\KatlaSpace{}Axiomatic\KatlaSpace{}steps}
\end{SaveVerbatim}

%% file: evaluation-listings.tex
\newcommand\FrexifyRHS[1][]{\UseVerb[#1]{FrexifyRHS}}
\begin{SaveVerbatim}[commandchars=\\\{\}]{FrexifyRHS}
\IdrisKeyword{((}\IdrisBound{c_s}\KatlaSpace{}\IdrisFunction{+}\KatlaSpace{}\IdrisBound{val_s}\IdrisKeyword{)}\KatlaSpace{}\IdrisFunction{+}\KatlaSpace{}\IdrisBound{val_s}\IdrisKeyword{)}\KatlaSpace{}\IdrisFunction{+}\KatlaSpace{}\IdrisKeyword{(}\IdrisData{2}\IdrisFunction{*}\IdrisKeyword{((}\IdrisData{2}\KatlaSpace{}\IdrisFunction{`power`}\KatlaSpace{}\IdrisBound{width}\IdrisKeyword{)}\IdrisFunction{*}\IdrisBound{c0}\IdrisKeyword{))}
\end{SaveVerbatim}
\newcommand\FrexifyLHS[1][]{\UseVerb[#1]{FrexifyLHS}}
\begin{SaveVerbatim}[commandchars=\\\{\}]{FrexifyLHS}
\IdrisBound{c_s}\KatlaSpace{}\IdrisFunction{+}\KatlaSpace{}\IdrisData{2}\IdrisFunction{*}\IdrisKeyword{(}\IdrisBound{val_s}\KatlaSpace{}\IdrisFunction{+}\KatlaSpace{}\IdrisKeyword{((}\IdrisData{2}\KatlaSpace{}\IdrisFunction{`power`}\KatlaSpace{}\IdrisBound{width}\IdrisKeyword{)}\IdrisFunction{*}\IdrisBound{c0}\IdrisKeyword{))}
\end{SaveVerbatim}

%% file: def.tex
\usepackage{suffix}
\usepackage[T1]{fontenc}
\usepackage{footmisc}
\usepackage{wrapfig}
\newcommand\idris{Idris2}
\newcommand\Haskell{Haskell}
\newcommand\agda{Agda}

\usepackage{colortbl}

\newcommand\seclabel[1]{\label{sec:#1}}
\newcommand\secref[1]{Section \ref{sec:#1}}
\WithSuffix\newcommand\secref*[1]{\S\ref{sec:#1}}
\newcommand\secsref[2]{Sections~\ref{sec:#1} and~\ref{sec:#2}}
\newcommand\applabel[1]{\label{app:#1}}
\newcommand\appref[1]{Appendix~\ref{app:#1}}
\WithSuffix\newcommand\appref*[1]{\ref{app:#1}}
\newcommand\subseclabel[1]{\label{subsec:#1}}
\newcommand\subsecref[1]{\S\ref{subsec:#1}}

\newcommand\figlabel[1]{\label{fig:#1}}
\newcommand\figref[1]{Fig.~\ref{fig:#1}}
\newcommand\Figref[1]{Figure~\ref{fig:#1}}

\newcommand{\citepos}[1]{\citeauthor{#1}'s~\citeyearpar{#1}}
\WithSuffix\newcommand\citepos*[2]{\citeauthor{#1}'s #2~\citeyearpar{#1}}
\WithSuffix\newcommand\citepos-[2]{\citeauthor{#1}#2~\citeyearpar{#1}}
\makeatletter

\newcommand\ibid[2][]{%
  \citetext{%
    \hyper@natlinkstart{#2}%
    \textit{ibid.}%
    \hyper@natlinkend%
    \ifx\ibid@empty#1\ibid@empty%
    \else%
    , #1%
    \fi%
  }%
}

\newcommand\citloc[1]{%
  \citetext{%
    \hyper@natlinkstart{#1}%
    \textit{cit.~loc.}%
    \hyper@natlinkend%
  }%
}
\newcommand\citlocpos[1]{%
  \citeauthor{#1}'s~\citloc{#1}%
}
\WithSuffix\newcommand\citlocpos*[2]{%
  \citeauthor{#1}'s #2~\citloc{#1}%
}
\let\FV@ProcessLine\relax
\makeatother

\newcommand\newmvar[3][]{%
  \newcommand#2[1][1]{#1{\ifcase ##1\or #3\or\else@ctrerr\fi}}%
}

\newcommand\setoid[1]{\text{\IdrisBound{#1}}}
\newcommand\idrisvar[1]{\text{\IdrisBound{#1}}}
\newcommand\idrisfun[1]{\text{\IdrisFunction{#1}}}
\newcommand\idrisdata[1]{\text{\IdrisData{#1}}}
\newmvar{\X}{\setoid{X}\or\setoid{Y}\or\setoid{Z}}
\newmvar{\x}{\idrisvar{x}}
\newmvar{\y}{\idrisvar{y}}
\newmvar{\f}{\idrisvar{f}}
\newmvar{\A}{\setoid{A}\or\setoid{B}\or\setoid{C}}
\let\C\undefined
\newmvar{\C}{\mathcal{C}\or\mathcal{B}\or\mathcal{A}}
\def\U{\text{\IdrisFunction{U}}}
\newcommand\UH{\text{\IdrisFunction{H}}}

\WithSuffix\newcommand\~~{\mathrel{\text{\setoidEqualityRelationInfix}}}
\WithSuffix\newcommand\~>{\mathrel{\text{\SetoidArrowInline}}}

\newcommand\Frexlib{\textsc{Frex}}
\newcommand\Fragmentlib{\textsc{Fragment}}
\newcommand\SetoidRewrite{\textsc{setoid\_rewrite}}

\newcommand\FrexLemma{\IdrisType{Lemma}}
\newcommand\Sig{\Sigma}
\newcommand\Pres{\mathcal{T}}
\newcommand\Axiom[1][\!]{#1\text{\IdrisFunction{.Axiom}}}
\newcommand\Op{\mathop{\text{\IdrisType{Op}}}}
\newcommand\arity{\mathop{\text{\IdrisFunction{arity}}}}
\newcommand\pair[2]{(#1,#2)}
\newcommand\triple[3]{(#1,#2,#3)}
\newcommand\Monoid{\mathbf{Monoid}}
\newcommand\CMonoid{\mathbf{CommutativeMonoid}}

\newcommand\definedby{:=}
\newcommand\set[1]{\left\{#1\right\}}

\usepackage{oksemantics}
\newcommand\sem\scottbrackets

\newcommand\IntroPrinterStepCount{24}

\newcommand\TwoColDIY[5][\hfill]{%
\begin{center}
\begin{tabular}{@{}l|l@{}}
\begin{minipage}[t]{#2\columnwidth}%
  #3%
\end{minipage}%
&
\begin{minipage}[t]{#4\columnwidth}%
  #5%
\end{minipage}%
\end{tabular}
\end{center}
}

\newcommand\TwoColDIYprime[5][\hfill]{%
\begin{center}
\begin{tabular}{@{}l@{\hspace{1em}}l@{}}
\begin{minipage}[t]{#2\columnwidth}%
  #3%
\end{minipage}%
&
\begin{minipage}[t]{#4\columnwidth}%
  #5%
\end{minipage}%
\end{tabular}
\end{center}
}

\newcommand\ThreeColDIY[1][\hfill]{\ThreeColDIY*{#1}{#1}}
\WithSuffix\newcommand\ThreeColDIY*[8]{%
\begin{center}
\begin{tabular}{@{}l|l|l@{}}
\begin{minipage}[t]{#3\columnwidth}%
  #4%
\end{minipage}%
&
\begin{minipage}[t]{#5\columnwidth}%
  #6%
\end{minipage}%
&
\begin{minipage}[t]{#7\columnwidth}%
  #8%
\end{minipage}%
\end{tabular}
\end{center}
}

\usepackage{tikz,pgfplots,pgfplotstable}
\pgfplotsset{compat=1.3}
\usetikzlibrary{shapes,backgrounds,patterns,shapes.misc,positioning,calc}

\newcommand\isomorphic\cong

\newcommand\FrexlibLoC{9,500}

\newcommand\+{\mathbin{\text{\IdrisFunction{+}}}}

\newcommand\involve[1]{\overline{#1}}
\newcommand\iinvolve[1]{\involve{\involve{#1}}}
\newcommand\iiinvolve[1]{\overline{\involve{\involve{#1}}}}
\newcommand\smashinvolve[1]{\smash{\involve{#1}}}
\WithSuffix\newcommand\involve*[1]{\vphantom{#1}\smashinvolve{#1}}
\WithSuffix\newcommand\iinvolve*[1]{\vphantom{#1}\smashinvolve{\involve{#1}}}
\WithSuffix\newcommand\iiinvolve*[1]{\vphantom{#1}\smashinvolve{\involve{\involve{#1}}}}

\newcommand\compose\circ

\newcommand\xto\xrightarrow
\newcommand\unit\eta
\newcommand\counit\varepsilon

\newcommand\Var{\text{\IdrisFunction{Var}}}
\newcommand\Embed{\text{\IdrisFunction{Embed}}}

\newcommand\injection\iota

\newcommand\env{\text{\IdrisBound{e}}}

\newcommand\PutListing[2]{%
  \begin{minipage}[t]{#1}
  #2
  \end{minipage}
}
\newcommand\MyCodeListing[2][.8\baselineskip]{%
\\[.8\baselineskip]\begin{minipage}{\textwidth}\centering%
#2
\end{minipage}
\\[#1]
}
\newcommand\naturals{\mathbb{N}}
\newcommand\integers{\mathbb{Z}}

\newcommand\AModule[2][]{{\IdrisHlightStyleModule{#2}}#1:}
\WithSuffix\newcommand\AModule*[2][]{{\IdrisHlightStyleModule{#2}}#1}
\newcommand\length{\text{\IdrisFunction{length}}}
\newcommand\List{\text{\IdrisType{List}}}
\WithSuffix\newcommand\++{\+\+}
\newcommand\Empty{\text{\IdrisData{[]}}}
\newcommand\aembed[1][]{\UseVerb[#1]{aembed}}
\begin{SaveVerbatim}[commandchars=\\\{\}]{aembed}
\IdrisBound{a}\IdrisFunction{.Embed}
\end{SaveVerbatim}

\begin{SaveVerbatim}[commandchars=\\\{\}]{bembed}
\IdrisBound{b}\IdrisFunction{.Embed}
\end{SaveVerbatim}

\definecolor{shade}{RGB}{223,223,223}
\definecolor{unshade}{RGB}{255,255,255}
\newcommand\shade[1]{\setlength{\fboxsep}{0pt}\colorbox{shade}{\ensuremath{#1}}}

\newcommand\CertificateType{\IdrisFunction{units}\KatlaSpace{}\IdrisKeyword{:}\KatlaSpace{}\IdrisKeyword{(}\IdrisBound{x}\KatlaSpace{}\IdrisKeyword{:}\KatlaSpace{}\IdrisFunction{U}\KatlaSpace{}\IdrisBound{m}\IdrisKeyword{)}\KatlaSpace{}\IdrisKeyword{\KatlaDash{}>}\KatlaSpace{}\IdrisFunction{O1}\KatlaSpace{}\IdrisFunction{.+.}\KatlaSpace{}\IdrisKeyword{(}\IdrisBound{x}\KatlaSpace{}\IdrisFunction{.+.}\KatlaSpace{}\IdrisFunction{O1}\IdrisKeyword{)}\KatlaSpace{}\IdrisFunction{.+.}\KatlaSpace{}\IdrisFunction{O1}\KatlaSpace{}\IdrisFunction{=\KatlaTilde{}=}\KatlaSpace{}\IdrisBound{x}}

%% file: abstract.tex
We present a new design for an algebraic simplification library
structured around concepts from universal algebra: theories,
models, homomorphisms, and universal properties of free algebras and
free extensions of algebras.
The library's dependently typed interface guarantees that
both built-in and user-defined simplification modules are terminating,
sound, and complete with respect to a well-specified class of
equations.
We have implemented the design in the Idris 2 and Agda dependently
typed programming languages and shown that it supports modular
extension to new theories, proof extraction and certification, goal
extraction via reflection, and interactive development.

%% file: ccsxml.tex
\ccsdesc[500]{Theory of computation~Type theory}
\ccsdesc[500]{Theory of computation~Constructive mathematics}
\ccsdesc[500]{Theory of computation~Equational logic and rewriting}
\ccsdesc[500]{Theory of computation~Automated reasoning}
\ccsdesc[300]{Theory of computation~Categorical semantics}
\ccsdesc[300]{Theory of computation~Algebraic semantics}
\ccsdesc[100]{Software and its engineering~Formal software verification}
\ccsdesc[100]{Software and its engineering~Functional languages}
\ccsdesc[100]{Mathematics of computing~Solvers}
\ccsdesc[100]{Computing methodologies~Representation of polynomials}

%% file: new-intro.tex
\section{Introduction}
Algebraic simplification involves using algebraic laws to normalise
expressions with unknowns. For example, the commutative monoid
axioms---associativity, neutrality, and commutativity---over
integers with addition serve to
simplify the left hand expression to the right hand expression below:
\[-6 + (x+3)+(y+x)
\;{}\xmapsto{\text{simplify}}{}\;
-3+2x+y
\]
%
Many application domains make use of this kind of simplification.
For example, algebraic simplification is often a useful first step in
a program optimiser, to avoid the need to analyse and transform
distinct but equivalent programs.
The present work focuses on another application domain: theorem
provers and programming languages based on type theory.
In these systems, users often need to prove to a type checker that
terms are equivalent, but constructing the proofs often involves rote
algebraic simplification steps that users resent having to produce
manually.

To free users from the need to construct rote algebraic proofs,
de\-pen\-dent\-ly ty\-ped languages and their ecosystems often
include simplifiers for common algebraic structures such as
monoids, semi-rings, and rings.
With these simplifiers, users need only establish the structures'
axioms, such as neutrality, associativity and commutativity, and can
then call the simplifiers to discharge rote simplification steps.
Implementation strategies for simplifiers vary:  some,
such as the Rocq ring solver~\cite{coq-ref-man-8:ring-solver},
use tactics to simplify  algebraic terms in typing goals,
while others, such as the Agda ring solver~\cite{kidney:ring-solver},
use proof-by-reflection to construct propositions to discharge
equations.
Some simplifiers, such as the Rocq ring solver, group together several algebraic
structures, while others, such as the system by \citet{gregoire-mahboubi:commutative-ring-done-right}, generalise several distinct structures to one
structure.
The
state-of-the-art are standalone simplifiers that, through heuristics
and long-term development, can deal with common cases.

\subsection{Representation Theorems}

Universal algebra has a long tradition concerning algebraic
simplification under the collective name `representation
theorems'. Each such representation theorem characterises canonical
representatives of algebraic expressions in terms of (typically inductive)
constructions such as reduced-words and formal polynomials. Such
characterisations often
reuse existing representation theorems of simpler algebraic
structures or familiar algebraic structures such as the integers or
the natural numbers.

The present work makes use of two kinds of representation theorems.
For a free algebra (abbreviated \emph{fral}), a representation theorem
amounts to an algebraic structure that interprets the algebraic
operations to produce a single canonical representative for all input
terms that are equivalent according to the algebraic laws.
For a free extension (abbreviated \emph{frex}), a representation
theorem chooses a canonical representative using the algebraic laws
while also evaluating concrete elements.
Free algebras and free extensions are related: each free extension is
also a free algebra of a theory specialised to a concrete algebra by
adding axioms over the concrete elements.

For example, for commutative monoids the fral representation theorem
states that the free commutative monoid over $n$ variables is
represented by the set of $n$-tuples of naturals $\naturals^n$.
We simplify with this representation by evaluating
a term in the fral and then reifying it back as a term:
\[
-6 + (x+3) + (y+x)
{}\xmapsto{\text{evaluate}(x_0 + (x_2 + x_1) + (x_3 + x_2))}{}
(1, 1, 2, 1)
{}\xmapsto{\text{reify}(x_0 \mapsto -6, x_1 \mapsto 3, x_2 \mapsto x, x_3 \mapsto y)}{}
-6 +3 + 2x + y
\]\label{new-intro fral example}
In the fral there is no concept of concrete elements, so fral
simplification must treat constants such as $-6$ and $3$ in the same
way as variables such as $x$ and $y$, as abstract and distinct
indeterminates.

The frex representation theorem for commutative monoids
states that the free extension of a commutative monoid over $C$ by $n$
variables is represented by the set $C \times \naturals^n$, where
the element $(c, a_1,
\ldots, a_n)$ represents the expression $c + a_1x_1 + \ldots +
a_nx_n$.
Simplifying a term using the frex representation involves evaluating a
term in the frex and then reifying it back as a term:
\[
-6 + (x+3) + (y+x)
{}\xmapsto{\text{evaluate}_{\integers}}{}
(-3, 2, 1)
{}\xmapsto{\text{reify}}{}
-3 + 2x + y
\]
The frex representation distinguishes variables from concrete
elements, gathering the latter together and evaluating them using the
operations of the concrete commutative monoid in use.

Both kinds of representation theorem allow us to choose a unique
syntactic representative, i.e., a normal form.
Such normalization by evaluation and then reification
also applies to more sophisticated notions of algebra that include the
equational theories of $\lambda$-calculi, and is familiar in those
settings as \emph{normalization-by-evaluation}~\cite{berger:nbe}.  Its first systematic
applications were in the formal study of various simply typed
calculi~\cite{altenkirch-et-al:cat-reconstruction,%
  cubric-dybjer-scott:normalization-and-the-yoneda-embedding,%
  altenkirch-et-al:nbe-stlc-sums}
and category theoretic
constructions~\cite{beylin-dybjer:extracting-coherence-for-monoidal-cats}.
It has served as a conversion-checking technique during the
type-checking of dependently typed calculi from their
inception~\cite{martin-lof:an-intuitionistic-theory-of-types},
gaining adoption after the seminal works
of~\citet{abel-aehlig-dybjer:nbe-ml-tt-universe,abel-coquand-dybjer:nbe-mltt-eq},
even for sophisticated
calculi~\cite{DBLP:journals/pacmpl/0001VW17,sterling-angiuli:normalization-for-cubical-type-theory,DBLP:journals/jfp/HuJP23}.

This article describes an extensible dependently typed library for
algebraic simplifiers based on fral and frex representation theorems,
building on our previous work on free extensions for
partial evaluation~\cite{yallop-et-al:partially-static-data-as-frex}.
Current implementations of algebraic simplifiers,
even in dependently typed settings, are restricted to
implementing the computational representation---i.e.~the data-structures
needed for the normal form together with the
normalisation-by-evaluation algorithm---alongside a formalisation of
the soundness proofs.
This work investigates what can be gained by encoding, in addition,
the full meta-theory of these representation
theorems, including generic representations of theories, their
algebras and algebra homomorphisms, and the universal properties of
the fral and frex.
For simplicity of development and exposition we apply this generic
machinery to a handful of familiar monoid varieties.
Moreover, we ensure all of these concepts remain computational by avoiding
the temptation of postulating axioms that could hinder reduction of closed
terms.
This last task is challenging to combine with
interactive performance, as formalising universal properties tends to
produce large terms that slow
type-checkers~\cite{gross-chlipala-spivak:performant-cat-theory-in-coq}.

\subsection{Paper Outline and Contributions}

\secref{new-overview} present background material: a
review of the mathematical foundation for \Frexlib{},
and a brief \idris{} tutorial that reviews
setoid-based equational reasoning.

\secsref{frexlib}{core} present our central contribution, a
fundamentally new approach to building algebraic simplifiers.
The standard existing approach is to write an ad-hoc simplifier for
some particular algebraic structure such as rings.
We take a different approach, teaching the implementation the basic concepts of
universal algebra --- signatures, theories and models, homomorphisms,
and universal properties (\secref{frexlib}) --- then build a
completely generic solver based on free algebras and free extensions
that can be instantiated with a particular algebra to discharge
concrete proof obligations (\secref{core}).  This new approach is
inherently modular and extensible, and delivers solvers that are sound
and complete by construction.  We have created implementations of our design in
two dependently typed languages: \Fragmentlib{} in \agda{},
and \Frexlib{}
in \idris{}\footnote{
{Available here:
  \url{https://github.com/frex-project/agda-fragment}
and here:
  \url{https://github.com/frex-project/idris-frex}}}.

\secref{certification} explains the completeness guarantees of the
library, and covers proof extraction, simplification, pretty-printing
and certification. The technical \secsref{core}{certification}
are aimed at library designers, and may be skimmed or skipped at
first reading.

\secref{reflection} considers a natural question: can one use
reflection to invoke \Frexlib{} automatically? The answer is a
qualified `yes', requiring much library-developer effort, but leading
to real advantages in \agda{} and limited advantages in \idris{}.

\secref{evaluation} reports some supplementary evaluation of
\Frexlib{}.  The key properties of our design are guaranteed by the
type theories of the languages in which we realise it: it delivers
sound and complete solvers in a completely generic way, with support
for proof extraction, certification, etc.  However, the practical
questions of usability and viability for interactive development
cannot be established by theorems and so we have also carried out some
experiments.  These experiments focus on the varieties of monoids that
also serve as our running example, and establish that the generic
solver is comfortably fast enough for interactive use, and can be
extended with new algebras in a modular way and without enormous
effort.

\secref{lessons} discusses system design issues that we encountered
with \Frexlib{}, and \secsref{related}{conclusion} conclude with
related and further work.

Appendices~\ref{app:first-appendix}--\ref{app:last-appendix}
have more information about \Frexlib's
\FrexlibLoC{}-line \idris{} codebase, and example code extraction.


%% file: overview.tex
\section{Mathematical Overview}
\seclabel{new-overview}
\emph{Universal algebra} concerns the generic description of algebraic
structures and relationships between them~\cite[Chp.~II]{burris-sankappanavar:universal-algebra}.
We summarise
the concepts relevant to \Frexlib{}.

\subsection{Presentations of Algebraic Structures}\seclabel{presentations}
A \emph{finitary signature} $\Sig = \pair{\Op\Sig}{\arity}$ consists
of a set $\Op\Sig$ of \emph{operators},
and an assignment $\arity : \Op\Sig \to \Nat$ of a natural
number to each operator called its \emph{arity}.
For example, the signature $\Sig_{(2,0)}$ often used for monoids
has two operators $\Op(\Sig_{(2,0)}) \definedby \set{(+),0}$ with
respective arities $2$ and $0$.
It is common to use a succinct notation that groups operators
and their arities: $\Sig_{(2,0)} \definedby \set{(+):2,0:0}$.

Signatures determine an algebraic language, whose semantic models
are \emph{algebras}.
An \emph{algebra} $\A = \pair{\U\,\A}{\A\sem-}$
  for a signature $\Sig$
  consists of
a set $\U\,\A$ called the \emph{carrier} and
an assignment $\A\sem f : (\U\,\A)^n \to \U\,\A$ of an \emph{$n$-ary operation} over this carrier for every $n$-ary operator $f : n$ in $\Sigma$.
Continuing the example above, $\Sig_{(2,0)}$-algebras amount to
triples $\triple{X}{\sem{(+)} : X^2 \to X}{\sem{0} \in X}$.
There may be many different algebras for a given signature and carrier set:
for instance, we can equip the natural
numbers $\naturals$ with the $\Sig_{(2,0)}$-algebra structures of
arithmetic addition $\triple{\naturals}{(+)}{0}$,
arithmetic multiplication $\triple\naturals{(\cdot)}1$,
or $\triple{\naturals}{\max}{0}$.
Examples abound:
subtraction over the integers has a $\Sig_{(2,0)}$-algebra
$\triple{\integers}{(-)}{0}$;
$n\times n$ matrices over $\naturals$ have
$\Sig_{(2,0)}$-algebra structures given by matrix addition and
multiplication with the zero and identity matrix respectively, and so
on.

Each signature determines a language consisting of \emph{terms}.
Given a set $\X$ of variables, the
\emph{$\Sig$-terms over $\X$} are given inductively as either a
variable in $\X$ or an application $f(t_1, \ldots, t_n)$ of an
operator $f : n$ from $\Sig$ to $n$ terms over $\X$.
The primary role of terms is to designate \emph{equations in context}
$\X \vdash t = s$, i.e.~triples consisting of a set $\X$ of
variables and two terms in context $\X$. For example, the
associativity equation, expressed over the signature $\Sig_{(2,0)}$,
is $x,y,z \vdash x + (y + z) = (x + y) + z$.

An \emph{environment} for a context (=set) $\X$ in an algebra $\A$ is
a function $\env : \X \to \U\,\A$.
An algebra $\A$ determines, for
each term in context $\X \vdash t$, an \emph{interpretation} function
$\A\sem t : (\U\,\A)^{\X} \to \U\,\A$.
Given an environment $\env$, $\A\sem t$ interprets
each variable in the environment $\A\sem{\x}\env \definedby \env\,\x$,
and structurally interprets each operator
application:
$\A\sem{f(t_1, \ldots, t_n)}\env\definedby \A\sem f(\A\sem{t_1}\env, \ldots, \A\sem{t_n}\env)$.
For example, the interpretation of the left-hand-side (LHS) of the
associativity axiom in the $\Sig_{(2,0)}$-algebra $\triple{\naturals}{\max}{0}$
given the environment $\{x \mapsto 5, y \mapsto 3, z \mapsto 8\}$ is
$\max(5, \max(3, 8)) = 8$.

We say that an equation is \emph{valid} in an algebra $\A$, writing
$\A \models (\X \vdash t = s)$, when $\A\sem t\env = \A\sem s\env$
for all environments $\env : \X \to \U\,\A$. 
For example, the $\Sig_{(2,0)}$-algebras
presented so far validate the associativity axiom, but interpreting
the binary operation as subtraction over the integers $\triple{\integers}{(-)}{0}$
does not: taking the
environment $\{x \mapsto 0, y \mapsto 0, z \mapsto 1\}$, we have
$0 - (0 - 1) = 1 \neq -1 = (0 - 0) - 1$.

A \emph{presentation} of an algebraic theory $\Pres = \pair{\Sig_{\Pres}}{\Pres\Axiom}$
consists of a signature $\Sig_{\Pres}$ and a set $\Pres\Axiom$ of
$\Sig_{\Pres}$-equations in context, which we call \emph{axioms}. A
\emph{$\Pres$-model} $\A$ is a $\Sig_{\Pres}$-algebra $\A$ validating
all $\Pres$-axioms.  For example, the axioms of the $\Monoid$
presentation consist of associativity and neutrality ($x \vdash x
+ 0 = x, 0 + x = x$) over the signature $\Sig_{(2,0)}$. The
axioms for the $\CMonoid$ presentation additionally include commutativity $x,y
\vdash y + x = x + y$. We can now generically manipulate classes of
algebraic structures using these concepts, while generalising the
usual examples: $\Monoid$-models are monoids, $\CMonoid$-models are
commutative monoids, etc.

\subsection{Homomorphisms, Free Models/Algebras, and Free Extensions}\seclabel{simplification}
For a presentation $\Pres$, we consider two classes of simplification problem of interest, which we call \emph{purely abstract} and \emph{partially concrete}, respectively.
In the purely abstract case, the input to the problem consists of a set of variables $\X$ and a $\Sigma_{\Pres}$-term over $\X$.
Usually, $\X = \Finn$ is a set of De Bruijn levels.
In the partially concrete case, the input instead consists of a $\Pres$-model $\A$, a set of variables $\X$ and a $\Sigma_{\Pres}$-term over the disjoint union $\U\,\A \uplus \X$.
Here, we do not treat the `variables' in the term coming from the inclusion of $\U\,\A$ as variables in the usual sense.
Indeed, in both cases, the goal of the simplification problem is to find a representative modulo the presentation's axioms and the rules of deduction.
However, in the partially concrete case, simplification must also fold the elements of $\U\,\A$ using the operations of $\A$.
When the set $\X$ is empty, simplification and evaluation should coincide.

The foregoing description is imprecise, but can be made precise using the notions of free algebra and free extension, whose definitions are formulated in terms of the unique existence of certain structure-preserving maps.
We dedicate the remainder of this section to introducing these concepts, drawing connection to the frex and fral simplification examples from
page \pageref{new-intro fral example}. We start with the fral:
\begin{equation}
-6 + (x+3) + (y+x)
{}\xmapsto{\text{evaluate}(x_0 + (x_2 + x_1) + (x_3 + x_2))}{}
(1, 1, 2, 1)
{}\xmapsto{\text{reify}\left(\begin{array}{@{}>{\scriptstyle}l@{}}
      x_0 \mapsto -6, x_1 \mapsto 3, \\
      x_2 \mapsto x, x_3 \mapsto y
    \end{array}
\right)}{}
-6 +3 + 2x + y
\qquad
\label{nbe-examples-fral}
\end{equation}

Let $\Pres$ be a presentation and $\X$ a set of variables.  We define a \emph{$\Pres$-model\/}
\emph{$\avar{}=\pair{\aModel}{\aEnv}$ over $\X$} to be a
$\Pres$-model $\aModel$ equipped with an $\X$-environment in this
algebra, i.e.~a function $\env : \X \to \U(\aModel)$.  This concept
formalises the inputs to the fral simplification process from
\eqref{nbe-examples-fral}.
The starting term is the expression $-6 + (x+3) + (y+x)$
in the meta-language involving the
meta-level variables $x$ and $y$.
The argument
$x_0 + (x_2 + x_1) + (x_3 + x_2)$
in the label on the left arrow is
a $\Sig_{(2,0)}$-term with object-level variables from
$\X \definedby \set{x_0, \ldots, x_3}$. The argument to the label on
the right arrow, $(x_0 \mapsto -6, x_1 \mapsto 3, x_2 \mapsto x, x_3
\mapsto y)$, is an $\X$-environment mapping the object-level variables $x_i \in \X$
to expressions in the meta-level such as the constants $-6$ and $3$ and the
meta-level variable $y$. To the fral simplification process
these three expressions have the same status.
The frex simplification process we describe later uses the concrete
nature of $-6$ and $3$, simplifying the meta-level expression further through evaluation.

Let $\A[1]$, $\A[2]$ be $\Sig$-algebras for a signature
$\Sig$.  A \emph{homomorphism} $h : \A[1] \to \A[2]$ of
$\Sig$-algebras is a semantics-preserving function $h : \U\,\A[1]
\to \U\,\A[2]$ between their carriers. Explicitly, for all operators
$f : n$ in $\Sig$ and elements $a_1, \ldots, a_n$ in $\U\,\A$, we have:
\( h(\A[1]\sem f\!(a_1, \ldots, a_n)) = \A[2]\sem f\!(h\,a_1,
\ldots, h\,a_n) \). A homomorphism between $\Pres$-models
is a homomorphism between the underlying signature algebras.
For example, the list-length function is a homomorphism from the
monoid of concatenation over lists to the monoid
of addition over naturals: $\length : \triple{\List \X}{(\++)}{\Empty}
\to \triple\naturals{(+)}0$.

\begin{wrapfigure}[4]{r}{2cm}
  \vspace{-1\baselineskip}
  \includegraphics{diagram-01.mps}
\end{wrapfigure}
A \emph{morphism} $\hName : \avar \to \bvar$ of $\Pres$-models over
$\X$ is a $\Pres$-model homomorphism that moreover makes the diagram
on the right commute.  A \emph{free $\Pres$-model over $\X$} is
then a $\Pres$-model over $\X$ from which there is a unique such
morphism to every other $\Pres$-model over $\X$. This
existence-and-uniqueness property is called the \emph{universal
property} of free $\Pres$-models.  For example, the free commutative
monoid over $\X = \set{x_0, \ldots, x_3}$ is the $\Sig_{(2,0)}$-algebra
over $\naturals^4$ given by componentwise arithmetic addition, and
equipped with the $\X$-environment sending $x_0$ to $(1, 0, 0, 0)$,
etc. The unique morphism out of this algebra into any commutative
monoid $\A$, equipped with an environment $\env$, sends $(a_0, \ldots,
a_3)$ to $a_0(\env\,x_0) + \ldots + a_3(\env\,x_3)$.

Confusingly, the literature sometimes refers to $\Pres$-models as
$\Pres$-algebras~\cite[\S{}V.6]{mac-lane:categories-2nd-ed}. To avoid
this confusion, we will use the terminology `$\Pres$-models' as much as
possible, with one exception. When $\Pres$ is implicit, we will
sometimes use the portmanteau `fral' of `\textbf{fr}ee
\textbf{al}gebra'. We prefer it over alternatives (`frmo' or `frem').

In summary, fral simplification provides a guiding principle
for designing the data structure in the middle of
\eqref{nbe-examples-fral}: it is an implementation of the fral.
\label{methodology-anchor}%
The success criterion for the design of the simplification code is the
universal property, which singles the fral up to a unique isomorphism
of models over $\X$. The different components of the universal
property provide the following methodology for this creative process.  Failure in
each step in this process may provide insight into how the current
data structure fails, and how it may be revised:
\begin{enumerate}
\item\label{methodology-step-structure} \textit{Equip the data structure with
  a $\Sig_{\Pres}$-algebra structure.} Failure in this step typically exposes
  inputs this simplifier cannot deal with. Revise the data structure to
  support these operations.
\item\label{methodology-step-validation} \textit{Prove it validates the
  axioms, i.e., forms a $\Pres$-model.}  Failure in this step
  exposes equations the structure does not use, i.e., completeness
  issues. Revise the data structure to use the additional equations, or
  relax the presentation, giving up on using these equations.
\item\label{methodology-step-existence} \textit{Define a function out of
  it to any other $\Pres$-model over $\X$.} Failure in this step
  typically exposes `junk' in the data structure: components that are
  not captured by the presentation such as missing operators.
  Revise the data structure to exclude the junk, or extend the presentation
  with more operators to account for the additional components.
\item\label{methodology-step-homomorphic}
  \textit{Show this function is a homomorphism.}
  Failure in this step typically exposes unprovable
  equations that the data structure validates.
  Revise the data structure/presentation.
\item\label{methodology-step-4-uniqueness} Show this homomorphism is
  unique.  Failure in this step similarly exposes implicit axioms in
  the data structure, revise data structure/presentation.
\end{enumerate}
While each step requires mathematical creativity, the overall structure
breaks the task down into smaller steps, a checklist. This checklist
organises the simplification code along these steps, making simplifier
development more methodical.

We now turn to frex simplification, through the second example from
page~\pageref{new-intro fral example}:
\begin{equation}
-6 + (x+3) + (y+x)
{}\xmapsto{\text{evaluate}_{\integers}}{}
(-3, 2, 1)
{}\xmapsto{\text{reify}}{}
-3 + 2x + y
\label{nbe-examples-frex}
\end{equation}
Let $\A$ be a $\Pres$-model and $\X$ a set whose elements we treat as representing
variables. An \emph{extension}
of $\A$ by $\X$ is a triple
$\avar =
\triple{\aModel}{\aVar}{\aEmbed}$
consisting of a $\Pres$-model $\aModel$, a function $\aVar : \X \to
\U(\aModel)$, and a $\Pres$-homomorphism $\aembed : \A \to \aModel$.
This concept lets us phrase the distinction between meta-level constants such as
$3$ and $6$, and meta-level variables such as $y$ in the meta-level expression
$-6 + (x+3) + (y+x)$ from \eqref{nbe-examples-frex}.
Taking $\A := \integers$ and $\X \definedby \set{x,y}$, both the
concrete elements of the algebra $\A$ (e.g., $3$ and $-6$) and the
abstract variables from $\X$ are part of the vocabularly of the
simplification process. It uses the evaluation equation
$-6 + 3 = -3$.

\begin{wrapfigure}[4]{r}{3.2cm}
  \vspace{-1\baselineskip}
  \includegraphics{diagram-02.mps}
\end{wrapfigure}
A \emph{morphism}\/ $\hName : \avar \to \bvar$ of extensions of $\A$
by $\X$ is a $\Pres$-homomorphisms $\hName : \aModel \to \bModel$ that
makes the two triangles in the diagram on the right commute.  A
\emph{free extension} of $\A$ by $\X$ is then an extension from which
there is a unique such morphism to every other extension of $\A$ by
$\X$.  For example, the free extension of a commutative monoid $\A$ by
two variables $\X \definedby \set{x,y}$ is the $\Sig_{(2,0)}$-algebra
over $\U\,\A\times \naturals^2$, given componentwise, equipped with
the $\Sig_{(2,0)}$-homomorphism that sends $u \in \U\,\A$ to $(u, 0,
0)$ and the function that sends $x$ to $(\A\sem0, 1, 0)$ and $y$ to
$(\A\sem0, 0, 1)$. The unique morphism to any
extension $(\A[2], \env, b)$ sends $(u,a_0,a_1)$ to
$b u + a_0(\env\,x) + a_1(\env\,y)$.

\begin{figure*}
  \centering
  \input{table-of-frex}
  \caption{Frals and frexes for varieties of monoids}
  \figlabel{monoid frexlets}
\end{figure*}
\Figref{monoid frexlets} summarises the frals and frexes
in this article. All but the last frex are well-known representations.
Our contribution is to implement, alongside these representations,
constructive proofs that they represent the fral or the frex. These proofs
require the universal algebraic concepts we introduced
so far. Other simplification modules can reuse these concepts and the representation
theorems for existing simplifiers. This design enables extensible and modular
user and library code.

\subsection{Setoids}

Designing \Frexlib{} around the concepts from
\secref{presentations} and~\secref{simplification} supports
term simplification in a uniform way.
Generalising the design further to use
\emph{setoids}~\cite{hofmann:thesis,Bishop1967-BISFOC-3} instead of
sets enables the same abstractions to offer more flexible
functionality: printing terms and proofs, proof simplication, code
generation, and more, which
\secsref{certification}{reflection}
explore.  The ability to extract such
intensional information from terms violates algebraic equality, and
differentiates this design from work involving quotients, for example
quotient inductive-inductive types
(QIITs)~\cite{altenkirch-et-al:qiits,kaposi-et-al:constructing-qiits,altenkirch-et-al:tt-in-tt-with-qiits}.
This
section presents some mathematical background about setoids and their
realisation in \idris{}.

A \emph{setoid} $\X = (\U\,\X,(\equiv_{\X}))$ consists of a set
$\U\,\X$ and an equivalence relation $(\equiv_{\X})$ over $\U\,\X$.  A
\emph{setoid homomorphism} $\f : \X[1] \~> \X[2]$ is a
relation-preserving function between sets $\X[1]$ and $\X[2]$:
$\f\,x \equiv_{\X[2]} \f\,y$ whenever $x \equiv_{\X} y$.
We think of elements in $\U\,\X$ as representatives of the equivalence
classes of $(\equiv_{\X})$, and so every setoid homomorphism induces
a (unique) function between the quotients $\X/(\equiv_{\X}) \to \X[2]/(\equiv_{\X[2]})$.
One way to construct a setoid is to equip a set $\X$ with its
equality relation to give $(\X, (=))$, but using setoids
also allows us to explicate sophisticated equivalence relations and define
operations on them that are not supported by the corresponding
quotients. For example, given a presentation $\Pres$ and a set $\X$,
we can define the \emph{provability} relation $\X \vdash - = -$ over
terms $\X \vdash t$ (cf.~\figref{provability}).  Related elements in
the setoid $(\Term\X, \X\vdash -=-)$ represent different, but provably
equal, terms. In contrast, elements in the quotient
$\Term\X/(\X\vdash-=-)$ represent equivalence classes of provably equal
terms. \Figref{provability} also includes the evaluation axiom (\textsc{eval}),
which we will use to present the frex.
\begin{figure}
  \input{provability}
  \caption{Provability in (a) equational logic (unshaded) and (b) the evaluation rule}
  \figlabel{provability}
\end{figure}

Setoids and their homomorphisms form a common technique to complete an
intensional type theory with extensional functions and quotients,
requiring users to establish that every defined function is a setoid
homomorphism.  However, in \Frexlib{} we make essential use of setoids
that quotient types~\cite{hofmann-phd} do not allow.
For example, the terms-up-to-provability
setoid supports operations such as $\mathrm{vars} : \Term\X
\to \List\;\X$ which extract the list of variables appearing in a given
term in-order. This function is not a setoid homomorphism.
The quotient can only support such extraction following a canonicalisation.
Both flavours of
function are useful in applications, but setoids, and not quotients,
support both.

In \Frexlib{}, we implement universal algebra \emph{internal} to setoids:
carriers are setoids; algebraic operations are setoid homomorphisms;
algebra homomorphisms and environments must also be setoid
homomorphisms; and the unique homomorphisms in the universal
properties of the fral and the frex must be setoid homomorphisms.
With this generalisation, the setoid $\Term\X/(\X\vdash-=-)$ also
satisfies the fral universal property constructively. Therefore, there
is a unique canonical setoid isomorphism to every other fral.  Similarly,
by including the evaluation equations (\textsc{eval}), we obtain a
setoid frex together with a setoid isomorphism to every other frex.
These
isomorphisms let us extract simplification proofs generically out of
user-defined simplifiers. We discuss the decision to use setoids
in \secref{conclusion} where we refer to the concrete benefits setoids offer.

%% file: table-of-frex.tex
\begingroup\small
\begin{tikzpicture}

\node[minimum width=0.3\textwidth] (monoid)
  at (0,0) { \bfseries monoids };
\node[minimum width=0.3\textwidth,anchor=north west] (cmonoid)
  at (monoid.north east) { \bfseries \textit{commutative} monoids };
\node[minimum width=0.3\textwidth,anchor=north west] (imonoid)
  at (cmonoid.north east) { \bfseries \textit{involutive} monoids };

\node[anchor=north] 
  at ($(monoid.north) + (0,-0.8)$)
     { \parbox{0.28\textwidth}{\centering lists of\\ variables} };
\node[anchor=north]
  at ($(cmonoid.north) + (0,-0.8)$)
     { \parbox{0.28\textwidth}{\centering origin-intercepting\\ linear polynomials} };
\node[anchor=north]
  at ($(imonoid.north) + (0,-0.8)$)
     { \parbox{0.28\textwidth}{\centering lists over ordinary \& \\ singly-involuted variables } };

\node[anchor=north]
  at ($(monoid.north) + (0,-1.8)$)
     { $yxxyx$ };
\node[anchor=north]
  at ($(cmonoid.north) + (0,-1.8)$)
     { \parbox{0.28\textwidth}{\centering $a_1x_1\! +\! \ldots\! +\! a_nx_n$ \\ $(a_i : \Nat)$}};
\node[anchor=north] 
  at ($(imonoid.north) + (0,-1.8)$)
     { $y\overline xxx\overline yx$ };

\node[anchor=north]
  at ($(monoid.north) + (0,-2.9)$)
     { \parbox{0.28\textwidth}{\centering ~\\[-0.5ex] alternating lists\\in $\mathbb{M}_{2\times 2}(\Nat{})[y]$\\~} };
\node[anchor=north]
  at ($(cmonoid.north) + (0,-2.9)$)
     { \parbox{0.28\textwidth}{\centering ~\\[-0.5ex]  linear polynomials \\ in $\A{}[x_1, \ldots, x_n]$\\~} };
\node[anchor=north]
  at ($(imonoid.north) + (0,-2.9)$)
     { \parbox{0.28\textwidth}{\centering alternating lists\\ with tagged variables \\ in $\String{}[x,y]$ } };

\node[anchor=north]
  at ($(monoid.north) + (0,-4.1)$)
     {   $\begin{pmatrix} 1 & 3 \\ 0 & 2 \end{pmatrix}y\begin{pmatrix} 0 & 1 \\ 1 & 0 \end{pmatrix}y$ };
\node[anchor=north]
  at ($(cmonoid.north) + (0,-4.3)$)
     { \parbox{0.28\textwidth}{\centering $c\! +\! a_1x_1\! +\! \ldots\! +\! a_nx_n$ \\ ($a_i : \Nat{}, c : \A$)}};
\node[anchor=north] (imonoid-frex-repr-eg)
  at ($(imonoid.north) + (0,-4.3)$)
     { $\EmptyStr{}x\helloStr{}y\ollehStr{}\overline x\EmptyStr{}$ };

\node[rectangle,rounded corners,draw=black,minimum width=0.26\textwidth,minimum height=5.3cm,anchor=north west]
  at ($(monoid.north west) + (0.01\textwidth,0)$) { };
\node[rectangle,rounded corners,draw=black,minimum width=0.26\textwidth,minimum height=5.3cm,anchor=north west]
  at ($(cmonoid.north west) + (0.01\textwidth,0)$) { };
\node[rectangle,rounded corners,draw=black,minimum width=0.26\textwidth,minimum height=5.3cm,anchor=north west]
  at ($(imonoid.north west) + (0.01\textwidth,0)$) { };

\node[rectangle,fill=red,fill opacity=0.1,minimum width=0.96\textwidth,minimum height=2cm,anchor=north west] (frals)
  at ($(monoid.north west) + (-0.06\textwidth,-0.7cm)$) { };

\node[rectangle,fill=green,fill opacity=0.1,minimum width=0.96\textwidth,minimum height=2.2cm,anchor=north west] (frexes)
  at ($(monoid.north west) + (-0.06\textwidth,-2.9cm)$) { };

\node[anchor=north,rotate=90] at ($(frals.west) + (0.03\textwidth,0)$) { free algebra};
\node[anchor=north,rotate=90] at ($(frexes.west) + (0.03\textwidth,0)$) { free extension};

\end{tikzpicture}
\endgroup

%% file: provability.tex
\begin{gather*}
\inferrule
    {~}
    {
      \X \vdash t = t
    }
    \textsc{refl}
\qquad
\inferrule
    {\X \vdash s = t}
    {
      \X \vdash t = s
    }\textsc{sym}
\qquad
\inferrule
    {\X \vdash t = s\quad
      \X \vdash s = r
    }
    {
      \X \vdash t = r
    }\textsc{trans}
\qquad
\inferrule{
  (\X \vdash t = s) \in \Pres\Axiom
}{
  \X \vdash t = s
}\textsc{ax}
\\
 \inferrule
    {\X[2] \vdash t=s\quad
      \theta_1,\theta_2 : \X[2] \to \Term\X[2]\quad
     (\X \vdash \theta_1y = \theta_2y)_{y \in \X[2]}
    }
    {
      \X \vdash t[\theta_1] = s[\theta_2]
    }\textsc{cong}
\qquad
\shade{
\inferrule{
  \U\A \vdash t
}{
  \X \vdash t = \A\sem{t}
}\textsc{eval}
}
\end{gather*}

%% file: background.tex
\begin{figure}
\TwoColDIY%
    {.6}{\EquivalenceDef{}
          \SetoidDef{}}%
    {.35}{%
      \SetoidSugar{}
      \SetoidCarrierSugar{}
      \SetoidRelationSugar{}
    }
    \caption{(a) Equivalence relations and setoids as records and
      (b) example desugaring into a GADT and projections}
\figlabel{setoid}
\end{figure}
\paragraph{Setoids in \idris{}.}
We represent equivalence
relations and setoids in \idris{} with \emph{records} in \figref{setoid}a.
\idris{} records are syntactic sugar for a single-constructor
\IdrisKeyword{data} declaration and automatically generated
\emph{field projections}, as in \figref{setoid}b.  \idris{} also
automatically generates the post-fix projections for each
field using a dotted notation, writing
\IdrisBound{b}\IdrisFunction{.equivalence.relation} for the
nested projection.
The annotation \IdrisKeyword{0} preceding the definition of the field $\U$ is a
\emph{quantity} \cite{mcbride:qtt,atkey:qtt} indicating
that the field is not represented at runtime, but may be used in
types. Readers can safely ignore these annotations. There is only a handful of them in
this article.
We maintain them to demonstrate and emphasise that \Frexlib{} does not
require us to retain, for example, runtime representations of types.
If you are reading this article in colour,
our listings include semantic highlighting, designating the semantic class
of each lexeme: \IdrisData{data} constructor, \IdrisType{type}
constructor, \IdrisFunction{defined} function or value, and
\IdrisBound{variable} in a binding/bound occurence.
We define setoid homomorphisms:
\MyCodeListing{
  \PutListing{.49\textwidth}\SetoidHomo
  \PutListing{.39\textwidth}\SetoidArrow
}
The Appendix includes expanded examples for setoids of functions and quotient
setoids.

This technique is affectionately dubbed `setoid hell', since we
need to prove that all our functions are setoid homomorphisms.
Following \citeauthor{hu-carette:agda-ct}~\citeyearpar{hu-carette:agda-ct}, we manage
setoid hell by structuring code categorically, organising results
into homomorphisms between appropriate setoids.

%% file: frexlib.tex
\section{Universal Algebra in \Frexlib{}}
\seclabel{frexlib}
To define an interface to algebraic simplifiers, we first
specify and represent algebraic structures.
We implement signatures and their operators in \Frexlib{} as follows
(below, left and middle):
\ThreeColDIY[]{.27}{
  \SignaturesPartA{}
  \SignaturesPartB{}
}{.35}{\OpDecl{}}
 {.35}{\MonoidOperationIntro{}}
The implementation uses \idris{}'s implicit record field for
\ArityName{}.  Users define concrete instances of \SignatureName{},
such as the signature \MonoidSig{} for monoids, by defining a
type family for the indexed field
\OpWithArityName{} (above, right).

\begin{figure*}
  \TwoColDIY{.375}{
      \Tuple{}
  \algebraOverDecl{}
  \AlgebraDecl{}
  }{.6}{
\CongruenceWRT{}
\SetoidAlgebra{}
  }
\caption{Algebras and setoid algebras in \Frexlib{}}
\figlabel{algebras}
\end{figure*}

\Frexlib{} represents the domain of an $n$-ary operation
with an $n$-ary vector (\figref{algebras}). (As in
\Haskell{},
backticks turn any name into an infix operator.)  For
example, the additive natural numbers form
an algebra for the monoid
signature as follows:
\MyCodeListing[0\baselineskip]{
  \PutListing{.5\textwidth}\NatAdditive{}
  }
Instead of specifying the \IdrisBound{Semantics} field directly,
this code uses the smart constructor \IdrisFunction{MkAlgebra},
which has a \IdrisBound{Sem} argument, instead of \IdrisBound{Semantics}.
This smart constructor transfers its \IdrisBound{Sem} argument into
\IdrisData{MakeAlgebra}'s \IdrisBound{Semantics} field by uncurrying
each \nvar{}-ary function into a function taking an \nvar{}-ary vector of arguments.
The departure from our usual naming scheme in which it is the constructor of
a record \IdrisType{R} that is called \IdrisData{MkR} indicates that the
smart constructor \IdrisFunction{MkAlgebra}, not the record constructor
\IdrisData{MakeAlgebra},
is the preferred way to construct
\IdrisType{Algebra} instances.
The \IdrisKeyword{\textbackslash{}case} keyword is an anonymous
function that immediately pattern-matches its argument.  \emph{Setoid
  algebras} further require an equivalence relation that forms a
congruence w.r.t.~the operations~(\figref{algebras}).

We implement terms over a signature as follows, mirroring their mathematical definition:
\MyCodeListing{
  \PutListing{.7\textwidth}\TermDecl}
Terms form an algebra, the \emph{free} algebra, with symbols denoting term formers:
\MyCodeListing{
\PutListing{.7\textwidth}\FreeDecl
}
Terms also form a monad, with \IdrisData{Done} as its unit and substitution
as its sequencing operation.

\begin{figure*}[t]
\MonoidAxiom{}

\MonoidTheory{}

\MonoidStructure{}

\MonoidModel{}
\caption{Axiomatising monoids in \Frexlib{}}
\figlabel{monoids}
\end{figure*}

Turning to equations, \Frexlib{} only needs
equations in a finite context, and we call its cardinality
the \emph{support} of the equation. We implement equations
and presentations as follows:
\ThreeColDIY{.26}{\EquationDecl[gobble=2]{}}{.3}{\PresentationDecl{}}
            {.38}{\AssociativityScheme[numbers=right,numbersep=-.1cm]{}}
\noindent\\[.3\baselineskip] For example, the monoid presentation
$\Monoid$ has three axioms: left and right neutrality, and
associativity. \Frexlib{} defines a generic collection of axiom
schemes (above, right).  The second argument to \IdrisFunction{EqSpec}
lists the arities \IdrisData{[n1, ..., nk]} of the operations in a
scheme, so on lines 1--2 the argument \IdrisData{[2]} in the type
\SchemeOps{} states that the scheme involves a single binary
operation.  The first argument \IdrisData{3} in \MkEquation{} (lines 5--6)
indicates that the declaration of the scheme involves three variables
\IdrisFunction{X}\;\IdrisData{0}, \IdrisFunction{X}\;\IdrisData{1}, and
\IdrisFunction{X}\;\IdrisData{2}. We use these
generic axiom schemes to construct, for example, the theory of monoids in the
middle of \figref{monoids}.

\begin{figure*}[t]
\Models{}

\medskip

\Entails{}

\medskip

\ValidatesEquation{}

\medskip

\Validates{}
  \caption{Equational validity in an algebra}
  \figlabel{equational validity}
\end{figure*}

\figref{equational validity} shows
\Frexlib{}'s representation of what it means for an algebra
to validate an equation. We use \idris{}'s dependent
pairing construct to pair an algebra with an environment in the
standard entailment syntax \ModelEnv{}.
The following code validates the monoid axioms
for our running example:
\MyCodeListing{
  \PutListing{.5\textwidth}\AdditiveNatValid{}
}
We define models for a presentation:
\MyCodeListing{
  \PutListing{.5\textwidth}\ModelDecl{}
}
We can now define a monoid to be a $\Monoid$-model, as in
\figref{monoids}. For another example, now putting everything together,
we validate the monoid structure of multiplication as follows:
\MyCodeListing{
  \PutListing{.55\textwidth}{\NatMultitive[numbers=right]}
}
\label{cast explanation}
The coercion function %
(\IdrisFunction{cast} \IdrisKeyword{:} \IdrisBound{from} \IdrisKeyword{->} \IdrisBound{to}) %
on Line 3 is a method in the standard Idris type-class/interface
\IdrisType{Cast}
\IdrisBound{from} \IdrisBound{to}.  \Frexlib{} exports a
\IdrisType{Cast}-instance that converts an algebra into a setoid
algebra whose equivalence relation is propositional equality
(\IdrisKeyword{=}).  Lines 7--9 use results about the natural numbers
from \idris{}'s standard library.

\begin{figure}
  \ListInvMonoid{}
  \caption{The involutive monoids of list reversal}
  \figlabel{involutive lists}
\end{figure}

\paragraph{Using \Frexlib{}.}
Unless they are already working abstractly with an algebraic structure,
we expect that in practice users start by recognising that their concrete
algebra validates the axioms of an existing simplification
module---\emph{frexlet} for short. Such modules export a presentation,
convenient notation suites for its signature, and fral and/or frex simplifiers
for this presentation.

As a concrete example, we will take computations with lists
that also involve the \IdrisFunction{reverse} function. These form an
\emph{involutive} monoid: a monoid $\A$ equipped with a unary
\emph{involution} operator $x\mapsto\overline x : \U\A \to \U\A$
satisfying two axioms $\overline{\overline \x} = \x$ and
$\overline{\x\y} = \overline\y\,\overline\x$. We then equip our type
of interest, lists, with an involutive structure as in
\figref{involutive lists}. We can use this algebra and the involutive
monoid to discharge equations containing list variables and concrete
lists: \InvMonoidFrexExample[numbers=right, numbersep=-.125cm]

The \IdrisFunction{solve} function takes as argument the number of
variables (\IdrisBound{n}=\IdrisData{2} on line~2) in the algebraic
term to simplify, and an algebraic simplifier from the frexlet
(\IdrisFunction{Involutive.Frex.Frex} on line~4). The final argument
is a pair of terms with \IdrisBound{n}=\IdrisData{2} variables
(\IdrisFunction{Dyn} \IdrisData{0} and \IdrisFunction{Dyn}
\IdrisData{1}) and concrete values from the algebra. By importing
notation modules the frexlet provides, we can use infix multiplicative
notation such as (\IdrisFunction{.*.}). The type-checker then infers
the terms to substitute for each variable.

In this example, we used \IdrisFunction{solve} to define a stand-alone
lemma, but we may also call \IdrisFunction{solve} directly from a
chain of equational reasoning steps. When we extract lemmas, we often
want to prove them more abstractly, for \emph{all} involutive
monoids. In that case we use a fral:
\InvMonoidFralExample[numbers=right,numbersep=-.125cm] Lines 2--3 and 8--9 overload the
infix and postfix notation using the frexlet's built-in notation
suites. Concretely, the projection \IdrisFunction{Notation1} brings
into scope the functions
\IdrisKeyword{(}\IdrisFunction{.*.}\IdrisKeyword{)} and
\IdrisKeyword{(}\IdrisFunction{.inv}\IdrisKeyword{)} when writing
algebraic terms.  The \IdrisFunction{solve} function takes the number
of free variables and a corresponding fral simplifier (line~10), as
well as the two terms representing the equation of interest.  The
variables \IdrisBound{x}, \IdrisBound{y}, \IdrisBound{z} (bound in line 7) are
implicitly used in this call. \subsecref{solver implementation}
covers the type of \IdrisFunction{solve} in more detail.

%% file: core.tex
\section{Free Extensions and Free Models/Algebras}
\seclabel{core}

Before delving into the details of \Frexlib{}'s core, we revisit
our frexlet representations using examples for elements in the
fral and the frex for ordinary, commutative, and involutive monoids~(see~\figref{monoid frexlets}).

The elements in the free monoid are lists of the variables appearing
in the term. The elements in the free extension of a
monoid are lists alternating between concrete elements in the given
monoid, and freely adjoined variables. The figure shows an element
in the free extension by 1 variable ($y$) of the multiplicative monoid
of $2\times2$ matrices with natural-number components. The matrix $y$
is unknown, and so its occurrence separates the elements in the list.

Further assuming commutativity equates more terms, resulting
in a representation of the free commutative monoid over \nvar{}
variables as an \nvar-vector of coefficients, representing a linear
polynomial. Freely extending a commutative monoid \A{} by
\nvar{} variables can be represented by a concrete coefficient $c :
\A$ together with an \nvar{}-vector of coefficients, representing a
linear polynomial over $\A$.

If we instead include an involutive operation $x \mapsto
\overline x$ over the monoid, we get lists of variables and alternating lists
whose letters may be tagged as involuted. The figure demonstrates
the free extension of the monoid structure of \String{} concatenation,
with string reversal for the involution.

\subsection{Universal Properties}
We now show the \idris{} realisation of the universal algebraic
concepts from \secref{new-overview}. In the previous \secref{frexlib},
we introduced these definitions:
\begin{itemize}
\item An \IdrisType{Algebra} for a signature, which consists of a carrier
  and an interpretation of operations (\figref{algebras}). It does not
  have to satisfy any equation.
\item A \IdrisType{SetoidAlgebra} is simultaneously a setoid structure
  (an equivalence relation) and an algebra structure over the same
  carrier, whose operations are setoid homomorphisms (\figref{algebras}).
\item A (setoid) \IdrisType{Model} for a presentation is a setoid
  algebra validating the presentation's axioms.
\end{itemize}
In this section, we introduce these definitions:
\begin{itemize}
\item Homomorphisms of setoid algebras.
\item A \IdrisType{ModelOver} a setoid \X{} is a model \A{}  equipped with a setoid
  homomorphism \X{} to \A{}.
\item An \IdrisType{Extension} of a model \A{} by a setoid \X{} is another
  model \A[2]{} together with a setoid homomorphism from \X{} to \A[2]{} and a
  model homomorphism from \A{} to \A[2].
\item Their morphisms.
\end{itemize}
We will then show how we define in \idris{} both the free model
(fral) over a setoid \X{} and the free extension (frex) of \A{} by
\X{} as the initial ones.

\subseclabel{UPs}
\begin{figure}
  \TwoColDIY{.31}{%
    \OpPreservation{}
    }{.65}{%
    \IsAlgHomomorphism{}
    \vspace{1\baselineskip}
  \AlgHomomorphism{}
  }
\caption{Setoid algebra homomorphisms in \Frexlib{}}
\figlabel{homo}
\end{figure}

\Frexlib{} defines homomorphisms of setoid algebras in \figref{homo}, by
requiring the underlying function to be a setoid homomorphism between
the corresponding setoids. The code uses an appropriate
\IdrisFunction{cast} function (see page~\pageref{cast explanation})
that assembles these setoids from the
data in each setoid algebra.
Each $\SigAlgebra$ defines a homomorphic extension operator
\AlgBind{}\AlgBindType{} by structural induction over the
term (i.e.~folding). The following term valuates to \BindExResult{}, e.g.:
\begin{center}
\BindEx{}
\end{center}

Similarly, \figref{algebras and extensions} presents the declarations
for models over a setoid and extension of a model by a setoid. It
expresses the equations in the commuting diagrams through
the extensionality relation on the setoid of functions from
\figref{quotients-and-functions}b in \appref{setoid example} and the power of a
model by a setoid (see \subsecref{powers}). As \idris{}
supports type-directed disambiguation, we overload the record name
\IdrisType{($\~>$)}.

\begin{figure*}
  \TwoColDIY%
      {.43}{%
        \ModelOver{}
        \PreservesEnvDecl{}
        \ModelOverHomo{}
      }
      {.55}{%
        \Extension{}
        \vspace{-.5\baselineskip}
        \ExtensionHomo{}
      }
  \caption{Structure and its preservation for (a) models
    over a setoid, and (b) extensions of a model}
  \figlabel{algebras and extensions}
\end{figure*}

The \emph{free} model over a set (fral) and the \emph{free}
extension (frex) of an algebra by a set is then the initial such
structure: there is a unique structure-preserving map from the free
structure to every structure.
This succinct definition, while
standard, packs much structure. By way of introduction, we will unpack
it for the free commutative monoid over $\Finn$, the finite set with
$\nvar$ elements.

First, we designate a commutative monoid for the
the fral. This structure is the data structure our simplifier will use
to represent the equivalence classes of terms.
In \figref{monoid frexlets}, we mentioned the
carrier consists of origin-intercepting linear polynomials with \Nat{}
coefficients $p = a_1x_1 + \ldots + a_{\nvar}x_{\nvar}$, which we
represent with $\nvar$-tuples of natural numbers and pointwise addition:
\begin{center}
\begin{minipage}{.35\textwidth}
  \ModelCarrier{}
\end{minipage}
\begin{minipage}{.6\textwidth}
  \[
  \begin{aligned}[t]
    0 \definedby{}& 0x_1 +\ldots+0x_n & \quad p \Plus q \definedby{}& (a_1 + b_1)x_1 + \ldots + (a_{\nvar} + b_{\nvar})x_{\nvar} \\
    \definedby{}& \ReplicateZeroPartA \ldots \ReplicateZeroPartB
    & \definedby{}& \OpenBracket a_1 \Plus b_2 \Comma \ldots \Comma a_{\nvar} \Plus b_{\nvar} \CloseBracket
    \\
    ={}& \ReplicateZero & ={}& \ZipIt{}
  \end{aligned}
  \]
\end{minipage}
\end{center}

Denote the resulting $\AppliedFreeCMonoidModelType$ by $\AppliedFreeCMonoidModel$.  For the  $\EnvName$ component, use tabulation to define
$ \FreeCMonoidUnit\,\nvar : \Finn \to \ModelCarrierName$, with $\idrisdata 1$ in
the argument position and $\idrisdata 0$ elsewhere:
\renewcommand\tabulateDirac[1][]{\UseVerb[#1]{tabulateDirac}}
\begin{SaveVerbatim}[commandchars=\\\{\}]{tabulateDirac}
\IdrisFunction{tabulate}\KatlaSpace{}$\KatlaSpace{}\IdrisFunction{kronecker}\KatlaSpace{}\IdrisBound{i}
\end{SaveVerbatim}
\renewcommand\dirac[1][]{\UseVerb[#1]{dirac}}
\begin{SaveVerbatim}[commandchars=\\\{\}]{kronecker}
\IdrisFunction{kronecker}
\end{SaveVerbatim}
\renewcommand\diracij[1][]{\UseVerb[#1]{diracij}}
\begin{SaveVerbatim}[commandchars=\\\{\}]{diracij}
\IdrisFunction{kronecker}\KatlaSpace{}\IdrisBound{i}\KatlaSpace{}\IdrisBound{j}
\end{SaveVerbatim}
\[
\begin{array}[t]{@{}l@{\qquad}l@{}}
\begin{aligned}[t]
\FreeCMonoidUnit\ \nvar\ i \definedby{}& 1x_i \\
\definedby{}& \ReplicateZeroPartA \ldots\KroneckerMiddle\ldots\ReplicateZeroPartB \\
={}& \tabulateDirac
\end{aligned}
&
\begin{array}[t]{@{}l@{}}
\text{where}\\ \diracij \definedby
\begin{cases}
  \diracsi = \diracsj: & \idrisdata{1} \\
  \diracsi \neq \diracsj:  & \idrisdata{0}
\end{cases}
\end{array}
\end{array}
\]
The initiality of this structure follows from the normal form property
--- every origin-intersecting linear polynomial $p$ can be represented
as $p = \sum_{i=1}^n a_i\cdot \FreeCMonoidUnit\ \nvar\ \diracsi$:\\
\[
\begin{array}{@{}l@{}}
  \AppliedCMonoidNormalFormTypePartB
  \\
  \multicolumn{1}{r}{%
    \AppliedCMonoidNormalFormTypePartC\ \AppliedCMonoidNormalFormTypePartA
  }
\end{array}
\]
Since monoid homomorphisms preserve the summation and
multiplication-by-a-natural, the unique structure
preserving map $\hName : \pair{\AppliedFreeCMonoidModel}{\FreeCMonoidUnit\,\nvar} \to \avar$
is this homomorphism:
\[
\hName\ \idrisvar{xs}= \CMonoidUniqueMap{}
\]

This standard argument lies behind many simplifiers, as well as more
advanced techniques like normalisation-by-evaluation. \Frexlib{} takes the same approach, but also explores how to use
general-purpose constructions involving frals and frexes, and bespoke
facts about algebraic structures, to construct new frals and frexes.

To summarise, to implement a fral/frex simplifier, the developer
follows these steps:
\begin{itemize}
\item Design a data-structure for the carrier of the frex/fral's
  algebra, e.g.~for commutative monoids:
  \IdrisType{Vect} \nvar{} \IdrisType{Nat} for the fral
  and \IdrisKeyword{(}\U{} \avar{}, \IdrisType{Vect} \nvar{}
  \IdrisType{Nat}\IdrisKeyword{)} for the frex.
\item Equip it with a setoid algebra structure: pointwise
  operations with propositional equality.
\item Equip it with the appropriate additional structure, e.g.~the
  unit for the fral and the \IdrisFunction{Var}iable function and the
  \IdrisFunction{Embed}ding homomorphism for the frex.
\item Define the function underlying the homomorphism into any other
  algebra over the variable setoid or extension, e.g.~linear
  combination for commutative monoids.
\item Prove that this function is a homomorphism and its uniqueness.
\end{itemize}
These steps realise the simplifier development methodology
we described in \secref{simplification} (page~\pageref{methodology-anchor}).

\subsection{Solver Interface}\subseclabel{solver implementation}
We can now explain the interface \Frexlib{} gives to the \IdrisFunction{solve}
functions. We describe the frex-based interface in detail; the
fral-based interface is similar. We implement the core functionality in the
auxiliary function \IdrisFunction{solveVect} in \figref{solveVect}.
\begin{figure}
  \begin{minipage}{.75\textwidth}
    \solveVectType[numbers=left]{}
  \end{minipage}
  \caption{Core frex-based simplification routine}\figlabel{solveVect}
\end{figure}

The argument \IdrisBound{frex} (line 2) is an implementation of a frex
simplifier for some \IdrisBound{pres}-model \avar{}, extended with
\nvar{} free variables (line 1). We erase the number of variables at
runtime, and so we also erase the type \IdrisData{Fin} \IdrisBound{n}.
We cast an erased type to a setoid instead of an unerased type, i.e.:
\\[.4\baselineskip]
\begin{tabular}{lcr}
  \begin{minipage}[b]{.4\textwidth}
    \irrelevantCast
  \end{minipage}
  &
  instead of
  &
  \begin{minipage}[b]{.4\textwidth}
    \relevantCast
  \end{minipage}
\end{tabular}
\\[.4\baselineskip]
\noindent
The function also takes an \IdrisBound{env}ironment of terms to
substitute for the free variables in the simplification equation (line
2). In this auxiliary function, we present the environment using an
\nvar-ary vector of terms over the algebra's carrier. Next comes the
\IdrisBound{eq}uation we want to discharge (line 3), involving either
concrete values (of type \U{} \avar) and any of the \nvar{} available
variables. Both the \IdrisBound{frex} and the \avar{}lgebra with its
\IdrisBound{env}ironment give rise to extensions in the formal sense,
which we can use to give an environment
for the \IdrisBound{eq}uation in question, namely a setoid
homomorphism from the joint setoid of constants and free variables to
the carrier of the model underlying the extension:
\extEnv{}
where:
\begin{minipage}{.8\textwidth}
  \setoidEither{}
\end{minipage}

We use these environments to interpret the equation, once in the
\IdrisBound{frex} (lines 6--7) and once in the given algebra
(lines~9--10). If the equation holds in all extensions, it will hold
in the \IdrisBound{frex} and in \avar{}, and, moreover, homomorphisms
of extensions will preserve this interpretation.  Interpreting this
equation in the \IdrisBound{frex} may have better decidability
properties over equivalence in \avar{}.

We use \idris{}'s \IdrisKeyword{auto}-implicits mechanism to search
for the equivalence of the \IdrisBound{frex} interpretations. This
mechanism will try to find terms that resolve the implicit argument
\IdrisBound{prf}, using a heuristic informed by unification, that will
also attempt to apply data constructors.

Typically, \idris{}'s judgemental
equality decides
the \IdrisBound{frex} setoid's
equivalence relation, and the number of variables we extend by is known
statically.  This case can happen when the equivalence relation
on the setoid algebra \avar{} is decidable by judgemental
equality. Then, the type of the \IdrisBound{prf} argument (line~5) is
a propositional equality between closed terms. Judgemental
equality decides this relation between the interpretations in the
type of \IdrisBound{prf}.  In \idris{}, the \IdrisKeyword{auto}-search
heuristic tries to use \IdrisData{Refl}, and promotes the required
equation to a judgemental equality constraint.  Even when the setoid relation
is not decidable by judgemental equality,
making \IdrisBound{prf} an \IdrisKeyword{auto}-implicit may provide
more functionality in the future. We may be able to
freely extend algebras whose propositional equality is only partially
decidable by judgemental equality (e.g.~function types in a
type theory with function extensionality), or by a
sophisticated decision procedure (e.g.~multiset equality).

\begin{figure}
  \begin{minipage}{.75\textwidth}
    \Visibility{}
    \metaPi{}
    \metaPI{}
  \end{minipage}
  \caption{Metaprogramming abstractions for curried $\Pi$-types}\figlabel{meta-pi}
\end{figure}

We use metaprogramming abstractions (\figref{meta-pi}) to
simplify the user-facing interface. The \IdrisFunction{PI} combinator
produces an \nvar-ary telescope of
\IdrisData{Visible}/\IdrisData{Hidden}/\IdrisData{Auto} arguments,
packaged as an \nvar-ary vector which it passes this its argument \IdrisBound{b}.
Using this abstraction to reduce
\IdrisFunction{solve} (\figref{solve}) to \IdrisFunction{solveVect} (\figref{solveVect}).
Compare their types: \IdrisFunction{solve} curries the \IdrisBound{env}ironment
$n$-vector into $n$ implicit arguments.
Unification can resolve these arguments to the free
variables in the discharged equation.

\begin{figure}
  \begin{minipage}{.75\textwidth}
    \solveType[numbers=left]{}
  \end{minipage}
  \caption{User-facing frex-based simplification routine}\figlabel{solve}
\end{figure}

\goodbreak
\subsection{Powers}
The commutative monoid structure \AppliedFreeCMonoidModel{} instantiates a general
construction: $\Pres$-models have
powers by setoids. The \emph{power} of an
algebra $\A$ by a set(oid) $\X$ is the terminal
\emph{parameterisation}.
\subseclabel{powers}
\begin{wrapfigure}[4]{r}{2.6cm}
  \vspace{-1\baselineskip}
  \includegraphics{diagram-03.mps}
\end{wrapfigure}
Parameterisations, shown succinctly on the right, are an $\X$-indexed
collection of algebra homomorphisms $\aEval\ \f : \aModel \to \A$.
Requiring $\aEval\ \f$ to be homomorphic implies that operations are
given pointwise. Structure preservation uses the contravariant
action $\idrisfun{pre}\ \UH \hName$ precomposing a homomorphism $\UH
\hName : \aModel \to \bModel$. Universality singles out
the carrier of the power as the function-space \PowerConstruction{}.
For $\X = \Finn$, we can represent it by $\nvar$-tuples from
$\U\,\A$.

\input{generic}

%% file: generic.tex
\subsection{Frex via Coproducts with Fral}
\subseclabel{frex-via-coproducts}
\begin{wrapfigure}[4]{r}{3cm}
  \vspace{-1\baselineskip}
\includegraphics{diagram-06.mps}
\end{wrapfigure}
The fral and the frex are related: the free extension of $\A$ by $\X$ is
the \emph{coproduct} of $\A$ with the free algebra over $\X$.
A \emph{cospan} $\avar{}$ between $\A_1$ and $\A_2$ consists of two
homomorphisms with a shared codomain: $\A_1
\xto{\aLft}$\aSink$\xleftarrow{\aRgt} \A_2$.  A \emph{cospan
homomorphism} $h : \avar \to \bvar$ is a homomorphism $\UH : \aSink
\to \bSink$ preserving the cospan as in the diagram on the right.  The
coproducts $\A_1\oplus\A_2$ of two models $\A_1$ and $\A_2$ is then
the initial \emph{cospan}.  All models have coproducts, but these may
be difficult to represent. However, in cases such as commutative
monoids, the coproduct is straightforward to represent: its carrier is
the cartesian product of the components carriers.

The universal property of the frex $\A{}[\X]$ combines
those of the fral $\FreeName\,\Pres\!\X$ and
its coproduct with $\A$. Consider the following two diagrams:
\[
\hfill\includegraphics{diagram-05.mps}
\qquad
\hfill\includegraphics{diagram-04.mps}
\]
The \Var-arrows and the \Lft-arrows correspond through the universal
property of the fral: \Lft{} is the unique homomorphic extension of
the \Var{}-arrows. The \Embed-homomorphisms are exactly
the \Rgt-homomorphism.
This identification lets us construct:
\MyCodeListing{
\FrexByCoprodPartA{}\IdrisKeyword{ :}\FrexByCoprodPartB{}\\
\FrexByCoprodPartC{}
}
For commutative monoids it gives the
commutative monoid of linear polynomials with natural numbers as degree-1
coefficients whose carrier is represented by $\CMonoidFrexCarrier{}$.

\subsection{Fral via a Frex}
\subseclabel{fral via frex}
We use the relationship between the fral and frex, described in \subsecref{frex-via-coproducts}, to derive a fral from a corresponding frex.
In particular, the following calculation shows that every free algebra arises as a free extension of $\FreeName\,\Pres\emptyset$. We use $\uplus$ for the disjoint union, i.e., the coproduct of sets:
\begin{align*}
\FreeName\,\Pres\X
&\isomorphic \FreeName\,\Pres(\X\uplus\emptyset) & (\X \isomorphic \X \uplus \emptyset)\\
&\isomorphic (\FreeName\,\Pres\X)\oplus(\FreeName\,\Pres\emptyset) & (\text{$\FreeName\,\Pres$ left adjoint, left adjoints preserve coproducts})\\
&\isomorphic (\FreeName\,\Pres\emptyset){}[\X]
\end{align*}
Therefore, we may construct a fral from an initial
algebra and its frex:
\MyCodeListing{\PutListing{.5\textwidth}\ByFrexFull{}}
This generic construction can produce suboptimal
representations.  For example, the initial monoid is easy to
construct: its carrier is the unit type. Freely extending this
initial monoid produces alternating lists, that interleave the unit
value. Taking lists of variables instead leads to a simpler representation but
requires more complicated proofs.  However, generic constructions such as the foregoing allow us to trade efficiency for rapid prototyping.

\subsection{Reusing Frexlets}
\subseclabel{reusing frexlets}
The final example demonstrates reuse of one simplifier when
constructing another. Recall the presentation of involutive monoids
from the end of \secref{frexlib}.
\begin{proposition}[Jacobs]\label{prop:involution}
  The carrier set and monoid operations of the free involutive monoid on \X{}
  are those of the free monoid on the product\/ $\BoolX$.
  Similarly, the frex of an involutive monoid by $\X$ is the
  frex of its underlying monoid by\/ $\BoolX$, extended with an involution operation.
\end{proposition}
We can prove this proposition directly, establishing the involutive
axioms, and have taken this strategy in \Frexlib{}.  However, it is
possible to prove this result without referring to the specific
representation of the monoid fral/frex, resulting in a
simplifier-transformer reusing the code implementing simplifiers for
monoids to implement simplifiers for involutive monoid.  The potential
for this kind of reuse extends beyond monoids.  We can phrase the
result in much greater generality, and give a higher-level proof,
using \citepos{jacobs:involution} axiomatisation of involutions. This
more abstract proof generalises to other notions of involutive
algebras, and we plan to exploit it in the future for generic frexlet
reuse. The more abstract proof goes beyond the scope of this
article, involving more abstract category theoretic notions.

%% file: printing.tex

\section{Completeness and Certification}
\seclabel{certification}
\Frexlib{} uses setoids to extract intensional information: it
automatically extracts printable representations of simplification proofs, and moreover formalises
the completeness of fral/frex simplifiers. Here is how it works:
\begin{itemize}
\item It is easy to construct a frex and a fral using
      a quotient setoid: terms quotiented by provability.
\item The equivalence relation in this setoid is not necessarily
  \emph{effective}/\emph{decidable}.
\item An effective fral/frex must be constructed manually/creatively.
\item Any fral/frex is canonically isomorphic to the quotient frex, and an
  effective fral/frex with a constructive universality proof has an effective isomorphism to
  the quotient setoid.
\item In this way, a simplifier using an effective frex is
  constructively sound and complete.
\item The universal property of the effective fral/frex lets us
  interpret equations it simplifies to the same value as related
  values in every model/extension.
\item
  So the effective fral/frex lets us extract the data structure that represents
  equivalence in the setoid. For the
  quotient fral/frex, it is a representation of the equational proof.
\end{itemize}
Through this mechanism, there is no need to write any proof extraction
code for our simplifiers. The universal properties of the fral/frex
have done all the presentation-specific heavy-lifting. The remainder
of this section expands this description in more technical detail
(\subsecref{completeness}), how \Frexlib{} we represent certificates
(\subsecref{extracting certificates}), and how we simplify proofs
(\subsecref{proof simplification}).

\subsection{Completeness}
\subseclabel{completeness}
It is straightforward to represent the deduction rules of equational
logic from \figref{provability} as a datatype
\IdrisType{Provability}$\,t\,s$ relating pairs of terms $t$ and $s$.
There are two variations on this provability relation: for free models
and for free extensions.

For free models over \X{}, we index the relation by $t,s \in \Term
\X$.  Its definition consists of the unshaded rules in
\figref{provability}. The quotient setoid
$Q \definedby \Term\X/\text{\IdrisType{Provability}}$ then satisfies the universal
property of the free model over \X{}.
For the free extension of a model $\avar$ by \X{}, we index the
relation by $t,s \in \Term(\X \uplus \U\,\avar)$, i.e., terms with
variables in the disjoint union $\X \uplus \U\,\avar$. The
left-injected values represent variables from \X{}. For each $c \in
\U\,\avar$, the corresponding right-injected variable, which we denote
by $\underline{c}$, is a term representing $c$.  The provability
relation then adds the shaded evaluation equation from
\figref{provability}. It amounts to datatype constructor, for every
operator $f : n$ and constants $c_1, \ldots, c_n \in \U\,\avar$,
representing the equation \( (\underline c_1, \ldots, \underline c_n)
= \underline{\text\avar\sem f(c_1, \ldots, c_n)} \).  The quotient
setoid $Q\definedby \Term(\X \uplus \U\,\avar) /
\text{\IdrisType{Provability}}$ then satisfies the universal property
of the frex of \avar{} by \X.

The interpretation of a term $t$ in $Q$ is $t$ itself, and so the quotient
setoid does not help at all with simplification.  Since $Q$ satisfies
the universal property, it is canonically isomorphic to every other
fral/frex. So the interpretation of every two terms $s,t$ in every
fral/frex is related iff the datatype
$\text{\IdrisType{Provability}}\,t\,s$ is inhabited. Therefore
evaluation in fral/frex is sound and complete for algebraic
simplification.
After designing a fral/frex $S$ whose equivalence relation is
\emph{effective}, we can use $Q$ as follows. The input type to the
function \IdrisFunction{solve} is exactly the carrier of $Q$.  By
appealing to the universal property of the fral/frex $S$, we obtain an
inhabitant of \IdrisType{Provability}, and we extracted the
simplification proof, which we call the \emph{certificate}.

\subsection{Representing Certificates}
\subseclabel{extracting certificates} \Frexlib{} provides the
\FrexLemma{} \IdrisBound{pres} type, consisting of a
\IdrisBound{pres}\IdrisFunction{.signature}-equation and derivation
for it in the quotient fral for \IdrisBound{pres}.
Such lemmata are sound: every \FrexLemma{} for a theory holds
in all models of this theory.
\Frexlib{} provides a \IdrisFunction{mkLemma} smart constructor which
runs the given fral simplifier, constructs a
proof that a stated equivalence holds, and returns a valid
\IdrisType{Lemma}.
This mechanism allows users to build up a library of
lemmata for their theories. Users can then seamlessly invoke
these lemmata in any model, avoiding further \Frexlib{} calls.

This approach forces the user's project to depend on most of
\Frexlib{} indirectly through such modules.
If users do not want to introduce that dependency, \Frexlib{} gives
them the option not to depend on it.
\Frexlib{} also supports proof-certificate extraction,
allowing users to produce standalone libraries of ordinary \idris{} functions
rather than \FrexLemma{} independent of
the \Frexlib{} library. For example, the fral simplifier for monoids
generates a module exporting the following \idris{} function:
\begin{center}
\CertificateType{}
\end{center}
together with an explicit equational proof that does not call any \Frexlib{}
simplification steps (see \figref{certification} in the \appref{printing}).
We now explain the certification process in technical detail.

Our goals for this extraction are to (1) produce libraries from lemmata,
and (2) produce somewhat idiomatic \idris{} code.
The derivation found by \Frexlib{} may not be what a human would have
chosen but it should nonetheless be possible for a sufficiently patient
human to follow the reasoning steps.

The main challenge was to go from the rich type of derivations trees,
i.e., the quotient setoid relation, to a representation that we can
print in a relatively readable way. The derivation trees have
arbitrarily nested transitivity, symmetry, and $n$-ary congruence
steps.  We transform them into a type of linear/flat derivations that
could be pretty-printed using combinators for setoid
reasoning.

We represent derivations in layers:
\begin{center}
  \begin{tabular}{@{}l@{\qquad}l@{}}
(a) the reflexive-transitive closure of &
(b) the symmetric closure of\\%
(c) the unary congruence closure of &
(d) axiomatic reasoning steps.
  \end{tabular}
\end{center}
We now detail each of these layers, implementing each layer by a
dedicated data structure~(cf.~\figref{linear-derivations} in
\appref{printing} for the full details):\\
\textit{(a) Reflexive-transitive closure (\IdrisType{RTList}):}
type-aligned~\cite{DBLP:conf/haskell/PloegK14} lists of steps in
the closed-over relation: the target element of each list position
is the source element of the next.\\
\textit{(b) Symmetric closure (\IdrisType{Symmetrise}):}{ either the relation or its opposite}.\\
\textit{(c) Unary congruence closure (\IdrisType{Locate}):} It suffices to pair a term
with a distinguished variable for the contextual hole, together with a
step in the closed-over relation. To ease our pretty-printing code,
we distinguish between using the closed-over relation in an empty
context, and using it in a context with a distinguished variable
represented by the Idris value \IdrisData{Nothing}.\\
\textit{(d) Axiomatic steps (\IdrisType{Step}):} An atomic step is either a setoid equivalence,
or an axiom.\\
Putting these together gives the type \IdrisType{Derivation} of
linear derivations~(cf.~\figref{linear-derivations}(e) in \appref{printing}).

Every derivation tree decomposes into a value in this layered
representation. The modular definition of \IdrisType{Derivation} as a
composition of the relation-transformers \IdrisType{RTList},
\IdrisType{Symmetrise}, \IdrisType{Locate} and \IdrisType{Step} makes
decomposition straightforward.  We use generic combinators for each
closure relation-transformers. Closure under congruence is the
trickiest part, decomposing an $n$-ary congruence in the derivation tree
into $n$ separate
unary congruences, pushing them under the reflexive-transitive and
symmetric closure layers, and erasing any congruence steps with the
identity context.

\subsection{Proof Simplication}
\subseclabel{proof simplification}

Certification also allows us to inspect \Frexlib{}-generated proofs.
Frexlet developers can check whether data-structures and
proofs are suboptimal, spurring code refactoring.
Concretely, when developing \Frexlib{}, we noticed proofs with loops:
multi-step derivations that start and end in the same term.
Such loops come from internal data structures that optimise
simplifier-development effort, but insert semantically irrelevant
subterms that can be simplified away.
\Frexlib{} implements a generic proof simplifier that automatically removes
all such loops.
This mechanism suggests further work, developing
modules for simplifying these proofs further.

%% file: reflection.tex
\section{Goal Extraction via Reflection}
\seclabel{reflection}
\newcommand\prove[1][]{\UseVerb[#1]{prove}}
\begin{SaveVerbatim}[commandchars=\\\{\}]{prove}
\IdrisFunction{Elab}
\end{SaveVerbatim}

\newcommand\Elab[1][]{\UseVerb[#1]{Elab}}
\begin{SaveVerbatim}[commandchars=\\\{\}]{Elab}
\IdrisType{Elab}
\end{SaveVerbatim}

\newcommand\TC[1][]{\UseVerb[#1]{TC}}
\begin{SaveVerbatim}[commandchars=\\\{\}]{TC}
\IdrisType{TC}
\end{SaveVerbatim}

\newcommand\TTImp[1][]{\UseVerb[#1]{TTImp}}
\begin{SaveVerbatim}[commandchars=\\\{\}]{TTImp}
\IdrisType{TTImp}
\end{SaveVerbatim}

Thus far, our examples have illustrated interaction with \Frexlib{}
using \solve{}. The \solve{} function provides a similar interface to
the simplifiers in (e.g.)  Agda's standard library: it takes the fral
or frex simplifier, the number of free variables and the abstract
syntax of a goal. However, these simplifiers additionally provide
ergonomic goal extraction with Agda's \emph{proof reflection}
mechanism.

Proof reflection is a metaprogramming paradigm, available in
proof assistants and dependently typed programming languages, that
supports bi-directional communication between a language and its
implementation. The language provides
a representation of its terms,
operators that construct, manipulate and destruct term representations, and
primitives \emph{quote} and \emph{unquote} that respectively
\emph{reify} terms into the representation and
\emph{reflect} back encoded terms as ordinary terms.

Given mechanisms for querying unsolved proof obligations, proof
reflection enables the implementation of verified decision procedures
for automatically discharging such obligations without
boilerplate~\cite{boutin:reflection,christiansen-et-al:elab-reflection}.
Coupled with the meta-theoretic properties that
dependently typed implementations of decision procedures can enforce
(e.g.~relative soundness and completeness), reflection-driven interfaces
yield easy-to-use tactics with strong guarantees. It is then natural to
ask: is it possible to construct an interface to \Frexlib{}
that uses proof reflection to avoid the need to explicitly supply the
equation to discharge?

\begin{figure}
  \begin{tabular}{@{}c@{}|c@{}}
  \begin{minipage}[b]{.55\textwidth}
    \MagicExOne{}
    \vspace{-1\baselineskip}
  \end{minipage}
  &
  \begin{minipage}[b]{.44\textwidth}
    \vspace{-1\baselineskip}
    \begin{code}
    \>[0]\AgdaFunction{agdaEx}\AgdaSpace{}%
    \AgdaSymbol{:}\AgdaSpace{}%
    \AgdaSymbol{∀}\AgdaSpace{}%
    \AgdaSymbol{\{}\AgdaBound{x}\AgdaSpace{}%
    \AgdaBound{y}\AgdaSymbol{\}}\AgdaSpace{}%
    \AgdaSymbol{→}\AgdaSpace{}%
    \AgdaSymbol{(}\AgdaNumber{2}\AgdaSpace{}%
    \AgdaOperator{\AgdaPrimitive{+}}\AgdaSpace{}%
    \AgdaBound{x}\AgdaSymbol{)}\AgdaSpace{}%
    \AgdaOperator{\AgdaPrimitive{+}}\AgdaSpace{}%
    \AgdaSymbol{(}\AgdaBound{y}\AgdaSpace{}%
    \AgdaOperator{\AgdaPrimitive{+}}\AgdaSpace{}%
    \AgdaNumber{3}\AgdaSymbol{)}\AgdaSpace{}%
    \AgdaOperator{\AgdaDatatype{≡}}\AgdaSpace{}%
    \AgdaBound{x}\AgdaSpace{}%
    \AgdaOperator{\AgdaPrimitive{+}}\AgdaSpace{}%
    \AgdaSymbol{(}\AgdaBound{y}\AgdaSpace{}%
    \AgdaOperator{\AgdaPrimitive{+}}\AgdaSpace{}%
    \AgdaNumber{5}\AgdaSymbol{)}\<%
    \\
    \>[0]\AgdaFunction{agdaEx}\AgdaSpace{}%
    \AgdaSymbol{=}\AgdaSpace{}%
    \AgdaMacro{fragment}\AgdaSpace{}%
    \AgdaFunction{CSemigroupFrex}\AgdaSpace{}%
    \AgdaFunction{+-csemigroup}\<%
  \end{code}
  \end{minipage}
  \end{tabular}
  \caption{Goal extraction in (a) \idris{} \Frexlib{}'s elaborator reflection script;
    (b) the \textbf{fr}ex \textbf{Ag}da aug\textbf{ment}ation lib.}
  \figlabel{magic}
\end{figure}
As the example code in \figref{magic} illustrates, using proof
reflection to provide such an interface to \Frexlib{} is possible in
both \idris{} and \agda{}.
Rather than designing custom reflection-based drivers for individual
simplifiers, we combine proof reflection with \Frexlib{}'s design
philosophy of extensibility and common core reuse and provide a single
generic metaprogram parameterized by a signature and a model of a
presentation.
The metaprogram can be instantiated for any algebraic simplifier,
built-in or user-defined.
\figref{magic} (a) shows the invocation of the \idris{} elaboration
script \IdrisFunction{frexMagic}, and \figref{magic} (b) shows the
\agda{} proof reflection macro \AgdaMacro{fragment}.
Both implementations aim to infer the abstract syntax of the goal
equation based on the expected type.


The drivers have no information about the structure of the algebraic
signature argument ahead of time. \Frexlib{}'s inductive \Term{}
representation means that relevant abstract operator names can be
extracted from the presentation.
However, the process of matching goal fragments against the abstract
syntax of the algebraic interpretation is tightly coupled to the
language's reflection primitives. Implementing \Frexlib{} in both
\idris{} and \agda{} allows us to compare differences in behaviour.

The differences between the \idris{} and \agda{} implementations can
be seen by considering the normalisation of arithmetic expressions
such as \ExOne{}.
In \idris{}, the reflected syntax passed to the driver represents the
normalized expression \ExTwo{}.
As far as the theory of monoids is concerned, \ReflectionSY{} is an atomic
expression and so it is treated as another free variable, distinct
from \ReflectionY{}. The \idris{} driver then incorrectly infers the
invalid equation \ExThree{}, and fails to discharge the goal.
In contrast, Agda does not normalise quoted expressions before
reflection, and so the Agda driver successfully finds the equation,
allowing \Frexlib{} to solve the example.
Agda's approach is not always superior: there are
similar examples for which the \agda{} driver fails.  The extent to which
the implementation can avoid such pathologies ultimately depends on
the engineering effort available to develop heuristics.

As these problems indicate, this rather naive approach to automation
requires significant developer resources to deal with edge cases or
construct bespoke solutions under simplifying assumptions.
In large, mature ecosystems it may be possible to maintain practical
heuristics despite these challenges.
However, there might be better mechanisms for specifying
algebraic contexts, allowing the solver to extract the required
information automatically; we suggest such directions in
\secref{conclusion}.

%% file: evaluation.tex
\section{Supplementary Evaluation: Usability and Interactive Development}
\seclabel{evaluation}

The implementation of \Frexlib{} is still in its early stages, and
offers many opportuntities for further engineering work to extend its
functionality, expressiveness, ergonomics, and efficiency.
However, we have already carried out some small experiments to assess
user experience and frexlet developer experience to establish that the
approach is feasible, and to identify further directions.

\subsection{Quantitative evaluation}
\idris{} encourages interactive, type-driven development, so it is
important that the checker is responsive when the user modifies the program.
Following~\citet{nielsen:usability}, our \idris{}
implementation aims for response times under one second, and we treat
a response time of over 10 seconds when type checking a modification
to \Frexlib{} client code as a bug.

\pgfplotstableread[col sep = comma]{csv/1vars_commutative.csv}\OneVarCommutativeData
\pgfplotstableread[col sep = comma]{csv/1vars_noncommutative.csv}\OneVarNoncommutativeData
\pgfplotstableread[col sep = comma]{csv/5vars_commutative.csv}\FiveVarCommutativeData
\pgfplotstableread[col sep = comma]{csv/5vars_noncommutative.csv}\FiveVarNoncommutativeData
\pgfplotstableread[col sep = comma]{csv/10vars_commutative.csv}\TenVarCommutativeData
\pgfplotstableread[col sep = comma]{csv/10vars_noncommutative.csv}\TenVarNoncommutativeData
\pgfplotstableread[col sep = comma]{csv/15vars_commutative.csv}\FifteenVarCommutativeData
\pgfplotstableread[col sep = comma]{csv/15vars_noncommutative.csv}\FifteenVarNoncommutativeData

For typical small equalities that arise incidentally in dependently
typed programs, Frex's performance falls very comfortably
within \citeauthor{nielsen:usability}'s limits.  For example, the
checking time\footnote{We use a dated
AMD FX-8320 machine with 16GB memory, running Idris 2 version
0.5.1-1011cc616 on Debian Linux.} is under 0.1s for terms of size six or below with the commutative solver
and terms of size 14 or below with the non-commutative solver, creating an impression of instantaneous response.

As the term size increases, Frex eventually crosses the one second
interactivity threshold.  \figref{timings} shows how type-checking
times grow with term size and with the number of free variables in a
randomly generated term for the commutative and non-commutative monoid
solvers.
As the figure shows, Frex's type-checking time generally remains below
the interactivity threshold up to terms of around size 30, and only
exceeds the 10 second threshold (beyond which users' attention is
lost) for a few terms of size 45 or above.
Our experience with Frex development suggests that the anomalously
high checking times for these terms is likely to arise from a
performance bottleneck in \idris{}'s evaluator (\secref{lessons})
and that the ongoing development of \idris{} may eventually eliminate
the problem, bringing the type-checking time for most terms up to size
60 down to a few seconds.
\begin{figure}[ht]

\begin{tikzpicture}
  \begin{axis}[only marks,width=0.85\columnwidth,height=5.3cm,
    xlabel={term size (leaves)},
    ylabel={run time (s)},
    ymax = 45,
    xmin = 0,
    xmax = 60,
    xlabel style={font=\small},
    ylabel style={font=\small},
    ymode=log,
    log ticks with fixed point,
    scaled ticks=false,
    tick label style={/pgf/number format/fixed},
    title={\footnotesize commutative solver},
    every axis title/.style={below right,at={(0,1)}},
    legend style={anchor=north west,at={(1.05,1.0)},draw=black,fill=white,align=left}
    ]
    \addplot[blue!50!black,mark=triangle,mark size=1pt] table[x index = {0}, y expr=(\thisrow{time})]{\OneVarCommutativeData};
    \addlegendentry{\scriptsize 1 free var};
    \addplot[red!50!black,,mark=otimes,mark size=1pt] table[x index = {0}, y expr=(\thisrow{time})]{\FiveVarCommutativeData};
    \addlegendentry{\scriptsize 5 free vars};
    \addplot[green!50!black,mark=star,mark size=1pt] table[x index = {0}, y expr=(\thisrow{time})]{\TenVarCommutativeData};
    \addlegendentry{\scriptsize 10 free vars};
    \addplot[yellow!50!black,mark=diamond,mark size=1pt] table[x index = {0}, y expr=(\thisrow{time})]{\FifteenVarCommutativeData};
    \addlegendentry{\scriptsize 15 free vars};

    \draw [dashed,green!50!black] (axis cs:0,0.1) -- node[anchor=east,fill=white,pos=0.98, inner sep=1pt] { \footnotesize instantaneity threshold } (axis cs:60,0.1);
    \draw [dashed,green!50!black] (axis cs:0,1) -- node[anchor=west,fill=white,pos=0.02,inner sep=1pt] { \footnotesize interactivity threshold } (axis cs:60,1);
    \draw [dashed,green!50!black] (axis cs:0,10) -- node[anchor=west,fill=white,pos=0.02, inner sep=1pt] { \footnotesize attention threshold } (axis cs:60,10);
  \end{axis}
\end{tikzpicture}

\begin{tikzpicture}
  \begin{axis}[only marks,width=0.85\columnwidth,height=5.3cm,
    xlabel={term size (leaves)},
    xmin=0,
    xmax=60,
    ymax = 30,
    ymode=log,
    log ticks with fixed point,
    ylabel={run time (s)},
    xlabel style={font=\small},
    ylabel style={font=\small},
    scaled ticks=false,
    tick label style={/pgf/number format/fixed},
    title={\footnotesize non-commutative solver},
    every axis title/.style={below right,at={(0,1)}},
    legend style={anchor=north west,at={(1.05,1.0)},draw=black,fill=white,align=left}
    ]
    \addplot[blue!50!black,mark=triangle,mark size=1pt] table[x index = {0}, y expr=(\thisrow{time})]{\OneVarNoncommutativeData};
    \addlegendentry{\scriptsize 1 free var};
    \addplot[red!50!black,mark=otimes,mark size=1pt] table[x index = {0}, y expr=(\thisrow{time})]{\FiveVarNoncommutativeData};
    \addlegendentry{\scriptsize 5 free vars};
    \addplot[green!50!black,mark=star,mark size=1pt] table[x index = {0}, y expr=(\thisrow{time})]{\TenVarNoncommutativeData};
    \addlegendentry{\scriptsize 10 free vars};
    \addplot[yellow!50!black,mark=diamond,mark size=1pt] table[x index = {0}, y expr=(\thisrow{time})]{\FifteenVarNoncommutativeData};
    \addlegendentry{\scriptsize 15 free vars};

    \draw [dashed,green!50!black] (axis cs:0,0.1) -- node[anchor=east,fill=white,pos=0.98, inner sep=1pt] { \footnotesize instantaneity threshold } (axis cs:60,0.1);
    \draw [dashed,green!50!black] (axis cs:0,1) -- node[anchor=west,fill=white,pos=0.02, inner sep=1pt] { \footnotesize interactivity threshold } (axis cs:60,1);
    \draw [dashed,green!50!black] (axis cs:0,10) -- node[anchor=west,fill=white,pos=0.02, inner sep=1pt] { \footnotesize attention threshold } (axis cs:60,10);
  \end{axis}
\end{tikzpicture}

   \caption{\Frexlib{} monoid simplifiers type-checking times}
  \figlabel{timings}
\end{figure}

\subsection{Qualitative evaluation}
To evaluate \Frexlib{}'s usage, we have reproduced
\citeauthor{brady-mckinna-et-al:indexed-binary}'s \citeyearpar{brady-mckinna-et-al:indexed-binary}
dependently typed representation of binary
arithmetic. \citeauthor{brady-mckinna-et-al:indexed-binary} index
binary representations by the natural numbers that they represent, and
so the programmer needs to prove arithmetic operations correct.
Such proofs interleave insightful
equational reasoning steps with rote calculational steps
such as the following:
\begin{center}
\FrexifyLHS{} \IdrisFunction{=} \FrexifyRHS{}
\end{center}
which may be discharged with \IdrisFunction{solve}.
We do not use our reflection capabilities since
these kinds of examples, in which the binding-time analysis is challenging,
are beyond their reach at the moment.
With early \Frexlib{} implementations the task was arduous due to
performance bottlenecks in \idris{} that are now eliminated.
The only other significant obstacle we encountered was the usual pain
point involved in invoking an algebraic simplifier without a goal
extraction mechanism: the need to repeat the equation and its relevant
rewriting-context when calling \Frexlib{}.

%% file: lessons.tex
\section{System Design Lessons}
\seclabel{lessons}
\Frexlib{} uses generic and dependently typed programming techniques
extensively, requiring significant type level computation that taxes
the capabilities of the host language implementation.
In developing \Frexlib{} in \agda{} and \idris{} we have eliminated some
performance bottlenecks in \idris{}'s type checker, and learned valuable lessons
about practical dependently typed language implementation. We share these
lessons here, in the hope that developers of other systems will find them useful.

\input{idris2}
\input{agda}

%% file: idris2.tex
\subsection{Idris2}
\seclabel{idris2}
At its heart, the type checker implements: dynamic
pattern
unification~\cite{miller:unification,reed:unification,gundry:thesis},
which instantiates implicit arguments; and conversion checking
whether two terms evaluate to the same reduct.
Both components require an evaluator.  \idris{} uses
normalisation by evaluation~\cite{berger:nbe} with a \emph{syntactic representation}
(terms) and a
\emph{semantic representation} (values in weak head normal form). The
static evaluator is call-by-name and produces a weak head normal form from a term, and
\idris{} implements a quotation mechanism which reconstructs a term from a semantic
representation of a weak head normal form.

Profiling the \idris{} executable reveals that most performance
bottlenecks encountered in developing \Frexlib{} are caused by
the evaluator.
We have experimented with alternative implementations of the evaluator
that compile terms using Scheme's backend and a glued
representation of values~\cite{DBLP:journals/mscs/CoquandD97,Chapman2005epigram} rather
than interpreting terms directly.
These implementations give
modest performance
gains, but in the end the most effective way is to avoid evaluation in the first place! We have therefore also experimented with the following ways
to avoid evaluating terms: preserving subterm sharing, choosing appropriate
data representation in unification, and taking advantage of the typical
structure of unification and conversion problems.

\subsubsection*{Preserving Sharing}

Instantiating implicit arguments in dependently typed programs often
leads to significant \emph{sharing} of subterms.
For example,
\IdrisData{[}\IdrisData{True}\IdrisData{,} \IdrisData{False}\IdrisData{]}
\IdrisKeyword{:} \IdrisType{Vect} \IdrisData{2} \IdrisType{Bool}
elaborates to the following term:
\IdrisKeyword{(}\IdrisData{::}\IdrisKeyword{)} \IdrisKeyword{(}\IdrisData{S} \IdrisData{Z}\IdrisKeyword{)} \IdrisType{Bool} \IdrisData{True} \IdrisKeyword{(}\IdrisKeyword{(}\IdrisData{::}\IdrisKeyword{)} \IdrisData{Z} \IdrisType{Bool} \IdrisData{False} \IdrisKeyword{(}\IdrisData{Nil} \IdrisType{Bool}\IdrisKeyword{)}\IdrisKeyword{)},
sharing the subexpressions \IdrisData{Z} and \IdrisType{Bool}. As the vector
gets longer, sharing increases.
Following~\citet{kovacs:smalltt}, we preserve sharing by introducing a
metavariable for \emph{every} implicit argument, inlining only when it is
guaranteed that the definition cannot break sharing. Consequently,
we inline a metavariable whose definition is itself a metavariable applied
to local variables. Otherwise, we do not substitute metavariable solutions
into terms at all until they are required for unification or display
purposes.

\subsubsection*{Unification}

Unification operates on \emph{values}, not \emph{terms}, but sometimes \idris{}
needs to postpone a unification problem if it is blocked due to an unsolved
metavariable. When the metavariable is solved, \idris{} needs to re-evaluate the
terms being unified. Previously, \idris{} stored postponed problems as a pair
of (syntactic) terms in an environment, re-evaluated once the
blocking metavariable is solved.
However, \Frexlib{} produces some large postponed problems, for which quotation
to syntax is expensive.  Now, in addition to the evaluator and quotation, we
have introduced a \emph{continue} operation, which re-evaluates the
metavariable at the head of a blocked value, and avoids unnecessary quotation.

\subsubsection*{Conversion Checking}

Types in \Frexlib{} can be large, and sometimes a unification problem that arises
while type checking \Frexlib{} is postponed due to an unsolved
metavariable which blocks evaluation. In this case, we might have a
unification problem of the form
\IdrisFunction{f} \IdrisBound{x1} \IdrisKeyword{...} \IdrisBound{xn}
\IdrisKeyword{=?=}
\IdrisFunction{f} \IdrisBound{y1} \IdrisKeyword{...} \IdrisBound{yn}
where the \IdrisBound{xi}, \IdrisBound{yi} etc may be very large subterms,
and the terms unify if they are convertible.
If most corresponding terms are equal after evaluation, but one differs, it may
take a long time to find the differing subterm which blocks unification,
especially since checking the convertibility of subterms involves evaluation.
Fortunately, terms in blocked unification problems
tend to differ at the heads rather than at deeply nested subterms. Therefore,
we always check the heads of the values of corresponding \IdrisBound{xi} and
\IdrisBound{yi} first, postponing the unification
problem if any are unequal. This heuristic significantly improves performance,
preventing a lot of unnecessary evaluation.

\subsubsection*{Influence on Language Design and Ecosystem}

The development of \Frexlib{} has led to the implementation of a number of
desirable language features in \idris{}.
Many of these have been minor changes to the treatment of implicit
arguments and \texttt{parameters} blocks.
More significantly, \Frexlib{} makes extensive use of \texttt{auto}
implicit arguments, which are solved by a search procedure which uses
constructors and functions marked as search hints.  To assist
the development of \Frexlib{} and improve the readability of its code we have added the
ability to mark \emph{local} functions as search hints, allowing
us to restrict the scope of the hints and so avoid excessive
growth of the search space.
\Frexlib{} is now part of the \idris{} test suite, ensuring that it will
remain consistent with any updates to \idris{}.

%% file: agda.tex
\subsection{Agda}
Agda is a well-established dependently typed interactive proof
environment. \idris{} and Agda and their communities have different
goals, leading to subtle implementation differences.

The key differences between the two languages arise from Agda's focus on proving versus \idris{}'s focus on programming.
\idris{} currently uses a single \emph{universe}~\cite{palmgren:universes}, allowing
\IdrisType{Type}\,\IdrisKeyword{:}\,\IdrisType{Type}, and is hence inconsistent
by Girard's paradox. In contrast, Agda's well-developed predicative
theory of universes avoids Girard's paradox. Agda also
protects users from other logical paradoxes of its more experimental
features with its `\texttt{-{}-safe}' compiler flag.
In the spirit of \citeauthor{hu-carette:agda-ct}~\citeyearpar{hu-carette:agda-ct} and \citeauthor{agda-stdlib}~\citeyearpar{agda-stdlib}, we adopt a
conservative set of compiler options (\texttt{-{}-without-K -{}-safe}). All our definitions are
universe-polymorphic. This conservativity broadens the applicability of
\Frexlib{} in the Agda ecosystem by guaranteeing compatibility with
all of Agda's various configurations, and
further assures us about the correctness of \Frexlib{} itself.
\anon[{[Reference redacted for anonymous review]}]{\citeauthor{corbyn:agda-fragment}~\citeyearpar{corbyn:agda-fragment}}
discusses these ideas in greater detail.

%% file: conclusion.tex
\section{Related Work}
\seclabel{related}
Within the Rocq ecosystem, an abundance of tactics enable algebraic simplification.
\citepos{boutin:reflection} ring
and field tactics\footnote{See the Rocq documentation:
  \url{https://rocq-prover.org/distrib/current/refman/addendum/ring.html} .}
let programmers discharge proof obligations involving (and requiring!)
addition, multiplication, and division operations.
\citepos{strub:coqmt} CoqMT extends Rocq's Calculus of
Inductive Constructions, allowing users to extend the conversion rule with
arbitrary decision procedures for first order theories (e.g. Presburger
arithmetic).
To ensure preservation of good meta-theoretical
properties, Strub extends only term level conversion.
Implementations of Hilbert's Nullstellensatz theorem
(\citepos{harrison:grobner} in HOL Light
and \citepos{pottier:grobner} in Rocq)
help users discharge proof obligations involving
polynomial equalities on a commutative integral domain.

Rocq's \SetoidRewrite{} is an advanced tactic library for setoid rewriting.%
\footnote{See the Rocq documentation:
  \url{https://rocq-prover.org/refman/addendum/generalized-rewriting.html}
  .} Disregarding the difference between the direct manipulation of
proof-terms in \idris{} and the tactic-based manipulation in Rocq,
\SetoidRewrite{} provides abstractions for manipulating parameterised
relations (covariant and contravariant), and users can register
setoids of interest and custom `morphisms' --- horn-like equational
clauses --- with the library. The various tactics in the library apply
these user-defined axioms to the goal. Users may also register
tactics, and the library includes an expressive collection of
term-traversal primitives (climbing up and down the syntax tree,
repeating sub-tactics, and so on). While \SetoidRewrite{} does not deal
with algebraic simplification directly, it may help in generalising
equality-based simplifiers to setoid-based simplifiers. In comparison,
\Frexlib{}'s setoid reasoning is minimal, implementing only the
necessary features for the library.

In Idris1,
\citeauthor{slama-brady:automatically-proving-equivalence-by-type-safe-reflection}~\citeyearpar{slama-brady:automatically-proving-equivalence-by-type-safe-reflection} and
\citeauthor{slama:thesis}~\citeyearpar{slama:thesis} implement a hierarchy of rewriting procedures for
algebraic structures of increasing complexity.  Though the procedures'
completeness is not enforced by type as in \Frexlib{}, these
simplifiers are based on a Knuth-Bendix resolution of critical pairs,
and so are likely to be complete.  \Frexlib{} also investigates a hierarchy of
rewriting procedures, but: (1)~frexlets are complete by
construction, (2) \Frexlib{} is based on normalisation-by-evaluation
(like Boutin's tactic, and unlike Slama and Brady's), and (3) \Frexlib{} is
extensible, with support for sufficiently motivated users to extend the library
with bespoke solvers.

Normalisation-by-evaluation is an established technique for
simplifying terms in a concrete equational theory, often involving
function types. One compelling example is \citepos{allais:nu-rules}
work, which demonstrates by a careful model
construction that the equational theory decided by
nor\-ma\-li\-sa\-tion-by-eval\-u\-a\-tion can be enriched with additional
rules. They implement a simply typed language internalising the
functorial and fusion laws
for list {\tt fold}, {\tt map}, and {\tt append} and prove
their construction sound and complete with respect to the extended
equational theory.

In Agda, \citepos{cockx:rewriting} and \citepos{cockx-et-al:taming-of-the-rew} `\texttt{-{}-rewriting}'
flag allows users to enrich the existing
reduction relation with new rules. Their implementation goes beyond Allais et al.'s:
it may restart stuck computations.
They leave to future work the soundness of user-provided reduction
rules, i.e.~ensuring rules neither introduce nontermination nor break
canonicity.
Unlike our approach, neither Allais et al's nor
Cockx et al's technique currently supports commutativity.

Existing formalisations for specific algebraic structures (monoids,
groups, rings, actions, etc.)
abound~\cite{mahboubi-tassi:mathcomp,agda-stdlib,hu-carette:agda-ct,lean-mathlib}. \Frexlib{}
is set apart from these by formalising the fragments of universal
algebra needed for its architecture. However, in \Frexlib{} we do not aspire
to formalise universal
algebra for its own sake.
Formalising more complete
fragments of the theory is an active area of research, with recent
contributions by \citet{GUNTHER2018147} and
\citet{DBLP:journals/corr/abs-2111-07936} in Agda.
\citet{carette-farmer-sharoda:leveraging-the-information-contained-in-theory-presentations}
generate \agda{} code for a comprehensive collection of multi-sorted
algebraic theories and their associated machinery via a pre-processing
phase from a much smaller description.
\citet{fiore-szamozvancev:soas-in-agda} similarly generate definitions
for the significantly more general second-order abstract syntax in
\agda{} via a preprocessing step, while formalising more of the
meta-theory as library code.  We consider it an open problem in this
domain to include the concise information as first-class data in the
meta-language while nonetheless enjoy the full ergonomics of
hand-written, inlined, definitions.

\agda{}'s category theory library~\cite{hu-carette:agda-ct} uses
setoids extensively. Each category includes both its homsets and their
dualisation. This choice allows for the operation `take the opposite
category' to be judgementally involutive. In turn, one can define
dualisation of a property purely formally, e.g. coproducts can be
defined as products in the opposite category. Unfortunately, this
design makes essential use of $\eta$-equality on records, yet
unsupported by \idris{}.

The Meta-F$\star$ language~\cite{martinez-et-al:metafstar} provides
normalisation tactics for commutative monoids and semi-rings through
its metaprogramming facilities. \Frexlib{}'s usage resembles
these tactics' usage, and we hope a
\Frexlib{} port to F$\star$ will make use of F$\star$'s metaprogramming
facilities to reduce some syntactic noise during goal extraction.

\section{Conclusions and Further Work}
\seclabel{conclusion}

We have presented a novel, mathematically structured, design for
algebraic simplification suites that guarantees sound and complete
simplification, even of user-defined simplifiers. Preliminary
evaluation shows that, despite a high level of abstraction, the
resulting library is responsive, with comparable functionality
to other libraries, in a combination of features that no single existing
library provides. \Frexlib{}'s unique design---the frex and
the fral---offer new prospects and questions.

Several computational type theories
support extensional function types and quotients, such as
observational type theory~\cite{altenkirch-et-al:ott-now,pujet-tabareau:ott-now-for-good,pujet-tabareau:impredicative-ott,pujet-leray-tabareau:ott-meets-cic} and cubical type theory~\cite{vezzosi-et-al:cubical-agda,cohen-et-al:cubical-tt}.
Extensional function types and quotient types are also typical reasons
for using setoids.  So it is natural to consider the tradeoffs for
implementing a similar library in a type theory with extensional
function types and quotient types.

Even in such rich type theories, we would still want to implement
setoid algebras.  Setoid algebras let us implement intensional aspects
such as proof printing, certification, and simplification, which we
view as useful functionality of a simplification library. Quotient
types cannot replace this functionality: eliminators of the quotient
must respect the quotient.

However, a richer type theory may make other aspects of the library
easier. For example, the commutative monoid fral and frex use the
power of an algebra by a setoid. If we worked in a type theory with
function extensionality, we could use the power of an algebra by a set
and do away with some of the setoid machinery involved.  A more
speculative direction involves using the improvements in the type
theory to implement category theoretic abstractions (presheaf
categories, ends and coends, 2-functors) which may allow us to `teach'
Idris not just universal algebra, but categorical algebra. We expand
on this idea further below.

Cubical type theories offer another potential implementation strategy
through non-trivial identity types.  Our Agda implementation
(\Fragmentlib) is compatible with Cubical Agda, as it type-checks with
the pragmas \texttt{-{}-safe} and \texttt{-{}-without-K}. Cubical identity types
turn types
into untruncated setoids, so our machinery applies unchanged.
However, cubical identity types cannot replace setoids.
Extracting intensional information is desirable and relevant  for cubical type theory too, so
we will still
implement setoid algebras. So cubical type theory will not
improve \Frexlib{}'s core design.

Implementing some of the internal data-structures may be
simplified by a more powerful type theory. For example, we can
implement the power of a model by an $n$-element set either as a
function or as an $n$-vector. It might be possible to use univalence
to ease transporting structure between these two different
implementations. However, using cubical abstractions---rather than
merely being compatible with a cubical type-theory---is a lot more
sophisticated. Such transport might lead to unexpected challenges, and
would require further experimentation.

\citepos{yallop-et-al:partially-static-data-as-frex} partial
evaluators include additional frexlets (e.g.~abelian groups,
semirings, distributive lattices). We plan to follow suit and port the
remaining simplifiers, then conduct larger evaluation and comparison studies.
The main challenge is that, unlike
\citeauthor{yallop-et-al:partially-static-data-as-frex}, we must
mechanising proofs showing these frexlets complete, which is costly.
One elegant motivation for including more
simplifiers is the following. The frex generalises the `ring of
polynomials over a ring' to that of an algebra of polynomials over an
algebra. By porting
\citeauthor{yallop-et-al:partially-static-data-as-frex}'s family of
representations, we will fully realise this generalisation.

Our experiment with reflection-based goal extraction as well as the reflection-based
interfaces of existing solvers show that with enough engineering
efforts, library designers can extract the goal equation from the goal
type. However, software engineers for dependently typed
languages are a scarce resource. We plan to explore other principled
approaches. In practice, when invoked inside a chain of equational
steps, the goal equation already appears in the source-code, albeit in
a context. Programmers seem willing to type the goal
equation once, since it documents the reasoning steps, but seem
unhappy to do so \emph{twice}. Generic
programming with holes\footnote{See Brad Hardy's Agda-Holes library:
  \url{https://github.com/bch29/agda-holes} .} may help.

Another promising direction is \emph{bootstrapping} of
the \Frexlib{} library using simplifier certification.
Bootstrapping might start with a hierarchy of inefficient simplifiers
that are easy to implement.
Next, these simplifiers may then be used to develop a hierarchy of
more efficient simplifiers and proof-simplifiers.
Finally, the certification mechanism can extract proofs to complete
the bootstrap.

We would also like to extend \Frexlib{}'s design beyond algebraic
structures. More general notions of theories abound: multisorted,
second-order/parameterised, and essentially algebraic.
Supporting these may allow \Frexlib{} to cover much more complex situations, such as decision
procedures for first order theories (e.g.~Presburger arithmetic,
cf.~\citepos{strub:coqmt} CoqMT) normalisation-by-evaluation
for fusion laws~\cite{allais:nu-rules}, and equational manipulation of
big-operators~\cite{betot-et-al:big-operators,markert:big-operators,lau:big-operators}.
Note that \Frexlib{} can already deal with big-operators such as \IdrisFunction{sum} so
long as the argument list is a concrete collection of constants and
variables such as
\IdrisFunction{sum}\ \IdrisData{[2,}\ \IdrisBound{x}\IdrisData{]}. We
only need the more sophisticated theories when the length of the lists
is abstract.

\Frexlib{} uses many
category-theoretic concepts, but the library itself is oblivious to
category theory. 
We hope that making use of a rich category theory library like \citepos{hu-carette:agda-ct} {\tt agda-categories} might
lead to a sleeker and even more modular \Frexlib{} implementation. More
specifically, we plan to explore a general treatment of involutive
algebras following \citeauthor{jacobs:involution}~\citeyearpar{jacobs:involution}, and Power's
\emph{distributive tensor} of equational
theories~\cite{power:discrete-lawvere-theories,%
  hyland-power:discrete-lawvere-theories}
for a uniform treatment of semi-ring varieties.
Instantiating each of the $6$ semi-group varieties makes it possible to cover
each instance of the following combinations:
\[
\left(
\begin{array}{@{}l@{}}\hphantom{\cup{}}
  \set{\text{ordinary}}\times\set{\text{ordinary}, \text{involutive},
        \text{non-reversing involutive}}
\\\cup
\set{\text{commutative}}\times\set{\text{ordinary}, \text{involutive}
}
\end{array}
\right)\times\set{\begin{array}{@{}c@{}}
    \text{commutative}\\
    \text{semigroup}, \text{monoid}, \text{group}
  \end{array}
}
\]
and modularly construct $(2+3)\times 3 = 15$ semi-ring varieties,
including rings and semirings. As this example shows, this kind of
modular treatment can provide a multiplicative development boost.

\newpage















%% file: automated-appendix.tex
\section{Module Structure of \Frexlib{}}\label{app:first-appendix}
We summarises the implementation's code modules
and their relationship to this article:
\input{deps}

\section{Extensional Function and Quotient Setoids}
\label{sec:setoid example}
\applabel{setoid example}
\begin{figure*}
  \TwoColDIYprime{.45}{
 \QuotientDef[numbers=right, numbersep=0.2cm]{}
 \FunSpaceDef[numbers=right, firstnumber=17,numbersep=0.2cm]{}
  }{.54}{%
    \VectEquality{}%
    \VectSetoid{}%
    \VectMap{}%
  }
  \caption{(a) Quotient, function-space, and (b) vector setoids (top) and a higher-order homomorphism (bottom)}
  \figlabel{quotients-and-functions}
  \figlabel{vect-setoid}
\end{figure*}
\Figref{quotients-and-functions}a defines the quotient of a type by a
function \QuotientingMap{}, taking two elements to be equal when their
images under the function \IdrisBound{q} are equal, and
the setoid of homomorphisms
between two setoids together with extensional equality.  This example
also demonstrates \idris's local definitions (lines 4--6, e.g.), possibly
with quantities, named-argument function calls (lines 8--15, e.g.), application operator \IdrisBound{\$},
and
anonymous functions (lines 9--10, e.g.). \idris, like \Haskell, implicitly
quantifies (with quantity \IdrisKeyword{0}) over unbound variables in
type-declarations such as the type \ImplicitArg{} in \QuotientName{}.
These underscores mean elaboration must fill-in the blanks
uniquely using unification.

\Figref{vect-setoid}b presents a setoid over $\nvar$-length
vectors over a given setoid. The vector functorial action
\VectMapName{} has a setoid homomorphism structure
between the two setoids of homomorphisms: (1) \VectMapF{} is a homomorphism (lines 19--25), and
that (2) it maps extensionally equal homomorphisms to extensionally
equal homomorphisms (26--30). These proofs use \idris's equational
reasoning notation for setoids (lines 20--25 and 27--30), a
deeply embedded chain of equational steps. Each step
\IdrisData{(\KatlaTilde\KatlaTilde)} appeals to
transitivity, and requires a justification. The last two dots in the
thought bubble operator \BasicThoughtBubble{} modify the reason:
plain usage (line~23) appeals to a setoid equivalence;
an
equals in the middle dot, e.g.~\EqualThoughtBubble{}, appeals to reflexivity via propositional equality (lines 22, 25, 28, 30); and
a comparison symbol in the end, e.g.~\EqualSymThoughtBubble{}, appeals to
symmetry (lines 25, 30).

\section{Proof Printing and Certification}
\applabel{last-appendix}
\applabel{printing}
\Figref{linear-derivations} presents the layered representation of
linear derivations.
\begin{figure}
\TwoColDIY{.4}{%
  \RTList
  \caption*{(a) reflexive-transitive closure}
  \Symmetrise
  \caption*{(b) symmetric closure}
  \vspace{2\baselineskip}
  \Derivation%
  \caption*{(e) linear derivations}
}{.57}{%
  \Locate
  \caption*{(c) unary congruence closure}
  \Step
  \caption*{(d) axiomatic steps}
}
\caption{Layered (a--d) representation of linear derivations (e)}
\figlabel{linear-derivations}
\end{figure}

\figref{proof-extraction} shows an automatically extracted proof for
the equation $(x•3)•2 = 5•x$ in the additive monoid structure $(\Nat,
0, (\Plus))$. The extracted proof has $\IntroPrinterStepCount$ steps ---
 far from the shortest proof possible.
Extraction removes reflexivity and transitivity steps, and the pointed
bracket tells whether the step uses the axiom directly (angle points
right) or using symmetry (angle points left). Square brackets mean
appealing to congruence, where the context is the congruence's
context, and the term in the hole is the equation's LHS.
\begin{figure*}
  \IntroPrintedListing{}
  \caption{\Frexlib-extracted proof of $(x•3)•2 = 5•x$
    in the additive monoid over \Nat{}}
  \figlabel{proof-extraction}
\end{figure*}
\figref{certification} shows an automatically extracted certificate
for the equation $0 + (x + 0) + 0 = x$ in a generic monoid $\mvar{} =
(\Um{}, \OI{}, (\DotPlus))$.  The certificate is generated inside a
module that parameterises over the generic monoid \mvar{} and
introduces the various notations and reasoning functions.

\begin{figure*}
  \Certificate{}
  \caption{\Frexlib-certificate for the  of $0 + (x + 0) + 0 = x$ in a generic monoid \mvar{}}
  \figlabel{certification}
\end{figure*}

%% file: deps.tex
\begin{center}
\begin{tabular}{ll}
  \begin{minipage}[t]{.44\textwidth}
\begin{itemize}[leftmargin=*]
\item[\textbullet] \AModule[ (\secref*{frexlib}--\subsecref{frex-via-coproducts})]{Frex}
  core definitions
\begin{itemize}[leftmargin=*]
\item[\textbullet] \AModule[ (\secref*{frexlib})]{Signature}
  operations \& arities
\item[\textbullet] \AModule[ (\secref*{frexlib},\subsecref{UPs})]{Algebra}
  algebraic structures and terms, homomorphisms
\item[\textbullet] \AModule[ (\secref*{frexlib})]{Presentation}
  axioms, equational theories
\item[\textbullet] \AModule[ (\secref*{frexlib})]{Axiom}
  common axiom schemes
\item[\textbullet] \AModule[ (\secref*{frexlib})]{Model}
  axiom-validating algebras
\item[\textbullet] \AModule[ (\subsecref{powers})]{Powers}
  parameterised algebras
\item[\textbullet] \AModule[ (\secref*{frexlib})]{Free}
  simplification in all algebras
\begin{itemize}[leftmargin=*]
\item[\textbullet] \AModule[ (\subsecref{UPs})]{Definition}
  universal property
\item[\textbullet] \AModule[ (\secref*{certification})]{Construction} a
  non-effective quotient construction used for extraction, printing,
  and certification
\begin{itemize}[leftmargin=*]
\item[\textbullet] \AModule[ (\subsecref{fral via frex})]{ByFrex}
  reuse a frex simplifier to define a fral simplifier
\item[\textbullet] \AModule[ (\subsecref{extracting
    certificates}--\subsecref{proof simplification})]{Linear}
  generic proof simplification and printing
\item[\textbullet] \AModule[ (\secref*{certification})]{Idris}
  generic certification
\end{itemize}
\end{itemize}
\item[\textbullet] \AModule[ (\subsecref{frex-via-coproducts})]{Coproduct}
  universal property
\item[\textbullet] \AModule[ (\subsecref{UPs}--\subsecref{frex-via-coproducts})]{Frex}
  universal property, reuse coproduct and fral simplifier to define a frex simplifier
\begin{itemize}[leftmargin=*]
\item[\textbullet] \AModule[ (\secref*{certification})]{Construction}
  non-effective quotient construction used for extraction, printing,
  and certification
\end{itemize}
\item[\textbullet] \AModule[ (\secref*{certification})]{Lemma}
  auxiliary representation for auxiliary lemmata discharged by fral
  simplifiers, printed, or certified
\item[\textbullet] \AModule[ (\secref*{reflection})]{Magic}
  generic reflection code for ergonomic invocation
\end{itemize}
\end{itemize}
\end{minipage}
&
\begin{minipage}[t]{.52\textwidth}
\begin{itemize}[leftmargin=*]
\item[\textbullet] \AModule{Frexlet.Monoid}
  modules concerning varieties of monoids and their simplifiers
\begin{itemize}[leftmargin=*]
\item[\textbullet] \AModule[ (\secref*{frexlib})]{Theory}
  signature, axioms, pretty printing for the theory of ordinary monoids
\item[\textbullet] \AModule[ (\secref*{frexlib})]{Notation}
  shared infix notation (additive and multiplicative) for monoid varieties
\item[\textbullet] \AModule[ (\figref{monoid frexlets})]{Frex}
  frex simplifier for monoids
\item[\textbullet] \AModule[ (\subsecref{fral via frex})]{Free}
  fral simplifier, reuses frex simplifier
\item[\textbullet] \AModule[ (\secref*{frexlib})]{Nat}
  additive and multiplicative monoid structure of the natural numbers
\item[\textbullet] \AModule{Pair} types with the cartesian product as
  a proof-relevant monoid structure
\item[\textbullet] \AModule{List}
  monoid structure of lists with catenation
\item[\textbullet] \AModule{Commutative}
  commutative monoids modules
\begin{itemize}[leftmargin=*]
\item[\textbullet] \AModule{Theory}
  commutativity axiom
\item[\textbullet] \AModule[ (\subsecref{UPs}]{Free})
  fral simplifier
\item[\textbullet] \AModule[ (\subsecref{UPs})]{NatSemiLinear}
  auxiliary definitions for fral simplifier
\item[\textbullet] \AModule[ (\subsecref{frex-via-coproducts})]{Frex}
  simplifier, reuses fral via coproducts
\item[\textbullet] \AModule[ (\subsecref{frex-via-coproducts})]{Coproduct}
  coproduct of commutative monoids
\item[\textbullet] \AModule{Nat}
  addition and multiplication of naturals
\end{itemize}
\item[\textbullet] \AModule{Involutive}
  modules concerning monoids equipped with an involution
\begin{itemize}[leftmargin=*]
\item[\textbullet] \AModule[ (\secref*{frexlib})]{Theory}
  signature and axioms
\item[\textbullet] \AModule{Free}
  simplifier, reuses frex simplifier
\item[\textbullet] \AModule[ (\subsecref{reusing frexlets})]{Frex}
  simplifier, reuses monoid frex
\item[\textbullet] \AModule[ (\secref*{frexlib})]{List}
  involutive monoid structure of list reversal
\end{itemize}
\end{itemize}
\end{itemize}
\end{minipage}
\end{tabular}
\end{center}